\documentclass[12pt]{article}
\usepackage{
		graphicx,		
		amsmath,		
		amssymb,		
		color,xcolor			
        }
\usepackage{comment}

\def\dAlembert{\kern1pt\vbox{\hrule height 1.2pt\hbox{\vrule width 1.2pt\hskip 3pt
   \vbox{\vskip 6pt}\hskip 3pt\vrule width 0.6pt}\hrule height 0.6pt}\kern1pt}

\usepackage{cancel}

\begin{document}


\begin{center}
{\bf Gravitational Noether-Ward identities for scalar field}\\
\end{center}
\bigskip
\begin{center}
{\bf Tomislav Prokopec}
\end{center}

\begin{center}
{\rm ITP, EMMEPH and Spinoza Institute, Utrecht University, 
Princetonplein 5, 3584 CC Utrecht, The Netherlands}
\end{center}

\bigskip
\begin{center}
{\bf Abstract}
\end{center}
We consider the gravitational Noether-Ward identities for the evolution of general metric perturbations
on quantum matter backgrounds. In this work we consider Einstein's gravity covariantly coupled to 
a massive, non-minimally coupled, quantum scalar field in general curved backgrounds. 
We find that each term in the equation of motion for gravitational perturbations satisfies its own 
Noether-Ward identity. 
Even though each term is non-transverse, the whole equation of motion maintains transversality. 
In particular, each counterterm needed to renormalize the graviton self-energy 
satisfies its own Noether identity, and we derive the explicit form for each.
Finally, in order to understand how the Noether-Ward identities are affected by the definition of the metric perturbation, we consider two inequivalent definitions of metric perturbations
and derive the Noether-Ward identities for both definitions. This implies that 
there are Noether-Ward identities for every definition of the metric perturbation.

\bigskip

\section{Introduction}
\label{Introduction}

Understanding gravitational symmetries has gained in importance,
especially in cosmology. Imposing constraints on cosmological perturbations implied by 
large scale diffeomorphisms during inflation has led to a deeper understanding of higher $n$-point functions of cosmological correlators generated in inflation~\cite{Maldacena:2002vr,Creminelli:2004yq,Cheung:2007sv,Martin:2012pe,Hui:2018cag,Goodhew:2021oqg}.
Furthermore, understanding soft photon and gravitational states~\cite{Weinberg:1965nx}
has led to the development of soft scattering amplitudes and
a connection between asymptotic states (characterised by BMS transformations~\cite{Bondi:1962px})
and quantum information theory of black holes~\cite{Strominger:2013jfa,Arkani-Hamed:2017jhn,Strominger:2017zoo}. Recently it was shown that there is an intricate connection beween the asymptotic symmetries,
Ward identity and  gravitational memory effects~\cite{DeLuca:2024cjl,DeLuca:2024asq}.

On curved backgrounds the Ward (or Slavnov-Taylor) identities can play an important role even 
for the tree-level gauge field two-point functions, by imposing additional constraints 
on the two-point functions which cannot be derived from the equations of motion.
They were used in Refs.~\cite{Glavan:2022nrd,Glavan:2022dwb,Glavan:2023lvw} 
 to show that the gauge sector of the photon two-point function in de Sitter violate de Sitter symmetries 
in the general covariant gauges, and their importance was emphasized 
for the construction of the photon two-point functions on more general cosmological spaces~\cite{Glavan:2022pmk,Domazet:2024dil,Glavan:2025iuw}. While can surmise that 
the Ward identities are equally important for the graviton two-point functions,
no systematic analysis of the importance of the Ward identities for 
construction of the tree-level graviton two-point functions has been attempted.~\footnote{It is known that Ward identities do not impose additional conditions on two-point functions in exact gauges. For example,
the photon two-point function in the exact Lorenz gauge~\cite{Tsamis:2006gj} 
automatically satisfies the Slavnov-Taylor identity. The same is true for the 
graviton two-point functions in the exact de Donder gauges~\cite{Miao:2011fc,Mora:2012zi}.
However, we do expect that the Ward identity plays an important role 
in the graviton two-point function in de Sitter in 
the general covariant gauges~\cite{Frob:2016hkx}.
}

There has been little work regarding the role of the gravitational Ward identities in curved spacetimes.
Fr\"ob~\cite{Frob:2017gez} studied how the Ward identity constrains the graviton two-point function on general curved spaces, however a little is known about the role of the  gravitational Ward identities that go beyond the tree level, for example for the graviton 
self-energy~\cite{Burns:2014bva,Liu:2024utl} and for higher-order vertex functions. It was observed that after renormalization
the one-loop graviton self-energy of a massless scalar scalar field in de Sitter 
remains to be non-transverse, unless one adds a finite cosmological constant 
counterterm~\cite{Park:2011ww,Miao:2024atw}, 
which removes  the quantum (one-loop) contribution from the energy-momentum tensor
by adding the suitably tuned finite cosmological constant counterterm contribution, which also removes the non-transverse part of the graviton self-energy. 
Which self-energy one uses affects the physical results. For example, it is known that 
the scalar gravitational potentials exhibit a secular growth generated by the quantum fluctuations
of the massless scalar in de Sitter inflation, but the precise form of that growth depends on 
the self-energy one uses~\cite{Park:2015kua,Miao:2024atw,Miao:2024nsz}.
However, if one wants to understand the graviton self-energy in more general settings, for example 
in more realistic inflationary models beyond de Sitter, in subsequent radiation, matter or dark energy epochs,
or in other curved backgrounds such as black holes, one needs general understanding 
of the transversality properties of the graviton self-energy. 
At this moment there is no clarity regarding 
what are the transversality properties of the graviton self-energy induced by quantized matter 
fields of different spin on different curved gravitational backgrounds.
The purpose of this work is to shed light on this important question.

In this work we consider the Noether-Ward identities in general curved spaces, for the case in which matter is quantum, but gravity is left classical. For simplicity we focus here on a real, massive, nonminimally 
coupled scalar field.
This study can be viewed as a prelude for the Noether-Ward identities in which both gravity and matter are quantized.
The principal applications we have in mind is the propagation of 
classical and quantum gravitational perturbations through the quantum
matter medium of the early Universe. Without understanding this question we cannot really 
claim that we understand the mechanism for generation of cosmological perturbations 
needed to explain the cosmic microwave background data and the origin of Universe's 
large scale structure.

\section{Preliminaries}
\label{Preliminaries}

\medskip
\noindent
{\bf The action.} The classical bare action consists of a gravitational action $S^{\tt b}_{\rm g}$ and 
a scalar field action $S^{\tt b}_{\rm m}$, 
\begin{equation}
S^{\tt b}[g_{\mu\nu},\phi]
 = S^{\tt b}_{\rm g}[g_{\mu\nu}]+S^{\tt b}_{\rm m}[g_{\mu\nu},\phi]
\label{gravitational plus matter action}
\end{equation}
where for the gravitational action 
we take the Hilbert-Einstein action,
\begin{equation}
S^{\tt b}_{\rm g}[g_{\mu\nu}]=\frac{1}{\kappa_{\tt b}^2}	\int\! {\rm d}^{D\!} x \, \sqrt{-g} \,
	\Bigl[
	R - (D\!-\!2)\Lambda_{\tt b}\Bigr]
\,,
\label{Hilbert-Einstein bare action}
\end{equation}
where $\kappa_{\tt b}$ and $\Lambda_{\tt b}$ are the bare gravitational couplings, which 
can be expressed in terms of the classical quatities $\kappa$ and $\Lambda$ as, 
\begin{equation}
\frac{1}{\kappa^2_{\tt b}} = \frac{1}{\kappa^2} + \delta\Big(\frac{1}{\kappa^2}\Big)
	\, ,\quad
	\frac{\Lambda_{\tt b}}{\kappa^2_{\tt b}} = \frac{\Lambda}{\kappa^2} 
	        + \delta\Big( \frac{\Lambda}{\kappa^2}\Big)
\,,\qquad
\label{bare gravitational couplings}	
\end{equation}
where $\kappa^2 = 16\pi G$ is the loop counting parameter of quantum gravity,
$G$ is the classical Newton constant and $\Lambda$ is the classical (geometric) cosmological constant, 
and $\delta\Big(\frac{1}{\kappa^2}\Big)$ and $ \delta\Big( \frac{\Lambda}{\kappa^2}\Big)$
belong to the quantum loop contributions $\kappa^2$ and $\Lambda$. With these remarks in mind,
the bare gravitational action~(\ref{Hilbert-Einstein bare action}) naturally splits into the classical action 
and the counterterm action,
\begin{eqnarray}
S_{\rm g}[g_{\mu\nu}]  &\!\!=\!\!&
   \frac{1}{\kappa^2}	\int\! {\rm d}^{D\!} x \, \sqrt{-g} \,
	\Bigl[
	R - (D\!-\!2)\Lambda\Bigr]
\,,
\label{Hilbert-Einstein classical action}\\
S_{\rm g}^{\tt ct}[g_{\mu\nu}]  &\!\!=\!\!&
  \int\! {\rm d}^{D\!} x \, \sqrt{-g} \,
	\biggl[
	 \delta\Big(\frac{1}{\kappa^2}\Big)R 
	 - (D\!-\!2) \delta\Big(\frac{\Lambda}{\kappa^2}\Big)\biggr]
\,.\qquad
\label{Hilbert-Einstein counterterm action}
\end{eqnarray}
Since gravity is a nonrenormalizable theory~\cite{tHooft:1974toh,Goroff:1985th,vandeVen:1991gw},
 the counterterm action $S_{\rm g}^{\tt ct}$ is in general insufficient to renormalize the theory, so 
the renormalization of the one-loop matter contributions to the effective action requires additional 
four-derivative gravitational counterterms,~\footnote{It is sometimes convenient to replace 
the counterterm action in Eq.~(\ref{gravitational counterterm action}) by the following equivalent counterterm
action,
\begin{eqnarray}
\tilde S_{\rm g}^{\prime\tt ct}[g_{\mu\nu}]  &\!\!=\!\!&
  \int\! {\rm d}^{D\!} x \, \sqrt{-g} \,
	\Bigl[\alpha_{R^2}R^2 + \alpha_{\rm Weyl^2}W_{\mu\nu\rho\sigma}W^{\mu\nu\rho\sigma}
	            + \alpha_{\rm G\!B}{\rm G\!B}
              \Bigr]
\,,\qquad
\label{gravitational counterterm action2}
\end{eqnarray}
where 
\begin{eqnarray}
{\rm G\!B} &\!\!\equiv\!\!& R^2 \!-\! 4 R_{\mu\nu}R^{\mu\nu} 
                                         \!+\!  R_{\mu\nu\rho\sigma}R^{\mu\nu\rho\sigma}
\,,\quad
\label{Gauss-Bonnet}\\
W_{\mu\nu\rho\sigma}W^{\mu\nu\rho\sigma} &\!\!\equiv\!\!& R_{\mu\nu\rho\sigma}R^{\mu\nu\rho\sigma} 
 \!-\! \frac{4}{D\!-\!2} R_{\mu\nu}R^{\mu\nu} 
    \!+\! \frac{2}{(D\!-\!1)(D\!-\!2)} R^2 
\,.\qquad
\label{Weyl squared}
\end{eqnarray}
The Gauss-Bonnet term ${\rm G\!B}$ is topological in $D=4$, and therefore its variation produces 
a result that is $\propto (D\!-\!4)$, and thus it vanishes in $D=4$ 
(for a non-singular $\alpha_{\rm G\!B}$). Sometimes a singular Gauss-Bonnet counterterm term
is used to model the Weyl anomaly in the energy-momentum tensor, also known as 
conformal anomaly or trace anomaly.
The Weyl tensor $W^{\mu}_{\;\;\nu\rho\sigma}$,
which is defined by, 
\begin{equation}
W_{\mu\nu\rho\sigma} \equiv R_{\mu\nu\rho\sigma}
   - \frac{2}{D\!-\!2}\left(g_{\mu[\rho}R_{\sigma]\nu}
   -g_{\nu[\rho}R_{\sigma]\mu}\right)
    +\frac{2}{(D\!-\!1)(D\!-\!2)}g_{\mu[\rho}g_{\sigma]\nu}R
\,,\qquad
\label{Weyl curvature tensor: intro}
\end{equation}
is also known as the conformal tensor, as it is invariant under Weyl transformations,
under which the metric tensor transforms as, $g_{\mu\nu}\rightarrow \Omega^2(x) g_{\mu\nu}$,
where  $\Omega^2(x)$ is a nonsingular function, and  $\Omega^2(x)\neq 0$.
} 
\begin{eqnarray}
\tilde S_{\rm g}^{\tt ct}[g_{\mu\nu}]  &\!\!=\!\!&
  \int\! {\rm d}^{D\!} x \, \sqrt{-g} \,
	\Bigl[\alpha_{R^2}R^2 + \alpha_{\rm Ric^2}R_{\mu\nu}R^{\mu\nu}
	            + \alpha_{\rm Riem^2}R_{\mu\nu\rho\sigma}R^{\mu\nu\rho\sigma}
              \Bigr]
\,.\qquad
\label{gravitational counterterm action}
\end{eqnarray}

 The bare matter action $S^{\tt b}_{\rm m}[g_{\mu\nu},\phi]$ 
we consider in this paper is that of a non-minimally coupled real scalar field $\phi$,
\begin{equation}
S^{\tt b}_{\rm m}\big[\phi,g_{\mu\nu}\big] = \int\! {\rm d}^{D\!} x \, \sqrt{-g} \,
	\biggl[-\frac{Z_{\tt b}}{2} g^{\mu\nu} (\partial_\mu \phi) (\partial_\nu \phi)
	-\frac{1}{2} m_{\tt b}^2  \phi^2
		-  \frac{1}{2}\xi_{\tt b} R \phi^2
	\biggr]	
	\, ,\quad
\label{bare scalar matter action}	
\end{equation}
where $g_{\mu\nu}$ and $g^{\mu\nu}$ are the metric tensor and its inverse, respectively,
$g={\rm det}[g_{\mu\nu}]$, $R=R(g_{\mu\nu})$ is the Ricci curvature scalar. 
Similarly to the case of gravitational action in Eq.~(\ref{bare gravitational couplings}), 
the bare coupling parameters in Eq.~(\ref{bare scalar matter action})
$Z_{\tt b}$, $m_{\tt b}^2$ and $\xi_{\tt b}$ can be split into their classical and quantum parts,
\begin{equation}
Z_{\tt b} = 1 + \delta Z
	\, ,\quad
	m_{\tt b}^2 = m^2 + \delta m^2
	\, ,\quad
	\xi_{\tt b} = \xi + \delta \xi
\,,\qquad
\label{bare gravitational couplings}	
\end{equation}
after which one can rewrite the bare matter action as a sum of the classical and counterterm action,
\begin{eqnarray}
S_{\rm m}\big[\phi,g_{\mu\nu}\big]
&\!\!=\!\!&\int\! {\rm d}^{D\!} x \, \sqrt{-g} \,
	\biggl[-\frac{1}{2} g^{\mu\nu} (\partial_\mu \phi) (\partial_\nu \phi)
	-\frac{1}{2} m^2  \phi^2
		-  \frac{1}{2}\xi R \phi^2
	\biggr]	
	\, ,\quad
\label{classical scalar matter action}\\	
S^{\tt ct}_{\rm m}\big[\phi,g_{\mu\nu}\big]
&\!\!=\!\!&\int\! {\rm d}^{D\!} x \, \sqrt{-g} \,
	\biggl[-\frac{\delta Z}{2} g^{\mu\nu} (\partial_\mu \phi) (\partial_\nu \phi)
	-\frac{1}{2} \delta m^2  \phi^2
		-  \frac{1}{2} \delta\xi R \phi^2
	\biggr]	
	\, ,\qquad
\label{counterterm scalar matter action}
\end{eqnarray}
where $m^2$ and $\xi$ denote the classical mass-squared and the nonminimal coupling, respectively, and
$\delta Z$,  $\delta m^2$ and  $\delta m^2$ are the couplings needed for renormalization
of the matter loop diagrams, such that the counterterm matter action 
in Eq.~(\ref{counterterm scalar matter action})
naturally belongs to the loop contributions.
 Note that our conventions are such that the conformal nonminimal coupling is given by 
 $\xi = \xi_c\equiv (D\!-\!2)/[4(D\!-\!1)]$.
 The action in Eq.~(\ref{classical scalar matter action})
is already written in terms of a canonically normalized field $\phi$, which is related to the
bare field  $\phi_{\tt b}$ as, $\phi_{\tt b}=\sqrt{Z_{\tt b}}\phi$.

\medskip
\noindent
{\bf The equations of motion.}
Scalar field $\phi$ obeys a classical equation of motion obtained by varying 
the classical action~(\ref{classical scalar matter action}),
\begin{equation}
\big(\dAlembert -m^2 -\xi R\big)\phi(x) =0
\,,\qquad
\dAlembert = \frac{1}{\sqrt{-g}}\partial_\mu \sqrt{-g}g^{\mu\nu}\partial_\nu
\,.\qquad
\label{classical equation of motion}
\end{equation}
The solution of this equation is the classical on-shell field $\phi(x)=\langle\hat\phi(x)\rangle$,
which is equal to the expectation value of the quantum field $\hat\phi(x)$ 
with respect to a state defined by the density operator $\hat\rho(t)$, {\it i.e.} 
$\big\langle\hat {\cal O}(t)\big\rangle\equiv{\rm Tr}\big[\hat\rho(t)\hat {\cal O}(t)\big]$,
where $\hat {\cal O}(t)$ is some operator of quantum theory.

The quantum fluctuations $\delta\hat\phi(x)\equiv \hat\phi(x)-\langle\hat\phi(x)\rangle$ 
of the scalar field around its 
expectation value $\langle\hat\phi(x)\rangle$ can be described by the corresponding two-point functions.
The corresponding scalar propagator, defined as the expectation value of a time ordered product, 
\begin{equation}
i\Delta_\phi(x;x') \equiv \big\langle T\big[\delta\hat\phi(x)\delta\hat\phi(x')\big] \big\rangle
\,,
\label{propagator: definition}
\end{equation}
then obeys the following tree-level equation of motion,
\begin{equation}
\big(\Box_x -m^2 -\xi R(x)\big)i\Delta_\phi(x;x') = i\hbar\frac{\delta^D(x\!-\!x')}{\sqrt{-g}}
\,.
\label{propagator equation of motion}
\end{equation}
%
In addition, the propagator obeys an equation of motion at the second leg $x'$,
\begin{equation}
\big(\Box_{x'} -m^2 -\xi R(x')\big)i\Delta_\phi(x;x') = i\hbar\frac{\delta^D(x\!-\!x')}{\sqrt{-g}}
\,,
\label{propagator equation of motion x'}
\end{equation}
which is usually implemented by imposing the following symmetry condition on the propagator, 
\begin{equation}
 i\Delta_\phi(x;x')= i\Delta_\phi(x';x) 
\,,
\label{propagator: symmetry x and  x'}
\end{equation}
which follows immediately from its definition~(\ref{propagator: definition}).
In general curved backgrounds which are the focus of this work, it is useful to define
the positive and negative frequency Wightman functions,
\begin{equation}
i\Delta_\phi^{(+)}(x;x') \equiv \big\langle\delta\hat\phi(x)\delta\hat\phi(x')\big\rangle
\,,\qquad
i\Delta_\phi^{(-)}(x;x') \equiv \big\langle\delta\hat\phi(x')\delta\hat\phi(x)\big\rangle 
\,,\quad
\label{Wightman functions}
\end{equation}
in terms of which one can represent the time-ordered (Feynman)
 propagator in Eq.~(\ref{propagator: definition}) as, 
\begin{equation}
i\Delta_\phi(x;x')= \Theta(\Delta x^0)i\Delta_\phi^{(+)}(x;x')
 + \Theta(-\Delta x^0)i\Delta_\phi^{(-)}(x;x')
\,.
\label{propagator: definition 2}
\end{equation}
The Wightman functions defined in Eq.~(\ref{Wightman functions}) obey homogeneous differential equations,
\begin{eqnarray}
\big(\Box_{x} -m^2 -\xi R(x)\big)i\Delta^{(\pm)}_\phi(x;x') &\!\!=\!\!& 0
\,,
\nonumber\\
\big(\Box_{x'} -m^2 -\xi R(x')\big)i\Delta^{(\pm)}_\phi(x;x') &\!\!=\!\!& 0
\,,\qquad
\label{equation of motion Wightman}
\end{eqnarray}
and they are related by a complex conjugation,
\begin{equation}
\big[i\Delta^{(+)}_\phi(x;x')\big]^* = i\Delta^{(-)}_\phi(x;x')
\,,\qquad
\label{Wightman functions: conjugation}
\end{equation}
and they transform under the exchange of the two legs into each other,
\begin{equation}
i\Delta^{(+)}_\phi(x';x) = i\Delta^{(-)}_\phi(x;x')
\,.\qquad
\label{Wightman functions: exchange of x and x}
\end{equation}
In order to complete the two-point functions needed for the Schwinger-Keldysh 
non-equilibrium formalism,
one can define the Dyson (anti-time ordered) propagator as,
\begin{eqnarray}
i\Delta^D_\phi(x;x') &\!\!=\!\!& \Theta(\Delta x^0)i\Delta_\phi^{(-)}(x;x')
 + \Theta(-\Delta x^0)i\Delta_\phi^{(+)}(x;x')
\nonumber\\
 &\!\!=\!\!& \big[i\Delta_\phi(x;x')\big]^ *
\,,
\label{Dyson propagator: definition}
\end{eqnarray}
which obeys, 
\begin{equation}
\big(\Box_x -m^2 -\xi R(x)\big)i\Delta^ D_\phi(x;x') 
= -i\hbar\frac{\delta^D(x\!-\!x')}{\sqrt{-g}}
\,.
\label{Dyson propagator equation of motion}
\end{equation}
Finally, in Keldysh notation we have, $i\Delta_\phi(x;x' ) \equiv i\Delta^{++}_\phi(x;x' )$,
 $i\Delta^D_\phi(x;x' ) \equiv i\Delta^{--}_\phi(x;x' )$,
$i\Delta^ {(+)}_\phi(x;x' ) \equiv i\Delta^{-+}_\phi(x;x' )$,
and $i\Delta^ {(-)}_\phi(x;x' ) \equiv i\Delta^{+-}_\phi(x;x' )$.

\bigskip
\noindent
{\bf Isometries.}
One is typically interested in how gravitational perturbations evolve on some simple gravitational background, given by a background metric $\overline{g}_{\mu\nu}$, 
which possesses isometries. This means that there exists a set of Killing vector fields
$K^{(i)} (i=1,\cdots , N)$ along which the metric is invariant,
\begin{equation}
  \mathcal{L}_{K^{(i)}}\overline{g}_{\mu\nu} \equiv - 2 \overline{\nabla}_{(\mu}K^{(i)}_{\nu)} =0
  \,,\qquad
\label{Killing vectors}
\end{equation}
where $\mathcal{L}_K$ denotes a Lie derivative along the vector field $K$.
For example, when $\overline{g}_{\mu\nu}=\eta_{\mu\nu}$ 
is the Minkowski metric on a $D$ dimensional flat Lorentzian spacetime, the set of 
Killing vectors $K^{(i)}$ spans the Poincar\'e algebra $\mathit{iso}(1,D-1)$ in $D$ dimensions,
for de Sitter and anti-de Sitter spaces $K^{(i)}$ span $\mathit{so}(1,D)$ and  $so(2,D-1)$ algebra,
respectively, for Schwarzschild black holes $K^{(i)}$ constitute the $D(D\!-\!1)/2$ 
rotational vector fields plus time translations, {\it etc.}

\bigskip
\noindent
{\bf Metric perturbations.} We are interested in understanding the dynamics of
gravitational perturbations $\delta g_{\mu\nu}$ on some background 
$\overline{g}_{\mu\nu}$. There is no unique way of defining metric perturbations,
so here we shall consider two commonly used definitions:
\begin{eqnarray}
{\rm  Case~A:}\qquad g_{\mu\nu} &\!\!=\!\!& \overline{g}_{\mu\nu} + \kappa\delta g_{\mu\nu}
\,, \qquad 
\label{metric perturbations: case A}\\
{\rm Case~B:}\qquad  g^{\mu\nu} &\!\!=\!\!& \overline{g}^{\mu\nu}  +  \kappa\delta g^{\mu\nu} 
\,. \qquad
\label{metric perturbations: case B}
\end{eqnarray}
A third definition, that is often used in cosmology~\cite{Maldacena:2002vr} is,
$g_{\mu\nu} = \big[\exp\big(\overline{g}+\kappa\delta g\big)\big]_{\mu\nu}$, which 
is convenient, as its inverse is simple, 
$g^{\mu\nu} = \big[\exp\big(\overline{g}-\kappa\delta g\big)\big]^{\mu\nu}$.
For the sake of brevity  in this work we do not consider that definition. 
 The rescaling of the metric perturbation $\kappa\delta g_{\mu\nu}$ by $\kappa$ in 
 Eqs.~(\ref{metric perturbations: case A})--(\ref{metric perturbations: case B})
 introduces a canonically normalized gravitational field perturbation  $\delta g_{\mu\nu}$, where 
\begin{equation}
 \kappa^2 = 16\pi G
 \,,\quad
 \label{kappa 2}
\end{equation}
is the loop-counting parameter of quantum gravity, such that the $n$-th loop contributes as 
$\sim(\hbar\kappa^2)^{n}$.

From the definitions~(\ref{metric perturbations: case A})--(\ref{metric perturbations: case B})
it follows that the (inverse) metric and the (inverse) metric determinant can be perturbed as, 
\begin{eqnarray}
{\rm Case~A:}\quad g^{\mu\nu} &\!\! =\!\!& 
      \overline{g}^{\mu\nu}\!-\! \kappa\delta g^{\mu\nu}
      \!+\! \kappa^2
      \delta g^{\mu\rho}\delta g_\rho^{\;\nu}
       + {\mathcal O}(\delta g_{\mu\nu}^3) \,,
\label{metric perturbations: case A inverse and g}\\
  &&\hskip -2.3cm
  \sqrt{-g}  = \sqrt{-\overline{g}}\left[1
    \!+\! \frac{\kappa}2\overline{g}^{\mu\nu}\delta g_{\mu\nu}
    \!+\!\frac{\kappa^2}{8}\!\left(\!\Big(\overline{g}^{\mu\nu}\delta g_{\mu\nu}\Big)^{\!2}
     \!\!-\!2\delta g_{\mu\nu}\delta g^{\mu\nu}\right)
     \!+\! {\mathcal O}(\delta g_{\mu\nu}^3)\right]
,
\nonumber\\
{\rm Case~B:}\quad  g_{\mu\nu} &\!\! =\!\!& 
      \overline{g}_{\mu\nu}\!-\!  \kappa\delta g_{\mu\nu}
      \!+\!  \kappa^2\delta g_{\mu\rho}\delta g^\rho_{\;\nu}
       + {\mathcal O}\Big(\big(\delta g^{\mu\nu}\big)^3\Big) 
      \,, 
\label{metric perturbations: case B inverse and g}\\
  &&\hskip -2.9cm
  \sqrt{-g}  = \sqrt{-\overline{g}}\left[1
    \!-\!\frac{\kappa}2 \overline{g}_{\mu\nu}\delta g^{\mu\nu}
    \!+\!\frac{\kappa^2}{8}\!\left(\!\Big(\overline{g}_{\mu\nu}\delta g^{\mu\nu}\Big)^{\!2}
     \!\!+\!2\delta g_{\mu\nu}\delta g^{\mu\nu}\right)
     \!+\! {\mathcal O}\Big(\big(\delta g^{\mu\nu}\big)^3\Big)\right]
.
\nonumber
\end{eqnarray}
One of the questions we discuss in this work is what are the ramifications -- if any --
of the differences  in the expansions in 
Eqs.~(\ref{metric perturbations: case A inverse and g})--(\ref{metric perturbations: case B inverse and g}).



\begin{figure}[h!]
\centering
\vskip -0.2cm
\includegraphics[width=4.5cm,height=2.cm,clip]{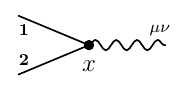}
\hskip 1cm
\includegraphics[width=4.5cm,height=2.cm,clip]{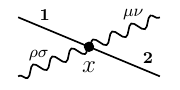}
\vskip-3mm
\caption{\small The three- and four-point vertices. The solid straight lines represent the scalar 
propagators, the wavy lines correspond to the graviton propagators.}
\label{fig: three and four point vertices}
\end{figure}

\bigskip
\noindent
{\bf Feynman rules.}
In what follows we present the Feynman rules needed for 
the one-loop graviton self-energy. First we need are
the three- and four-point vertex functions shown in 
Figure~\ref{fig: three and four point vertices}.
It is convenient to split the vertices to the minimal coupling and the non-minimal coupling contributions,
 \begin{equation}
 V_{(n)} = V_{(n,0)} + \delta_\xi V_{(n)}
 \,.\qquad
 \label{vertices: minimal vs the non-minimal coupling}
\end{equation}
A simple variation of the scalar action~(\ref{classical scalar matter action}) with respect to 
$\delta g_{\mu\nu}$ and $\delta g^{\mu\nu}$, respectively, gives 
for the ($\xi=0$, amputated) three-point vertex functions, 
\begin{eqnarray}
\Big[V_{(3,0)}^{\mu\nu}(x)\Big]_A 
&\!\!=\!\!& -\frac{i\kappa}{\hbar}\sqrt{-\overline{g}}
  \left[-\overline{\nabla}^{(\mu}_1\overline{\nabla}^{\nu)}_2
   +\frac{\overline{g}^{\mu\nu}}{2}
          \left(\overline{\nabla}_1\!\cdot\!\overline{\nabla}_2+m^2\right)\right]
\,,\qquad
\label{3 point vertex: A}\\
\Big[V^{(3,0)}_{\mu\nu}(x) \Big]_B
&\!\!=\!\!&- \frac{i\kappa}{\hbar}\sqrt{-\overline{g}}
  \left[\overline{\nabla}_{(\mu}^1\overline{\nabla}_{\nu)}^2
   -\frac{\overline{g}_{\mu\nu}}{2}
          \left(\overline{\nabla}_1\!\cdot\!\overline{\nabla}_2+m^2\right)
          \right]
\,,
\label{3 point vertex: B}
\end{eqnarray}
and the non-minimal coupling contributions are,
\begin{eqnarray}
\Big[\delta_\xi V_{(3)}^{\mu\nu}(x) \Big]_A
&\!\!=\!\!& -\frac{i\kappa}{\hbar}\sqrt{-\overline{g}}\xi
  \left[-\overline{G}^{\mu\nu} +\overline{\nabla}^{\mu}\overline{\nabla}^{\nu}
   - \overline{g}^{\mu\nu}\overline{\dAlembert}\right]
\,,\qquad
\label{3 point vertex: A xi}\\
\Big[\delta_\xi V^{(3)}_{\mu\nu}(x) \Big]_B
&\!\!=\!\!&- \frac{i\kappa}{\hbar}\sqrt{-\overline{g}}\xi
  \left[\overline{G}_{\mu\nu} -\overline{\nabla}_{\mu}\overline{\nabla}_{\nu}
   + \overline{g}_{\mu\nu}\overline{\dAlembert}\right]
\,,
\label{3 point vertex: B xi}
\end{eqnarray}
where $\overline{\nabla}_{\mu}$ is the background covariant derivative,
$\overline{\dAlembert} =\overline{g}^{\mu\nu} 
 \overline{\nabla}_{\mu}\overline{\nabla}_{\nu}$ and
$\overline{G}_{\mu\nu}$ is the background Einstein tensor (defined with respect to
the background metric $\overline g_{\mu\nu}$).
The structure of vertices~(\ref{3 point vertex: A xi})--(\ref{3 point vertex: B xi})
suggests to introduce the following operator,
\begin{equation}
\overline{\mathcal{G}}_{\mu\nu}
\equiv\overline{G}_{\mu\nu} -\overline{\nabla}_{\mu}\overline{\nabla}_{\nu}
   + \overline{g}_{\mu\nu}\overline{\dAlembert}
\,.\qquad
\label{mathcal Gmn}
\end{equation}
Apart from the trivial difference in the position of indices, 
the three-point vertices differ by a sign, which is immaterial as 
the three-point vertices appear quadratically in the three-point self-energy diagram.~\footnote{This sign difference is well known. For example, it occurs 
in the graviton tadpole (energy-momentum tensor), and it is usually 
absorbed in the definition of the energy-momentum tensor,
\begin{equation}
T^{\mu\nu}(x) = \frac{2}{\sqrt{-g}}\frac{\delta S_{\rm m}}{\delta g_{\mu\nu}(x)}
\,,\qquad
T_{\mu\nu}(x) = - \frac{2}{\sqrt{-g}}\frac{\delta S_{\rm m}}{\delta g^{\mu\nu}(x)}
\,.\quad
\nonumber
\end{equation}
 }

The four-point vertex functions for the minimal couplings are, 
\begin{eqnarray}
\Big[V_{(4,0)}^{\mu\nu\rho\sigma}(x;x')  \Big]_A
&\!\!=\!\!& -\frac{i\kappa^2}{\hbar}\sqrt{-\overline{g}}
  \bigg[\left(\overline{\nabla}^{(\mu}_{1}\overline{g}^{\nu)(\rho}
                    \overline{\nabla}^{\sigma)}_{2}
  +\overline{\nabla}^{(\mu}_{2}\overline{g}^{\nu)(\rho}
             \overline{\nabla}^{\sigma)}_{1}\right)
\nonumber\\
&&\hskip 1.7cm
  -\,\frac12\left(\overline{g}^{\mu\nu}\overline{\nabla}^{(\rho}_1 
                    \overline{\nabla}^{\sigma)}_{2}
  +\overline{g}^{\rho\sigma}\overline{\nabla}^{(\mu}_{1}
              \overline{\nabla}^{\nu)}_{2}\right)
\nonumber\\
&&\hskip -0.4cm
  +\,\frac14\left(\overline{g}^{\mu\nu}\overline{g}^{\rho\sigma}
           -2\overline{g}^{\mu(\rho}\overline{g}^{\sigma)\nu}\right)
   \left(\overline{\nabla}_1\!\cdot\overline{\nabla}_2\!+\!m^2\right)\!\bigg]
              \delta^D(x\!-\!x')
,
\qquad
\label{4 point vertex: A}\\
\Big[V^{(4,0)}_{\mu\nu\rho\sigma}(x;x')  \Big]_B
&\!\!=\!\!&-\frac{i\kappa^2}{\hbar}\sqrt{-\overline{g}}
  \bigg[\!-\frac12\left(\overline{g}_{\mu\nu}\overline{\nabla}_{(\rho}^1
                  \overline{\nabla}_{\sigma)}^{2}
  \!+\!\overline{g}_{\rho\sigma}\overline{\nabla}_{(\mu}^{1}
                \overline{\nabla}_{\nu)}^{2}\right)
\nonumber\\
&&\hskip -0.4cm
     +\,\frac14\left(\overline{g}_{\mu\nu}\overline{g}_{\rho\sigma}
                \!+\!2\overline{g}_{\mu(\rho}\overline{g}_{\sigma)\nu}
               \right)
          \left(\overline{\nabla}_1\!\cdot\overline{\nabla}_2
                    \!+\!m^2\right)\!\bigg]\delta^D(x\!-\!x')
\,.\qquad
\label{4 point vertex: B}
\end{eqnarray}
The non-minimal coupling contributions to the four-point vertices
can be extracted from the considerations in Appendix~B. From 
Eqs.~(\ref{second variation of the Hilbert-Einstein action: Lichnerowicz}) 
and~(\ref{second variation of the Hilbert-Einstein action: Lichnerowicz: case B})
we have,
\begin{eqnarray}
\left[\delta_\xi V_{(4)}^{\mu\nu\rho\sigma}(x;x') \right]_A
&\!\!=\!\!& -\frac{i\kappa^2}{\hbar}\xi \sqrt{-\overline{g}}
  \bigg\{\!\!-\!\frac12\big( \overline{g}^{\mu\nu}\overline{G}^{\rho\sigma}
                           \!+\!\overline{g}^{\rho\sigma} \overline{G}^{\mu\nu}\big)
                               \!+\!2\overline{g}^{\mu)(\rho} \overline{G}^{\sigma)(\nu}
\quad
 \nonumber\\
&&\hskip 1.85cm 
                         -\, \frac14  \big(\overline{g}^{\mu\nu}\overline{g}^{\rho\sigma} 
           \!-\!2\overline{g}^{\mu(\rho}\overline{g}^{\sigma)\nu}\big)\overline{R}
\label{4 point vertex: A xi}\\
&&\hskip -3.9cm  
       +\,\frac12 \Big( \overline{g}^{\mu\nu}
      \overline{\nabla}^\rho\overline{\nabla}^\sigma
             \!+\!\overline{g}^{\rho\sigma}\overline{\nabla}^\mu\overline{\nabla}^\nu
         \Big)
\!-\!\overline{g}^{\mu)(\rho} \overline{\nabla}^{\sigma)}\overline{\nabla}^{(\nu}                
          \!\!-\! \frac12\Big(\overline{g}^{\mu\nu} \overline{g}^{\rho\sigma}
          \!\!-\!\overline{g}^{\mu(\rho}\overline{g}^{\sigma)\nu}    
             \Big)\overline{\dAlembert}
                     \bigg\}  \delta^D(x\!-\!x') 
,
\nonumber
\end{eqnarray}
\begin{eqnarray}
\Big[\delta_\xi V^{(4)}_{\mu\nu\rho\sigma}(x;x')  \Big]_B
&\!\!=\!\!&- \frac{i\kappa^2}{\hbar}\xi \sqrt{-\overline{g}}
  \bigg\{\!\!-\!\frac12\big( \overline{g}_{\mu\nu}\overline{G}_{\rho\sigma}
                           \!+\!\overline{g}_{\rho\sigma} \overline{G}_{\mu\nu}\big)
\nonumber\\
&&\hskip 1.85cm   
                        -\, \frac14  \big(\overline{g}_{\mu\nu}\overline{g}_{\rho\sigma} 
           \!-\!2\overline{g}_{\mu(\rho}\overline{g}_{\sigma)\nu}\big)\overline{R}
\label{4 point vertex: B xi}\\
&&\hskip -3.8cm  
       +\,\frac12 \Big( \overline{g}_{\mu\nu}\!
      \overline{\nabla}_\rho\overline{\nabla}_\sigma
             \!+\!\overline{g}_{\rho\sigma}\!\overline{\nabla}_\mu\overline{\nabla}_\nu
         \Big)
\!-\!\overline{g}_{\mu)(\rho}\! \overline{\nabla}_{\sigma)}\overline{\nabla}_{(\nu}                
          \!-\! \frac12\Big(\overline{g}_{\mu\nu} \overline{g}_{\rho\sigma}  
          \!-\! \overline{g}_{\mu(\rho}\overline{g}_{\sigma)\nu}
             \Big)\overline{\dAlembert}
                     \bigg\}\delta^D(x\!-\!x')
\,.
\nonumber
\end{eqnarray}
Apart from some presumably unimportant sign differences,
the four-point vertices~(\ref{4 point vertex: A})--(\ref{4 point vertex: B}) 
significantly differ in that the first line in the four-point vertex in 
Eq.~(\ref{4 point vertex: A}) is absent in the 4-point vertex in Eq.~(\ref{4 point vertex: B}).
The question then arises whether the corresponding four-point diagram contributions to the 
graviton self-energy are physically equivalent. We emphasize that this question 
is different from the question of gauge independence, addressed in cosmological settings in 
Refs.~\cite{Miao:2017feh,Glavan:2020gal,Glavan:2020ccz,Glavan:2021adm,Glavan:2024elz},
 but may be equally important to resolve.
Similarly, there are potentially significant differences in the non-minimal coupling
vertex functions in Eqs.~(\ref{4 point vertex: A xi})--(\ref{4 point vertex: B xi}).
This completes our discussion of the vertex functions, which will be used in the one-loop diagrams.

To complete the Feynman rules we also need the propagators.  In the 1PI formalism the scalar matter 
propagator satisfies the tree-level equation of motion~(\ref{propagator equation of motion}),
and therefore it is a functional of the metric, 
\begin{equation}
i\Delta_\phi(x;x') \equiv i\Delta_\phi(x;x'|g_{\mu\nu}]
\,.
\label{propagator as a functional of the metric}
\end{equation}
The propagator is known only for a small class of metrics $\overline{g}_{\mu\nu}$, 
typically for those with a high level of symmetry,
such as Minkowski space, (anti)-de Sitter space and some cosmological spaces such as radiation era.
In fact, it is unlikely that we will ever know the explicit form of the propagator for general curved backgrounds.
Nevertheless, it is often of a great value to understand the evolution for the metric of the 
form $g_{\mu\nu}=\overline{g}_{\mu\nu} +\kappa \delta g_{\mu\nu}$, where $\overline{g}_{\mu\nu}$ 
is a highly-symmetric metric for which the propagator is known, and $\kappa\delta g_{\mu\nu}$
is a small perturbation on top of $\overline{g}_{\mu\nu}$. Such a problem is especially important 
in cosmology, in which we are interested in  
the evolution of small, but general, metric perturbations 
on cosmological backgrounds $\overline{g}_{\mu\nu}$. Cosmological backgrounds break time translation
invariance, which means that the evolution on such backgrounds is non-equilibrium.
For such non-equilibrium evolutions it is not enough to know the propagator, but in addition the knowledge 
of the two Wightman functions, defined in Eq.~(\ref{Wightman functions}), is also required.
These two-point functions obey homogeneous equations of motion~(\ref{equation of motion Wightman})
and -- just like the propagator -- are not known for general gravitational backgrounds $g_{\mu\nu}$.
The formalism of choice for studying the evolution of gravitational perturbations on various metric backgrounds
is the Schwinger-Keldysh formalism, also known as the {\it in-in} formalism. 
In order to simplify our notation and condense
the paper, in this work we shall present the {\it in-out} formalism, for which the Feynman propagator suffices, 
and discuss at appropriate places how to generalise our considerations to the {\it in-in} formalism.

\medskip

\section{Effective action}
\label{Effective action}

\bigskip
\noindent
{\bf Effective action.} The classical action~(\ref{classical scalar matter action}) 
is quadratic in the scalar field $\phi$,
and therefore the quantum effective action obtained by integrating the scalar matter perturbation
can be formally written as the 1PI one-loop effective action, which in the background field method
and {\it in-out} formalism~\footnote{For notational simplicity we use here the {\it in-out} formalism.
Later we shall note how to generalize the analysis presented here to the {\it in-in} formalism,
which is the formalism of choice in curved backgrounds.}
has the form,
\begin{eqnarray}
{\rm e}^{\frac{i}{\hbar}\Gamma[g_{\mu\nu},\phi]}
  &=& {\rm e}^{\frac{i}{\hbar}S_{\rm g}[g_{\mu\nu}]}
  \int \mathcal{D}\varphi \langle \Psi[\varphi_-]|\hat \rho|\Psi[\varphi_+]\rangle
     {\rm e}^{\frac{i}{\hbar}S_{\rm m}[g_{\mu\nu},\phi+\varphi]}
\,,\qquad
\label{1PI one-loop effective action 0}
\end{eqnarray}
from where we infer,
\begin{eqnarray}
 \Gamma[g_{\mu\nu},\phi]&=&S_{\rm g}[g_{\mu\nu}]+S_{\rm m}[g_{\mu\nu},\phi]
    +\frac{i\hbar}{2}{\rm Tr}\ln\left[\frac{1}{\mu^{D+2}}
     \frac{\delta^2S_{\rm m}[g_{\mu\nu},\phi]}{\delta \phi(x)\delta \phi(x')}\right]
\nonumber\\
  &&\hskip -2.2cm
  =\, S_{\rm g}[g_{\mu\nu}]+S_{\rm m}[g_{\mu\nu},\phi]
    +\frac{i\hbar}{2}{\rm Tr}\ln\left[\frac{\sqrt{-g}\big[\dAlembert\!-\!m^2\!-\!\xi R\big]
                  \delta^D(x\!-\!x')}{\mu^{D+2}}\right]
\,,\qquad
\label{1PI one-loop effective action}
\end{eqnarray}
where $\varphi_-$ and $\varphi_+$ are the fields at the {\it in} and {\it out} states which are also integrated over 
in part integral measure $\mathcal{D}\varphi$, and $\mu$ is an arbitrary energy scale, introduced for dimensional reasons (to make the 
argument of the logarithm dimensionless).
Formally, Eq.~(\ref{1PI one-loop effective action}) is an exact effective action which includes 
all of the quantum matter effects in general gravitational backgrounds, provided one adds the counterterms that regularize it. A convenient way of rewriting the 
one-loop contribution $\Gamma^{(1)}$ in Eq.~(\ref{1PI one-loop effective action})
is to use the scalar propagator $i\Delta_\phi(x;x')$, which 
satisfies Eq.~(\ref{propagator equation of motion}) with some given boundary conditions,
in terms of which the one-loop contribution to the effective action becomes,
\begin{eqnarray}
 \Gamma^{(1)} = - \frac{i\hbar}{2}{\rm Tr}\ln\left[\frac{\Delta_\phi(x;x')}{\hbar\mu^{D-2}}\right]
\,,\qquad
\label{1PI one-loop effective action 2}
\end{eqnarray}
where $\Delta_\phi(x;x')\equiv\Delta_\phi(x;x'|g_{\mu\nu},\phi]$ is in general an unknown functional of the metric tensor $g_{\mu\nu}$ and the field $\phi$, which themselves are general off-shell variables in the effective 
action.
Indeed, the scalar propagator is known only for a handful of curved spacetimes.
To make progress, we shall expand the $\Gamma[g_{\mu\nu},\phi]$ in 
Eq.~(\ref{1PI one-loop effective action}) 
around some symmetric background $\overline{g}_{\mu\nu}$, for which the solutions of the 
semi-classical equations of motion for the field $\overline{\phi}$, the metric $\overline{g}_{\mu\nu}$ and
the propagator $i\Delta_{\bar{\phi}}(x;x')\equiv i\Delta_{\phi}(x;x'|\overline{g}_{\mu\nu}]$ are all known.

Writing the metric and the field as ({\it cf.} Eq.~(\ref{metric perturbations: case A})),
\begin{equation}
{\rm Case~A:} \quad g_{\mu\nu} = \overline{g}_{\mu\nu} +\kappa\delta g_{\mu\nu}
\,,\qquad \phi = \overline{\phi}
\,,\qquad
\label{expansion of the metric and field A}
\end{equation}
motivates the following expansion of the effective action, 
\begin{eqnarray}
  \Gamma[g_{\mu\nu},\phi]&=& \Gamma\big[\overline{g}_{\mu\nu},\overline{\phi}\,\big]
   +\kappa \int\! {\rm d}^Dx \left(\frac{\delta\Gamma[g_{\mu\nu},\phi]}
    {\delta g_{\mu\nu}(x)}\right)_{\overline{g}_{\mu\nu},\overline{\phi}}\delta g_{\mu\nu}(x)
\nonumber\\
  &&\hskip -2.15cm
+\,\frac{\kappa^2}2 \int\! {\rm d}^Dx{\rm d}^D{x'} \left(\frac{\delta^2\Gamma[g_{\mu\nu},\phi]}
    {\delta g_{\mu\nu}(x)\delta{g}_{\rho\sigma}(x')}\right)_{\overline{g}_{\mu\nu},\overline{\phi}}
      \!\delta g_{\mu\nu}(x)\delta{g}_{\rho\sigma}(x')
      +\mathcal{O}\big(\kappa^3\delta g_{\mu\nu}^3\big)
\,,\qquad
\label{1PI one-loop effective action: expansion A}
\end{eqnarray}
where functional derivatives are evaluated on the background fields.
In this work we are primarily interested in the evolution of the metric perturbations,
and thus we restrict the classical field to that which obeys the symmetries of the background,
and for that purpose the scalar field expansion in 
Eq.~(\ref{expansion of the metric and field A}) suffices.~\footnote{A more general expansion 
includes the field expansion, $\phi = \overline{\phi} + \kappa \delta\phi$, with $\delta\phi(x)$ being 
a small -- but arbitrary -- field perturbation. In this more general case, one would have to vary the action 
in Eq.~(\ref{1PI one-loop effective action: expansion A}) also with respect to $\delta \phi(x)$.}

For the metric perturbation defined in Eq.~(\ref{metric perturbations: case B}),
\begin{equation}
{\rm Case~B:} \quad g^{\mu\nu} = \overline{g}^{\mu\nu} +\kappa\delta g^{\mu\nu}
\,,\qquad \phi = \overline{\phi}
\,,\qquad
\label{expansion of the metric and field B}
\end{equation}
the effective action expansion looks like,
\begin{eqnarray}
  \Gamma[g_{\mu\nu},\phi]&=& \Gamma\big[\overline{g}_{\mu\nu},\overline{\phi}\,\big]
   +\kappa \int\! {\rm d}^Dx \left(\frac{\delta\Gamma[g_{\mu\nu},\phi]}
    {\delta g^{\mu\nu}(x)}\right)_{\overline{g}_{\mu\nu},\overline{\phi}}\delta g^{\mu\nu}(x)
\nonumber\\
  &&\hskip -2.5cm
+\,\frac{\kappa^2}2\! \int\! {\rm d}^Dx{\rm d}^D{x'} \left(\frac{\delta^2\Gamma[g_{\mu\nu},\phi]}
    {\delta g^{\mu\nu}(x)\delta{g}^{\rho\sigma}(x')}\right)_{\overline{g}_{\mu\nu},\overline{\phi}}
      \!\delta g^{\mu\nu}(x)\delta{g}^{\rho\sigma}(x')
      +\mathcal{O}\big(\kappa^3{(\delta g_{\mu\nu})}^3\big)
.\qquad\!
\label{1PI one-loop effective action: expansion B}
\end{eqnarray}
The neglected terms denoted as $\mathcal{O}\big(\kappa^3{(\delta g_{\mu\nu})}^3\big)$ 
consist of the classical three- and higher-point graviton vertex contributions 
and the corresponding quantum (loop) vertex corrections, whose consideration is beyond the scope of this work.




\bigskip
\noindent
{\bf Energy-momentum tensor.} The first variation of
the effective action in Eqs.~(\ref{1PI one-loop effective action: expansion A}) and 
(\ref{1PI one-loop effective action: expansion B}) involves 
the classical and quantum contributions to the energy momentum tensor.
The classical energy-momentum tensor 
is obtained by varying the classical action~(\ref{classical scalar matter action}) once.
For Case~A we have, 
\begin{eqnarray}
\overline{T}_A^{\mu\nu}(x) &\!\!=\!\!& \frac{2}{\sqrt{-\overline{g}}}
 \left(\frac{\delta S_{\rm m}}{\kappa\delta g_{\mu\nu}(x)}\right)_{\overline{g}_{\mu\nu},\overline{\phi}}
 = \big(\overline{\nabla}^\mu\overline{\phi}\,\big)\big( \overline{\nabla}^\nu\overline{\phi}\,\big)
\label{energy momentum tensor classical A}\\
&&\hskip -.45cm
+ \,\,\overline{g}^{\mu\nu}\left[-\frac12 \overline{g}^{\alpha\beta}\overline{\nabla}_\alpha\overline{\phi}
 \overline{\nabla}_\beta\overline{\phi}-\frac12m^2\overline{\phi}^2\right]
 +\xi \left[\,\overline{G}^{\mu\nu}-\overline{\nabla}^\mu\overline{\nabla}^\nu
 + \overline{g}^{\mu\nu} \overline{\dAlembert}\,\right]\big(\,\overline{\phi}^2\,\big)
 \,,\!\!
\nonumber
\end{eqnarray}
and for Case B,
\begin{eqnarray}
\overline{T}_{\mu\nu}^B(x) &\!\!\!=\!\!\!& \frac{-2}{\sqrt{-\overline{g}}}
 \left(\frac{\delta S_{\rm m}}{\kappa\delta g^{\mu\nu}(x)}\right)_{\overline{g}_{\mu\nu},\overline{\phi}}
= \big(\overline{\nabla}_\mu\overline{\phi}\,\big)\big( \overline{\nabla}_\nu\overline{\phi}\,\big)
\label{energy momentum tensor classical B}\\
&&\hskip -.45cm
+\, \overline{g}_{\mu\nu}\!\left[-\frac12 \overline{g}^{\alpha\beta}\overline{\nabla}_\alpha\overline{\phi}
 \overline{\nabla}_\beta\overline{\phi}-\frac12m^2\overline{\phi}^2\right]
 +\,\xi \left[\,\overline{G}_{\mu\nu}-\overline{\nabla}_\mu\overline{\nabla}_\nu
 + \overline{g}_{\mu\nu} \overline{\dAlembert}\,\right]\big(\,\overline{\phi}^2\,\big)
 \,,
\nonumber
\end{eqnarray}
where we follow the literature and 
absorb the sign difference in the cubic vertices~(\ref{3 point vertex: A})--(\ref{3 point vertex: B xi}) 
by the minus sign 
difference in the definition of the energy-momentum tensor.
The subscripts $\overline{g}_{\mu\nu},\overline{\phi}$ in Eqs.~(\ref{energy momentum tensor classical A}) and~(\ref{energy momentum tensor classical B})
mean that $\overline{T}_A^{\mu\nu}(x)$ and $\overline{T}^B_{\mu\nu}(x)$ are defined on a background 
$\overline{g}_{\mu\nu},\overline{\phi}$, which obey the equations of motion,
\begin{equation}
\sqrt{-\overline{g}}\left(\overline{\dAlembert} - m^2 - \xi\overline{R}\,\right)\overline{\phi}(x) = 0
\,.\quad
\label{eom for bar phi}
\end{equation}
The quantum one-loop contribution to the energy-momentum tensor -- also known as the graviton tadpole -- 
is obtained by varying the action once, and evaluating it on the quantum perturbations of the field.
One can express it in terms of the scalar field propagator $\Delta_{\bar\phi}(x;x')$ as,
\begin{eqnarray}
\Big\langle\Delta\hat{\overline{T}}_A^{\mu\nu}(x)\Big\rangle &\!=\!&  \bigg\{
\overline{\nabla}^\mu {\overline{\nabla}'}^\nu i\Delta_{\bar\phi}(x;x')
 + \overline{g}^{\mu\nu}\bigg[-\frac12 \overline{g}^{\alpha\beta}\overline{\nabla}_\alpha 
 \overline{\nabla}'_\beta i\Delta_{\bar\phi}(x;x')
 \nonumber\\
&&\hskip -1.45cm
 -\frac12m^2 i\Delta_{\bar\phi}(x;x')\bigg]
 \bigg\}_{x'\rightarrow x}
 +\,\xi \left[\overline{G}^{\mu\nu}-\overline{\nabla}^\mu\overline{\nabla}^\nu
 + \overline{g}^{\mu\nu} \overline{\dAlembert}\right]\big[ i\Delta_{\bar\phi}(x;x)\big]
 \,,\qquad
\label{energy momentum tensor: 1 loop A}\\
\Big\langle\Delta\hat{\overline{T}}^B_{\mu\nu}(x)\Big\rangle &\!=\!&  \bigg\{
\overline{\nabla}_\mu {\overline{\nabla}'}_\nu i\Delta_{\bar\phi}(x;x')
 + \overline{g}_{\mu\nu}\bigg[-\frac12 \overline{g}^{\alpha\beta}\overline{\nabla}_\alpha 
 \overline{\nabla}'_\beta i\Delta_{\bar\phi}(x;x')
 \nonumber\\
&&\hskip -1.45cm
 -\frac12m^2 i\Delta_{\bar\phi}(x;x')\bigg]
 \bigg\}_{x'\rightarrow x}
 +\,\xi \left[\overline{G}_{\mu\nu}-\overline{\nabla}_\mu\overline{\nabla}_\nu
 + \overline{g}_{\mu\nu} \overline{\dAlembert}\right]\big[ i\Delta_{\bar\phi}(x;x)\big]
 \,,\qquad
\label{energy momentum tensor: 1 loop B}
\end{eqnarray}
\begin{figure}[t]
\centering
\vskip -1.2cm
\includegraphics[width=4.4cm]{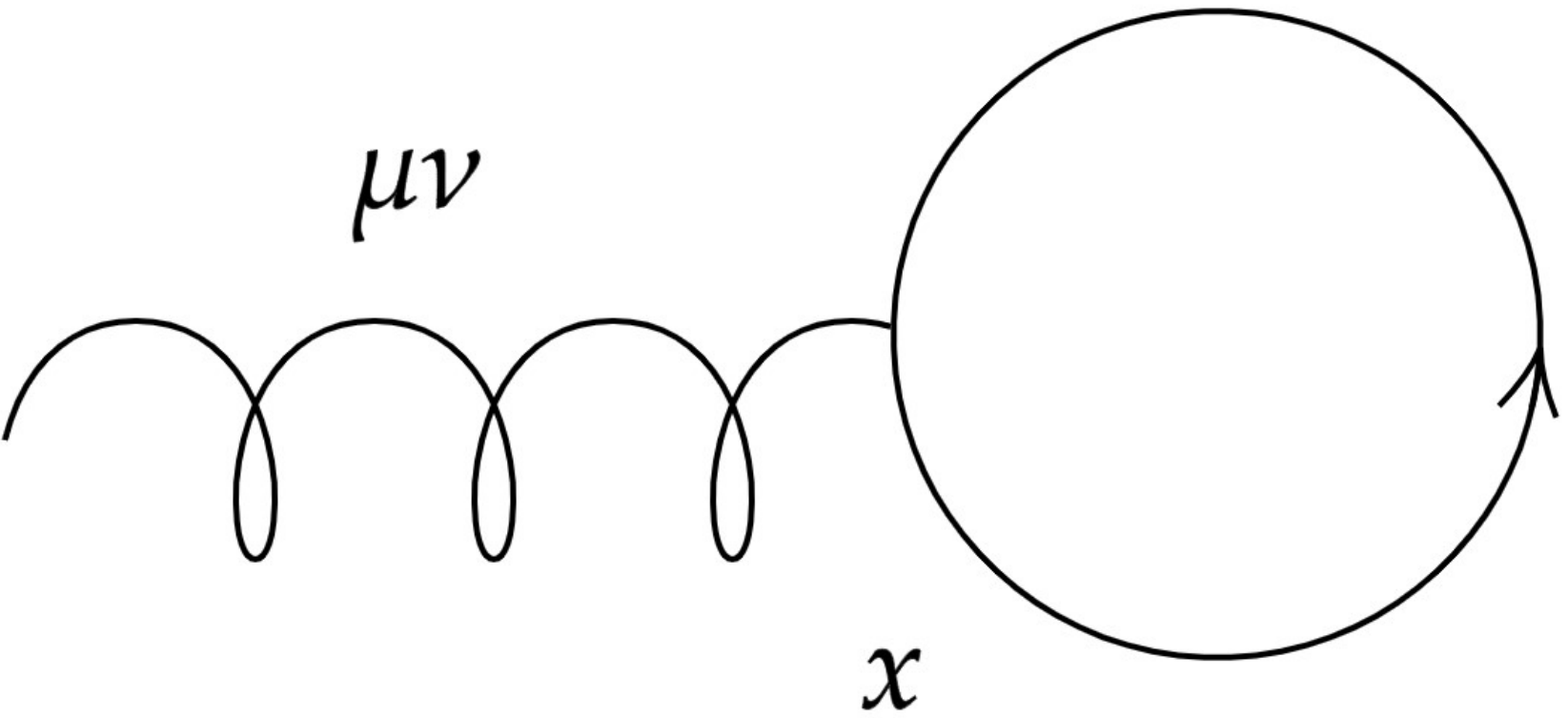}
\vskip -0.5cm
\caption{\footnotesize The one-loop graviton tadpole diagram contributing to the background equation of motion~(\ref{SC Einstein equation}).
This diagram contributes to the semiclassical Einstein equation~(\ref{SC Einstein equation})
such that the external graviton leg ought to be amputated.}
\label{figure: tadpole diagram}
\end{figure}
where now the propagator $ i\Delta_{\bar\phi}(x;x')$ obeys the background equation,~\footnote{The 
equation of motion~(\ref{eom for bar iDelta(x;x')}) for the propagator on the background metric,
$i\Delta_{\bar\phi}(x;x')\equiv i\Delta_\phi(x;x'|\overline{g}_{\mu\nu}]$
differs from the more general equation~(\ref{propagator equation of motion}) 
for which the propagator is a functional of the general metric tensor $g_{\mu\nu}$,
$i\Delta_\phi(x;x')\equiv i\Delta_\phi(x;x'|g_{\mu\nu}]$. 
}
\begin{equation}
\sqrt{-\overline{g}}\left(\overline{\dAlembert} - m^2 - \xi\overline{R}\,\right)i\Delta_{\bar\phi}(x;x')
 = i\hbar \delta^D(x\!-\!x')
\,.\quad
\label{eom for bar iDelta(x;x')}
\end{equation}
and therefore it shares the symmetries of the background metric $\overline{g}_{\mu\nu}$, provided one chooses the initial state that respects these symmetries.
This can be achieved by selecting a state that obeys the symmetries of the background.
The background metric $\overline{g}_{\mu\nu}$ then obeys a semi-classical Einstein equation,
\begin{equation}
\overline{G}_A^{\mu\nu}\big(\overline{g}_{\alpha\beta}\big)
 =\frac{\kappa^2}{2} \left[\overline{T}_A^{\mu\nu}(x)
               +\Big\langle\Delta\hat{\overline{T}}^{\mu\nu}_A(x)\Big\rangle
               + \overline{T}_{\tt ct}^{\mu\nu}(x)\right]
\,,\quad
\label{SC Einstein equation}
\end{equation}
where 
\begin{equation}
\overline{T}_{\tt ct}^{\mu\nu}(x) = \frac{2}{\sqrt{-\overline{g}}} \left(\frac{\delta S_{\rm ct}}{\kappa\delta g_{\mu\nu}(x)}\right)_{\overline{g}_{\mu\nu},\overline{\phi}}
\,,\quad
\label{Tmn counterterm}
\end{equation}
originates from varying the counterterm 
action given by the sum of the actions in Eqs.~(\ref{Hilbert-Einstein counterterm action}), 
(\ref{gravitational counterterm action}) and~(\ref{counterterm scalar matter action}),
\begin{equation}
S_{\rm ct}\big[\phi,g_{\mu\nu}\big] = S_{\rm g}^{\tt ct}[g_{\mu\nu}] 
                                                               + \tilde S_{\rm g}^{\tt ct}[g_{\mu\nu}] 
                                                               +  S^{\tt ct}_{\rm m}\big[\phi,g_{\mu\nu}\big]
\,,\qquad
\label{total counterterm action}
\end{equation}
and $\Big\langle\Delta\hat{\overline{T}}^{\mu\nu}_A(x)\Big\rangle$ is the one-loop energy-momentum tensor
in Eq.~(\ref{energy momentum tensor: 1 loop A}), whose diagrammatic representation is shown in 
figure~\ref{figure: tadpole diagram}. The explicit form of the counterterm action in Eq.~(\ref{Tmn counterterm}) is given in Appendices~A and~B.
 The value of the coupling constants in the counterterm
action needed to renormalize 
$\Big\langle\Delta\hat{\overline{T}}^{\mu\nu}_A(x)\Big\rangle$ is well 
known~\cite{tHooft:1974toh,Marunovic:2012pr}, 
and it can be obtained on selected as well as on general gravitational backgrounds~\cite{Birrell:1982ix},
as only near-coincident (ultraviolet) properties of the propagator are needed to determine the value of these constants.
Since the details of the renormalization procedure are not relevant for this work, we shall not dwell on
determining the precise value of the counterterms. Instead we shall assume that the value has been 
determined and that 
\begin{equation}
\Big\langle\Big(\Delta\hat{\overline{T}}^{\mu\nu}_{\rm ren}\Big)_A(x)\Big\rangle
= \Big\langle\Delta\hat{\overline{T}}^{\mu\nu}_A(x)\Big\rangle
               + \overline{T}_{\tt ct}^{\mu\nu}(x)
\,,\qquad
\label{renormalized energy-momentum tensor}
\end{equation}
is the renormalized energy-momentum tensor, which is thus finite in $D=4$.
Finally, we note that in this work we assume that the symmetries of the state 
are the same as those of the background metric $\overline{g}_{\mu\nu}$, such that 
$\big\langle(\Delta\hat{\overline{T}}^{\mu\nu}_{\rm ren})_A(x)\big\rangle$ respects 
the isometries of the background metric.

\begin{figure}[t]
\centering
\vskip -2.9cm
\includegraphics[width=13cm]{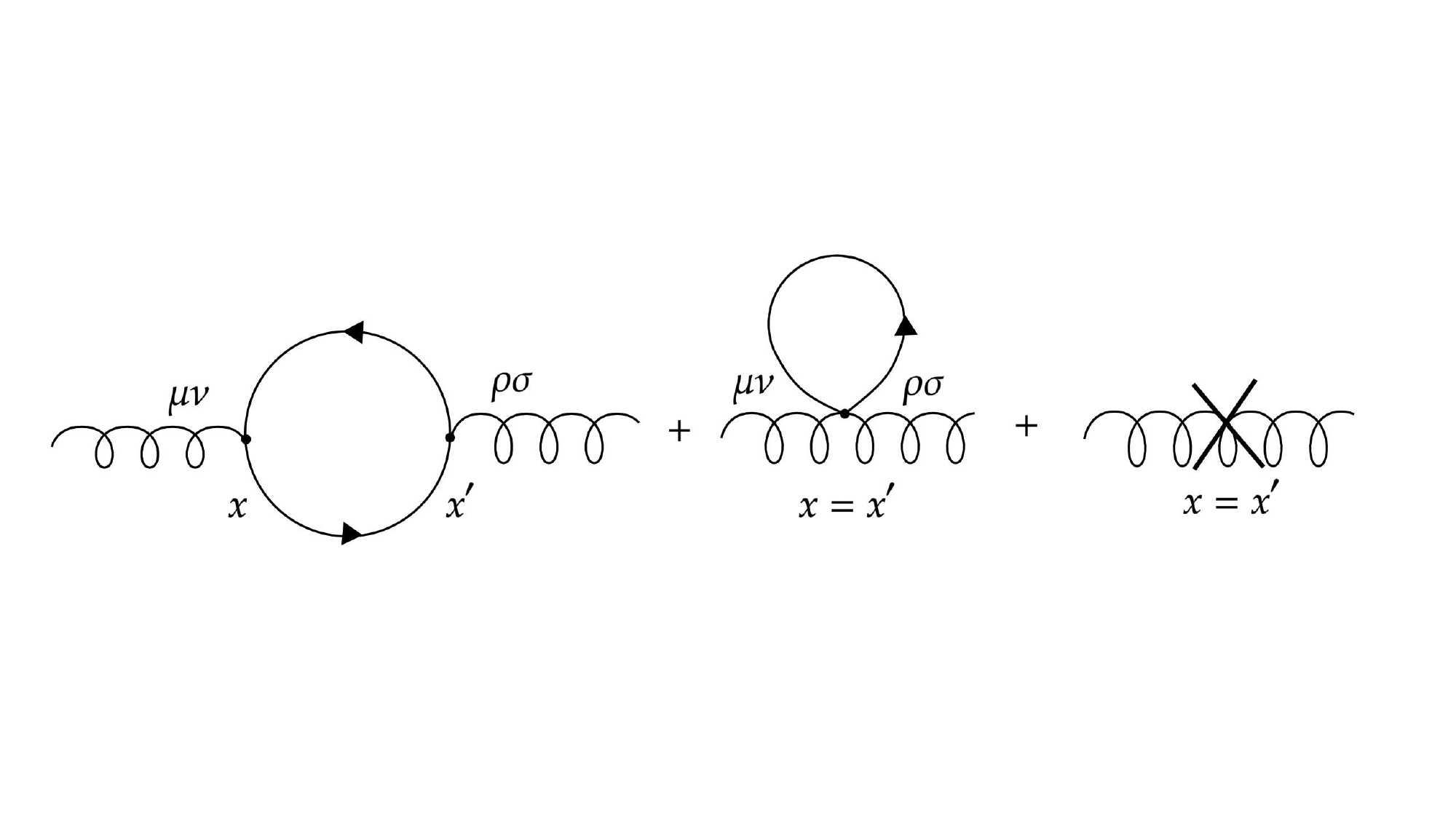}
\vskip -2.1cm
\caption{\footnotesize The one-loop diagrams contributing to the graviton 
self-energy~(\ref{total self-energy}) 
induced by scalars. The first two diagrams are the three- and four-point contributions,
and the last diagram represents the local counterterms.}
\label{figure: 1 loop diagrams}
\end{figure}

\bigskip
\noindent
{\bf The graviton self-energy.}
The second variation in 
the effective action expansion in Eqs.~(\ref{1PI one-loop effective action: expansion A}) and 
(\ref{1PI one-loop effective action: expansion B}) involves 
the classical contribution
-- which can be easily obtained from the four-point vertex functions 
in Eqs.~(\ref{4 point vertex: A})--(\ref{4 point vertex: B xi}) --
 and the quantum contribution, which is known as the graviton self-energy.
 The total self-energy consists of the three- and four-point diagrams, 
supplemented by the counter-term diagram,
\begin{equation}
\!\, -i\big[{}^{\mu\nu}\overline{\Sigma}_A^{\rho\sigma}\big](x;x')
 =  -i\big[{}^{\mu\nu}\overline{\Sigma}_{\mbox{\tiny 3pt}}^{\rho\sigma}\big]_A(x;x')
           -i\big[{}^{\mu\nu}\overline{\Sigma}_{\mbox{\tiny 4pt}}^{\rho\sigma}\big](x;x')
        -i\big[{}^{\mu\nu}\overline{\Sigma}_{\mbox{\tiny ct}}^{\rho\sigma}\big]_A(x;x')
\,,
\label{total self-energy}
\end{equation}
which can be also represented in terms of a second variation of 
the one-loop contribution to the effective action in Eq.~(\ref{1PI one-loop effective action: expansion A}),
\begin{eqnarray}
 \frac{1}{\sqrt{-\overline{g}}\sqrt{-\overline{g}'}}
  \left(\frac{\delta^2\Gamma^{(1)}[g_{\alpha\beta},\phi]}
    {\delta g_{\mu\nu}(x)\delta{g}_{\rho\sigma}(x')}\right)_{\overline{g}_{\mu\nu},\overline{\phi}}
     &\!\!=\!\!& 
    \!\, -\big[{}^{\mu\nu}\overline{\Sigma}_A^{\rho\sigma}\big](x;x') 
\,,\qquad
\label{1PI one-loop self energy: in terms of effective action A}\\
\frac{1}{\sqrt{-\overline{g}}\sqrt{-\overline{g}'}}
  \left(\frac{\delta^2\Gamma^{(1)}[g_{\alpha\beta},\phi]}
    {\delta g^{\mu\nu}(x)\delta{g}^{\rho\sigma}(x')}\right)_{\overline{g}_{\mu\nu},\overline{\phi}}   
 &\!\!=\!\!&    \!\, -\big[{}_{\mu\nu}\overline{\Sigma}^B_{\rho\sigma}\big](x;x') 
\,.\qquad
\label{1PI one-loop self energy: in terms of effective action B}
\end{eqnarray}
The Feynman diagrammatic representation of the different one-loop contributions is shown in 
figure~\ref{figure: 1 loop diagrams}, and these can be naturally split into the 
three-point, four-point and counterterm contributions.
The counterterm contributions ought to be chosen such that the total self-energy
$-i\big[{}^{\mu\nu}\overline{\Sigma}_A^{\rho\sigma}\big](x;x')$
is finite in $D=4$. 
Just like in the case of the energy-momentum tensor discussed above,
this can be achieved by suitable choice of the couplings in the counterterm action. Moreover,
one can show that the couplings that renomalize the one-loop energy-momentum tensor must simultaneously
renormalize the one-loop graviton self-energy. 

Making use of the vertex functions~(\ref{3 point vertex: A})--(\ref{3 point vertex: B xi})
and~(\ref{4 point vertex: A})--(\ref{4 point vertex: B xi}) 
one can easily obtain the three- and four-point contributions to the self-energy,~\footnote{Alternative expressions for the self-energy contributions are, 
\begin{eqnarray}
 -i\big[ {}^{\mu\nu}\overline{\Sigma}^{\rho\sigma}_{\mbox{\tiny 3pt}}\, \big]_A(x;x')
 &\!\!=\!\!& \frac{1}{\sqrt{\overline{g}(x)\overline{g}(x')}} 
 \frac{1}{\hbar}\left\langle\frac{i\delta S_{\rm m}^{(3)}}{\delta g_{\mu\nu}(x)}
  \frac{i\delta S_{\rm m}^{(3)}}{\delta g_{\rho\sigma}(x')}\right\rangle_{\overline{g}_{\mu\nu},\rm 1PI}
\,,
\label{three-point self-energy: general}\\
 -i\big[ {}^{\mu\nu}\overline{\Sigma}^{\rho\sigma}_{\mbox{\tiny 4pt}}\, \big]_A(x;x')
 &\!\!=\!\!& \frac{1}{\sqrt{\overline{g}(x)\overline{g}(x')}}\left\langle\frac{i\delta S_{\rm m}^{(4)}}{\delta g_{\mu\nu}(x)\delta g_{\rho\sigma}(x')}
  \right\rangle_{\overline{g}_{\mu\nu},\rm 1PI}
  \,,\qquad
\label{four-point self-energy: general}
\end{eqnarray}
where $S_{\rm m}^{(3)}$ is the cubic tree-level action 
(linear in $\delta g_{\mu\nu}$ and quadratic in the fields $\phi$), $S_{\rm m}^{(4)}$ is the quartic
tree-level action 
(quadratic in gravitational perturbations $\delta g_{\mu\nu}$ and quadratic in the fields $\phi$), and
$\big\langle\;\cdot\;\big\rangle_{\rm 1PI}$ denotes the two-point Wick contractions that include only one-particle 
irreducible diagrams.
}~\footnote{Our definition of the self-energy in Eqs.~(\ref{three-point self-energy: 1 loop})--(\ref{four-point self-energy: 1 loop}) is such that it is a bitensor. Had we not divided by 
$\sqrt{\overline{g}(x)\overline{g}(x')}$ one would have 
obtained bi-tensor densities, which is the convention that is often used in the literature.}
\begin{eqnarray}
\! -i\big[ {}^{\mu\nu}\overline{\Sigma}^{\rho\sigma}_{\mbox{\tiny 3pt}}\, \big]_A(x;x')
\! &\!\!=\!\!&\!
  \frac{\hbar}{2\sqrt{\overline{g}(x)\overline{g}(x')}} V_{(3)}^{\mu\nu}(x)\big[i\Delta_{\bar\phi}(x;x')\big]_1
         \big[i\Delta_{\bar\phi}(x;x')\big]_2 
                   V_{(3)}^{\rho\sigma}(x') 
\,,
\nonumber\\
\label{three-point self-energy: 1 loop}\\
 -i\big[ {}^{\mu\nu}\overline{\Sigma}^{\rho\sigma}_{\mbox{\tiny 4pt}}\, \big]_A(x;x')
 &\!\!=\!\!&  \frac{1}{\sqrt{\overline{g}(x)\overline{g}(x')}} \frac{\hbar}{2}V_{(4)}^{\mu\nu\rho\sigma}(x;x')
            \big[i\Delta_{\bar \phi}(x;x')\big]_{1,2}
 \,,\qquad
\label{four-point self-energy: 1 loop}
\end{eqnarray}
where numbers $1$ and $2$ on the scalar propagators indicate the leg at which
 the vertex derivatives act, see figure~\ref{fig: three and four point vertices}.
More explicitly, 
Eqs.~(\ref{three-point self-energy: 1 loop})--(\ref{four-point self-energy: 1 loop}) yield,
\begin{eqnarray}
\!\!\!
    -i\big[ {}^{\mu\nu}\overline{\Sigma}^{\rho\sigma}_{\mbox{\tiny 3pt}} \big](x;x')
       &\!\!=\!\!& \frac{\kappa^2}{4\hbar}
         \bigg\{\!
            \!-\! 2\overline{\nabla}^{\mu)}{\overline{\nabla}'}^{(\rho}
                i\Delta_{\bar\phi}(x;x')
                {\overline{\nabla}'}^{\sigma)}\overline{\nabla}^{(\nu}
                i\Delta_{\bar\phi}(x;x')
\nonumber\\
 &&\hskip -3.4cm
           +\,
                     \overline{g}^{\mu\nu}\Big(\overline{\nabla}^{\alpha}{\overline{\nabla}'}^{(\rho}
                    i\Delta_{\bar\phi}(x;x')
                    {\overline{\nabla}'}^{\sigma)}\overline{\nabla}_{\alpha}
                    i\Delta_{\bar\phi}(x;x')\!+\!m^2
                    {\overline{\nabla}'}^{\rho}i\Delta_{\bar\phi}(x;x')
                    {\overline{\nabla}'}^{\sigma}i\Delta_{\bar\phi}(x;x')\Big)
\nonumber\\
 &&\hskip -3.4cm
                 \!+\, {\overline{g}'}^{\rho\sigma}\Big({\overline{\nabla}'}^{\gamma}\overline{\nabla}^{(\mu}\!
                    i\Delta_{\bar\phi}(x;x')
                    {\overline{\nabla}'}_{\gamma}\overline{\nabla}^{\nu)}
                    i\Delta_{\bar\phi}(x;x')\!+\!m^2
                   \overline{\nabla}^{\mu}i\Delta_{\bar\phi}(x;x')
                    \overline{\nabla}^{\nu}i\Delta_{\bar\phi}(x;x')\Big)
\nonumber\\
 &&\hskip -3.4cm
            -\,\frac{1}{2}\overline{g}^{\mu\nu}{\overline{g}'}^{\rho\sigma}
               \Big(\overline{\nabla}^{\alpha}{\overline{\nabla}'}^{\gamma}
                i\Delta_{\bar\phi}(x;x')
                \overline{\nabla}_{\gamma}'\overline{\nabla}_{\alpha}
                i\Delta_{\bar\phi}(x;x')
                \!+\!m^2{\overline{\nabla}'}^{\gamma}i\Delta_{\bar\phi}(x;x')
                   \overline{\nabla}'_{\gamma}i\Delta_{\bar\phi}(x;x')
\nonumber\\
 &&\hskip -3.cm
                    +\,m^2\overline{\nabla}^{\alpha}i\Delta_{\bar\phi}(x;x')
                   \overline{\nabla}_{\alpha}i\Delta_{\bar\phi}(x;x')
                  \!+\!m^4
                \big[i\Delta_{\bar\phi}(x;x')\big]^2  
                    \Big)
\nonumber\\
&&\hskip -3.4cm
            +\,\xi\Big[\!-2{\overline{\mathcal{G}}'}^{\rho\sigma}
                    \big(\overline{\nabla}^{\mu}i\Delta_{\bar\phi}(x;x')
                       \overline{\nabla}^{\nu}i\Delta_{\bar\phi}(x;x')\big)
 \notag\\
&&\hskip -.cm
                -\, 2\overline{\mathcal{G}}^{\mu\nu}
                    \big({\overline{\nabla}'}^{\rho}i\Delta_{\bar\phi}(x;x')
                        {\overline{\nabla}'}^{\sigma}i\Delta_{\bar\phi}(x;x')\big)
\nonumber\\
&&\hskip -2.5cm
                +\overline{g}^{\mu\nu}{\overline{\mathcal{G}}'}^{\rho\sigma}
                 \Big(\overline{\nabla}^{\alpha}i\Delta_{\bar\phi}(x;x')
                    \overline{\nabla}_{\alpha}i\Delta_{\bar\phi}(x;x')
                    \!+\!m^2\big[i\Delta_{\bar\phi}(x;x')\big]^{2}\,\Big)
\nonumber\\
&&\hskip-2.5cm
                      +\,{\overline{g}'}^{\rho\sigma}\overline{\mathcal{G}}^{\mu\nu}
                      \Big({\overline{\nabla}'}^{\gamma}i\Delta_{\bar\phi}(x;x')
                    \overline{\nabla}'_{\gamma}i\Delta_{\bar\phi}(x;x')\big)
                     \!+\!m^2\big[i\Delta_{\bar\phi}(x;x')\big]^{2}\,\Big)\Big]
\nonumber\\
&&\hskip -2.cm
       -\, 2\xi^2\overline{\mathcal{G}}^{\mu\nu}{\overline{\mathcal{G}}'}^{\rho\sigma}
                \left[i\Delta_{\bar\phi}(x;x')\right]^2
        \!\bigg\}
,\;\;\;
\label{3pt contribution to self-energy: total}
\end{eqnarray}
where we used a shorthand notation,
\begin{equation}
\overline{\mathcal{G}}^{\mu\nu} = \overline{G}^{\mu\nu}(x)
   \!-\!\overline{\nabla}^\mu\overline{\nabla}^\nu
    \!+\!\overline{g}^{\mu\nu}\overline{\dAlembert}
    \,,\qquad
{\overline{\mathcal{G}}'}^{\rho\sigma} = \overline{G}^{\rho\sigma}(x')
   \!-\!{\overline{\nabla}'}^\rho{\overline{\nabla}'}^\sigma
    \!+\!{\overline{g}'}^{\rho\sigma}\overline{\dAlembert}'
\,,\quad\!\!
\label{cal Gmn}
\end{equation}
and
\vskip -0cm
\begin{eqnarray}
 \Big\{\!\! -\!i\big[{}^{\mu\nu}
                \overline{\Sigma}_{\mbox{\tiny 4pt}}^{\rho\sigma}\big](x;x')\Big\}_{\!A}
        &\!\!=\!\!& \frac{i\kappa^2}{4}
        \bigg\{\!
            \bigg[ \!-\!2\overline{g}^{\mu)(\rho}\Big(\overline{\nabla}^{\sigma)}{\overline{\nabla}'}^{(\nu}
                        \!+\! {\overline{\nabla}'}^{\sigma)}\overline{\nabla}^{(\nu}
                \Big) i\Delta_{\bar\phi}(x;x')
\nonumber\\
&&\hspace{-2cm}     
           +\,  \Big(
                   \overline{g}^{\mu\nu}
                       \overline{\nabla}^{(\rho}{\overline{\nabla}'}^{\sigma)}
                            i\Delta_{\bar\phi}(x;x')
                    \!+\! \overline{g}^{\rho\sigma}
                       \overline{\nabla}^{(\mu}{\overline{\nabla}'}^{\nu)}
                            \Delta_{\bar\phi}(x;x')
                \Big)
\nonumber\\
&&\hspace{-3.9cm}
           -\, \frac{1}{2}\Big(
                    \overline{g}^{\mu\nu}\overline{g}^{\rho\sigma}
                   \!-\! 2\overline{g}^{\mu(\rho}\overline{g}^{\sigma)\nu}\!
                \Big)\Big(\overline{g}^{\alpha\beta}
                   \overline{\nabla}_{\alpha}\overline{\nabla}'_{\beta}i\Delta_{\bar\phi}(x;x') 
                      \! +\!m^2i\Delta_{\bar\phi}(x;x)\Big)\bigg]\frac{\delta^D(x\!-\!x')}{\sqrt{-\overline{g}}}
\nonumber\\
&&\hspace{-3.9cm}              
            +\,\xi \bigg[
                \Big(\overline{g}^{\mu\nu}\overline{G}^{\rho\sigma} 
                   \!+\!\overline{g}^{\rho\sigma}\overline{G}^{\mu\nu}\Big)
                     \!-\!4{\overline{g}}^{\mu)(\rho}\overline{G}^{\sigma)(\nu}
                     \!+\! \frac12\Big(\overline{g}^{\mu\nu} \overline{g}^{\rho\sigma}
                    \!-\!2\overline{g}^{\mu)(\rho}\overline{g}^{\sigma)(\nu}\Big)\overline{R}
\label{4-point contribution self-energy: case A}\\
&&\hspace{-3cm}  
                    \!-\,
                   \Big(\overline{g}^{\rho\sigma} \overline{\nabla}^\mu\overline{\nabla}^\nu
                                        \! \!\!+\!\overline{g}^{\mu\nu}\overline{\nabla}^\rho\overline{\nabla}^\sigma\Big)
                  \!+\!2{\overline{g}}^{\mu)(\rho}\overline{\nabla}^{\sigma)}\overline{\nabla}^{(\nu}
                  \!+\!\Big(\overline{g}^{\mu\nu}\overline{g}^{\rho\sigma}
                    \!-\!{\overline{g}}^{\mu)(\rho}\overline{g}^{\sigma)(\nu}\Big)\overline{\dAlembert}
           \bigg]
 \nonumber\\
&&\hspace{-3cm} 
                 \times
               \bigg[i\Delta_{\bar\phi}(x;x)\frac{\delta^D(x\!-\!x')}{\sqrt{-\overline{g}}}\bigg]
        \bigg\}
,\;
\nonumber
\end{eqnarray}
and (from Eqs.~(\ref{4 point vertex: A}) and~(\ref{4 point vertex: B})),
\begin{eqnarray}
 \Big\{\!\! -\!i\big[{}_{\mu\nu}
                \overline{\Sigma}^{\mbox{\tiny 4pt}}_{\rho\sigma}\big](x;x')\Big\}_{\!B}
        &\!\!\!=\!\!\!& \frac{i\kappa^2}{4}
        \bigg\{ \bigg[\Big(
                    \overline{g}_{\mu\nu}
                        \overline{\nabla}_{\!(\rho}\overline{\nabla}_{\sigma)}'
                            i\Delta_{\bar\phi}(x;x')
                    \!+\! \overline{g}_{\rho\sigma}
                        \overline{\nabla}_{\!(\mu}\overline{\nabla}_{\nu)}'
                            i\Delta_{\bar\phi}(x;x')
                \Big)
\notag\\
&&\hspace{-3.9cm}         
            -\,\frac{1}{2}\Big(
                   \overline{g}_{\mu\nu}\overline{g}_{\rho\sigma}
                   \!+\! 2\overline{g}_{\mu(\rho}\overline{g}_{\sigma)\nu}\!
                \Big)\Big(\overline{g}^{\alpha\beta}
                   \overline{\nabla}_{\alpha}\overline{\nabla}_{\beta}'i\Delta_{\bar\phi}(x;x')
                   \!+\!m^2i\Delta_{\bar\phi}(x;x) \Big)
                                       \bigg]\frac{\delta^D(x\!-\!x')}{\sqrt{-\overline{g}}}         
\nonumber\\
&&\hspace{-3.9cm}              
            +\,\xi \bigg[
               \Big(\overline{g}_{\mu\nu}\overline{G}_{\rho\sigma}
                \!+\!\overline{G}_{\mu\nu}\overline{g}_{\rho\sigma}\Big)
                    \!+\!\frac12\Big( \overline{g}_{\mu\nu}\overline{g}_{\rho\sigma}
                   \!-\! 2{\overline{g}}_{\mu)(\rho}\overline{g}_{\sigma)(\nu}\Big)\overline{R}
\label{4-point contribution self-energy: case B}\\
&&\hspace{-3.cm}  
                    -\,\Big(\overline{g}_{\mu\nu}\overline{\nabla}_{\!\rho}\overline{\nabla}_{\!\sigma}
                    \!+\!\overline{g}_{\!\rho\sigma}\overline{\nabla}_{\!\mu}\overline{\nabla}_{\!\nu}\Big)
                   \!+\! 2\overline{g}_{\mu)(\rho}\overline{\nabla}_{\!\sigma)}\overline{\nabla}_{\!(\nu}
                    \!+\!\Big(\overline{g}_{\mu\nu}\overline{g}_{\!\rho\sigma}
                    \!-\!{\overline{g}}_{\mu)(\rho}\overline{g}_{\sigma)(\nu}\Big)\overline{\dAlembert}
           \bigg]
\nonumber\\
&&\hspace{-3.cm}
              \times \bigg[i\Delta_{\bar\phi}(x;x)\frac{\delta^D(x\!-\!x')}{\sqrt{-\overline{g}}}\bigg]
        \bigg\}
,
\nonumber
\end{eqnarray}
where the four-point diagrams~(\ref{4-point contribution self-energy: case A})--(\ref{4-point contribution self-energy: case B}) refer to cases A and B for metric perturbations 
defined in Eqs.~(\ref{metric perturbations: case A})--(\ref{metric perturbations: case B}), respectively.
We have not made any distinction between Cases A and B 
in the three-point contribution~(\ref{3pt contribution to self-energy: total}),
as they are trivially related by lowering of all of the indices in Eq.~(\ref{3pt contribution to self-energy: total}).

Finally, the counterterm action contribution 
in Eq.~(\ref{total self-energy}) can be written as a second variation of 
the counterterm action in Eq.~(\ref{total counterterm action}),
\begin{equation}
 -i\big[{}^{\mu\nu}\overline{\Sigma}_{\mbox{\tiny ct}}^{\rho\sigma}\big]_A(x;x')
  = \frac{1}{\sqrt{-\overline{g}}\sqrt{-\overline{g}'}}
  \left(\frac{\delta^2S_{\rm ct}[g_{\alpha\beta},\phi]}
    {\delta g_{\mu\nu}(x)\delta{g}_{\rho\sigma}(x')}\right)_{\overline{g}_{\mu\nu},\overline{\phi}}
\,,\qquad
\label{counterterm self-energy}
\end{equation}
and the precise form of the contributions from various terms in $S_{\rm ct}$ can be found in 
Appendices~A and~B. Note that the value of the counterterms that renormalize the energy-momentum tensor
in Eq.~(\ref{SC Einstein equation}) must be identical to those that renormalize the self-energy in 
Eq.~(\ref{total self-energy}), 
which follows from the considerations in the following section. Finally, for completeness we note that 
the contribution of the counterterm action to the self-energy in Case~B can be obtained as, 
\begin{equation}
 -i\big[{}_{\mu\nu}\overline{\Sigma}^{\mbox{\tiny ct}}_{\rho\sigma}\big]^B(x;x')
  = \frac{1}{\sqrt{-\overline{g}}\sqrt{-\overline{g}'}}
  \left(\frac{\delta^2S_{\rm ct}[g_{\alpha\beta},\phi]}
    {\delta g^{\mu\nu}(x)\delta{g}^{\rho\sigma}(x')}\right)_{\overline{g}_{\mu\nu},\overline{\phi}}
\,.\qquad
\label{counterterm self-energy: case B}
\end{equation}
%


\section{Noether-Ward identities for gravitational perturbations}
\label{Noether-Ward identities for gravitational perturbations}

We are interested in how the effective action~(\ref{1PI one-loop effective action: expansion A})
 changes under linear coordinate transformations,
\begin{equation}
x^\mu \rightarrow \widetilde{x}^\mu = x^\mu+\xi^\mu
\,.
\label{coordinate transformations}
\end{equation}
The metric and the field transform as,
\begin{eqnarray}
 g_{\mu\nu}(x) &\!\!\rightarrow\!\!& \widetilde g_{\mu\nu}(x)= g_{\mu\nu}(x) - \mathcal{L}_\xi g_{\mu\nu}
   +\mathcal{O}(\xi^2)
    = g_{\mu\nu}(x) - 2\nabla_{(\mu} \xi_{\nu)} + \mathcal{O}(\xi^2)
\,,\qquad \nonumber\\
 \phi(x) &\!\!\rightarrow\!\!& \widetilde\phi(x) = \phi(x) - \mathcal{L}_\xi \phi(x)
   +\mathcal{O}(\xi^2)
    = \phi(x) - \xi\!\cdot\!\nabla\phi(x)  + \mathcal{O}(\xi^2)
\,,
\label{expansion of the metric and field: gauge}
\end{eqnarray}
where $\xi\!\cdot\!\nabla = \xi^\mu \partial_\mu$, $\xi_\nu = g_{\nu\mu}\xi^\mu$ and 
$\nabla_{\mu}$ is the full covariant derivative for a general metric tensor. The inverse metric
tensor transforms as, 
\begin{eqnarray}
 g^{\mu\nu}(x) &\!\!\rightarrow\!\!& \widetilde g^{\mu\nu}(x)= g^{\mu\nu}(x) + \mathcal{L}_\xi g^{\mu\nu}
   +\mathcal{O}(\xi^2)
\nonumber\\    
&&\hskip -0.5cm
   =\, g^{\mu\nu}(x) + 2\nabla^{(\mu} \xi^{\nu)}  + \mathcal{O}(\xi^2)
\,,\qquad 
\label{linear transformation: Lie derivative inverse metric}
\end{eqnarray}
such that the unity tensor $\delta^\mu_{\;\nu} = g^{\mu\rho}g_{\rho\nu}$ remains invariant.
The effective action in Eqs.~(\ref{1PI one-loop effective action: expansion A})
and~(\ref{1PI one-loop effective action: expansion B})
is invariant under these transformations,
\begin{equation}
\Gamma[g_{\mu\nu},\phi] 
  = \Gamma\big[g_{\mu\nu}-\mathcal{L}_\xi g_{\mu\nu},\phi-\xi\!\cdot\!\nabla\phi(x)\big]
\,.\quad
\label{invariance of effective action}
\end{equation}
Let us first consider what Eq.~(\ref{invariance of effective action}) implies for the case of a general metric and field.  
Inserting the one-loop effective action in Eq.~(\ref{1PI one-loop effective action})
into Eq.~(\ref{invariance of effective action}) gives, 
\begin{eqnarray}
\int {\rm d}^Dx \left\{ \frac{\delta \Gamma[g_{\alpha\beta},\phi]}{\delta g_{\mu\nu}(x)}
                     2 \nabla_{\mu}\xi_{\nu}(x) 
                 + \frac{\delta \Gamma[g_{\alpha\beta},\phi]}{\delta \phi(x)} 
                   \xi_{\mu} g^{\mu\nu} \nabla_{\nu}\phi(x)
    \right\}  + \mathcal{O}\big(\xi_\alpha^2\big)   = 0
\,,\quad
\label{1PI one-loop effective action: towards NW}
\end{eqnarray}
from which one can derive the fundamental Noether-Ward identity by integrating by parts
the derivative in $\nabla_{\mu}\xi_{\nu}$.  But before we do that, it is instructive to evaluate the functional
derivatives in Eq.~(\ref{1PI one-loop effective action: towards NW}). Varying 
Eq.~(\ref{invariance of effective action}) one obtains,
\begin{eqnarray}
\frac{\delta \Gamma[g_{\alpha\beta},\phi]}{\delta g_{\mu\nu}(x)}
  &\!\!=\!\!& \int {\rm d}^Dy \sqrt{-g(y)}\bigg\{\! -\frac{1}{\kappa^2}
                \Big[G^{\mu\nu}(y) - \nabla^\mu_y\nabla^\nu_y + g^{\mu\nu}(y)\dAlembert_y
              \Big]\delta^D(y\!-\!x)
\nonumber\\
&&\hskip -2.1cm
        -\, \frac{D\!-\!2}{2}\frac{\Lambda}{\kappa^2}g^{\mu\nu}\delta^D(y\!-\!x)
              \!+\!\frac12\Big[T^{\mu\nu}(x)\!+\!\big\langle\Delta \hat T^{\mu\nu}(x)\big\rangle
                 \!+\!T_{\rm ct}^{\mu\nu}(x)\Big]
                             \delta^D(y\!-\!x)\bigg\}
  \,,\quad\;
\label{first variation of effective action 1}\\
\frac{\delta \Gamma[g_{\alpha\beta},\phi]}{\delta\phi(x)}
  &\!\!=\!\!& \int {\rm d}^Dy \sqrt{-g(y)}\Big\{\! 
  -g^{\alpha\beta}(y)\big(\nabla^y_\alpha\phi(y)\big)\nabla^y_\beta
                  - \big(m^2+\xi R(y)\big)\phi(y)\Big\}
\nonumber\\
&&\hskip 3.0cm                  
                  \times\,\delta^D(y\!-\!x)
  \,,\qquad
\label{first variation of effective action 2}
\end{eqnarray}
where the derivative $\nabla^y_\beta$ can be partially integrated,
to give a d'Alembertian acting on $\phi(y)$, and 
the classical and one-loop energy-momentum tensor in Eq.~(\ref{first variation of effective action 1})
are given by,
\begin{eqnarray}
T^{\mu\nu}(x) &\!\!=\!\!& 
\frac{2}{\sqrt{-g}}\frac{\delta S_{\rm m}[g_{\alpha\beta},\phi]}{\delta g_{\mu\nu}(x)}
    = \nabla^\mu\phi\nabla^\nu\phi+g^{\mu\nu}
       \bigg[\!-\frac12 g^{\alpha\beta}\nabla_\alpha\phi\nabla_\beta\phi-\frac12m^2\phi^2\bigg]
\nonumber\\
&&\hskip 3.0cm  
     +\,\xi\big[G^{\mu\nu}-\nabla^\mu\nabla^\nu+g^{\mu\nu}\dAlembert\big]\phi^2(x)
\,,\quad
\label{energy momentum tensor: classical}
\end{eqnarray}
\begin{eqnarray}
\big\langle\Delta \hat T^{\mu\nu}(x)\big\rangle
   &\!\!=\!\!&  \frac{2}{\sqrt{-g}}\frac{\delta \Gamma^{(1)}[g_{\alpha\beta},\phi]}{\delta g_{\mu\nu}(x)}
\nonumber\\
&&\hskip .0cm      
    =\, \bigg\{\bigg[\nabla^\mu{\nabla'}^\nu    +g^{\mu\nu}
       \bigg(\!-\frac12 g^{\alpha\beta}\nabla_\alpha\nabla'_\beta-\frac12m^2\bigg)\bigg]i\Delta_\phi(x;x')
\nonumber\\
&&\hskip 1.0cm  
     +\,\xi\big[G^{\mu\nu}-\nabla^\mu\nabla^\nu+g^{\mu\nu}\dAlembert\big]i\Delta_\phi(x;x)
      \bigg\}_{x'\rightarrow x}
\,,\qquad
\label{energy momentum tensor: quantum}
\end{eqnarray}
and $T_{\rm ct}^{\mu\nu}(x)$ is the energy-momentum tensor associated with the counterterm 
action~(\ref{total counterterm action}),
needed to renormalize $\big\langle\Delta \hat T^{\mu\nu}(x)\big\rangle$.
Here we give a couple of steps showing how to obtain the result in 
Eq.~(\ref{energy momentum tensor: quantum}). Varying the one-loop part of 
the effective action in Eq.~(\ref{1PI one-loop effective action}) gives,
\begin{eqnarray}
\big\langle\Delta \hat T^{\mu\nu}(x)\big\rangle
   &\!\!\!=\!\!\!&  \frac{i\hbar}{\sqrt{-g}}{\rm Tr}_y\Bigg\{
      \int {\rm d}^Dy' 
 \nonumber\\
&&\hskip -1.45cm    
    \times\frac{\delta}{\delta g_{\mu\nu}(x)}
          \Big[\sqrt{-g}\Big(\!-g^{\alpha\beta}\nabla_\alpha\nabla_\beta'-m^2-\xi R\Big)
          \delta^D(y\!-\!y')\Big]
         \frac{\Delta_\phi(y',y)}{\hbar}
\Bigg\}
\nonumber\\
&&\hskip -2.3cm      
   =   \frac{1}{\sqrt{-g}}{\rm Tr}_y\sqrt{-g}\Bigg\{
      \int {\rm d}^Dy' 
    \delta^D(x\!-\!y)
          \Bigg[\bigg[\nabla^{(\mu}{\nabla'}^{\nu)}
  \!+\!g^{\mu\nu}\bigg(
              \!\!-\!\frac{1}{2}g^{\alpha\beta}\nabla_\alpha\nabla_\beta'  \!-\!\frac12m^2 \bigg)
 \nonumber\\
&&\hskip 0.45cm             
           +\,\xi\Big(G^{\mu\nu}-\nabla^\mu\nabla^\nu +g^{\mu\nu}\dAlembert\Big)       
                  \bigg]
          \delta^D(y\!-\!y')\Bigg]
         i\Delta_\phi(y',y)
\Bigg\}
  \,.\qquad
\label{energy momentum tensor: quantum: step 1}
\end{eqnarray}
Integrating this over $y'$ and taking a trace, ${\rm Tr}_y\rightarrow \int\! {\rm d}^Dy$,
yields Eq.~(\ref{energy momentum tensor: quantum}). 
Next, inserting Eqs.~(\ref{energy momentum tensor: classical})--(\ref{energy momentum tensor: quantum})
into Eq.~(\ref{1PI one-loop effective action: towards NW}),
integrating by parts the derivative in $ \nabla_{\mu}\xi_{\nu}$
and dropping the boundary term~\footnote{That boundary term can be dropped when 
one integrates over a volume $V$ at whose boundary $\partial V$ the coordinate shift
 $\xi_\alpha(x)$ vanishes, which is the case with {\it proper gauge transformations}. The class of transformations
 for which  $\xi(x)$ does not vanish at the boundary $\partial V$ are known as {\it large gauge transformations}.
 Such transformations can induce physical changes in theories of topological nature and are therefore 
interesting by themselves. In this work we focus on 
proper gauge transformations, and thus do not consider the effects induced by large gauge transformations. We refer the reader to the references mentioned in 
section~\ref{Introduction}, many of which exploited large gauge transformations.
\label{footnote: boundary terms}
}
one obtains, 
\begin{eqnarray}
&&\hskip -0.5cm
\int {\rm d}^Dx\sqrt{-g}\, \xi_\nu(x) \Bigg\{\frac{2}{\kappa^2}\nabla_{\mu}G^{\mu\nu}
  \!+\! (D\!-\!2)\frac{\Lambda}{\kappa^2}\nabla_\mu g^{\mu\nu}
                        - \nabla_\mu\left(T^{\mu\nu}(x)+\big\langle\Delta \hat T^{\mu\nu}(x)\big\rangle\right)
\nonumber\\
 &&\hskip 0.5cm               
 +\,\big(\nabla^\nu\phi\big)\Big[\dAlembert
                  - \big(m^2+\xi R(x)\big)\Big]\phi(x)
    \Bigg\} 
 \nonumber\\
 &&\hskip 0.5cm   
     +\,\int {\rm d}^Dx\sqrt{-g}\frac{2}{\kappa^2}\big[\nabla^\mu\nabla^\nu-g^{\mu\nu}\dAlembert\big]
            \nabla_{(\mu}\xi_{\nu)}(x)
     + \mathcal{O}\big(\xi_\alpha^2\big)   = 0
\,.\qquad
\label{1PI one-loop effective action: towards NW 2}
\end{eqnarray}
Consider first the terms in the last line, which can be rewritten as, 
\begin{eqnarray}  
    \frac{2}{\kappa^2}\int {\rm d}^Dx\sqrt{-g}\big[\nabla^\mu\nabla^\nu-g^{\mu\nu}\dAlembert\big]
            \nabla_{(\mu}\xi_{\nu)}
            = \frac{2}{\kappa^2}\int {\rm d}^Dx\sqrt{-g}\,\nabla_\mu\big[R^{\mu\nu}\xi_\nu(x)\big]
\,.\;
\label{1PI one-loop effective action: towards NW 3}
\end{eqnarray}
This is a boundary term which can be dropped for proper gauge transformations $\xi_\alpha$ 
which vanish at the boundary $\partial V$ 
of the volume integral $V$, see footnote~\ref{footnote: boundary terms} for the justification.
The term in the second line of Eq.~(\ref{1PI one-loop effective action: towards NW 2}) 
vanishes on the account of the equation of motion~(\ref{classical equation of motion}) for $\phi$. 
And finally, the first line in Eq.~(\ref{1PI one-loop effective action: towards NW 2}) 
contains three terms. The first term vanishes by the contracted Bianchi identity, 
$\nabla_\mu G^{\mu\nu}=0$, and the second ($\nabla_\mu g^{\mu\nu}=0$) is 
the metric compatibility condition,
which are both geometric identities that hold true for an arbitrary metric tensor.
Finally, the third term in the first line is the covariant energy-momentum conservation,
\begin{equation}
\nabla_\mu\left(T^{\mu\nu}(x)+\big\langle\Delta \hat T^{\mu\nu}(x)\big\rangle
+T^{\mu\nu}_{\rm ct}(x)\right) = 0
\,.\qquad
\label{covariant energy-momentum conservation}
\end{equation}
While it is quite easy to prove that the classical energy-momentum 
tensor~(\ref{energy momentum tensor: classical}) is covariantly conserved, 
it is instructive to give a couple of 
steps towards proving covariant conservation of the quantum contribution 
in~(\ref{energy momentum tensor: quantum}),
\begin{eqnarray}
\nabla_\mu\big\langle\Delta \hat T^{\mu\nu}(x)\big\rangle    
   &\!\!\!=\!\!\!&  \bigg\{\bigg[\dAlembert{\nabla'}^\nu \!+\! \nabla^\mu\nabla_\mu'{\nabla'}^\nu    
   \!+\!\nabla^\nu
       \bigg(\!\!-\!\frac12 g^{\alpha\beta}\nabla_\alpha\nabla'_\beta
           \!-\!\frac{m^2}{2}\bigg)\bigg]i\Delta_\phi(x;x')
\nonumber\\
&&\hskip 1.0cm  
     +\,\xi\big[G^{\mu\nu}\nabla_\mu-\dAlembert\nabla^\nu+\nabla^\nu\dAlembert\big]i\Delta_\phi(x;x)
      \bigg\}_{x'\rightarrow x}
  \,,\qquad
\label{energy momentum tensor: quantum: conservation}
\end{eqnarray}
where we made use of, $\nabla_\mu G^{\mu\nu}=0$ and 
 $\big[\nabla_\mu i\Delta(x;x')\big]_{x'\rightarrow x} 
\!=\big[\nabla_\mu' i\Delta(x;x')\big]_{x'\rightarrow x} $
 $\!= \frac12\nabla_\mu \big[ i\Delta(x;x)\big]$.
 One can use the equation of motion~(\ref{eom for bar iDelta(x;x')}) in the first term 
 in Eq.~(\ref{energy momentum tensor: quantum: conservation}), extract ${\nabla'}^\nu$
 from the second term (picking up a factor $1/2$) can commute the derivative $\nabla^\nu$ 
 in the penultimate term ($\dAlembert\nabla^\nu = \nabla^\nu \dAlembert +R^{\mu\nu}\nabla_\mu$)
to obtain, 
\begin{eqnarray}
\nabla_\mu\big\langle\Delta \hat T^{\mu\nu}(x)\big\rangle    
   &\!\!\!=\!\!\!&\bigg[\frac12(m^2+\xi R){\nabla}^\nu i\Delta(x;x)
 \!+\!\frac12 {\nabla}^\nu \Big[ \big(\nabla^\mu\nabla_\mu' i\Delta_\phi(x;x') \big)_{x'\rightarrow x} \Big]
\nonumber\\
&&\hskip -0.4cm  
   +\,\nabla^\nu
  \bigg(\!-\!\frac12 g^{\alpha\beta}\big(\nabla_\alpha\nabla'_\beta i\Delta_\phi(x;x')\big)_{x'\rightarrow x}
           \!-\!\frac{m^2}{2} i\Delta_\phi(x;x)\bigg)\bigg]
\nonumber\\
&&\hskip -0.4cm  
     +\,\xi\big[G^{\mu\nu}\nabla_\mu-R^{\mu\nu}\nabla_\mu\big]i\Delta_\phi(x;x)
     +\left({\nabla'}^\nu \frac{i\hbar \delta^D(x\!-\!x')}{\sqrt{-g}}\right)_{\!x'\rightarrow x}
\nonumber\\
&&\hskip -0.4cm
    =\,\, 0
  \,,\qquad
\label{energy momentum tensor: quantum: conservation B}
\end{eqnarray}
where the local term in the penultimate line vanishes in dimensional regularization.~\footnote{It is well known that all power-law divergences vanish automatically (by analytic extension) in dimensional regularization, 
and thus the coincident delta function, $\delta^D(x\!-\!x)\rightarrow 0$, which scales as the volume of spacetime,
as well as all of its derivatives --
$\bigl\{\nabla_\mu\bigl[\delta^D(x\!-\!x')/\sqrt{-g}\bigr]\bigr\}_{x'\rightarrow x}\rightarrow 0$, {\it etc.} 
--  vanish in dimensional regularization.}
This also means that in semi-classical gravity, the classical and quantum energy-momentum
tensor are {\it separately} conserved, the fact that is usually not appreciated.~\footnote{This holds when
covariant derivatives are taken with respect to the metric $g_{\mu\nu}$, which 
solves the full equation of motion of semiclassical gravity,
\begin{equation}
G^{\mu\nu}(g_{\alpha\beta}) + \frac{D\!-\!2}{2}\Lambda g^{\mu\nu} = 8\pi G\left(T^{\mu\nu}(x) 
         + \left\langle\Delta \hat{T}^{\mu\nu}(x)\right\rangle
         + T_{\rm ct}^{\mu\nu}(x)\right)
\,.
\label{eom SCG}
\end{equation}
}
Furthermore, this also implies that the counterterm energy-momentum tensor in 
Eq.~(\ref{covariant energy-momentum conservation}) must be by itself conserved. 
In fact, as it was shown in Appendices~A and~B, this is true
independently on the choice of the counterterm coefficients, as the energy-momentum 
generated by any term in the counterterm action is covariantly conserved, 
see Appendices~A and~B. For example, the energy-momentum tensor associated with the 
Ricci scalar squared counterterm is (see Eq.~(\ref{counterterm action R2 c1})), 
\begin{eqnarray}
T_{R^2}^{\mu\nu}
 &\!\!=\!\!&  \alpha_{R^2}
    \bigg\{\!\!-\!g^{\mu\nu}\,\overline{R}^2
              \!-\! 4
               \big(G^{\mu\nu}\!\!-\!\nabla^{\mu}\nabla^{\nu}
            \!\!+\! g^{\mu\nu}\dAlembert
                 \big)R
    \bigg\}
\,,\quad
\label{Tmn: R 2}
\end{eqnarray}
which can be easily shown to be covariantly conserved ({\it cf.} Eq.~(\ref{counterterm action R2 d1})), 
\begin{eqnarray}
\nabla_\mu T_{R^2}^{\mu\nu}  = 0
\,.
\quad
\label{Tmn: R 2 b}
\end{eqnarray}
To summarize, we have shown that not only the total energy momentum tensor is 
conserved as indicated in Eq.~(\ref{covariant energy-momentum conservation}), but also each of the individual parts is separately conserved,
\begin{equation}
\nabla_\mu T^{\mu\nu}(x) = 0
\,,\qquad 
\nabla_\mu \big\langle\Delta \hat T^{\mu\nu}(x)\big\rangle = 0
\,,\qquad 
\nabla_\mu T^{\mu\nu}_{\rm ct,i}(x) = 0\quad (\forall i)
\,,\quad
\label{covariant energy-momentum conservation: parts}
\end{equation}
where $T^{\mu\nu}_{\rm ct,i}(x)$ is the energy-momentum tensor associated to an individual counterterm
action, 
discussed in detail in Appendices~A and~B, and $T^{\mu\nu}_{\rm ct}(x)=\sum_i T^{\mu\nu}_{\rm ct,i}(x)$.

\medskip

The above considerations -- done for the metric tensor $g_{\mu\nu}$ -- can be 
easily repeated for the inverse metric $g^{\mu\nu}$, resulting 
in the energy-momentum conservation,
\begin{equation}
\nabla^\mu\left(T_{\mu\nu}(x)+\big\langle\Delta \hat T_{\mu\nu}(x)\big\rangle
+ T^{\rm ct}_{\mu\nu}(x)\right) = 0
\,,\qquad
\label{covariant energy-momentum conservation 2}
\end{equation}
with suitably adapted definitions of the classical and  quantum energy-momen\-tum tensor
({\it cf.} Eqs.~(\ref{energy momentum tensor classical A})--(\ref{energy momentum tensor classical B})  
and~(\ref{energy momentum tensor: 1 loop A})--(\ref{energy momentum tensor: 1 loop B})).
The conservation law in Eq.~(\ref{covariant energy-momentum conservation 2}) 
is to be compared with that in Eq.~(\ref{covariant energy-momentum conservation}).
Just as above, each term in Eq.~(\ref{covariant energy-momentum conservation 2}) is individually conserved,
\begin{equation}
\nabla^\mu T_{\mu\nu}(x) = 0
\,,\qquad
\nabla^\mu\big\langle\Delta \hat T_{\mu\nu}(x)\big\rangle= 0
\,,\qquad
\nabla^\mu T^{\rm ct,i}_{\mu\nu}(x)= 0
\,,\qquad
\label{covariant energy-momentum conservation 2b}
\end{equation}
where $T^{\rm ct,i}_{\mu\nu}(x)$ denote contributions from individual counterterms, and \\
 $\sum_i T_{\mu\nu}^{\rm ct,i}(x)=T_{\mu\nu}^{\rm ct}(x)$.

Even though our derivation was done from an unusual angle, 
most of the material presented thus far in this section is known~\cite{Birrell:1982ix}, and therefore expected.
This analysis can be used as a sanity check. In what follows, we perform 
an equivalent derivation, but now for the case we are really interested in in this paper, namely 
for the case when the metric is broken into a background metric, and a small, but arbitrary, 
perturbation as in Eqs.~(\ref{metric perturbations: case A})--(\ref{metric perturbations: case B}), 
such that we know the solutions of the equations of motion~(\ref{eom for bar phi}),
(\ref{eom for bar iDelta(x;x')}) and~(\ref{SC Einstein equation}) 
for the background quantities $\overline\phi$, 
$i\Delta_{\bar\phi}(x;x')$ and $\overline{g}_{\mu\nu}$.

\bigskip
\noindent
{\bf Noether-Ward identity for the background quantities.}
Eq.~(\ref{invariance of effective action}) still holds, so we can take 
Eq.~(\ref{1PI one-loop effective action: towards NW}) as a starting point for our derivation.
Expanding Eq.~(\ref{1PI one-loop effective action: towards NW}) around $\overline{g}_{\mu\nu}$
and $\overline{\phi}$ gives, 
\begin{eqnarray}
&&\hskip -0.5cm
\int\! {\rm d}^Dx \Bigg\{ \Bigg[\!\left(\frac{\delta \Gamma[g_{\alpha\beta},\phi]}{\delta g_{\mu\nu}(x)}
           \right)_{\overline{g}_{\mu\nu},\overline{\phi}}
\nonumber\\
&&\hskip 0cm  
             \!\! +\!\int\! {\rm d}^Dx'
             \Bigg(\!\left(\frac{\delta^2 \Gamma[g_{\alpha\beta},\phi]}{\delta g_{\mu\nu}(x)
                 \delta g_{\rho\sigma}(x')}
           \right)_{\overline{g}_{\mu\nu},\overline{\phi}}\!\!\delta g_{\rho\sigma}(x')
           \!+\!\left(\frac{\delta^2 \Gamma[g_{\alpha\beta},\phi]}{\delta g_{\mu\nu}(x)\delta \phi(x')}
                \right)_{\overline{g}_{\mu\nu},\overline{\phi}}\!\delta\phi(x')
           \!\Bigg)
           \Bigg]
\nonumber\\
&&\hskip 1.5cm            
                    \times\, 2\Big( \overline{\nabla}_{(\mu}\xi_{\nu)}(x)
                       -\Big(\overline{\nabla}_{(\mu}\delta g_{\nu)}^{\;\alpha}
                       -\frac12\overline{\nabla}^{\;\alpha}\delta g_{\mu\nu}\Big) \xi_{\alpha}(x)\Big)
\nonumber\\
&&\hskip 0cm  
               +\,\Bigg[ \left(\frac{\delta \Gamma[g_{\alpha\beta},\phi]}{\delta \phi(x)}
                     \right) _{\overline{g}_{\mu\nu},\overline{\phi}}
               +\!\int\! {\rm d}^Dx'
             \Bigg(\!\left(\frac{\delta^2 \Gamma[g_{\alpha\beta},\phi]}{\delta \phi(x)
                 \delta g_{\rho\sigma}(x')}
           \right)_{\overline{g}_{\mu\nu},\overline{\phi}}\!\!\delta g_{\rho\sigma}(x')
 \nonumber\\
&&\hskip 0.cm 
          \!+\,\left(\frac{\delta^2 \Gamma[g_{\alpha\beta},\phi]}{\delta\phi(x)\delta \phi(x')}
                \right)_{\overline{g}_{\mu\nu},\overline{\phi}}\!\delta\phi(x')
           \!\Bigg)
           \Bigg]      
 \xi_{\mu} \big(\overline{g}^{\mu\nu}\!-\!\delta g^{\mu\nu}\big)\overline{\nabla}_{\nu}
                          \left(\overline{\phi}+\delta\phi\right)
    \Bigg\}
\nonumber\\
&&\hskip 1.5cm     
  +\, \mathcal{O}\big(\delta g_{\mu\nu}^2,\delta \phi^2,\xi_\alpha^2\big)   = 0
\,.\qquad
\label{1PI one-loop effective action: towards NW for bg}
\end{eqnarray}
This is the general identity accurate  to linear order in perturbations. Since we are primarily interested 
in gravitational perturbations, from now on we shall assume that $\delta \phi=0$ and 
drop the scalar perturbations~\footnote{The effects of quantum scalar perturbations can be 
self-consistently described in the framework of the 2PI {\it in-in} effective action for scalar perturbations,
whereby scalar perturbations are described by the scalar propagator and gravitational perturbations 
can be treated either as classical or as quantum~\cite{ProkopecVecchioni:2026}.
Note also that the analysis around Eq.~(\ref{1PI one-loop effective action: towards NW 2}), 
which includes the gravitational gauge transformation of the scalar field, 
did not lead to any interesting identity for the scalar field.}.
With this simplification Eq.~(\ref{1PI one-loop effective action: towards NW for bg}) becomes,
\begin{eqnarray}
&&\hskip -0.5cm
\int\! {\rm d}^Dx \Bigg\{ \Bigg[\!\left(\frac{\delta \Gamma[g_{\alpha\beta},\phi]}{\delta g_{\mu\nu}(x)}
           \right)_{\overline{g}_{\mu\nu},\overline{\phi}}
            +\!\int\! {\rm d}^Dx'
             \left(\frac{\delta^2 \Gamma[g_{\alpha\beta},\phi]}{\delta g_{\mu\nu}(x)
                 \delta g_{\rho\sigma}(x')}
           \right)_{\overline{g}_{\mu\nu},\overline{\phi}}\!\!\delta g_{\rho\sigma}(x')
           \Bigg]
\nonumber\\
&&\hskip 0.cm            
                    \times\, 2\bigg[ \overline{\nabla}_{(\mu}\xi_{\nu)}(x)
                       \!-\!\Big(\overline{\nabla}_{(\mu}\delta g_{\nu)}^{\;\alpha}
                       \!-\!\frac12\overline{\nabla}^{\;\alpha}\delta g_{\mu\nu}\Big) \xi_{\alpha}(x)\bigg]
    \Bigg\}  + \mathcal{O}\big(\delta g_{\mu\nu}^2,\xi_\alpha^2\big)   = 0
\,.\qquad
\label{1PI one-loop effective action: towards NW for bg 2}
\end{eqnarray}
The first term in Eq.~(\ref{1PI one-loop effective action: towards NW for bg 2})
can be analysed in the same way as it was done in 
Eq.~(\ref{1PI one-loop effective action: towards NW 2}) above, 
but with unbarred quantities replaced by barred quantities,
\begin{eqnarray}
&&\hskip -0.8cm
\int\! {\rm d}^Dx \!\left(\frac{\delta \Gamma[g_{\alpha\beta},\phi]}{\delta g_{\mu\nu}(x)}
           \right)_{\overline{g}_{\mu\nu},\overline{\phi}}2\nabla_{(\mu}\xi_{\nu)}
\nonumber\\
 &&\hskip -0.cm      
        =\,\int {\rm d}^Dx\sqrt{-\overline{g}}\, \xi_\alpha(x) \Bigg\{
   \bigg[\delta^\alpha_\nu\overline{\nabla}_\mu+\Big(\overline{\nabla}_{(\mu}\delta g_{\nu)}^{\;\alpha}
                       \!-\!\frac12\overline{\nabla}^{\;\alpha}\delta g_{\mu\nu}\Big) \bigg]
\nonumber\\
 &&\hskip -0.cm  
 \times\,\bigg[\frac{2}{\kappa^2}\overline{G}^{\mu\nu}
                 \!+\! (D\!-\!2)\frac{\Lambda}{\kappa^2}\overline{g}^{\mu\nu} \!-\! \left(\overline{T}^{\mu\nu}(x)
   \!+\!\Big[\big\langle\Delta\hat{\overline{T}}^{\mu\nu}(x)\big\rangle\Big]_{\overline{g}_{\mu\nu},\overline{\phi}}
   \!+\!\overline{T}_{\rm ct}^{\mu\nu}(x)
   \right)
          \bigg] 
    \Bigg\} 
 \nonumber\\
 &&\hskip 1.2cm   
     +\,\frac{2}{\kappa^2}\int {\rm d}^Dx\sqrt{-\overline{g}}
            \Big[\overline{\nabla}^\mu\overline{\nabla}^\nu-\overline{g}^{\mu\nu}\overline{\dAlembert}\Big]
  \nonumber\\
 &&\hskip 4cm   
        \times\,\Big[ \overline{\nabla}_{(\mu}\xi_{\nu)}(x)
                       \!-\!\Big(\overline{\nabla}_{(\mu}\delta g_{\nu)}^{\;\alpha}
                       \!-\!\frac12\overline{\nabla}^{\;\alpha}\delta g_{\mu\nu}\Big) \xi_{\alpha}(x)\Big]
\,.\qquad
\label{1PI one-loop effective action:  towards NW for bg 3}
\end{eqnarray}
The terms in the last two lines are boundary terms, which vanish for proper gauge transformations
$\xi_\alpha$, which are those that vanish at the boundary $\partial V$.
The action of the derivative $\overline{\nabla}_\mu$ in the second line gives 
the background conservation laws ({\it cf.} Eq.~(\ref{covariant energy-momentum conservation})) 
which vanish in semi-classical gravity, leaving us with,
\begin{eqnarray}
&&\hskip -0.7cm
\int\! {\rm d}^Dx \!\left(\frac{\delta \Gamma[g_{\alpha\beta},\phi]}{\delta g_{\mu\nu}(x)}
           \right)_{\overline{g}_{\mu\nu},\overline{\phi}}2\nabla_{(\mu}\xi_{\nu)}
\nonumber\\
 &&\hskip -0.3cm      
        =\,\int {\rm d}^Dx\sqrt{-\overline{g}}\, \xi_\alpha(x) \Bigg\{
     \Big(\overline{\nabla}_{(\mu}\delta g_{\nu)}^{\;\alpha}
                       \!-\!\frac12\overline{\nabla}^{\;\alpha}\delta g_{\mu\nu}\Big)
\nonumber\\
 &&\hskip -0.3cm  
 \times\,\bigg[\frac{2}{\kappa^2}\overline{G}^{\mu\nu}
                         \!+\! (D\!-\!2)\frac{\Lambda}{\kappa^2}\overline{g}^{\mu\nu}
                        \!-\! \left(\overline{T}^{\mu\nu}(x)
   \!+\!\Big[\big\langle\Delta \hat{\overline{T}}^{\mu\nu}(x)\big\rangle\Big]_{\overline{g}_{\mu\nu},\overline{\phi}}
     \!\!+\!\overline{T}_{\rm ct}^{\mu\nu}(x)\right)
          \bigg] 
    \Bigg\} 
.\qquad\!
\label{1PI one-loop effective action:  towards NW for bg 4}
\end{eqnarray}
The second order variation in Eq.~(\ref{1PI one-loop effective action: towards NW for bg 2})
contributes as, 
\begin{eqnarray}
&&\hskip -0.78cm
\int\! {\rm d}^Dx\,{\rm d}^Dx'
             \left(\frac{\delta^2 \Gamma[g_{\alpha\beta},\phi]}{\delta g_{\mu\nu}(x)
                 \delta g_{\rho\sigma}(x')}
           \right)_{\overline{g}_{\mu\nu},\overline{\phi}}\!\!\delta g_{\rho\sigma}(x')
          \times 2\overline{\nabla}_{(\mu}\xi_{\nu)}
\nonumber\\
&&\hskip -0.2cm            
       =    \int\!\! {\rm d}^Dx\,{\rm d}^Dx'\sqrt{-\overline{g}}\sqrt{-\overline{g}'}
            \bigg\{\bigg[
        \frac{1}{\kappa^2}\big[{}^{\mu\nu}\mathcal{L}^{\rho\sigma}\big]_{\overline{g}_{\mu\nu}}\!(x;x')
\nonumber\\
&&\hskip -0.2cm 
         -\,\frac{D\!-\!2}{4}\frac{\Lambda}{\kappa^2}\Big(\overline{g}^{\mu\nu}\overline{g}^{\rho\sigma}
           \!\!-\!2\overline{g}^{\mu(\rho}\overline{g}^{\sigma)\nu}\Big)
            \frac{\delta^D(x\!-\!x')}{\sqrt{-\overline{g}}}
        -\, \frac{1}{\kappa^2}\big[{}^{\mu\nu}\Sigma^{\rho\sigma}\big]_{\overline{g}_{\mu\nu}}\!(x;x')  
\nonumber\\
&&\hskip -0.2cm 
 -\,  \frac{1}{\kappa^2}\big[{}^{\mu\nu}\Sigma_{\rm ct}^{\rho\sigma}\big]_{\overline{g}_{\mu\nu}}\!(x;x')  
       \!+\! \big[{}^{\mu\nu}\mathcal{T}^{\rho\sigma}\big]_{\overline{g}_{\mu\nu},\overline{\phi}}(x;x') 
\bigg]\delta g_{\rho\sigma}(x') \bigg\}
    \!\times\! 2\overline{\nabla}_{(\mu}\xi_{\nu)}(x)
\,,\qquad
\label{1PI one-loop effective action: towards NW for bg 5}
\end{eqnarray}
where $\big[{}^{\mu\nu}\Sigma^{\rho\sigma}\big]_{\overline{g}_{\mu\nu}}(x;x')$
is the graviton self-energy, whose one-loop contribution is discussed in section~\ref{Effective action},
 $\big[{}^{\mu\nu}\Sigma_{\rm ct}^{\rho\sigma}\big]_{\overline{g}_{\mu\nu}}(x;x')$
 is the corresponding counterterm contribution needed to renormalize
  $\big[{}^{\mu\nu}\Sigma^{\rho\sigma}\big]_{\overline{g}_{\mu\nu}}(x;x')$,
$\big[{}^{\mu\nu}\mathcal{L}^{\rho\sigma}\big]_{\overline{g}_{\mu\nu}}(x;x')$
is the Lichnerowicz operator
(whose precise form is given in Eq.~(\ref{second variation of the Hilbert-Einstein action: Lichnerowicz})),
\begin{equation}
\big[{}^{\mu\nu}\mathcal{L}^{\rho\sigma}\big]_{\overline{g}_{\mu\nu}}(x;x')
    = \frac{\kappa^2}{\sqrt{-\overline{g}}\sqrt{-\overline{g}'}}
    \left(\frac{\delta^2S_{\rm g}\big[\overline{g}_{\alpha\beta}\big]}
              {\delta g_{\mu\nu}(x)\delta g_{\rho\sigma}(x')}\right)_{\overline{g}_{\mu\nu}}
\,,
\label{Lichnerowicz operator}
\end{equation}
and 
$\big[{}^{\mu\nu}\mathcal{T}^{\rho\sigma}\big]_{\overline{g}_{\mu\nu}}(x;x')$
is the corresponding contribution from the classical matter action,
\begin{eqnarray}
\big[{}^{\mu\nu}\mathcal{T}^{\rho\sigma}\big]_{\overline{g}_{\mu\nu},\overline{\phi}}(x;x')
    &\!\!=\!\!& \frac{1}{\sqrt{-\overline{g}}\sqrt{-\overline{g}'}}
    \left(\frac{\delta^2S_{\rm m}\big[\overline{g}_{\alpha\beta}\big]}
              {\delta g_{\mu\nu}(x)\delta g_{\rho\sigma}(x')}\right)_{\overline{g}_{\mu\nu},\overline{\phi}}
\,\qquad
\label{second variation of matter action: def}\\
 &\!\!=\!\!& \big[{}^{\mu\nu}\mathcal{T}^{\rho\sigma}\big]^{\xi=0}_{\overline{g}_{\mu\nu},\overline{\phi}}(x;x')
 + \big[{}^{\mu\nu}\mathcal{T}^{\rho\sigma}\big]^{\xi\neq 0}_{\overline{g}_{\mu\nu},\overline{\phi}}(x;x')
 \nonumber
\,,\qquad
\end{eqnarray}
where 
\begin{eqnarray}
\big[{}^{\mu\nu}\mathcal{T}^{\rho\sigma}\big]^{\xi=0}_{\overline{g}_{\mu\nu},\overline{\phi}}(x;x')  
 &\!\!=\!\!&\bigg\{\frac14\Big( 
   \overline{g}^{\mu\nu}\big(\overline{\nabla}^\rho\overline{\phi}\,\big)
                    \big(\overline{\nabla}^\sigma\overline{\phi}\,\big)
  +\overline{g}^{\rho\sigma}\big(\overline{\nabla}^\mu\overline{\phi}\,\big)
                 \big(\overline{\nabla}^\nu\overline{\phi}\,\big)
  \Big)
\nonumber\\
 &&\hskip -3.82cm 
  -\, \overline{g}^{\mu)(\rho}\big(\overline{\nabla}^{\sigma)}\overline{\phi}\,\big)
                 \big(\overline{\nabla}^{(\nu}\overline{\phi}\,\big)
  \!+\!\frac14 \Big(\overline{g}^{\mu\nu}\overline{g}^{\rho\sigma}\!\!-\!2\overline{g}^{\mu(\rho}\overline{g}^{\sigma)\nu}
           \Big)\bigg(\!\!\!-\!\frac12 
                    \big(\overline{\nabla}^\alpha\overline{\phi}\,\big)
                \big( \overline{\nabla}_\alpha\overline{\phi}\,\big)
               \!-\! \frac12 m^2\overline{\phi}^2\!\bigg)\!
 \bigg\}
  \nonumber\\
 &&\hskip -.45cm
  \times \frac{\delta^D(x\!-\!x')}{\sqrt{-g}}
\label{second variation of classical matter action: xi=0}
\\
\big[{}^{\mu\nu}\mathcal{T}^{\rho\sigma}\big]^{\xi\neq 0}_{\overline{g}_{\mu\nu},\overline{\phi}}(x;x')  
 &\!\!=\!\!& 
  -\, \frac{\xi}{2}\bigg\{\!\!-\!\frac12\big( \overline{g}^{\mu\nu}\overline{G}^{\rho\sigma}
                           \!+\!\overline{g}^{\rho\sigma} \overline{G}^{\mu\nu}\big)
                               \!+\!2\overline{g}^{\mu)(\rho} \overline{G}^{\sigma)(\nu}
\quad
 \nonumber\\
&&\hskip -2.cm 
                         -\, \frac14  \big(\overline{g}^{\mu\nu}\overline{g}^{\rho\sigma} 
           \!-\!2\overline{g}^{\mu(\rho}\overline{g}^{\sigma)\nu}\big)\overline{R}
       \!-\!\frac12 \Big( \overline{g}^{\mu\nu}
      {\overline{\nabla}'}^{(\rho}\overline{\nabla}^{\sigma)}
             \!+\!\overline{g}^{\rho\sigma}{\overline{\nabla}'}^{(\mu}\overline{\nabla}^{\nu)}
         \Big)
\label{second variation of classical matter action}\\
&&\hskip -3.7cm  
+\,\frac12\overline{g}^{\mu)(\rho}\Big( {\overline{\nabla}'}^{\sigma)}\overline{\nabla}^{(\nu}
 \!\!+\!
 {\overline{\nabla}}^{\sigma)}{\overline{\nabla}'}^{(\nu}\Big)                     
          \!+\! \frac12\Big(\overline{g}^{\mu\nu} \overline{g}^{\rho\sigma} 
          \!-\!  \overline{g}^{\mu(\rho}\overline{g}^{\sigma)\nu}    
             \Big)\overline{\nabla}\!\cdot\!\overline{\nabla}'
  \bigg\}   \bigg[\frac{\delta^D(x\!-\!x')}{\sqrt{-g}} \overline{\phi}^2\bigg]
\,,\!
\nonumber
\end{eqnarray}
which can be also obtained from the four-point vertex function in Eqs.~(\ref{4 point vertex: A})
and~(\ref{4 point vertex: A xi}),
\begin{equation}
\big[{}^{\mu\nu}\mathcal{T}^{\rho\sigma}\big]_{\overline{g}_{\mu\nu}}(x;x')
=\frac{\hbar}{2i\kappa^2}\frac{1}{\sqrt{-\overline{g}}\sqrt{-\overline{g}'}} V_{(4)}(x;x')
\overline{\phi}(x)
\overline{\phi}(x')
\,.\quad
\label{second variation of classical matter action 2}
\end{equation}

In semiclassical gravity the Noether-Ward identity was split into several parts: a contracted Bianchi identity,
metric compatibility,
 and a separate conservation of the classical, quantum and counterterm parts of the energy-momentum tensor.
Guided by that insight, in what follows we consider separately these five contributions.

\bigskip
\noindent
{\bf Gravitational action.}
The two gravitational contributions are, 
\begin{eqnarray}
&&\hskip -0.7cm
\frac{2}{\kappa^2}\int {\rm d}^Dx\sqrt{-\overline{g}}\,\bigg\{ \xi_\alpha(x) 
     \Big(\overline{\nabla}_{(\mu}\delta g_{\nu)}^{\;\alpha}
                       \!-\!\frac12\overline{\nabla}^{\;\alpha}\delta g_{\mu\nu}\Big)
               \Big(  \overline{G}^{\mu\nu} \!+\! \frac{D\!-\!2}{2}\Lambda\overline{g}^{\mu\nu}\Big)
 \nonumber\\
&&\hskip -0.1cm                             
     \!+\!\int\! {\rm d}^Dx'\sqrt{-\overline{g}'}\,
       \Big[ \big[{}^{\mu\nu}\mathcal{L}^{\rho\sigma}\big]_{\overline{g}_{\mu\nu}}\!(x;x')
  \nonumber\\
&&\hskip -0.1cm 
  -\,\frac{D\!-\!2}{4}\Lambda\Big(\overline{g}^{\mu\nu}\overline{g}^{\rho\sigma}
           \!\!-\!2\overline{g}^{\mu(\rho}\overline{g}^{\sigma)\nu}\Big)\Big]
        \delta g_{\rho\sigma}(x') \overline{\nabla}_\mu \xi_\nu(x)\bigg\}       
 \!+\!\mathcal{O}\big(\delta g_{\alpha\beta}^2,\xi_\gamma^2\big)
  = 0
\,.\qquad\;
\label{towards NW for bg: gravity}
\end{eqnarray}
This ought to hold for arbitrary $\xi_\alpha(x)$
and separately for the Einstein tensor and cosmological constant contributions, 
so we can remove the integral to obtain the following two identities, 
\begin{eqnarray}
&&\hskip -0.6cm 
     \Big(\overline{\nabla}_{(\mu}\delta g_{\nu)}^{\;\alpha}
                       \!-\!\frac12\overline{\nabla}^{\;\alpha}\delta g_{\mu\nu}\Big)
                            \overline{G}^{\mu\nu}
\,\qquad
\nonumber\\
&&\hskip 0.3cm            
       -\,\delta^ {\;\alpha}_{\mu}\overline{\nabla}_{\nu}
\bigg\{\bigg[\!\!-\!\frac12\big( \overline{g}^{\mu\nu}\overline{G}^{\rho\sigma}
                           \!+\!\overline{g}^{\rho\sigma} \overline{G}^{\mu\nu}\big)
                               \!+\!2\overline{g}^{\mu)(\rho} \overline{G}^{\sigma)(\nu}
\quad
 \nonumber\\
&&\hskip 0.3cm 
                         -\, \frac14  \big(\overline{g}^{\mu\nu}\overline{g}^{\rho\sigma} 
           \!-\!2\overline{g}^{\mu(\rho}\overline{g}^{\sigma)\nu}\big)\overline{R}
       \!+\!\frac12 \Big( \overline{g}^{\mu\nu}
      \overline{\nabla}^\rho\overline{\nabla}^\sigma 
           \!+\!\overline{g}^{\rho\sigma}\overline{\nabla}^{(\mu}\overline{\nabla}^{\nu)}
         \Big)
\nonumber\\
&&\hskip 0.3cm  
          -\,\overline{g}^{\mu)(\rho} \overline{\nabla}^{\sigma)}\overline{\nabla}^{(\nu}                
          \!-\! \frac12\Big(\overline{g}^{\mu\nu} \overline{g}^{\rho\sigma}    
          \!-\! \overline{g}^{\mu(\rho}\overline{g}^{\sigma)\nu}   
             \Big)\overline{\dAlembert}
                     \bigg]         \delta g_{\rho\sigma}\bigg\}
   + \mathcal{O}\big(\delta g_{\alpha\beta}^2\big) = 0
\,,\qquad
\label{towards NW for bg: gravity 2a}\\
&&\hskip -0.6cm 
     \Big(\overline{\nabla}_{(\mu}\delta g_{\nu)}^{\;\alpha}
                       \!-\!\frac12\overline{\nabla}^{\;\alpha}\delta g_{\mu\nu}\Big)
                            \overline{g}^{\mu\nu}
\,\qquad
\nonumber\\
&&\hskip 0.3cm    
   +\,\delta_\nu^\alpha\overline{\nabla}_\mu\left\{ \frac12
   \left(\overline{g}^{\mu\nu}\overline{g}^{\rho\sigma}
        - 2\overline{g}^{\mu(\rho}\overline{g}^{\sigma)\nu} \right)\delta g_{\rho\sigma}\right\}
           + \mathcal{O}\big(\delta g_{\alpha\beta}^2\big) = 0
\,,\qquad
\label{towards NW for bg: gravity 2}
\end{eqnarray}
where we made use of 
Eq.~(\ref{second variation of the Hilbert-Einstein action: Lichnerowicz 3})
in Eq.~(\ref{towards NW for bg: gravity 2a}).
One can immediately see that the second identity (which originates from the cosmological constant term) 
is true, and thus it provides no useful information.
The first identity~(\ref{towards NW for bg: gravity 2a}) looks nontrivial, and it
can be proved by expanding $\nabla_\mu G^{\mu\nu} = 0 $ in powers of 
$\delta g_{\mu\nu}$ around $\overline{g}_{\mu\nu}$,
\begin{eqnarray}
 0 &\!\!=\!\!& \nabla_\mu G^{\mu\alpha} = \overline{\nabla}_\mu \overline{G}^{\mu\alpha}
    + \Big(\delta \Gamma_{\mu\beta}^\mu \overline{G}^{\beta\alpha}
         +\delta \Gamma_{\mu\beta}^\nu \overline{G}^{\beta\mu} \Big)
         + \overline{\nabla}_\mu \delta{G}^{\mu\alpha}
          +\mathcal{O}\big(\delta g_{\alpha\beta}^2\big)
\nonumber\\
&\!\!=\!\!&
         \delta \Gamma_{\mu\nu}^\alpha \overline{G}^{\beta\nu}
         + \overline{\nabla}_\mu 
            \left(\frac{1}{\sqrt{-\overline{g}}}\delta\Big[\sqrt{-g}G^{\mu\nu}\Big]\right)
          +\mathcal{O}\big(\delta g_{\alpha\beta}^2\big)
\,,\qquad
\label{proof of gravitational NW identity}
\end{eqnarray}
where, to get to the second line, we used $\overline{\nabla}_\mu \overline{G}^{\mu\nu}=0$
and $ \delta \Gamma_{\mu\alpha}^\mu 
= \frac12  \overline{\nabla}_\alpha\delta g_{\gamma}^{\;\gamma} 
= \overline{\nabla}_\alpha \frac{\delta \sqrt{-g}}{\sqrt{-\overline{g}}}$.
Recalling Eq.~(\ref{second variation of the Hilbert-Einstein action: Lichnerowicz 3})
and $ \delta \Gamma_{\mu\nu}^\alpha =\overline{\nabla}_{(\mu}\delta g_{\nu)}^{\;\alpha}
                       \!-\!\frac12\overline{\nabla}^{\;\alpha}\delta g_{\mu\nu} $ 
we see that Eq.~(\ref{towards NW for bg: gravity 2a}) reduces to 
Eq.~(\ref{proof of gravitational NW identity}), completing the proof
of the linear Noether identity in Eq.~(\ref{towards NW for bg: gravity 2a})
for the gravitational part of the action. Including higher order terms
in $\delta g_{\alpha\beta}$ in 
Eqs.~(\ref{towards NW for bg: gravity 2a}) and~(\ref{towards NW for bg: gravity 2}) would 
couple the linear and higher order gravitational perturbations. 
The identity~(\ref{towards NW for bg: gravity 2a})
does  not provide useful constraint for classical gravitational perturbations, as it is a part of the 
identity $\nabla_\mu G^{\mu\nu}(g_{\alpha\beta}) = 0 $, which holds for a general metric tensor
$g_{\mu\nu}$ and thus also for general perturbations perturbations
$\delta g_{\mu\nu} = g_{\mu\nu} - \overline{g}_{\mu\nu}$ around some background metric
 $\overline{g}_{\mu\nu}$. Nevertheless, they can be useful to check accuracy of a calculation.

\bigskip
\noindent
{\bf Classical matter action.} Consider next the classical part of the matter action. From 
Eqs.~(\ref{1PI one-loop effective action:  towards NW for bg 4}) 
and~(\ref{1PI one-loop effective action: towards NW for bg 5}),
\begin{eqnarray}
&&\hskip -0.7cm
- \int {\rm d}^Dx\sqrt{-\overline{g}}\, \xi_\alpha(x) 
     \Big(\overline{\nabla}_{(\mu}\delta g_{\nu)}^{\;\alpha}
                       \!-\!\frac12\overline{\nabla}^{\;\alpha}\delta g_{\mu\nu}\Big)
\overline{T}^{\mu\nu}(x)
\,\qquad
\nonumber\\
&&\hskip -0.3cm
 + \int\! {\rm d}^Dx\,{\rm d}^Dx'\sqrt{-\overline{g}}\sqrt{-\overline{g}'}
            \Big\{
 \, \big[{}^{\mu\nu}\mathcal{T}^{\rho\sigma}\big]_{\overline{g}_{\mu\nu},\overline{\phi}}(x;x') 
\delta g_{\rho\sigma}(x') \Big\}    \!\times\! 2\overline{\nabla}_{(\mu}\xi_{\nu)}(x)
\,.\qquad\;
\label{towards NW for classical matter}
\end{eqnarray}
When Eqs.~(\ref{second variation of classical matter action: xi=0})--(\ref{second variation of classical matter action}) are inserted into 
Eq.~(\ref{towards NW for classical matter}) and one integrates over $x'$ (in some terms) one obtains, 
\begin{eqnarray}
&&\hskip -0.6cm
-\! \int {\rm d}^Dx\sqrt{-\overline{g}}\, \xi_\alpha(x) 
     \Big(\overline{\nabla}_{(\mu}\delta g_{\nu)}^{\;\alpha}
                       \!-\!\frac12\overline{\nabla}^{\;\alpha}\delta g_{\mu\nu}\Big)
\overline{T}^{\mu\nu}(x)
\,\qquad
\nonumber\\
&&\hskip -0.6cm
 +\! \int\!\! {\rm d}^Dx\sqrt{-\overline{g}}
\bigg[
  \frac14\Big(
   \overline{g}^{\mu\nu}\!\big(\overline{\nabla}^\rho\overline{\phi}\,\big)
                    \big(\overline{\nabla}^\sigma\overline{\phi}\,\big)
  \!+\!\overline{g}^{\rho\sigma}\!\big(\overline{\nabla}^\mu\overline{\phi}\,\big)
                 \big(\overline{\nabla}^\nu\overline{\phi}\,\big)
  \Big) 
  \!-\! \overline{g}^{\mu)(\rho}\big(\overline{\nabla}^{\sigma)}\overline{\phi}\,\big)
                 \big(\overline{\nabla}^{(\nu}\overline{\phi}\,\big)\!\!
\nonumber\\
 &&\hskip -.5cm 
  -\,\frac18\Big(\overline{g}^{\mu\nu}\overline{g}^{\rho\sigma}\!-\!2\overline{g}^{\mu(\rho}\overline{g}^{\sigma)\nu}
           \Big)\bigg( 
                    \big(\overline{\nabla}^\alpha\overline{\phi}\,\big)
                \big( \overline{\nabla}_\alpha\overline{\phi}\,\big)
               \!+\!  m^2\overline{\phi}^2\,\bigg)
\nonumber\\
&& \hskip -.5cm
-\,\frac{\xi}{2}\bigg(\!-\!\frac12\Big(\overline{g}^{\mu\nu}\overline{G}^{\rho\sigma}
      \!+\!\overline{g}^{\rho\sigma}\overline{G}^{\mu\nu}\Big)
                               \!+\!2\overline{g}^{\mu)(\rho} \overline{G}^{\sigma)(\nu}
                        \! -\! \frac14  \big(\overline{g}^{\mu\nu}\overline{g}^{\rho\sigma} 
           \!-\!2\overline{g}^{\mu(\rho}\overline{g}^{\sigma)\nu}\big)\overline{R}                         
             \bigg)  \bigg]
\nonumber\\
&& \hskip 1.5cm
\times\,\delta g_{\rho\sigma}(x) \times\, 2\overline{\nabla}_{(\mu}\xi_{\nu)}(x)
\nonumber\\
&& \hskip -.3cm
 -\,\frac{\xi}{2} \int\!\! {\rm d}^Dx\,{\rm d}^Dx'\sqrt{-\overline{g}}\sqrt{-\overline{g}'}
\,\bigg[\frac12 \Big( \overline{g}^{\mu\nu}\overline{\nabla}^{(\rho}\overline{\nabla}^{\sigma)}
                    \!\!-\!\overline{g}^{\rho\sigma}\overline{\nabla}^{(\mu}\overline{\nabla}^{\nu)}
         \Big)\!-\!\overline{g}^{\mu)(\rho} \overline{\nabla}^{\sigma)}\overline{\nabla}^{(\nu}  
\nonumber\\
&&\hskip -.cm                
          -\, \frac12\Big(\overline{g}^{\mu\nu} \overline{g}^{\rho\sigma}  
          \!-\!      \overline{g}^{\mu(\rho}\overline{g}^{\sigma)\nu}
             \Big)\overline{\dAlembert}\bigg]\bigg[\frac{\delta^D(x\!-\!x')}{\sqrt{-g}}\bigg]\overline{\phi}^2(x')
   \!\times\! \delta g_{\rho\sigma}(x')2\overline{\nabla}_{(\mu}\xi_{\nu)}(x)
\,.\quad
\label{towards NW for classical matter 2}
\end{eqnarray}
The derivatives in the last two lines can be integrated by parts,
and the boundary terms can be dropped for proper gauge transformations, upon which the remaining integral over $x'$ can be performed.
The identity~(\ref{towards NW for classical matter 2}) must hold for arbitrary $\xi_\alpha(x)$, which allows us to remove 
the integral over $x$ in Eq.~(\ref{towards NW for classical matter 2}), resulting in the following identity,
\begin{eqnarray}
&&\hskip -0.7cm
-\, \Big(\overline{\nabla}_{(\mu}\delta g_{\nu)}^{\;\alpha}
                       \!-\!\frac12\overline{\nabla}^{\;\alpha}\delta g_{\mu\nu}\Big)
\overline{T}^{\mu\nu}(x)
\,\qquad
\nonumber\\
&&\hskip -0.cm
 -\,2\delta_\nu^\alpha\overline{\nabla}_\mu
            \bigg\{\!\bigg[
  \frac14\Big(
   \overline{g}^{\mu\nu}\!\big(\overline{\nabla}^\rho\overline{\phi}\,\big)
                    \big(\overline{\nabla}^\sigma\overline{\phi}\,\big)
  \!+\!\overline{g}^{\rho\sigma}\!\big(\overline{\nabla}^\mu\overline{\phi}\,\big)
                 \big(\overline{\nabla}^\nu\overline{\phi}\,\big)
  \Big) 
  \!-\! \overline{g}^{\mu)(\rho}\big(\overline{\nabla}^{\sigma)}\overline{\phi}\,\big)
                 \big(\overline{\nabla}^{(\nu}\overline{\phi}\,\big)\!\!
\nonumber\\
 &&\hskip .cm 
  -\,\frac18\Big(\overline{g}^{\mu\nu}\overline{g}^{\rho\sigma}\!-\!2\overline{g}^{\mu(\rho}\overline{g}^{\sigma)\nu}
           \Big)\Big( 
                    \big(\overline{\nabla}^\alpha\overline{\phi}\,\big)
                \big( \overline{\nabla}_\alpha\overline{\phi}\,\big)
               \!+\!  m^2\overline{\phi}^2\,\Big)\bigg]\delta g_{\rho\sigma}\bigg\}
\nonumber\\
&& \hskip -.cm
 +\,\xi\delta_\nu^\alpha\overline{\nabla}_\mu\bigg\{
 \bigg[\!\!-\!\frac12\Big(\overline{g}^{\mu\nu}\overline{G}^{\rho\sigma}
      \!\!+\!\overline{g}^{\rho\sigma}\overline{G}^{\mu\nu}\Big)
                               \!+\!2\overline{g}^{\mu)(\rho} \overline{G}^{\sigma)(\nu}
                        \!\! -\! \frac14  \big(\overline{g}^{\mu\nu}\overline{g}^{\rho\sigma} 
           \!-\!2\overline{g}^{\mu(\rho}\overline{g}^{\sigma)\nu}\big)\overline{R}                         
\nonumber\\
&& \hskip -.cm
   +\,\frac12 \Big( \overline{g}^{\mu\nu}{\overline{\nabla}}^{(\rho}\overline{\nabla}^{\sigma)}
  \!+\! \overline{g}^{\rho\sigma}\overline{\nabla}^{(\mu} \overline{\nabla}^{\nu)}
         \Big) \!-\!\overline{g}^{\mu)(\rho} \overline{\nabla}^{\sigma)}{\overline{\nabla}}^{(\nu}
                   \!-\! \frac12\Big(\overline{g}^{\mu\nu} \overline{g}^{\rho\sigma} 
          \!-\!      \overline{g}^{\mu(\rho}\overline{g}^{\sigma)\nu}
             \Big)\overline{\dAlembert}\bigg]
\nonumber\\
&&\hskip 1.3cm   
\times\,\Big[\overline{\phi}^2\delta g_{\rho\sigma}(x)\Big]
 \bigg\} 
\,,\quad
\label{towards NW for classical matter 3}
\end{eqnarray}
which can be proven by expanding the classical energy-momentum conservation, $\nabla_\mu T^{\mu\nu}=0$,
around $\overline{g}_{\mu\nu}$ to linear order in $\delta g_{\mu\nu}=g_{\mu\nu}\!-\!\overline{g}_{\mu\nu}$.

\bigskip
\noindent
{\bf Quantum matter action.} Derivation of the Noether-Ward identity for the quantum part of 
the matter action follows the same steps as for the classical action, with the replacement $\overline{\phi}(x)\overline{\phi}(x')\rightarrow i\Delta_{\overline{\phi}}(x;x')$. As we pointed out when we discussed 
the Noether-Ward identity for semiclassical gravity (see Eqs.~(\ref{covariant energy-momentum conservation}) and~(\ref{covariant energy-momentum conservation: parts})), one can either consider the identity for the 
renormalized energy-momentum tensor, or one can consider separately the primitive 
(non-renormalized) 
quantum contribution to the energy momentum-tensor and the contribution from the counterterms.
Here we shall consider them separately. Let us begin with the primitive 
quantum contribution to the energy momentum-tensor,
\begin{eqnarray}
&&\hskip -0.6cm
-\, \frac{\kappa^2}{2}\Big(\overline{\nabla}_{(\mu}\delta g_{\nu)}^{\;\alpha}
                       \!-\!\frac12\overline{\nabla}^{\;\alpha}\delta g_{\mu\nu}\Big)
\Big[\big\langle\Delta \hat{\overline{T}}^{\mu\nu}(x)\big\rangle\Big]_{\overline{g}_{\mu\nu},\overline{\phi}}
\,\qquad
\nonumber\\
&&\hskip -0.cm
 +\,\int {\rm d}^Dx'\sqrt{-g'}\delta_\nu^\alpha\overline{\nabla}_\mu 
 \Big(\big[{}^{\mu\nu}\overline{\Sigma}_{\rm 3pt}^{\rho\sigma}\big](x;x')\!+\!\big[{}^{\mu\nu}\overline{\Sigma}_{\rm 4pt}^{\rho\sigma}\big](x;x')\Big)
       \delta g_{\rho\sigma}(x')
\nonumber\\
&&\hskip -0.cm
       +\,\mathcal{O}\big(\delta g_{\alpha\beta}^2\big)
       =0
\,,\quad
\label{towards NW for quantum matter}
\end{eqnarray}
where $-i\big[^{\mu\nu}\overline{\Sigma}_{\rm 3pt}^{\rho\sigma}\big](x;x')$ and 
$-i\big[^{\mu\nu}\overline{\Sigma}_{\rm 4pt}^{\rho\sigma}\big](x;x')$ are the three-point and four-point contributions to the
one-loop graviton self-energy given
in Eqs.~(\ref{3pt contribution to self-energy: total}) and~(\ref{4-point contribution self-energy: case A}), 
respectively.~\footnote{For a proof that 
variation of the one-loop energy momentum tensor gives the graviton self-energy see also 
Ref.~\cite{Marunovic:2012pr}.}
Since $\overline{\nabla}_\mu \big[\big\langle\Delta \hat{\overline{T}}^{\mu\nu}(x)\big\rangle\big]_{\overline{g}_{\mu\nu},\overline{\phi}}$
and $\nabla_\mu\big[\big\langle\Delta \hat{\overline{T}}^{\mu\nu}(x)\big\rangle\big]_{g_{\mu\nu},\phi}$
contain local singular contributions, which partially cancel in the difference, leaving 
ultra-local singular terms which are linear in metric perturbations and which vanish in dimensional regularization.
With this caveat in mind, the quantum Noether-Ward identity in Eq.~(\ref{towards NW for quantum matter}) 
can be considered as exact.


\bigskip
\noindent
{\bf Counterterms.} In order to renormalize the quantum energy-momntum tensor 
and the graviton self-energy at the one-loop level in general
one ought to add the following geometric counterterms,
\begin{enumerate}
\item Cosmological constant counterterm;
\item Newton constant counterterm;
\item Ricci scalar squared counterterm;
\item Ricci tensor squared counterterm;
\item Riemann tensor squared counterterm.
\end{enumerate}
In the presence of a scalar condensate, one may need to add to these the kinetic counterterm, the scalar 
mass counterterm and the (quartic) self-interaction counterterm.
In $D\!=\!4$ the counterterms 3 to 5 are not linearly independent,
as their linear combination  -- known as the Gauss-Bonnet counterterm action -- 
is  topological in $D\!=\!4$. 
This means that its variation produces a contribution to the energy-momentum tensor that are suppressed 
as $\propto (D\!-\!4)$, and thus cannot be used to renormalise the divergencies in $D\!=\!4$.~\footnote{The Gauss-Bonnet counterterm can be used to obtain one of the contributions from the Weyl anomaly to the energy-momentum tensor~\cite{Koksma:2008jn,Glavan:2019inb}.} 
Therefore, an alternative way of organizing the dimension four 
counterterms is to use the Ricci scalar squared and the Weyl tensor squared counterterms. 

Since the cosmological constant and Newton constant contributions were already considered in
subsection {\it Classical Noether identity} of this section, what remains to be considered are
the Ricci scalar squared, Ricci tensor squared and Riemann tensor squared counterterms,
\begin{eqnarray}
S_{\rm ct}[g_{\mu\nu}] &\!\!\!=\!\!\!& S_{R^2}[g_{\mu\nu}]
 \!+\!  S_{\rm Ric^2}[g_{\mu\nu}]
  \!+\! S_{\rm Riem^2}[g_{\mu\nu}]
 \label{counterterm action}\\
 &&\hskip -1.cm
  \equiv\,\alpha_{R^2}\!\int\! {\rm d}^Dx \sqrt{-g}\, R^2
 \!+\!  \alpha_{\rm Ric^2}\!\!\int\! {\rm d}^Dx \sqrt{-g}\, {\rm Ric}^2
 \!+\! \alpha_{\rm Riem^2}\!\!\int\! {\rm d}^Dx \sqrt{-g}\, {\rm Riem}^2
 \,,
\nonumber
\end{eqnarray}
where ${\rm Ric}^2 \equiv R_{\alpha\beta}R^{\alpha\beta}$ and 
${\rm Riem}^2 \equiv R_{\alpha\beta\gamma\delta}R^{\alpha\beta\gamma\delta}$. 
Generally speaking, two of these terms are enough to renormalize
the the quantum contributions as the following linear combination, known 
as Gauss-Bonnet term,
\begin{equation}
{\rm G\!B} \equiv R^2 \!-\! 4 {\rm Ric}^2 \!+\! {\rm Riem}^2
\,,\quad
\label{Gauss-Bonnet}
\end{equation}
is topological in $D=4$~\cite{Chern:1945,Yale:2011usf}.

As we pointed out above, each of the counterterms contributes to the energy-momentum tensor 
$T^{\mu\nu}_{\rm ct, i}$ a term which is covariantly conserved,
\begin{equation}
\nabla_\mu T^{\mu\nu}_{\rm ct, i}=0
\,,\qquad
\label{covariant energy-momentum conservation: part i}
\end{equation}
see Eq.~(\ref{covariant energy-momentum conservation: parts}).
In what follows we discuss in some detail how to prove the Noether identity for the 
Ricci scalar squared counterterm, and then outline how to the proof works for other counterterms.


\bigskip
\noindent
{\bf Noether identity for the Ricci scalar squared counterterm.}
Varying the Ricci scalar squared action in Eq.~(\ref{counterterm action}) gives 
({\it cf.}~Eq.~(\ref{counterterm action R2 c1})),
\begin{eqnarray}
T^{\mu\nu}_{R^2} \equiv \frac{2}{\sqrt{-g}}\frac{\delta S_{R^2}}{\delta g_{\mu\nu}(x)} &\!\!=\!\!&  \alpha_{R^2}
    \bigg\{
\!-g^{\mu\nu}R^2
              \!-\! 4
               \big(G^{\mu\nu}\!-\!\nabla^{\mu} \nabla^{\nu}
            \!+\! g^{\mu\nu}\dAlembert
                 \big)R
    \bigg\}
\,,\qquad\;
\label{counterterm action R2: variation}
\end{eqnarray}
which is covariantly conserved ({\it cf.} Eq.~(\ref{counterterm action R2 d1})), 
\begin{eqnarray}
\nabla_\mu T^{\mu\nu}_{R^2}= 0
\,.\qquad
\label{counterterm action R2: covariant conservation}
\end{eqnarray}
When this identity is expanded around the background metric $\overline{g}_{\mu\nu}(x)$,
and taking account of $\overline{\nabla}_\mu \overline{T}^{\mu\nu}_{R^2}= 0$
one obtains the following Noether identity for gravitational perturbations,
\begin{eqnarray}
&&\hskip -0.7cm
\left[\overline{\nabla}_{(\mu}\delta g_{\alpha)}^{\;\nu}
   \!-\!\frac12\overline{\nabla}^\nu\delta g_{\mu\alpha} \right]\overline{T}^{\mu\alpha}_{R^2}
\nonumber\\
&&\hskip 1.2cm
  +\,2\!\int\!\!{\rm d}^{D}x'\sqrt{-\overline{g}'}\,\overline{\nabla}_\mu\Bigg[
  \frac{1}{\sqrt{-\overline{g}}\sqrt{-\overline{g}'}}
   \left(\frac{\delta^2S_{R^2}}{\delta g_{\mu\nu}(x)\delta g_{\rho\sigma}(x')}
     \right)_{\!\overline{g}_{\alpha\beta}}\Bigg]
      \delta g_{\rho\sigma}(x')  
\nonumber\\
 &&\hskip 1.2cm     
      +\,{\mathcal{O}}\left(\delta g_{\mu\nu}^2\right)
   =0
   \,,
\label{Noether identity for R2}
\end{eqnarray}
where (see Eq.~(\ref{counterterm action R2 c2})),
\begin{eqnarray}
\frac{1}{\sqrt{-\overline{g}}\sqrt{-\overline{g}'}}
 \left(\frac{\delta^2 S_{R^2}}{\delta g^{\mu\nu}(x)\delta g^{\rho\sigma}(x')}
      \right)_{\!\overline{g}_{\alpha\beta}}\!\!
 &\!\!\!=\!\!\!&
  \alpha_{R^2}\bigg\{\!
    \frac14\overline{R}^2\big(\overline{g}^{\mu\nu} \overline{g}^{\rho\sigma}
     \!\!+\!6\overline{g}^{\mu(\rho}\overline{g}^{\sigma)\nu}\big)
\nonumber\\
&&\hskip -4.7cm
     +\,4\overline{R}\,\overline{G}^{\mu)(\rho}\overline{g}^{\sigma)(\nu}
  \!\!+\!2\big(\overline{R}^{\mu\nu}
           \!\!-\!\overline{\nabla}^{\mu}\overline{\nabla}^{\nu}
          \!\!+\!\overline{g}^{\mu\nu}\overline{\dAlembert}\big)
          \big({\overline{R}'}^{\rho\sigma} 
          \!\!-\!{\overline{\nabla}'}^{\rho}{\overline{\nabla}'}^{\sigma}
           \!\!\!+\!{\overline{g}'}^{\rho\sigma}\overline{\dAlembert}'\big)
\nonumber\\
&&\hskip -4.7cm
   -\, \overline{g}^{\mu\nu}\overline{R} \big(\overline{R}^{\rho\sigma} 
          \!\!-\!\overline{\nabla}^{\rho}\overline{\nabla}^{\sigma}
           \!\!+\!\overline{g}^{\rho\sigma}\overline{\dAlembert}\big)
       -\, {\overline{g}'}^{\rho\sigma}{\overline{R}'} \big({\overline{R}'}^{\mu\nu} 
          \!\!-\!{\overline{\nabla}'}^{\mu}{\overline{\nabla}'}^{\nu}
           \!\!+\!{\overline{g}'}^{\mu\nu}\overline{\dAlembert}'\big)     
\nonumber\\
&&\hskip -4.7cm
                 +\,2\overline{R}\big(\!-2\overline{g}^{\mu)(\rho}
                         \overline{\nabla}^{\sigma)}\overline{\nabla}^{(\nu}
                         \!\!+\!\overline{g}^{\rho\sigma}
                         \overline{\nabla}^{\mu}\overline{\nabla}^{\nu}
                          \!\!+\!\overline{g}^{\mu(\rho}\overline{g}^{\sigma)\nu}
                         \overline{\dAlembert}\big)
  \nonumber\\
&&\hskip -4.7cm
   +\,2\overline{R}'\big(\!-2{\overline{g}'}^{\mu)(\rho}
                         {\overline{\nabla}'}^{\sigma)}{\overline{\nabla}'}^{(\nu}
                         \!\!+\!{\overline{g}'}^{\rho\sigma}
                         {\overline{\nabla}'}^{\mu}{\overline{\nabla}'}^{\nu}
                          \!\!+\!{\overline{g}'}^{\mu(\rho}{\overline{g}'}^{\sigma)\nu}
                         \overline{\dAlembert}'\big)
  \nonumber\\
&&\hskip -4.7cm
+\,\Big[\!-\!\big(\overline{g}^{\mu\nu}\overline{g}^{\rho\sigma}
             \!-\!3\overline{g}^{\mu(\rho}\overline{g}^{\sigma)\nu}\big)
             \overline{\nabla}\!\cdot\!{\overline{\nabla}'}
\!+\!2\big(\overline{g}^{\mu\nu}\overline{\nabla}^{(\rho}{\overline{\nabla}'}^{\sigma)}
                  \!+\!\overline{g}^{\rho\sigma}\overline{\nabla}^{(\mu}{\overline{\nabla}'}^{\nu)}\big)
  \nonumber\\
&&\hskip -3.99cm
   -\,3\overline{g}^{\mu(\rho}\big(\overline{\nabla}^{\sigma)}{\overline{\nabla}'}^{(\nu}
                  \!+\!{\overline{\nabla}'}^{\sigma)}\overline{\nabla}^{(\nu}\big)           
       \Big]R      \bigg\}
         \frac{\delta^D(x\!-\!x')}{\sqrt{-\overline{g}}}
      \,.\qquad
\label{counterterm action R2: second variation}
\end{eqnarray}
The identity~ (\ref{Noether identity for R2}) is easily proved by expanding Eq.~(\ref{counterterm action R2: covariant conservation})
to linear oder in $\delta g_{\mu\nu}$. An intermediate step in the proof is,
\begin{eqnarray}
&&\hskip -0.6cm
\Big[\delta \Gamma_{\mu\alpha}^\mu \overline{T}_{R^2}^{\alpha\nu}
\!\!+\!\delta \Gamma_{\mu\alpha}^\nu \overline{T}_{R^2}^{\mu\alpha}\Big]
\!+\!\overline{\nabla}_\mu \int {\rm d}^Dx'
\left[\!-\frac{1}{2}\overline{g}^{\rho\sigma}\delta g_{\rho\sigma}(x)\delta^D(x\!-\!x')
      \overline{T}_{R^2}^{\mu\nu}(x)
     \right]
\nonumber\\
&&\hskip -0.cm
+\,2\overline{\nabla}_\mu \int {\rm d}^Dx'\sqrt{-\overline{g}'}
\left[
\frac{1}{\sqrt{-\overline{g}}\sqrt{-\overline{g}'}}
      \left(\frac{\delta S_{R^2}}{\delta g_{\rho\sigma}(x')\delta g_{\mu\nu}(x)}
          \right)_{\overline{g}_{\alpha\beta}}\!\!\delta g_{\rho\sigma}(x')
     \right]
\,,\qquad\;
\label{counterterm action R2: proof of conservation}
\end{eqnarray}
where $\delta \Gamma_{\mu\alpha}^\mu 
=\frac12 \overline{g}^{\rho\sigma}\overline{\nabla}_\alpha\delta g_{\rho\sigma}$.
Now noting that the first term in the first line cancels against the last term in the same line results in 
Eq.~(\ref{Noether identity for R2}), completing the proof.

\bigskip

\noindent
{\bf Noether identities for the remaining counterterms.}
The remaining counterterms needed to renormalize the one-loop graviton self-energy 
are the Ricci tensor squared counterterm and the Riemann tensor squared counterterms.
As mentioned above, one can choose a linear combination of these 
and use instead the Weyl tensor squared and Gauss-Bonnet counterterms~(\ref{Gauss-Bonnet}). Other 
(higher order) gravitational counterterms are needed to renormalize the graviton self-energy
caclulated at a higher loop order. The logic is always the same. If $S_{\rm ct,i}$ is the action of a gravitational
counterterm, then $\nabla_\mu T^{\mu\nu}_{\rm ct,i} =0$ holds true. Expanding this 
around $\overline{g}_{\mu\nu}$ to linear order in $\delta g_{\mu\nu}$ gives 
({\it cf.} Eq.~(\ref{Noether identity for R2})),
\begin{eqnarray}
&&\hskip -0.7cm
\left[\overline{\nabla}_{(\mu}\delta g_{\alpha)}^{\;\nu}
   \!-\!\frac12\overline{\nabla}^\nu\delta g_{\mu\alpha} \right]\overline{T}^{\mu\alpha}_{\rm ct,i}
\nonumber\\
&&\hskip 1.2cm
  +\,2\!\int\!\!{\rm d}^{D}x'\sqrt{-\overline{g}'}\,\overline{\nabla}_\mu\Bigg[
  \frac{1}{\sqrt{-\overline{g}}\sqrt{-\overline{g}'}}
   \left(\frac{\delta^2S_{\rm ct,i}}{\delta g_{\mu\nu}(x)\delta g_{\rho\sigma}(x')}
     \right)_{\!\overline{g}_{\alpha\beta}}\Bigg]
      \delta g_{\rho\sigma}(x')  
\nonumber\\
 &&\hskip 1.2cm     
      +\,{\mathcal{O}}\left(\delta g_{\mu\nu}^2\right)
   =0
   \,,\qquad
\label{Noether identity for ct i}
\end{eqnarray}
where  
\begin{equation}
\overline{T}^{\mu\nu}_{\rm ct,i} = \frac{2}{\sqrt{-\overline{g}}}
\left(\frac{\delta S_{\rm ct.i}}{\delta g_{\mu\nu}(x)}\right)_{\overline{g}_{\alpha\beta}}
\,.\quad
\label{energy momentum tensor ct i}
\end{equation}

\bigskip
\noindent
{\bf Summary.} In this section we have shown that various parts of the quantum equation of motion
(which include the self-energy correction from matter loops) obey individual Noether (or Noether-Ward)  identities for the gravitational perturbations. A natural question that then arises is whether all of these identities 
can be combined. The answer to this question is: yes. Indeed, it is quite easy to convince oneself that, when
all of the individual terms we have analyzed combine, one obtains 
({\it cf.} Eqs.~(\ref{1PI one-loop effective action:  towards NW for bg 4})--(\ref{1PI one-loop effective action: towards NW for bg 5})), 
\begin{equation}
\delta \Gamma^\nu_{\mu\alpha} ({\rm\overline{SCG}})^{\mu\alpha}
 + \overline{\nabla}_\mu ({\rm GP})^{\mu\nu}  = 0
\,,\quad
\label{Noether-Ward identity for gravitational perturbations A}
\end{equation}
and since the equation of motion of semiclassical gravity (for the background metric 
$\overline{g}_{\mu\nu}$) $({\rm\overline{SCG}})^{\mu\alpha}=0$ is satisfied, 
\begin{equation}
\overline{G}^{\mu\nu}
                         \!+\! \frac{D\!-\!2}{2}\Lambda\overline{g}^{\mu\nu}
                        \!=\! \frac{\kappa^2}{2}\bigg(\overline{T}^{\mu\nu}(x)
   \!+\!\Big[\big\langle\Delta \hat{\overline{T}}^{\mu\nu}(x)\big\rangle\Big]
     \!\!+\!\overline{T}_{\rm ct}^{\mu\nu}(x)\bigg)
\,,\qquad
\label{EoM SCG for bg metric}
\end{equation}
we obtain the following transversality condition for 
the equation of motion of gravitational perturbations $({\rm GP})^{\mu\nu}=0$,
\begin{eqnarray}
&&\hskip -0.6cm 
\overline{\nabla}_\mu\int {\rm d}^D x'\bigg[\big[{}^{\mu\nu}\overline{\mathcal{L}}^{\rho\sigma}\big](x;x')
\nonumber\\
&&\hskip 0.cm 
        -\, \big[{}^{\mu\nu}\overline{\Sigma}^{\rho\sigma}\big](x;x')  
 \!-\!  \big[{}^{\mu\nu}\overline{\Sigma}_{\rm ct}^{\rho\sigma}\big](x;x')  
       \!+\! \kappa^2\big[{}^{\mu\nu}\overline{\mathcal{T}}^{\rho\sigma}\big](x;x')
        \bigg]\delta g_{\rho\sigma}(x')
\nonumber\\
&&\hskip 0.8cm    
 -\,\frac{D\!-\!2}{4}\Lambda\overline{\nabla}_\mu
\Big[\Big(\overline{g}^{\mu\nu}\overline{g}^{\rho\sigma}
\!\!-\!2\overline{g}^{\mu(\rho}\overline{g}^{\sigma)\nu}\Big)\delta g_{\rho\sigma}(x)
\Big]
         =0
\,.\qquad\;
\label{Noether-Ward identity for gravitational perturbations B}
\end{eqnarray}
Since this equality holds for an arbitrary metric perturbation $\delta g_{\rho\sigma}(x')$,  
one can remove  $\delta g_{\rho\sigma}(x')$ and the integral over $x'$
from Eq.~(\ref{Noether-Ward identity for gravitational perturbations B}),~\footnote{This removal ought to be done with a due care, as the objects in Eq.~(\ref{Noether-Ward identity for gravitational perturbations B}) are bitensors, {\it i.e.} 2-index tensors at $x$ and 2-index tensors at $x'$, which 
act on $\delta g_{\rho\sigma}(x')$ at $x'$, making it a scalar at $x'$.} to obtain,
\begin{eqnarray}
&&\hskip -0.6cm 
\overline{\nabla}_\mu\bigg[\big[{}^{\mu\nu}\overline{\mathcal{L}}^{\rho\sigma}\big](x;x')
         \!-\!\frac{D\!-\!2}{4}\Lambda\Big(\overline{g}^{\mu\nu}\overline{g}^{\rho\sigma}
           \!\!-\!2\overline{g}^{\mu(\rho}\overline{g}^{\sigma)\nu}\Big)
             \frac{\delta^D(x\!-\!x')}{\sqrt{-\overline{g}}}
\nonumber\\
&&\hskip 0.8cm 
        -\, \big[{}^{\mu\nu}\overline{\Sigma}^{\rho\sigma}\big](x;x')  
 \!-\!  \big[{}^{\mu\nu}\overline{\Sigma}_{\rm ct}^{\rho\sigma}\big](x;x')  
       \!+\! \kappa^2\big[{}^{\mu\nu}\overline{\mathcal{T}}^{\rho\sigma}\big](x;x')
        \bigg] =0
\,,\qquad\;
\label{Noether-Ward identity for gravitational perturbations B2}
\end{eqnarray}
One may think that that is an empty statement, because 
the equation of motion for gravitational perturbations $({\rm GP})^{\mu\nu}=0$ is satisfied.
This is in fact not true, because Eq.~(\ref{Noether-Ward identity for gravitational perturbations B})
tells us that different non-transverse parts generated by each term in 
Eq.~(\ref{Noether-Ward identity for gravitational perturbations B})
consist of one-point function contributions which, when taken together, combine into the equation of motion 
of semiclassical gravity~(\ref{EoM SCG for bg metric}), which is satisfied for the background metric, 
thus representing a nontrivial check of the consistency and accuracy of the calculation.

\bigskip
\noindent
{\bf Case B.} For the metric pertubations 
defined in terms of the inverse metric~(\ref{metric perturbations: case B}), as in case A, each of the terms in the semiclassical
gravity equation ({\it cf.} Eq.~(\ref{eom SCG})),
\begin{equation}
G_{\mu\nu}(g_{\alpha\beta}) + \frac{D\!-\!2}{2}\Lambda g_{\mu\nu} 
= 8\pi G\bigg(T_{\mu\nu}(x) 
         + \big\langle\Delta \hat{T}_{\mu\nu}(x)\big\rangle
         + \sum_{\rm i} T^{\rm ct,i}_{\mu\nu}(x)\bigg)
\,,
\label{eom SCG: case B}
\end{equation}
 is separately conserved, 
\begin{eqnarray}
\nabla^\mu G_{\mu\nu}(g_{\alpha\beta}) &\!\!=\!\!& 0
\,,\qquad
\nabla^\mu g_{\mu\nu} =0
\,,\qquad
\nonumber\\
\nabla^\mu T_{\mu\nu} &\!\!=\!\!& 0
\,,\qquad
\nabla^\mu \big\langle\Delta \hat{T}_{\mu\nu}(x)\big\rangle=0
\,,\qquad
\nabla^\mu T^{\rm ct,i}_{\mu\nu}=0\quad (\forall i)
\,,\qquad
\label{eom SCG: case B separate conservation}
\end{eqnarray}
where $G_{\mu\nu}$ is the Einstein curvature tensor, 
\begin{eqnarray}
T_{\mu\nu} 
= -\frac{2}{\sqrt{-g}}\frac{\delta S_{\rm m}}{\delta g^{\mu\nu}}
\,,\qquad
\label{eom SCG: case B: energy momentum tensor}
\end{eqnarray}
is the classical energy momentum tensor 
(see Eq.~(\ref{energy momentum tensor classical B})),
$S_{\rm m}$ is the classical matter 
action in Eq.~(\ref{classical scalar matter action}),
$\big\langle\Delta \hat{T}_{\mu\nu}(x)\big\rangle$
are the quantum (one-loop) contribution to the energy-momentum
tensors (see Eq.~(\ref{energy momentum tensor: 1 loop B})), and 
$T^{\rm ct,i}_{\mu\nu} 
= -\frac{2}{\sqrt{-g}}\frac{\delta S_{\rm ct,i}}{\delta g^ {\mu\nu}}$
are the energy-momentum tensors induced by the 
counterterm actions $S_{\rm ct,i}$ discussed in Appendices~A and~B.
Each of the conservation laws in 
Eq.~(\ref{eom SCG: case B separate conservation}) can be expanded 
around $\overline{g}^{\mu\nu}$ in powers of $\delta g^{\mu\nu}$ to obtain the following two geometric Noether identities, 
\begin{eqnarray}
&&\hskip -0.5cm
\overline{\nabla}_\alpha
  \big[\,\overline{G}_{\mu\nu}\delta g^{\mu\alpha}\big]
  + \overline{G}^{\mu}_{\;\alpha}
         \Big(\overline{\nabla}_{(\mu}\delta g_{\nu)}^ {\;\alpha}
          - \frac12 \overline{\nabla}^\alpha\delta g_{\mu\nu}\Big) 
\nonumber\\
&&\hskip 1.cm
- 2\overline{\nabla}^\mu\!\!\int {\rm d}^D x' \sqrt{-\overline{g}'}
          \big[{}_{\mu\nu}\overline{\mathcal{L}}_{\rho\sigma}\big](x;x')
       \delta g^{\rho\sigma}(x') 
   +\mathcal{O}\big((\delta g^{\rho\sigma})^2\big) = 0
\,,\qquad
\label{Noether identities: case B: Gmn}\\
&&\hskip -0.5cm
\overline{\nabla}_\alpha
  \big[\overline{g}_{\mu\nu}\delta g^{\mu\alpha}\big]
  + \overline{g}^{\mu}_{\;\alpha}
         \Big(\overline{\nabla}_{(\mu}\delta g_{\nu)}^ {\;\alpha}
          - \frac12 \overline{\nabla}^\alpha\delta g_{\mu\nu}\Big) 
\nonumber\\
&&\hskip 1.cm
- \frac12\overline{\nabla}^\mu\Big[\big(\overline{g}_{\mu\nu}\overline{g}_{\rho\sigma}
\!+\!2\overline{g}_{\mu(\rho}\overline{g}_{\sigma)\nu}\big)
       \delta g^{\rho\sigma}(x) \Big]
   +\mathcal{O}\big((\delta g^{\rho\sigma})^2\big) = 0
\,,\qquad
\label{Noether identities: case B: gmn}
\end{eqnarray}
and the following Noether-Ward identities for the scalar matter field,
\begin{eqnarray}
&&\hskip -0.5cm
\overline{\nabla}_\alpha
  \big[\,\overline{T}_{\mu\nu}\delta g^{\mu\alpha}\big]
  + \overline{T}^{\mu}_{\;\alpha}
         \Big(\overline{\nabla}_{(\mu}\delta g_{\nu)}^ {\;\alpha}
          - \frac12 \overline{\nabla}^\alpha\delta g_{\mu\nu}\Big) 
\nonumber\\
&&\hskip 1.cm
- 2\overline{\nabla}^\mu\!\!\int {\rm d}^D x' \sqrt{-\overline{g}'}
          \big[{}_{\mu\nu}\overline{\mathcal{T}}_{\rho\sigma}\big](x;x')
       \delta g^{\rho\sigma}(x') 
   +\mathcal{O}\big((\delta g^{\rho\sigma})^2\big) = 0
\,,\qquad
\label{Noether identities: case B: Tmn}\\
&&\hskip -0.5cm
\overline{\nabla}_\alpha
  \big[\bigl\langle\Delta\hat{\overline{T}}_{\mu\nu}(x)\bigr\rangle
      \delta g^{\mu\alpha}\big]
  +\bigl\langle \Delta\hat{\overline{T}}^{\mu}_{\;\alpha}(x)\bigr\rangle
         \Big(\overline{\nabla}_{(\mu}\delta g_{\nu)}^ {\;\alpha}
          - \frac12 \overline{\nabla}^\alpha\delta g_{\mu\nu}\Big) 
\nonumber\\
&&\hskip 1.cm
- 2\overline{\nabla}^\mu\!\!\int {\rm d}^D x' \sqrt{-\overline{g}'}
          \big[{}_{\mu\nu}\overline{\Sigma}_{\rho\sigma}\big](x;x')
       \delta g^{\rho\sigma}(x') 
   +\mathcal{O}\big((\delta g^{\rho\sigma})^2\big) = 0
\,,\qquad
\label{Noether identities: case B: Delta Tmn quantum}\\
&&\hskip -0.5cm
\overline{\nabla}_\alpha
  \big[\,\overline{T}^{\rm ct,i}_{\mu\nu}\delta g^{\mu\alpha}\big]
  + \big({\overline{T}^{\rm ct,i}}\big)^{\mu}_{\;\alpha}
         \Big(\overline{\nabla}_{(\mu}\delta g_{\nu)}^ {\;\alpha}
          - \frac12 \overline{\nabla}^\alpha\delta g_{\mu\nu}\Big) 
\nonumber\\
&&\hskip 0.99cm
- 2\overline{\nabla}^\mu\!\!\int {\rm d}^D x' \sqrt{-\overline{g}'}
          \big[{}_{\mu\nu}\overline{\mathcal{T}}^{\rm \,ct,i}_{\rho\sigma}
             \big](x;x')
       \delta g^{\rho\sigma}(x') 
   +\mathcal{O}\big((\delta g^{\rho\sigma})^2\big) = 0
\,,\qquad
\label{Noether identities: case B: Tmn ct i}
\end{eqnarray}
where 
\begin{eqnarray}
 \big[{}_{\mu\nu}\overline{\mathcal{T}}_{\rho\sigma}\big](x;x')
  &\!\!=\!\!& \frac{1}{\sqrt{-\overline{g}}\sqrt{-\overline{g}'}}
\left(\frac{\delta^2 S_{\rm m}}{\delta g^{\mu\nu}(x)\delta g^{\rho\sigma}(x')}\right)_{\! \overline{g}_{\alpha\beta}}
\,,\qquad
\label{mnTrs: case B}\\
 \big[{}_{\mu\nu}\overline{\mathcal{T}}^{\rm ct,i}_{\rho\sigma}\big](x;x')
  &\!\!=\!\!& \frac{1}{\sqrt{-\overline{g}}\sqrt{-\overline{g}'}}
\left(\frac{\delta^2 S_{\rm ct,i}}{\delta g^{\mu\nu}(x)\delta g^{\rho\sigma}(x')}\right)_{\! \overline{g}_{\alpha\beta}}
\quad (\forall i)
\,,\qquad
\label{mnTrs: case B: ct i}
\end{eqnarray}
and $\big[{}_{\mu\nu}\overline{\Sigma}_{\rho\sigma}\big](x;x')$
consists of the three- and four-point contributions given in 
Eqs.~(\ref{3pt contribution to self-energy: total}) 
and~(\ref{4-point contribution self-energy: case B}).
A simple inspection shows that 
Eq.~(\ref{Noether identities: case B: gmn}) is satisfied;
we leave to the reader to show that the other four equations 
in Eqs.~(\ref{Noether identities: case B: Gmn})--(\ref{mnTrs: case B: ct i}) are satisfied. 

A simple comparison of Eqs.~(\ref{towards NW for bg: gravity 2a})--(\ref{towards NW for bg: gravity 2}),
(\ref{towards NW for classical matter 3}),
(\ref{towards NW for quantum matter}) and~(\ref{Noether identity for ct i}) which hold in case~A with 
Eqs.~(\ref{Noether identities: case B: Gmn})--(\ref{mnTrs: case B: ct i})
which hold in case~B shows that 
-- even at the linear level in perturbations 
-- both the equations of motion and the Noether-Ward identities 
for $\delta g_{\mu\nu}$ 
(case~A) and $\delta g^{\mu\nu}$ (case B) are different.


\section{De Sitter space example}
\label{De Sitter space example}

This work is primarily motivated by the evolution of gravitational perturbations in cosmology, and therefore 
in what follows we consider de Sitter space, which is a model space for cosmological inflation,
which is a quasi-exponentially expanding spacetime, {\it i.e.} in which the 
 principal slow roll parameter $\epsilon = -\dot H/H^2$ is small,  $|\epsilon|\ll 1$, where 
 $H(t)=\frac{{\rm d}}{{\rm d}t} \ln\big(a(t)\big)$ 
 is the Hubble expansion rate and $a(t)$ is the scale factor of the Universe.
On flat spatial sections of de Sitter the metric can be written as, 
\begin{equation}
 {\rm d}s^2 = - {\rm d}t^2 + a(t)^2 {\rm d} \vec{x}\!\cdot \!{\rm d} \vec{x}
 \,,\qquad 
 a(t)= a_0\exp\big(Ht\big)
\,.\qquad\;
\label{de Sitter metric}
\end{equation}
De Sitter space is a maximally symmetric space with a constant, positive Ricci
curvature scalar, $\overline{R} = D(D\!-\!1)H^2$, where $H$ is the (constant) Hubble rate,
$\dot H\equiv {\rm d}H/{\rm d}t =0$.
All of the background curvature tensors,
that is the Riemann tensor, Ricci tensor and Einstein tensor, can be expressed in terms of the background metric
tensor and the Ricci scalar as follows,
\begin{eqnarray}
\overline{R}^{\mu\nu\rho\sigma} &\!=\!& H^2\left(\overline{g}^{\mu\rho}\overline{g}^{\nu\sigma}
  - \overline{g}^{\mu\sigma}\overline{g}^{\nu\rho}\right)
      = \frac{\overline{R}}{D(D\!-\!1)}\left(\overline{g}^{\mu\rho}\overline{g}^{\nu\sigma}
  - \overline{g}^{\mu\sigma}\overline{g}^{\nu\rho}\right)
\,,\qquad
\nonumber\\
\overline{R}^{\mu\nu} &\!=\!& (D-1)H^2\overline{g}^{\mu\nu}=\frac{\overline{R}}{D}\overline{g}^{\mu\nu}
\,,\qquad  \overline{R} = D(D\!-\!1)H^2
\nonumber\\
\overline{G}^{\mu\nu} &\!=\!&-\frac{(D\!-\!1)(D\!-\!2)}{2}H^2 \,\overline{g}^{\mu\nu}
     = -\frac{(D\!-\!2)}{2D}\overline{R}\,\overline{g}^{\mu\nu}
\,.\qquad
\label{Ricci and Einstein tensor: de Sitter}
\end{eqnarray}

\bigskip
\noindent
{\bf Classical Noether identity.}
Making use of equalities~(\ref{Ricci and Einstein tensor: de Sitter}) the first line in Eq.~(\ref{towards NW for bg: gravity 2a}) can be recast as,
\begin{eqnarray}
 L_1 &\!\!\equiv\!\!&    \Big(\overline{\nabla}_{(\mu}\delta g_{\nu)}^{\;\alpha}
                       \!-\!\frac12\overline{\nabla}^{\;\alpha}\delta g_{\mu\nu}\Big)
                            \overline{G}^{\mu\nu}
                           = -\frac{(D\!-\!2)}{2D}\,\overline{R}\Big(\overline{\nabla}^\mu\delta g_{\mu}^{\;\alpha}
                       \!-\!\frac12\overline{\nabla}^{\;\alpha}\delta g_{\mu}^{\;\mu}\Big)
\,.\qquad
\label{NW in de Sitter: gravity: L1}
\end{eqnarray}
Lines 2-4  in Eq.~(\ref{towards NW for bg: gravity 2a}) combine as,
\begin{eqnarray}
 L_{2-4} &\!\!\equiv\!\!&  
       -\,
\bigg\{\frac{(D\!-\!4)}{4D}\overline{R}\Big( \overline{g}^{\rho\sigma}\overline{\nabla}^\alpha
                           \!-\!2\overline{g}^{\alpha(\rho}\overline{\nabla}^{\sigma)}\Big)
                           \!+\!\frac12\overline{\nabla}^{\alpha}\overline{\nabla}^\rho\overline{\nabla}^\sigma
  \!+\!\frac12\overline{g}^{\rho\sigma}\overline{\nabla}_\nu\overline{\nabla}^{\alpha}\overline{\nabla}^{\nu} 
\quad
 \nonumber\\
&&\hskip 0.1cm 
   -\,\frac12\overline{g}^{\alpha(\rho}\overline{\nabla}_{\nu}\overline{\nabla}^{\sigma)}\overline{\nabla}^{\nu}  
                  \!-\!\frac12\overline{\nabla}^{(\rho} \overline{\nabla}^{\sigma)}\overline{\nabla}^{\alpha}  
          \!+\!  \frac12\overline{g}^{\alpha(\rho}\overline{\nabla}^{\sigma)}\overline{\dAlembert}
          \!-\! \frac12\overline{g}^{\rho\sigma}  \overline{\nabla}^{\alpha}  \overline{\dAlembert}   
 \bigg\}  \delta g_{\rho\sigma}
\nonumber\\
&&\hskip 0.1cm 
   +\, \mathcal{O}\big(\delta g_{\alpha\beta}^2\big) = 0
\,,\qquad
 \nonumber\\
&&\hskip -0.3cm 
  = \frac{(D\!-\!2)}{2D}\,\overline{R}\Big(\overline{\nabla}^\mu\delta g_{\mu}^{\;\alpha}
                       \!-\!\frac12\overline{\nabla}^{\;\alpha}\delta g_{\mu}^{\;\mu}\Big)
\,,\qquad
\label{NW in de Sitter: gravity: L2-4}
\end{eqnarray}
where to get the last equality we made use of,
\begin{eqnarray}
\frac12\overline{g}^{\rho\sigma}\overline{\nabla}_\nu\overline{\nabla}^{\alpha}\overline{\nabla}^{\nu}
\delta g_{\rho\sigma}  &\!\!=\!\!&  \frac12\overline{g}^{\rho\sigma}\overline{\nabla}^{\alpha}\overline{\dAlembert}
\delta g_{\rho\sigma}
+\frac{\overline{R}}{2D}\overline{g}^{\rho\sigma}\overline{\nabla}^{\alpha}\delta g_{\rho\sigma}
\nonumber\\
  -\frac12\overline{g}^{\alpha\rho}\overline{\nabla}_{\nu}\overline{\nabla}^{\sigma}\overline{\nabla}^{\nu}
   \delta g_{\rho\sigma} &\!\!=\!\!& -\frac12\overline{g}^{\alpha\rho}\overline{\nabla}^{\sigma}
      \overline{\dAlembert}\delta g_{\rho\sigma}
       \!-\!\frac{\overline{R}}{2D(D\!-\!1)}\left(\overline{g}^{\rho\sigma}\overline{\nabla}^{\alpha}
       \!-\!\overline{g}^{\alpha(\rho}\overline{\nabla}^{\sigma)}\right) \delta g_{\rho\sigma} 
\nonumber\\    
- \frac12\overline{\nabla}^{\rho} \overline{\nabla}^{\sigma}\overline{\nabla}^{\alpha} \delta g_{\rho\sigma}  
 &\!\!=\!\!&- \frac12\overline{\nabla}^{\alpha} \overline{\nabla}^{\rho}\overline{\nabla}^{\sigma} \delta g_{\rho\sigma}
 \nonumber\\  
 &&\hskip 0cm
  +\,\frac{\overline{R}}{2D(D\!-\!1)}\left(\overline{g}^{\rho\sigma}\overline{\nabla}^{\alpha}
       \!-\!\big(2D\!-\!1\big)\overline{g}^{\alpha(\rho}\overline{\nabla}^{\sigma)}\right) \delta g_{\rho\sigma} 
       \,.\qquad
\label{NW in de Sitter: gravity: L2-4 b}
\end{eqnarray}
We can now clearly see that adding Eqs.~(\ref{NW in de Sitter: gravity: L1})
and~(\ref{NW in de Sitter: gravity: L2-4}) sums up to zero, up to corrections  
$\mathcal{O}\big(\delta g_{\alpha\beta}^2\big)$, showing that 
the identity~(\ref{towards NW for bg: gravity 2a}) holds in de Sitter.
Now adding to this $\frac{D\!-\!2}{2}\Lambda$ times Eq.~(\ref{towards NW for bg: gravity 2}),
and making use of the classical equation of motion for the background metric,
\begin{equation}
\overline{G}^{\mu\nu} + \frac{D\!-\!2}{2}\Lambda \overline{g}^{\mu\nu} =0
\,,\qquad
\label{classical EoM background metric: dS}
\end{equation}
gives the classical equation of motion for gravitational perturbations on de Sitter background,
\begin{eqnarray}
&&\hskip -0.8cm 
   \int {\rm d}^Dx'\sqrt{-\overline{g}(x')}
   \big[^{\mu\nu}\mathcal{L}^{\rho\sigma}\big](x;x')\delta g_{\rho\sigma}(x')
\nonumber\\
&&\hskip 2.0cm 
   -\, \frac{D\!-\!2}{4}\Lambda
   \left(\overline{g}^{\mu\nu}\overline{g}^{\rho\sigma}
        \!-\! 2\overline{g}^{\mu(\rho}\overline{g}^{\sigma)\nu} \right)\delta g_{\rho\sigma}(x) 
           \!+\! \mathcal{O}\big(\delta g_{\alpha\beta}^2\big) = 0
\,,\qquad
\label{classical EoM metric perturbations: dS}
\end{eqnarray}
which is -- by the Noether identity -- manifestly transverse, 
\begin{eqnarray}
&&\hskip -0.8cm 
   \int\! {\rm d}^Dx'\sqrt{-\overline{g}(x')}\,
   \overline{\nabla}_\mu [^{\mu\nu}\mathcal{L}^{\rho\sigma}]_{\overline{g}_{\alpha\beta}}
   (x;x')\delta g_{\rho\sigma}(x')
\nonumber\\
&&\hskip 2.0cm 
   -\, \frac{D\!-\!2}{4}\Lambda
   \overline{\nabla}_\mu\Big[\left(\overline{g}^{\mu\nu}\overline{g}^{\rho\sigma}
        \!-\! 2\overline{g}^{\mu(\rho}\overline{g}^{\sigma)\nu} \right)\delta g_{\rho\sigma}\Big]
           \!+\! \mathcal{O}\big(\delta g_{\alpha\beta}^2\big) = 0
\,.\qquad
\label{classical Noether for metric perturbations: dS}
\end{eqnarray}
We emphasize that the Lichnerowicz operator by itself is not transverse, but that the non-transverse 
pieces obtained from $\overline{\nabla}_\mu[^{\mu\nu}\mathcal{L}^{\rho\sigma}]_{\overline{g}_{\alpha\beta}}(x;x')$
combine with the non-transverse pieces from the cosmological constant part such that -- when one
makes use of the classical background equation~(\ref{classical EoM background metric: dS}) 
-- they cancel each other out.


The classical Noether identity discussed up to this point can be generalized to include classical 
and quantum matter contributions to the energy-momentum tensor. As long as the classical scalar condensate 
$\overline{\phi}$ and the quantum fluctuations around it do not break de Sitter symmetry, one can include 
them and still the background metric $\overline{g}_{\mu\nu}$ will be that of 
de Sitter space~(\ref{de Sitter metric}). We assume this to be the case in the rest of this paper.

\bigskip
\noindent
{\bf Quantum Noether-Ward identity.} The dynamics on de Sitter space can include 
classical and quantum scalar condensates. However, both must be such that 
the corresponding energy momentum tensors do not violate de Sitter symmetry. This means that 
both classical and quantum contributions must be proportional to the background metric.
This is realized by a space-time independent scalar condensates, for which the energy-momentum tensor
in (the background version of) Eq.~(\ref{energy momentum tensor: classical}) reduces to,
\footnote{
Since the classical scalar satisfies, $\big(\overline{\dAlembert}-m^2-\xi\overline{R}\big)\overline{\phi}=0$,
the condensate can be non-vanishing only when $m^2+\xi\overline{R}=0$, in which case the scalar is massless.
In this case however, the scalar propagator violates de Sitter symmetry, and thus this case 
requires extra care.
%
%
}
\begin{eqnarray}
\overline{T}^{\mu\nu}(x) &\!\!=\!\!& -\frac12
       \bigg[m^2 \!+\!\xi\frac{D\!-\!2}{D}\overline{R}\bigg] \overline{\phi}^2\overline{g}^{\mu\nu}
\,.\qquad
\label{energy momentum tensor: classical dS}
\end{eqnarray}
On the other hand, the (one-loop) quantum contribution can be calculated from the de Sitter invariant 
Chernikov-Tagirov massive scalar propagator~\cite{Chernikov:1968zm},
\begin{eqnarray}
i\Delta_{\overline{\phi}}(x;x') &\!\!=\!\!& \frac{\hbar H^{D-2}}{(4\pi)^{D/2}}
 \frac{\Gamma\big(\frac{D-1}{2}\!+\!\overline{\nu}\big)\Gamma\big(\frac{D-1}{2}\!-\!\overline{\nu}\big)}
     {\Gamma\big(\frac{D}{2}\big)}
\nonumber\\
 && \times\,_{2}F_1\left(\frac{D-1}{2}\!+\!\overline{\nu},\frac{D-1}{2}\!-\!\overline{\nu};
          \frac{D}{2};1\!-\!\frac{y(x;x')}{4}\right)
 \,,\qquad
\label{Chernikov-Tagirov propagator}
\end{eqnarray}
\begin{equation}
\hskip -2.5cm
\overline{\nu} = \sqrt{\left(\frac{D\!-\!1}{2}\right)^{\!2}\!-\!m^2 -\xi\overline{R}}
\,\,,\qquad   \big(m^2 + \xi\overline{R} > 0\big)
\label{nu on dS}
\end{equation}
where 
\begin{eqnarray}
y(x;x') = H^2 a(\eta)a(\eta')\Delta x^2(x;x')
\,,\quad 
\Delta x^2
  = -(|\eta\!-\!\eta'|\!-\!i\varepsilon)^2 \!+\!\|\vec x\!-\!\vec x^{\,\prime}\|^2
\,,\quad
\label{definition of y(x;x') and Delta x2}
\end{eqnarray}
with $\varepsilon>0$ an infinitesimal parameter (introduced to define the imaginary part of the propagator),  $a(\eta)=-1/(H\eta)\;(\eta<0)$
and $\eta$ is conformal time. The  quantum contribution is then,
\begin{eqnarray}
\!\big\langle\Delta \hat T^{\mu\nu}(x)\big\rangle_{\overline{g}_{\alpha\beta},\overline{\phi}}
   \!\!&\!\!=\!\!\!&  \frac{\hbar H^{D-2}}{(4\pi)^{D/2}}
 \frac{\Gamma\big(\frac{D-1}{2}\!+\!\overline{\nu}\big)\Gamma\big(\frac{D-1}{2}\!-\!\overline{\nu}\big)}
     {\Gamma\big(\frac{1}{2}\!+\!\overline{\nu}\big)\Gamma\big(\frac{1}{2}\!-\!\overline{\nu}\big)}
         \Gamma\Big(\!1\!-\!\frac{D}{2}\Big) \bigg[\!-\frac{m^2}{D}\overline{g}^{\mu\nu}
      \bigg]
,\quad\;\;\;
\label{energy momentum tensor: quantum dS}
\end{eqnarray}
where we made use of,
\begin{eqnarray}
i\Delta_{\overline{\phi}}(x;x)  &\!\!=\!\!&  \frac{\hbar H^{D-2}}{(4\pi)^{D/2}}
 \frac{\Gamma\big(\frac{D-1}{2}\!+\!\overline{\nu}\big)\Gamma\big(\frac{D-1}{2}\!-\!\overline{\nu}\big)}
     {\Gamma\big(\frac{1}{2}\!+\!\overline{\nu}\big)
       \Gamma\big(\frac{1}{2}\!-\!\overline{\nu}\big)}\Gamma\Big(1\!-\!\frac{D}{2}\Big)
\,,\qquad 
\label{energy momentum tensor: quantum dS 2a}
\\
\left[\overline{\nabla}^\mu{\overline{\nabla}'}^\nu\! i\Delta_{\overline{\phi}}(x;x')\right]_{x'\rightarrow x}  
\!&\!\!=\!\!& \frac{\hbar H^{D-2}}{(4\pi)^{D/2}}
 \frac{\Gamma\big(\frac{D-1}{2}\!+\!\overline{\nu}\big)\Gamma\big(\frac{D-1}{2}\!-\!\overline{\nu}\big)}
     {\Gamma\big(\frac{1}{2}\!+\!\overline{\nu}\big)\Gamma\big(\frac{1}{2}\!-\!\overline{\nu}\big)}
     \Gamma\Big(\!1\!-\!\frac{D}{2}\Big)
\nonumber\\
&&
  \times\,\bigg[-\frac{m^2\!+\!\xi \overline{R}}{D}\,\overline{g}^{\mu\nu}\bigg]
\,.\qquad\!
\label{energy momentum tensor: quantum dS 2b}
\end{eqnarray}
Notice that the one-loop contribution to the energy-momentum tensor~(\ref{energy momentum tensor: quantum dS})
does not (directly) depend on $\xi$, and that it diverges in $D=4$ since,
\begin{equation}
\Gamma\left(1\!-\frac{D}{2}\right) = \frac{2}{D\!-\!4} +\gamma_E - 1
\,,\qquad
\label{expanding Gamma 1-D/2}
\end{equation}
and it ought to be renormalized.
Here $\gamma_E = -\psi(1) = 0.577\cdots$ denotes the Euler-Mascheroni constant,
and $\psi(z) = \frac{{\rm d}}{{\rm d} z}\ln\left(\Gamma(z)\right)$ is the digamma function.

The semiclassical equation for the background metric is now,
\begin{equation}
\overline{G}^{\mu\nu} + \frac{D\!-\!2}{2}\Lambda \overline{g}^{\mu\nu}
   =\frac{\kappa^2}{2}\left(\overline{T}^{\mu\nu} \!\!+\! \left\langle\Delta \hat{T}^{\mu\nu}\right\rangle_{\overline{g}_{\alpha\beta},\overline{\phi}}
   \!+\!\overline{T}_{\rm ct}^{\mu\nu} 
 \right)
\,,
\label{semiclassical equation: dS}
\end{equation}
where $\overline{T}^{\mu\nu}$ is given in Eq.~(\ref{energy momentum tensor: classical dS}).
Since the quantum contribution in Eq.~(\ref{energy momentum tensor: quantum dS}) is proportional 
to $\overline{g}^{\mu\nu}$, with a coefficient which arespacetime independent, 
and the divergent contributions are $\propto H^0$ and $\propto H^ 2$,
 the counterterm action that renormalizes it is that of the Ricci scalar and cosmological constant,
\begin{equation}
S_{\rm ct}[g_{\mu\nu}] = \int {\rm d}^Dx\sqrt{-g} 
      \left[\delta\left(\frac{1}{\kappa^2}\right)R-(D\!-\!2)\delta\left(\frac{\Lambda}{\kappa^2}\right)\right]
\,,\qquad
\label{cosmological constant counterterm action}
\end{equation}
which contributes to the energy-momentum tensor as,
\begin{equation}
\overline{T}_{\rm ct}^{\mu\nu}
  = \frac{2}{\sqrt{-\overline{g}}} \left(\frac{\delta S_{\rm ct}[g_{\mu\nu}]}{\delta g_{\mu\nu}(x)}
    \right)_{\overline{g}_{\alpha\beta}}
     = - 2\delta\left(\frac{1}{\kappa^2}\right)\overline{G}^{\mu\nu}
      \!-\!(D\!-\!2)\delta\left(\frac{\Lambda}{\kappa^2}\right)\overline{g}^{\mu\nu}
\,.\quad
\label{cosmological constant counterterm energy-momentum tensor}
\end{equation}
Expanding Eq.~(\ref{energy momentum tensor: quantum dS}) around $D=4$ gives,
\begin{eqnarray}
\big\langle\Delta \hat T^{\mu\nu}(x)\big\rangle_{\overline{g}_{\alpha\beta},\overline{\phi}}
   \!&\!\!=\!\!&  -\frac{\hbar m^2}{32\pi^2 }\frac{\mu^{D-4}}{D\!-\!4}
   \times \left[m^2\!+\!\xi D(D\!-\!1)H^2-(D\!-\!2)H^2\right]\overline{g}^{\mu\nu}
\nonumber\\
&&\hskip -3.0cm
    \times \left[1\!+\!\frac{D\!-\!4}{2}\left(\ln\bigg[\frac{H^2}{4\pi\mu^2}\bigg]
    \!+\!\psi\Big(\frac12 \!+\!\overline{\nu}\Big)
    \!+\!\psi\Big(\frac12 \!-\!\overline{\nu}\Big)\!+\!\gamma_E\!-\!\frac32\right)\right]
  \,,\quad\;
\label{energy momentum tensor: quantum dS: expanded}
\end{eqnarray}
which in light of Eq.~(\ref{cosmological constant counterterm energy-momentum tensor}) 
gives for the counterterms in the minimal subtraction scheme (in which only the divergent parts
$\propto 1/(D\!-\!4)$ are subtracted),
\begin{eqnarray}
\delta\left(\frac{\Lambda}{\kappa^2}\right)
    &\!\!=\!\!& -\frac{\hbar m^4}{64\pi^2 }\!\times\!\frac{\mu^{D-4}}{D\!-\!4} 
 \nonumber\\ 
    \delta\left(\frac{1}{\kappa^2}\right)
        &\!\!=\!\!& \frac{\hbar m^2}{16\pi^2 } \left(\xi\!-\!\frac16\right)\!\times\!\frac{\mu^{D-4}}{D\!-\!4}
\,.\qquad
\label{energy momentum tensor: quantum dS: counterterms}
\end{eqnarray}
where we took account of $\overline{G}^{\mu\nu} =-(D\!-\!1)(D\!-\!2)H^2 \overline {g}^{\mu\nu}/2$.
The remaining finite parts can be evaluated in $D=4$, 
\begin{eqnarray}
\big\langle\Delta \hat T^{\mu\nu}(x)\big\rangle_{\overline{g}_{\alpha\beta},\overline{\phi}}
  +\overline{T}_{\rm ct}^{\mu\nu} &\!\!=\!\!&
   -\frac{\hbar m^2}{64\pi^2 }\bigg[m^2\!+\!12H^2\Big(\xi \!-\!\frac16\Big) \bigg] \!
 \nonumber\\ 
 &&\hskip -.1cm 
\times\left[\ln\bigg(\frac{H^2}{4\pi\mu^2}\bigg)
    \!+\!\psi\Big(\frac12 \!+\!\overline{\nu}\Big)
    \!+\!\psi\Big(\frac12 \!-\!\overline{\nu}\Big)\!+\!\gamma_E\!-\!2\right]\!\overline{g}^{\mu\nu}
\nonumber\\
 &&
 +\,\frac{\hbar m^2}{64 \pi^2 }\left(\frac{m^2}{2}\!-\!\frac{H^2}{3}\right)\overline{g}^{\mu\nu}
 .\qquad\;
\label{energy momentum tensor: quantum dS: expanded2}
\end{eqnarray}
Now, noting that the non-tachyonic condition on the mass in the the Chernikov-Tagirov propagator 
$m^2 + \xi\overline{R} > 0$ in Eq.~(\ref{nu on dS}), implies that the scalar condensate $\overline{\phi}$ on de Sitter must  vanish,~\footnote{A constant scalar condensate obeys, 
$\big(\overline{\Box} - m^2 - \xi\overline{R}\big)\overline{\phi}
= - \big(m^2+ \xi\overline{R}\big)\overline{\phi}=0$, which implies that the only solution 
of that equation is $\overline{\phi}=0$.
When $m^2+ \xi\overline{R}\leq 0$, then the propagator acquires de Sitter breaking 
parts~\cite{Onemli:2002hr,Onemli:2004mb},
and that situation must be reconsidered. That is however beyond the scope of this paper,
as our study of de Sitter focuses primarily on how the Noether-Ward 
identity works at the one-loop level in de Sitter space.}, which then implies that 
the classical contribution in Eq.~(\ref{energy momentum tensor: classical dS}) 
must vanish, $\overline{T}^{\mu\nu} =0$. 
With these remarks in mind, and taking account of the result in
 Eq.~(\ref{energy momentum tensor: quantum dS: expanded2}), yields the semiclassical 
Friedmann equation on de Sitter~(\ref{semiclassical equation: dS}) in the form,
\begin{eqnarray}
&&\hskip -0.5cm
H^2\left\{1\!+\! \frac{\hbar \kappa^2 m^2}{32\pi^2}
         \left[\Big(\xi\!-\!\frac16\Big)\left(
     \ln\bigg(\frac{H^2}{4\pi\mu^2}\bigg)
    \!+\!\psi\Big(\frac12 \!+\!\overline{\nu}\Big)
    \!+\!\psi\Big(\frac12 \!-\!\overline{\nu}\Big)\!+\!\gamma_E\!-\!2
    \right) +\frac{1}{36}\right]\right\}
\nonumber\\   
&&\hskip -0.cm
   =\, \frac{\Lambda}{3}
\!+\! \frac{\hbar \kappa^2 m^4}{384\pi^2}
 \left[\left(
     \ln\bigg(\frac{H^2}{4\pi\mu^2}\bigg)
    \!+\!\psi\Big(\frac12 \!+\!\overline{\nu}\Big)
    \!+\!\psi\Big(\frac12 \!-\!\overline{\nu}\Big)\!+\!\gamma_E\!-\!\frac52
    \right)\right]
\,,\quad
\label{semiclassical equation: dS: 2}
\end{eqnarray}
where we moved all the terms that are $\propto H^2$ to the left-hand-side.
Equation~(\ref{semiclassical equation: dS: 2}) is a transcendental equation for $H^2$ as a function of
$\Lambda, m^2,\xi$ and $\mu^2$, and cannot be solved in a closed form. However, when 
\begin{equation}
\frac{\hbar \kappa^2 m^2}{32\pi^2}\Big(\xi\!-\!\frac16\Big) \ll1 
\,,\qquad
\frac{\hbar \kappa^2 m^2}{32\pi^2}\frac{1}{36} \ll1 
\,,\qquad
\frac{\hbar \kappa^2 m^4}{128\pi^2} \ll \Lambda
\,,\quad
\label{cosmological constant counterterm energy-momentum tensor: validity condition}
\end{equation}
are all satisfied (basically when $m^2,\Lambda \ll M_{\rm P}^2$ (where $M_{\rm P}^2\equiv \hbar /(8\pi G)$), 
and $\xi$ is not too large),
one can do perturbative expansion of $H$ around its classical value, $H_{(0)}=\sqrt{\Lambda/3}$,
\begin{equation}
H_{(0)}^2  = H_{(0)}^2 \!+\! \hbar H_{(1)}^2 \!+\! \hbar^2 H_{(2)}^2 \!+\!  {\mathcal{O}}(\hbar^3)
\,,\quad
\label{perturbative solution for H}
\end{equation}
such that~(\ref{semiclassical equation: dS: 2}) gives at order $\hbar$,
\begin{eqnarray}
&&\hskip -0.65cm
H_{(1)}^2
   =  \frac{\kappa^2 m^4}{384\pi^2}
 \left[
     \ln\bigg(\frac{H_{(0)}^2}{4\pi\mu^2}\bigg)
    \!+\!\psi\Big(\frac12 \!+\!\overline{\nu}(H_{(0)})\Big)
    \!+\!\psi\Big(\frac12 \!-\!\overline{\nu}(H_{(0)})\Big)\!+\!\gamma_E\!-\!\frac52
   \right]
 \nonumber\\   
&&\hskip 0.35cm   
-\,\frac{\kappa^2 m^2H_{(0)}^2 }{32\pi^2}
         \Bigg[\frac{1}{36}
          \nonumber\\   
&&\hskip 0.35cm 
         +\Big(\xi\!-\!\frac16\Big)\left(
     \ln\bigg(\frac{H_{(0)}^2}{4\pi\mu^2}\bigg)
    \!+\!\psi\Big(\frac12 \!+\!\overline{\nu}(H_{(0)})\Big)
    \!+\!\psi\Big(\frac12 \!-\!\overline{\nu}(H_{(0)})\Big)\!+\!\gamma_E\!-\!2
    \right) \Bigg]
\,.
 \nonumber\\   
&&\hskip -0.cm 
\label{semiclassical equation: dS: 3}
\end{eqnarray}
Note that $H_{(1)}^2$ is dependent on scale  $\mu$, which was introduced through 
the counterterms~(\ref{energy momentum tensor: quantum dS: counterterms}) 
needed to renormalize the effective action. This scale dependence can be 
removed by RG-improving the effective action. Since RG improvement is not the main focus of this article,
here we satisfy ourselves by remarking that one can make a convenient choice of scale $\mu$.
Choosing, for example, $\mu^2 = H_{(0)}^2/(4\pi)$ removes the logarithms from
Eq.~(\ref{semiclassical equation: dS: 3}), making the calculation of $H_{(i)}\;(i\geq 1)$ easier.
Finally, we note that when $m^2 + 12\xi H^2 > 9/4$, $\overline{\nu}$ 
in Eqs.~(\ref{energy momentum tensor: quantum dS: expanded2}), (\ref{semiclassical equation: dS: 2})
and~(\ref{semiclassical equation: dS: 3}) becomes imaginary.
These expressions remain valid however,  since 
$\psi\Big(\frac{1}{2}\!+\!\overline{\nu}\Big)+\psi\Big(\frac{1}{2}\!-\!\overline{\nu}\Big)$ 
stays real for imaginary $\overline{\nu}$.

\medskip
To complete the analysis of the quantum case, we need to consider three more contributions
to the perturbations, namely the classical contribution~(\ref{towards NW for classical matter 3}),
quantum contributions~(\ref{towards NW for quantum matter}) and 
the counterterm contributions~(\ref{Noether identity for ct i})--(\ref{energy momentum tensor ct i}).

\medskip

Let us first consider the classical contribution in Eq.~(\ref{towards NW for classical matter 3}),
which after some manipulations reduces to,
\begin{eqnarray}
&&\hskip -0.6cm
 \Big(\overline{\nabla}^\mu\delta g_\mu^{\;\alpha}
               \!-\!\frac12\overline{\nabla}^{\alpha}\delta g_{\mu}^{\;\mu}\Big)
                     \frac12\bigg[m^2 \!+\!\xi\frac{D\!-\!2}{D}\,\overline{R}\bigg] \overline{\phi}^2
\qquad
\nonumber\\
&&\hskip 1.cm
 +\,\frac12\bigg[\Big(\frac12\overline{\nabla}^{\alpha}\delta g_{\mu}^{\;\mu}
           \!-\!\overline{\nabla}^{\mu}\delta g_{\mu}^{\;\alpha}
                      \Big)\bigg(\!m^2\!+\!\xi\frac{D\!-\!2}{D}\overline{R}\bigg)\overline{\phi}^2
  \bigg]     \!+\!\mathcal{O}\big(\delta g_{\alpha\beta}^2\big) \simeq 0
  \,,\qquad\;\;
\label{towards NW for classical matter 3: dS}
\end{eqnarray}
where to get this result we made use of $\overline{\phi}={\rm const.}$ and of 
Eqs.~(\ref{NW in de Sitter: gravity: L2-4 b}).

\medskip

Next we consider the quantum contribution in Eq.~(\ref{towards NW for quantum matter}), 
which should satisfy its own Noether-Ward identity,
\begin{eqnarray}
&&\hskip -0.6cm
-\, \frac{\kappa^2}{2}\Big(\overline{\nabla}_{(\mu}\delta g_{\nu)}^{\;\alpha}
                       \!-\!\frac12\overline{\nabla}^{\;\alpha}\delta g_{\mu\nu}\Big)
\Big[\big\langle\Delta \hat T^{\mu\nu}(x)\big\rangle\Big]_{\overline{g}_{\mu\nu},\overline{\phi}}
\,\qquad
\nonumber\\
&&\hskip -0.cm
 +\,\int {\rm d}^Dx'\sqrt{-g'}\delta_\nu^\alpha\overline{\nabla}_\mu 
 \Big(\big[^{\mu\nu} \overline{\Sigma}_{\rm 3pt}^{\rho\sigma}\big](x;x')
       \!+\!\big[^{\mu\nu} \overline{\Sigma}_{\rm 4pt}^{\rho\sigma}\big](x;x')\Big)
       \delta g_{\rho\sigma}(x')
\nonumber\\
&&\hskip -0.cm
       +\,\mathcal{O}\big(\delta g_{\alpha\beta}^2\big)
       = 0
\,,\quad
\label{towards NW for quantum matter: dS}
\end{eqnarray}
where $-i\big[^{\mu\nu}\Sigma_{\rm 3pt}^{\rho\sigma}\big](x;x')$
and $-i\big[^{\mu\nu}\Sigma_{\rm 4pt}^{\rho\sigma}\big](x;x')$ denote the three- and four-point contributions
to the graviton self-energy given in Eqs.~(\ref{3pt contribution to self-energy: total}) 
and~(\ref{4-point contribution self-energy: case A}), respectively.

Firstly, the divergence of the three-point diagram in general curved
spacetimes evaluates to, 
\begin{eqnarray}
\overline{\nabla}_\mu\big[{}^{\mu\nu}\overline{\Sigma}^{\rho\sigma}_{\mbox{\tiny 3pt}} \big](x;x')
       \!&\!\!\!=\!\!\!&\! \frac{\kappa^2}{4}
         \bigg\{ \bigg[2\Big(\overline{\nabla}^{\nu} {\overline{\nabla}'}^{(\sigma}\!i\Delta_{\overline{\phi}}(x;x')\Big)
               {\overline{\nabla}'}^{\rho)}
           \!\!\!-\! \overline{g}^{\rho\sigma}\Big( \overline{\nabla}^{\nu} {\overline{\nabla}'}^{\gamma}
               \! i\Delta_{\overline{\phi}}(x;x')\Big)
               {\overline{\nabla}'}_{\!\!\gamma}\!
\nonumber\\
 && \hskip -3.6cm
      -\, m^2\Big(\overline{g}^{\rho\sigma}\overline{\nabla}^{\nu}i\Delta_{\overline{\phi}}(x;x')\Big)\bigg]
           \frac{\delta^D(x\!-\!x')}{\sqrt{-\overline{g}}}
           \!+\! 2\xi {\mathcal{G}'}^{\rho\sigma}
     \bigg[\frac{\delta^D(x\!-\!x')}{\sqrt{-\overline{g}}}\overline{\nabla}^{\nu}\!i\Delta_{\overline{\phi}}(x;x')
       \bigg]\!
           \bigg\}
,\quad\;\;
\label{NW for quantum matter: divergence 3pt diagram: dS}
\end{eqnarray}
which in de Sitter simplifies to,
\begin{eqnarray}
\overline{\nabla}_\mu\big[ {}^{\mu\nu}\overline{\Sigma}^{\rho\sigma}_{\mbox{\tiny 3pt}} \big](x;x')
       &\!\!\!=\!\!\!& \kappa^2i\Delta_{\overline{\phi}}(x;x)
        \frac{m^2\!+\!\xi\overline{R}}{2D}\Big(\overline{g}^{\nu(\sigma}\overline{\nabla}^{\rho)}
           \!\!\!-\! \frac12\overline{g}^{\rho\sigma}\overline{\nabla}^{\nu}\Big)  
           \frac{\delta^D(x\!-\!x')}{\sqrt{-\overline{g}}}
\,.\qquad\;\;
\label{NW for quantum matter: divergence 3pt diagram: dS2}
\end{eqnarray}
The four point contribution~(\ref{4-point contribution self-energy: case A}) evaluates in de Sitter,
\begin{eqnarray}
 \Big\{\big[{}^{\mu\nu}
                \overline{\Sigma}_{\mbox{\tiny 4pt}}^{\rho\sigma}\big](x;x')\Big\}_{\xi=0, \rm dS}
        &\!\!=\!\!& \frac{\kappa^2}{4}
        \bigg\{\!
            \bigg[\frac{4m^2-(D\!-\!4)\xi\overline{R}}{2D}\Big(\overline{g}^{\mu\nu}\overline{g}^{\rho\sigma}
               \!-\!2\overline{g}^{\mu)(\rho}\overline{g}^{\sigma)(\nu}\Big)
\nonumber\\
&&\hspace{-0.cm}
          \times i\Delta_{\overline{\phi}}(x;x) \bigg]\frac{\delta^D(x\!-\!x')}{\sqrt{-\overline{g}}}
        \bigg\}
\,,\;\qquad\;
\label{4-point contribution self-energy: xi neq 0 dS}\\
 \Big\{\big[{}^{\mu\nu}
                \overline{\Sigma}_{\mbox{\tiny 4pt}}^{\rho\sigma}\big](x;x')\Big\}_{\xi\neq 0, \rm dS}
        &\!\!=\!\!& \frac{\kappa^2}{4}\xi
        \bigg\{\bigg[\frac{D\!-\!4}{2D}\overline{R}\Big(\overline{g}^{\mu\nu}\overline{g}^{\rho\sigma} 
                     \!-\!2\overline{g}^{\mu(\rho}\overline{g}^{\sigma)\nu}\Big)
\nonumber\\
&&\hspace{-3cm}  
                    \!+\,
                   \Big(\overline{g}^{\rho\sigma} \overline{\nabla}^\mu\overline{\nabla}^\nu
                                        \! \!\!+\!\overline{g}^{\mu\nu}\overline{\nabla}^\rho\overline{\nabla}^\sigma\Big)
                  \!-\!2{\overline{g}}^{\mu)(\rho}\overline{\nabla}^{\sigma)}\overline{\nabla}^{(\nu}
                  \!-\!\Big(\overline{g}^{\mu\nu}\overline{g}^{\rho\sigma}
                    \!-\!{\overline{g}}^{\mu)(\rho}\overline{g}^{\sigma)(\nu}\Big)\overline{\dAlembert}
           \bigg]
 \nonumber\\
&&\hspace{-3cm} 
                 \times
               \bigg[i\Delta_{\overline{\phi}}(x;x)\frac{\delta^D(x\!-\!x')}{\sqrt{-\overline{g}}}\bigg]
        \bigg\}
\,.\;\qquad\;
\label{4-point contribution self-energy: xi=0 dS}
\end{eqnarray}
Taking a covariant derivative then gives,
\begin{eqnarray}
 \Big\{\overline{\nabla}_\mu\big[{}^{\mu\nu}
                \overline{\Sigma}_{\mbox{\tiny 4pt}}^{\rho\sigma}\big](x;x')\Big\}_{\xi=0, \rm dS}
        &\!\!=\!\!& \frac{\kappa^2}{4}
        i\Delta_{\overline{\phi}}(x;x)
          \frac{-4m^2\!+\!(D\!-\!4)\xi\overline{R}}{D}
                \nonumber\\
&&\hspace{-0cm} 
\times \Big(
               \overline{g}^{\nu(\rho}\overline{\nabla}^{\sigma)}
               \!-\!\frac12\overline{g}^{\rho\sigma}\overline{\nabla}^{\nu}\Big)\frac{\delta^D(x\!-\!x')}{\sqrt{-\overline{g}}}
,\;
\label{4-point contribution self-energy: xi=0 dS B}\\
 \Big\{\overline{\nabla}_\mu\big[{}^{\mu\nu}
                \overline{\Sigma}_{\mbox{\tiny 4pt}}^{\rho\sigma}\big](x;x')\Big\}_{\xi\neq 0, \rm dS}
        &\!\!=\!\!& \frac{\kappa^2}{4} i\Delta_\phi(x;x)
        \bigg[-\frac{D\!-\!2}{D}\xi\overline{R}
           \bigg]
 \nonumber\\
&&\hspace{0.cm} 
                 \times
             \Big(
               \overline{g}^{\nu(\rho}\overline{\nabla}^{\sigma)}
               \!-\!\frac12\overline{g}^{\rho\sigma}\overline{\nabla}^{\nu}\Big)\frac{\delta^D(x\!-\!x')}{\sqrt{-\overline{g}}}
.\;
\label{4-point contribution self-energy: xi new 0 dS B}
\end{eqnarray}
Summing the two contributions and inserting the result into Eq.~(\ref{towards NW for quantum matter: dS}) yields, 
\begin{eqnarray}
\!\int\! {\rm d}^Dx'\sqrt{-g'}\delta_\nu^\alpha\overline{\nabla}_\mu 
\big[^{\mu\nu} \overline{\Sigma}_{\rm 4pt}^{\rho\sigma}\big](x;x')
       \delta g_{\rho\sigma}(x')
       &\!\!=\!\!&\kappa^2
 i\Delta_{\overline{\phi}}(x;x)
          \frac{-2m^2\!-\!\xi\overline{R}}{2D}
                \nonumber\\
&&\hspace{-.5cm} 
\times \Big(\overline{\nabla}^{\mu}\delta g_\mu^{\;\alpha}(x)
               \!-\!\frac12\overline{\nabla}^{\alpha}\delta g_\mu^{\;\mu}(x)\Big)
.\qquad\;
\label{towards NW for quantum matter: dS: 4pt contribution}
\end{eqnarray}
Adding to this the three-point contribution from
Eq.~(\ref{NW for quantum matter: divergence 3pt diagram: dS2}) yields 
the total non-transversality from the self-energy,
\begin{eqnarray}
&&\hskip -0.7cm
\!\int\! {\rm d}^Dx'\sqrt{-g'}\delta_\nu^\alpha
   \overline{\nabla}_\mu\big[ {}^{\mu\nu}\overline{\Sigma}^{\rho\sigma}_{\mbox{\tiny 3pt}} \big](x;x')
        \delta g_{\rho\sigma}(x')
                \nonumber\\
&&\hspace{3.5cm}
   =\, \kappa^2i\Delta_{\overline{\phi}}(x;x)
        \frac{-m^2}{2D} \Big(\overline{\nabla}^{\mu}\delta g_\mu^{\;\alpha}(x)
               \!-\!\frac12\overline{\nabla}^{\alpha}\delta g_\mu^{\;\mu}(x)\Big)
\,.\qquad\;\;
\label{NW for quantum matter: divergence total self-energy: dS}
\end{eqnarray}
When the first line in identity~(\ref{towards NW for quantum matter: dS}) is added 
to~(\ref{NW for quantum matter: divergence total self-energy: dS}),
\begin{eqnarray}
&&\hskip -0.6cm
-\, \frac{\kappa^2}{2}\Big(\overline{\nabla}_{(\mu}\delta g_{\nu)}^{\;\alpha}
                       \!-\!\frac12\overline{\nabla}^{\;\alpha}\delta g_{\mu\nu}\Big)
\Big[\big\langle\Delta \hat T^{\mu\nu}(x)\big\rangle\Big]_{\overline{g}_{\mu\nu},\overline{\phi}}
                \nonumber\\
&&\hspace{4.0cm} 
 =\kappa^2\frac{m^2}{2D}i\Delta_{\overline{\phi}}(x;x)\Big(\overline{\nabla}^{\mu}\delta g_{\mu}^{\;\alpha}
                       \!-\!\frac12\overline{\nabla}^{\;\alpha}\delta g_{\mu}^{\;\mu}\Big)
\,,\qquad\;
\label{towards NW for quantum matter: dS: line 1}
\end{eqnarray}
one gets {\it zero}, thus proving identity~(\ref{towards NW for quantum matter: dS}) for de Sitter.
Note that this proof holds in general $D$ dimensions for the primitive self-energy, which 
diverges as $\propto 1/(D-4)$ in $D=4$. Therefore, to complete the analysis we also need to consider 
the counterterms, each of which satisfies its own Noether identity.

\medskip
The counterterms needed to renormalize the one-loop contributions in de Sitter space 
are that the Ricci scalar (which renormalizes the Newton constant) and that of the cosmological constant. 
The counterterm action is given in Eq.~(\ref{cosmological constant counterterm action}) 
and~(\ref{energy momentum tensor: quantum dS: counterterms}) and the energy-momentum 
tensor is given in Eq.~(\ref{cosmological constant counterterm energy-momentum tensor}).
The corresponding Noether identites were already discussed in the beginning of this section
(when we discussed the classical Noether identity in de Sitter) 
in Eqs.~(\ref{NW in de Sitter: gravity: L1})--(\ref{classical Noether for metric perturbations: dS}),
just with different values of the coupling constants $1/\kappa^2$ and $\Lambda/\kappa^2$, 
and therefore we shall not repeat that analysis here. 
This completes our discussion of the Noether-Ward identities in de Sitter space.

\medskip

\section{Discussion}
\label{Discussion}

In this paper we consider the Noether-Ward identities for semiclassical gravity 
on general curved spaces. We limit our analysis to the case of Hilbert-Einstein gravity and 
our matter field is a massive non-minimally coupled 
real scalar field which interacts only gravitationally. The classical gravitational and matter 
actions are given in Eqs.~(\ref{Hilbert-Einstein classical action}) 
and~(\ref{classical scalar matter action})
and the corresponding counterterm actions are in
Eqs.~(\ref{Hilbert-Einstein counterterm action})--(\ref{gravitational counterterm action2}) 
and~(\ref{counterterm scalar matter action}), 
which are needed to renomalize the one-loop matter contributions
to the energy-momentum tensor~(\ref{energy momentum tensor: 1 loop A}) and the graviton 
self-energy~(\ref{3pt contribution to self-energy: total})--(\ref{4-point contribution self-energy: case A}).

The main results of this work are presented in 
section~\ref{Noether-Ward identities for gravitational perturbations}.
We begin the section by showing that for semiclassical gravity on a general gravitational
background $g_{\mu\nu}(x)$. The Noether theorem implies transversality (covariant conservation) 
of every classical term in the equation of motion of semiclassical gravity~(\ref{eom SCG}),
including each of the counterterms~(\ref{covariant energy-momentum conservation: parts})
 needed to renormalize the one-loop contribution~(\ref{energy momentum tensor: 1 loop A}),
and show that the quantum one-loop contribution is transverse, up to an ultra-local term
$\propto \left\{\nabla^\nu \left[\delta^2(x\!-\!x')/\sqrt{-g}\right]\right\}_{x'\rightarrow x}$
(see Eq.~(\ref{energy momentum tensor: quantum: conservation B})),
which vanishes in dimensional regularization. By expanding these individual Noether(-Ward)
identities around some background metric $\overline{g}_{\mu\nu}$ to linear order in general gravitational 
perturbations $\delta g_{\mu\nu} = g_{\mu\nu}-\overline{g}_{\mu\nu}$, we then show that 
each of the identities for semiclassical gravity generates the corresponding identity for 
gravitational perturbations, with non-transversal pieces that 
can be expressed as the perturbation of the connection $\delta\Gamma^\alpha_{\mu\nu}$
({\it i.e.} in terms of first derivatives of the metric perturbation)
and quantities that appear in the equation of motion of semiclassical gravity for the background metric, see 
Eqs.~(\ref{towards NW for bg: gravity 2a})--(\ref{towards NW for bg: gravity 2}),
(\ref{towards NW for classical matter 3}),
(\ref{towards NW for quantum matter}) and~(\ref{Noether identity for ct i}).
Even though we have shown that these identities hold to linear order in $\delta g_{\mu\nu}$, 
with some extra work they can be extended to include higher orders in $\delta g_{\mu\nu}$.
When the identities for each individual term are summed as they appear in the 
equation of motion for gravitational perturbations, 
the total non-transverse contribution can be expressed as a first order perturbation of the Christoffel connection
multiplied by the equation of motion of semiclassical gravity for the background metric, which vanishes, proving
that the equation of motion for gravitational 
perturbations~(\ref{Noether-Ward identity for gravitational perturbations B}) is covariantly transverse.
In section~\ref{De Sitter space example} we use our results from 
section~\ref{Noether-Ward identities for gravitational perturbations} and show by explicit calculations that 
they hold in de Sitter space.
Finally, in Appendices~A and~B we present the derivations of 
the first and second variation of the Hilbert-Einstein action and the 
geometric counterterm actions.

Even though in this work we have focused on the {\it in-out} formalism, in which 
the graviton self-energy is built from the the in-out vertices and the Feynman (time ordered) propagator,
our results are easily generalized to the {\it in-in} -- or Schwinger-Keldysh -- formalism, 
which is the formalism of choice for non-equilibrium situations such as cosmology. 
In this formalism, the Feynman propagator $i\Delta_\phi(x;x')$ is replaced by 
the Keldysh propagator $i\left[{}^a\Delta^b_\phi\right](x;x')$, where $a,b=\pm$ are Keldysh indices,
and each vertex gets a Keldysh polarity $a=\pm$, see 
Eqs.~(\ref{Wightman functions})--(\ref{Dyson propagator equation of motion}).
The classical transversality conditions are the conditions which relate transversality of classical two-point vertices
to classical one-point functions, and these remain unchanged.
Next, keeping in mind that 
the one-point functions satisfy,
$\langle\Delta\hat{T}_{+}^{\mu\nu}(x)\rangle=\langle\Delta\hat{T}_{-}^{\mu\nu}(x)\rangle
\equiv\langle\Delta\hat{T}^{\mu\nu}(x)\rangle$, one concludes that 
the quantum transversality conditions for the graviton self-energy remain identical for 
the $\{a,b\}=\{+,+\}$ self-energy $-i\left[^{+}_{\mu\nu}\Sigma^{+}_{\rho\sigma}\right](x;x')
\equiv -i\left[_{\mu\nu}\Sigma_{\rho\sigma}\right](x;x')$,
 for $-i\left[^{-}_{\mu\nu}\Sigma^{-}_{\rho\sigma}\right](x;x')$
the one-point contributions acquire an opposite sign, and finally
$-i\left[^{+}_{\mu\nu}\Sigma^{-}_{\rho\sigma}\right](x;x')$
and $-i\left[^{-}_{\mu\nu}\Sigma^{+}_{\rho\sigma}\right](x;x')$ are covariantly transverse.

The results of this work have a wide range of potential applications. They can be used to check
the validity of calculations of the one-loop graviton self-energy in expanding cosmological backgrounds
such as radiation era~\cite{Ota:2022xni,Ota:2023iyh,Frob:2025sfq,Sasaki:2025zao,Glavan:2026wwl,LiuProkopec:2026,FennemaProkopec:2026,BaghouzianProkopec:2026}
and matter era~\cite{ProkopecVecchioni:2026}. 
These calculations are important for our understanding of the evolution of cosmological perturbations
in the early Universe settings. For example, if cosmological gravitational perturbations are generated
in inflation, which predates the hot Big Bang expansion, then it is of an essential importance
to understand how these perturbations evolve in the background of quantum matter fields in an approximately 
thermal state. Furthermore, our analysis can be used in studies of how gravitational perturbations are 
affected by quantum fluctuating matter fields in black holes backgrounds.

A further important question we address in this paper 
is how different (inequivalent) definitions of the metric perturbations influence their evolution and the 
Noether-Ward identities that they must satisfy. 
To illustrate our point, we consider two definitions of the metric 
perturbations: $\kappa\delta g_{\mu\nu} = g_{\mu\nu}\!-\!\overline{g}_{\mu\nu}$ (case~A) and $\kappa\delta g^{\mu\nu} = g^{\mu\nu}\!-\!\overline{g}^{\mu\nu}$ (case~B).
We first observe that the corresponding Lichnerowicz 
operators in Eqs.~(\ref{second variation of the Hilbert-Einstein action: Lichnerowicz}) and~(\ref{second variation of the Hilbert-Einstein action: Lichnerowicz: case B}) and 
the graviton self-energies in Eqs.~(\ref{3pt contribution to self-energy: total}) and
(\ref{4-point contribution self-energy: case A})--(\ref{4-point contribution self-energy: case B})  
differ, which has as a consequence 
that the evolution of metric perturbations depends on how 
one defines them. Second, we find that the Noether-Ward identities 
obeyed by the metric perturbations for case~A 
are given in Eqs.~(\ref{towards NW for bg: gravity 2a})--(\ref{towards NW for bg: gravity 2}),
(\ref{towards NW for classical matter 3}),
(\ref{towards NW for quantum matter}) and~(\ref{Noether identity for ct i}). These identities differ from the Noether-Ward identities in case~B,
which are given in
Eqs.~(\ref{Noether identities: case B: Gmn})--(\ref{mnTrs: case B: ct i}).

 The analysis presented here can be generalized in several ways. Firstly, in this work 
 gravitational perturbations were assumed to be general, but classical. Instead, one can quantize gravitational
 perturbations, and characterize them by the gravitational two-point functions. In this case the starting 
 point would be the two- (or higher-)loop 2PI effective action, from which one would obtain
 the equation of motion for semiclassical gravity (which would contain both scalar and graviton one-
 and two-loop 
 contributions) and the equation of motion for the scalar matter-field and graviton two-point function.
 A proper analysis of the latter would have to include gauge fixing and ghost contributions.
 Finally, the analysis can be generalized to include other matter fields that contribute to the standard model,
 such as fermionic and (Abelian and non-Abelian) vector fields.


\section{Appendices}



\section*{Appendix A: Variation of the Hilbert-Einstein action}
\label{Appendix A: Variation of the Hilbert-Einstein action}

In this appendix we show the main steps leading to the second variation of the Hilbert-Einstein action,
\begin{equation}
S_{\rm HE}\big[g_{\mu\nu}\big] = \frac{1}{\kappa^2}\int {\rm d}^Dy\sqrt{-g(y)}
\big[ R(y) - (D\!-\!2)\Lambda\big]
\,,\quad (\kappa^2 = 16\pi G)
\,.\quad
\label{Hilbert-Einstein action}
\end{equation}
It is convenient to organize the calculation by introducing a small parameter 
$\epsilon$, which will be at the end of the calculation set to one. 
For notational simplicity in this Appendix and in Appendix~B
we define the metric perturbation as $\delta g_{\mu\nu}$, 
which is to be contrasted with the main text, in which 
the metric perturbation is defined as $\kappa \delta g_{\mu\nu}$.

\bigskip
\noindent
{\bf Case~A.} Thus we have for Case~A in Eq.~(\ref{metric perturbations: case A}),
\begin{eqnarray}
g_{\mu\nu} &\!\!=\!\!& \overline{g}_{\mu\nu}\!+\!\epsilon\delta g_{\mu\nu}
\,,\qquad
\label{variation gmn}\\
\sqrt{-g} &\!\!=\!\!&\sqrt{-\overline{g}}\left\{1\!+\!\epsilon\left[\frac12 \delta g_\alpha^{\;\alpha}\right]
                  \!+\!\epsilon^2\left[\frac18 \big(\delta g_\alpha^{\;\alpha}\big)^2
                   \!-\!  \frac14 \big(\delta g^{\mu\nu}\delta g_{\mu\nu}\big)\right]
                    \!+\!\mathcal{O}\big(\epsilon^3\big) \right\}
,\qquad
\label{variation root g}\\
g^{\mu\nu}  &\!\!=\!\!&\overline{g}^{\mu\nu}\!-\!\epsilon\delta g^{\mu\nu}
 \!+\!\epsilon^2\delta g^{\mu\rho}\delta g_{\rho}^{\;\nu}
  \!+\!\mathcal{O}\big(\epsilon^3\big) 
 \, ,\qquad
\label{variation inverse gmn}\\
R_{\mu\nu}  &\!\!=\!\!&\overline{R}_{\mu\nu}\!+\!\epsilon\delta R_{\mu\nu}
 \!+\!\epsilon^2\delta^2 R_{\mu\nu}
  \!+\!\mathcal{O}\big(\epsilon^3\big) 
  \,,\qquad
\label{variation Rmn}
\end{eqnarray}
where 
\begin{equation}
\overline{R}_{\mu\nu} = \partial_\rho   \overline{\Gamma}_{\nu\mu}^\rho
\!-\!\partial_\nu   \overline{\Gamma}_{\rho\mu}^\rho
\!+\!\overline{\Gamma}_{\rho\lambda}^\rho\overline{\Gamma}_{\nu\mu}^\lambda
\!-\!\overline{\Gamma}_{\nu\lambda}^\rho\overline{\Gamma}_{\rho\mu}^\lambda
\,,\qquad
\label{background Ricci tensor}
\end{equation}
denotes the background Ricci tensor. With these in mind we can write, 
\begin{eqnarray}
\sqrt{-g}R &\!\!=\!\!& \sqrt{-\overline{g}}\overline{R}
   \!+\!\epsilon \sqrt{-\overline{g}}\left[\frac12 \delta g_\rho^{\;\rho} \overline{R}
        \!-\! \delta g^{\mu\nu} \overline{R}_{\mu\nu}  
             \!+\!  \overline{g}^{\mu\nu}\delta R_{\mu\nu} \right]
\nonumber\\
&&\hskip -.0cm
 +\,\epsilon^2 \sqrt{-\overline{g}}\bigg[\Big(\frac18 (\delta g_\rho^{\;\rho}\big)^2
           \!-\!\frac14\delta g^{\mu\nu} \delta g_{\mu\nu}\Big)\overline{R}
        \!+\!\frac12 \delta g_{\rho}^{\;\rho}\Big( \overline{g}^{\mu\nu}\delta R_{\mu\nu}
               \!-\!\delta g^{\mu\nu}\overline{R}_{\mu\nu}\Big)
\nonumber\\
&&\hskip 2cm  
             +\, \Big(\delta g^{\mu\rho}\delta g_{\rho}^{\;\nu}\overline{R}_{\mu\nu}
                \!-\!\delta g^{\mu\nu}\delta R_{\mu\nu}
                   \!+\!\overline{g}^{\mu\nu}\delta^2 R_{\mu\nu}\Big)
                     \bigg]
\,.\quad
\label{expansion of sqrt g R}
\end{eqnarray}
The first and second variation of the cosmological constant term 
are obtained from variation of the integral measure $\sqrt{-g}$ 
given in Eq.~(\ref{variation root g}). Variation of the Ricci scalar term
is more complicated.
The first variation of the connection and Ricci tensor can be easily derived, 
\begin{eqnarray}
\delta \Gamma_{\alpha\beta}^\rho &\!\!=\!\!& 
           \overline{\nabla}_{(\alpha} \delta g_{\beta)}^\rho 
            -\frac12 \overline{\nabla}^\rho \delta g_{\alpha\beta}
\,,\qquad \delta \Gamma_{\rho\beta}^\rho = 
           \frac12\overline{\nabla}_{\beta}\big( \delta g_{\rho}^{\;\rho}\big) 
\,,\quad 
\label{connection: first variation}\\
\delta R_{\alpha\beta} &\!\!=\!\!&  
  \overline{\nabla}_{\rho} \overline{\nabla}_{(\alpha} \delta g_{\beta)}^\rho 
            -\frac12 \big(\overline{\dAlembert} \delta g_{\alpha\beta}
              +\overline{\nabla}_{(\alpha}\overline{\nabla}_{\beta)}
              \delta g_{\rho}^{\;\rho}
            \big)
\,.\quad
\label{Ricci tensor: first variation}
\end{eqnarray}
The second variation is more tricky. First observe that the second variation of the 
Ricci tensor can be written as, 
\begin{equation}
\delta^2 R_{\mu\nu} = \overline{\nabla}_\rho \delta^2 \Gamma_{\nu\mu}^\rho
  \!-\!\overline{\nabla}_\nu \delta^2 \Gamma_{\rho\mu}^\rho
   \!+\!\delta\Gamma_{\rho\lambda}^\rho\delta\Gamma_{\nu\mu}^\lambda
    \!-\!\delta\Gamma_{\nu\lambda}^\rho\delta\Gamma_{\rho\mu}^\lambda
\,,\qquad
\label{Ricci tensor: second variation}
\end{equation}
where the second variation of the connection reads,
\begin{equation}
\delta^2 \Gamma_{\alpha\beta}^\rho = 
           -\delta g^{\rho\sigma}\Big(\overline{\nabla}_{(\alpha} \delta g_{\beta)\sigma}
            \!-\!\frac12\overline{\nabla}_\sigma\delta g_{\alpha\beta}\Big)
\,,\qquad
\delta^2 \Gamma_{\rho\alpha}^\rho = 
           -\frac12\delta g^{\rho\sigma}\overline{\nabla}_{\alpha} \delta g_{\rho\sigma}
\,.
\label{connection: second variation}
\end{equation}
Inserting Eqs.~(\ref{connection: second variation}) and~(\ref{connection: first variation})
 into~(\ref{Ricci tensor: second variation}) gives the desired second variation of the Ricci tensor,
\begin{eqnarray}
\delta^2 R_{\mu\nu}  &\!\!=\!\!&  -\frac12 \delta g^{\rho\sigma}
\Big(\overline{\nabla}_\rho\overline{\nabla}_\nu\delta g_{\mu\sigma}
        \!+\!\overline{\nabla}_\rho\overline{\nabla}_\mu\delta g_{\nu\sigma}
        \!-\!\overline{\nabla}_\rho\overline{\nabla}_\sigma\delta g_{\mu\nu}
        \!-\!\overline{\nabla}_{(\mu}\overline{\nabla}_{\nu)}\delta g_{\rho\sigma}
   \Big)
\nonumber\\
&&\hskip 0cm   
 -\frac12 \big(\overline{\nabla}_\rho\delta g^{\rho\sigma}\big)
\Big(\overline{\nabla}_\nu\delta g_{\mu\sigma}
        \!+\!\overline{\nabla}_\mu\delta g_{\nu\sigma}
        \!-\!\overline{\nabla}_\sigma\delta g_{\mu\nu}\Big)
        \!+\!\frac12 \big(\overline{\nabla}_{\nu}\delta g^{\rho\sigma}\big)
         \big(\overline{\nabla}_{\mu}\delta g_{\rho\sigma}\big)
\nonumber\\
&&\hskip 0cm      
  +\,\frac14 \big(\overline{\nabla}_{\lambda}\delta g_{\rho}^{\;\rho}\big)
  \Big( \overline{\nabla}_{\nu} \delta g_{\mu}^\lambda
  + \overline{\nabla}_{\mu} \delta g_{\nu}^\lambda 
            -  \overline{\nabla}^{\lambda} \delta g_{\nu\mu}\Big)
\nonumber\\
&&\hskip 0cm 
   -\, \frac14\Big( \overline{\nabla}_{\nu} \delta g_{\lambda}^\rho
  + \overline{\nabla}_{\lambda} \delta g_{\nu}^\rho
            -  \overline{\nabla}^{\rho} \delta g_{\nu\lambda}\Big)
  \Big( \overline{\nabla}_{\rho} \delta g_{\mu}^\lambda
  \!+\! \overline{\nabla}_{\mu} \delta g_{\rho}^\lambda 
            \!-\!  \overline{\nabla}^{\lambda} \delta g_{\rho\mu}\Big)
\,,\qquad\;\;
\label{Ricci tensor: second variation 2}
\end{eqnarray}
which can be rewritten in a somewhat simplified form as, 
\begin{eqnarray}
\delta^2 R_{\mu\nu}  &\!\!=\!\!&  \frac12 \delta g^{\rho\sigma}
\Big(\overline{\nabla}_\rho\overline{\nabla}_\sigma\delta g_{\mu\nu}
        \!+\!\overline{\nabla}_{(\mu}\overline{\nabla}_{\nu)}\delta g_{\rho\sigma}
\!-\!\overline{\nabla}_\rho\overline{\nabla}_\nu\delta g_{\mu\sigma}
        \!-\!\overline{\nabla}_\rho\overline{\nabla}_\mu\delta g_{\nu\sigma}
   \Big)
\nonumber\\
&&\hskip -0.55cm   
 +\,\frac14 \big(\overline{\nabla}_{\mu}\delta g^{\rho\sigma}\big)
  \big( \overline{\nabla}_{\nu} \delta g_{\rho\sigma}\big)
 \!+\!\frac12 \big(\overline{\nabla}_\rho\delta g^{\sigma}_{\;(\mu}\big)
\big(\overline{\nabla}^\rho\delta g_{\nu)\sigma}\big)
 \!-\!\frac12 \big(\overline{\nabla}_\rho\delta g_{\sigma(\mu}\big)
\big(\overline{\nabla}^\sigma\delta g_{\nu)}^{\;\rho}\big)
\nonumber\\
&&\hskip 0cm 
   -\, \Big( \overline{\nabla}_{\rho} \delta g^{\rho\sigma}
            \!-\!\frac12 \overline{\nabla}^{\sigma} \delta g_{\rho}^{\rho}\Big)
  \Big( \overline{\nabla}_{(\mu} \delta g_{\nu)\sigma}
            \!-\! \frac12 \overline{\nabla}_{\sigma} \delta g_{\mu\nu}\Big)
\,,\qquad\;\;
\label{Ricci tensor: second variation 2}
\end{eqnarray}
which agrees with Eq.~(35.58b) in Ref.~\cite{Misner:1973prb}.

We have now all the needed elements to calculate variations of the Hilbert-Einstein 
action~(\ref{Hilbert-Einstein action}), which we can expand in a functional Taylor series as,
\begin{eqnarray}
S_{\rm HE}\big[\overline{g}_{\mu\nu}\!+\!\delta g_{\mu\nu}\big] 
 &\!\!=\!\!& \frac{1}{\kappa^2}\int\! {\rm d}^Dy\sqrt{-\overline{g}(y)}\overline{R}(y)
 \!+\!\frac{1}{\kappa^2}\int\! {\rm d}^Dx
\left(\frac{\delta S_{\rm HE}}{\delta g_{\mu\nu}(x)}\right)_{\overline{g}_{\alpha\beta}}
             \!\delta g_{\mu\nu}(x)
\nonumber\\
\nonumber\\
&&\hskip -2.99cm    
   +\,\frac12\frac{1}{\kappa^2}\!\int \!{\rm d}^Dx\,{\rm d}^Dx'\delta g_{\rho\sigma}(x')\left(\frac{\delta^2 S_{\rm HE}}
     {\delta g_{\rho\sigma}(x')\delta g_{\mu\nu}(x)}\right)_{\overline{g}_{\alpha\beta}}
           \!  \delta g_{\mu\nu}(x)
             \!+\! \mathcal{O}\big(\delta g_{\alpha\beta}^3\big)
\,.\qquad\;\;
\label{Hilbert-Einstein action: expanded}
\end{eqnarray}
Making use of the above results one can easily obtain the first variation,
\begin{equation}
\!\left(\frac{\delta S_{\rm HE}}{\delta g_{\mu\nu}(x)}\right)_{\overline{g}_{\alpha\beta}}
 \! = \frac{1}{\kappa^2}\!\int\! {\rm d}^Dy\sqrt{-\overline{g}(y)}
     \Big(\!-\overline{G}^{\mu\nu}(y)\!+\!\overline{\nabla}_y^\mu\overline{\nabla}_y^\nu
           \!-\!\overline{g}^{\mu\nu}(y)\overline{\dAlembert}_y\Big)
                  \delta^D(x\!-\!y)
\,.\!
\label{Hilbert-Einstein action: first variation}
\end{equation}
The derivative terms contribute boundary terms, which we neglect in this work,~\footnote{It is well known that on curved space-times boundary contributions can be important.
One such example is the York-Gibbons-Hawking boundary 
term~\cite{York:1972sj,Gibbons:1976ue}.
\label{York-Gibbons-Hawking}}
such that Eq.~(\ref{Hilbert-Einstein action: first variation}) simplifies to, 
\begin{equation}
\!\left(\frac{\delta S_{\rm HE}}{\delta g_{\mu\nu}(x)}\right)_{\overline{g}_{\alpha\beta}}
   = \frac{1}{\kappa^2}
     \big(\!-\overline{G}^{\mu\nu}(x)\big)
\,,\!\qquad
\label{Hilbert-Einstein action: first variation 2}
\end{equation}
which is the geometric side of the Einstein equation.
Varying the $\mathcal{O}(\epsilon^2)$ part of Eq.~(\ref{expansion of sqrt g R}) 
gives the second variation of the Hilbert-Einstein action~(\ref{Hilbert-Einstein action}),
\begin{eqnarray}
\left(\frac{\delta^2 S_{\rm HE}}{\delta g_{\mu\nu}(x)\delta g_{\rho\sigma}(x')} 
\right)_{\overline{g}_{\alpha\beta}}
&\!\!=\!\!& \frac{1}{\kappa^2}\int {\rm d}^Dy 
 \sqrt{-\overline{g}}
 \nonumber\\
&&\hskip -4.1cm 
 \times\bigg\{\frac14 \bigg(\overline{g}^{\mu\nu}\overline{g}^{\rho\sigma} 
           \!-\!2\overline{g}^{\mu(\rho}\overline{g}^{\sigma)\nu}\bigg)\overline{R}
                        \delta^D(x\!-\!y) \delta^D(x'\!-\!y)
\nonumber\\
&&\hskip -3.5cm  
       +\,\frac12 \overline{g}^{\alpha\beta}\bigg(\overline{g}^{\mu\nu}\delta^D(x\!-\!y)
    \left(
    \left(\frac{\delta R_{\alpha\beta}}{\delta g_{\rho\sigma}(x')}
             \right)_{\overline{g}_{\alpha\beta}}
      \right)_{\overline{g}_{\alpha\beta}}
               \!+\!\overline{g}^{\rho\sigma}\delta^D(x'\!-\!y)
                  \left(
  \left(\frac{\delta R_{\alpha\beta}}{\delta g_{\mu\nu}(x)}
       \right)_{\overline{g}_{\alpha\beta}}
\right)_{\overline{g}_{\alpha\beta}}\bigg)
\nonumber\\
&&\hskip -3.5cm                   
-\,\frac12\Big( \overline{g}^{\mu\nu}\overline{R}^{\rho\sigma}
                           \!+\!\overline{g}^{\rho\sigma} \overline{R}^{\mu\nu}\Big)
                            \delta^D(x\!-\!y) \delta^D(x'\!-\!y)
\nonumber\\
&&\hskip -3.5cm  
    +\,\Big( \overline{g}^{\alpha(\mu} \overline{g}^{\nu)(\rho} \overline{g}^{\sigma)\beta}\overline{R}_{\alpha\beta}   
       \!+\! \overline{g}^{\alpha(\rho} \overline{g}^{\sigma)(\mu} \overline{g}^{\nu)\beta}\overline{R}_{\alpha\beta}   
              \Big) \delta^D(x\!-\!y) \delta^D(x'\!-\!y)
\nonumber\\
&&\hskip -3.8cm  
   -\, \bigg(\!  \overline{g}^{\alpha(\mu}\overline{g}^{\nu)\beta} \delta^D(x\!-\!y)
              \left(\frac{\delta R_{\alpha\beta}}{\delta g_{\rho\sigma}(x')}
               \right)_{\overline{g}_{\alpha\beta}}
    \!+\!\overline{g}^{\alpha(\rho}\overline{g}^{\sigma)\beta}\delta^D(x'\!-\!y)
                           \left(\frac{\delta R_{\alpha\beta}}{\delta g_{\mu\nu}(x)}
              \right)_{\overline{g}_{\alpha\beta}}
                 \bigg)
\nonumber\\
&&\hskip -3.5cm                   
   +\,\overline{g}^{\alpha\beta}
     \left(\frac{\delta^2 R_{\alpha\beta}}{\delta g_{\mu\nu}(x)\delta g_{\rho\sigma}(x')}
          \right)_{\overline{g}_{\alpha\beta}}
                     \bigg\}
\nonumber\\
&&\hskip -4.2cm                                         
                   =\,  \frac{1}{\kappa^2}\int {\rm d}^Dy 
 \sqrt{-\overline{g}} \bigg\{
           \bigg[\!-\!\frac12\Big( \overline{g}^{\mu\nu}\overline{G}^{\rho\sigma}
                           \!+\!\overline{g}^{\rho\sigma} \overline{G}^{\mu\nu}\Big)
                               \!+\!2\overline{g}^{\mu)(\rho} \overline{G}^{\sigma)(\nu}
 \nonumber\\
&&\hskip -0.9cm 
                             -\, \frac14  \bigg(\overline{g}^{\mu\nu}\overline{g}^{\rho\sigma} 
           \!-\!2\overline{g}^{\mu(\rho}\overline{g}^{\sigma)\nu}\bigg)\overline{R}
                               \bigg]
                               \delta^D(x\!-\!y) \delta^D(x'\!-\!y)
\nonumber\\
&&\hskip -3.8cm  
       +\,\frac12\bigg(\overline{g}^{\mu\nu}
\delta^D(x\!-\!y) \overline{g}^{\alpha\beta}
\left(\frac{\delta R_{\alpha\beta}}{\delta g_{\rho\sigma}(x')}
         \right)_{\overline{g}_{\alpha\beta}}
               \!+\!\overline{g}^{\rho\sigma}\delta^D(x'\!-\!y)
          \overline{g}^{\alpha\beta} 
      \left(\frac{\delta R_{\alpha\beta}}{\delta g_{\mu\nu}(x)}
             \right)_{\overline{g}_{\alpha\beta}}
\bigg)
\nonumber\\
&&\hskip -3.8cm  
   -\,\bigg(\!  \overline{g}^{\alpha(\mu}\overline{g}^{\nu)\beta}\delta^D(x\!-\!y)
                           \left(\frac{\delta R_{\alpha\beta}}{\delta g_{\rho\sigma}(x')}
\right)_{\overline{g}_{\alpha\beta}}
    \!+\!\overline{g}^{\alpha(\rho}\overline{g}^{\sigma)\beta}\delta^D(x'\!-\!y)
                           \left(\frac{\delta R_{\alpha\beta}}{\delta g_{\mu\nu}(x)}
\right)_{\overline{g}_{\alpha\beta}}
                 \bigg)
\nonumber\\
&&\hskip -3.5cm 
   +\,\overline{g}^{\alpha\beta}
         \left(\frac{\delta^2 R_{\alpha\beta}}{\delta g_{\mu\nu}(x)\delta g_{\rho\sigma}(x')}
                 \right)_{\overline{g}_{\alpha\beta}}
                     \bigg\}                  
,\qquad\;\;
\label{second variation of the Hilbert-Einstein action}
\end{eqnarray}
where all metrics, curvature tensors and derivatives are evaluated at point $y$.
Inserting Eqs.~(\ref{Ricci tensor: first variation}) and~(\ref{Ricci tensor: second variation 2}) into Eq.~(\ref{second variation of the Hilbert-Einstein action})
one obtains, 
\begin{eqnarray}
\left(\frac{\delta^2 S_{\rm HE}}{\delta g_{\mu\nu}(x)\delta g_{\rho\sigma}(x')} 
 \right)_{\overline{g}_{\alpha\beta}}
&\!\!=\!\!&   \frac{1}{\kappa^2}\int {\rm d}^Dy 
 \sqrt{-\overline{g}} \bigg\{
           \bigg[\!-\!\frac12\big( \overline{g}^{\mu\nu}\overline{G}^{\rho\sigma}
                           \!+\!\overline{g}^{\rho\sigma} \overline{G}^{\mu\nu}\big)
\quad
 \nonumber\\
&&\hskip -3.9cm 
    +\,2\overline{g}^{\mu)(\rho} \overline{G}^{\sigma)(\nu}
    \!-\! \frac14  \big(\overline{g}^{\mu\nu}\overline{g}^{\rho\sigma} 
           \!-\!2\overline{g}^{\mu(\rho}\overline{g}^{\sigma)\nu}\big)\overline{R}
                               \bigg]
                               \delta^D(x\!-\!y) \delta^D(x'\!-\!y)
\nonumber\\
&&\hskip -3.cm  
       +\,\frac12\overline{g}^{\mu\nu}\delta^D(x\!-\!y)
       \big( \overline{\nabla}^\rho_y\overline{\nabla}^\sigma_y
          \!-\!\overline{\dAlembert}_y\overline{g}^{\rho\sigma}
               \big)\delta^D(x'\!-\!y)
\nonumber\\
&&\hskip -0.cm 
                      +\,\frac12\overline{g}^{\rho\sigma}\delta^D(x'\!-\!y)
       \big( \overline{\nabla}^\mu_y\overline{\nabla}^\nu_y
          \!-\!\overline{\dAlembert}_y\overline{g}^{\mu\nu}
               \big)\delta^D(x\!-\!y)
\nonumber\\
&&\hskip -3.cm  
       -\,\delta^D(x\!-\!y)
       \Big(2\overline{g}^{\mu)(\rho} 
       \overline{\nabla}^{\sigma)}_y\overline{\nabla}^{(\nu}_y
            \!-\! \overline{\nabla}^{\mu}_y\overline{\nabla}^{\nu}_y\overline{g}^{\rho\sigma}
              \!-\! \overline{g}^{\mu(\rho}\overline{g}^{\sigma)\nu}\overline{\dAlembert}_y
               \Big)\delta^D(x'\!-\!y)
\nonumber\\
&&\hskip -1.9cm 
                      -\,\delta^D(x'\!-\!y)
       \Big(2\overline{g}^{\mu)(\rho} 
       \overline{\nabla}^{\sigma)}_y\overline{\nabla}^{(\nu}_y
           \!-\! \overline{\nabla}^{\rho}_y\overline{\nabla}^{\sigma}_y\overline{g}^{\mu\nu}
              \!-\! \overline{g}^{\mu(\rho}\overline{g}^{\sigma)\nu}\overline{\dAlembert}_y
              \Big)\delta^D(x\!-\!y)
\nonumber\\
&&\hskip -3.cm  
       -\,\frac12\Big( \overline{\nabla}^{(\mu}_y\delta^D(x\!-\!y)\Big)
       \Big(2\overline{g}^{\nu)(\rho} 
       \overline{\nabla}^{\sigma)}_y
            \!-\! \overline{\nabla}^{\nu)}_y\overline{g}^{\rho\sigma}
               \Big)\delta^D(x'\!-\!y)
\nonumber\\
&&\hskip -3cm 
       -\,\frac12\Big( \overline{\nabla}^{(\rho}_y\delta^D(x'\!-\!y)\Big)
       \Big(2\overline{g}^{\sigma)(\mu} 
       \overline{\nabla}^{\nu)}_y
            \!-\! \overline{\nabla}^{\sigma)}_y\overline{g}^{\mu\nu}
               \Big)\delta^D(x\!-\!y)
\nonumber\\
&&\hskip 0.cm 
                +\, \Big(\overline{\nabla}^{\lambda}_y\delta^D(x'\!-\!y)\Big)
                    \overline{g}^{\mu(\rho}\overline{g}^{\sigma)\nu}
                    \Big(\overline{\nabla}_{\lambda}^y\delta^D(x\!-\!y)\Big)
\nonumber\\
&&\hskip -3cm 
                      +\,\frac12\overline{g}^{\mu\nu}
       \Big(\overline{\nabla}^{(\rho}_y\delta^D(x\!-\!y)\Big)
        \Big(\overline{\nabla}^{\sigma)}_y\delta^D(x'\!-\!y)\Big)
\nonumber\\
&&\hskip 0cm 
                      +\,\frac12\overline{g}^{\rho\sigma}
       \Big(\overline{\nabla}^{(\mu}_y\delta^D(x'\!-\!y)\Big)
        \Big(\overline{\nabla}^{\nu)}_y\delta^D(x\!-\!y)\Big)
\nonumber\\
&&\hskip 0cm 
           -\,\frac12\overline{g}^{\mu\nu}\overline{g}^{\rho\sigma}
                   \Big(\overline{\nabla}^{\lambda}_y\delta^D(x\!-\!y)\Big)
        \Big(\overline{\nabla}_{\lambda}^y\delta^D(x'\!-\!y)\Big)
\nonumber\\
&&\hskip -3cm 
                      -\,\overline{g}^{\mu)(\rho}
       \Big(\overline{\nabla}^{\sigma)}_y\delta^D(x\!-\!y)\Big)
        \Big(\overline{\nabla}^{(\nu}_y\delta^D(x'\!-\!y)\Big)
\nonumber\\
&&\hskip 0cm 
           +\,\frac12\overline{g}^{\mu(\rho}\overline{g}^{\sigma)\nu}
                   \Big(\overline{\nabla}^{\lambda}_y\delta^D(x\!-\!y)\Big)
        \Big(\overline{\nabla}_{\lambda}^y\delta^D(x'\!-\!y)\Big)
                     \bigg\}             
,\qquad\;\,
\label{second variation of the Hilbert-Einstein action 2}
\end{eqnarray}
where the last eight lines and a half from the previous two lines
 come from $\overline{g}^{\alpha\beta}\delta^2R_{\alpha\beta}$.
The easiest way of integrating this over $y$ is to move all of the derivatives from 
$y$ to $x$ or $x'$ (after which all of the derivatives can be pulled out of the integral).
This results in the following {\it nonlocal Lichnerowicz operator}, 
\begin{eqnarray}
\big[{\!\,}^{\mu\nu}\mathcal{\overline{L}}^{\rho\sigma}
  \big]_A(x;x')
&\!\!\equiv\!\!&  \frac{\kappa^2}{\sqrt{-g}\sqrt{-g'}}
   \left( \frac{\delta^2 S_{\rm HE}}{\delta g_{\mu\nu}(x)\delta g_{\rho\sigma}(x')} 
           \right)_{\overline{g}_{\alpha\beta}}
     \nonumber\\
&&\hskip -3.1cm 
         =\,  \bigg\{\!\!-\!\frac12\big( \overline{g}^{\mu\nu}\overline{G}^{\rho\sigma}
                           \!+\!\overline{g}^{\rho\sigma} \overline{G}^{\mu\nu}\big)
                               \!+\!2\overline{g}^{\mu)(\rho} \overline{G}^{\sigma)(\nu}
                           \!-\! \frac14  \big(\overline{g}^{\mu\nu}\overline{g}^{\rho\sigma} 
           \!-\!2\overline{g}^{\mu(\rho}\overline{g}^{\sigma)\nu}\big)\overline{R}
\nonumber\\
&&\hskip -2.7cm  
       +\,\Big(\frac12  \overline{g}^{\mu\nu}
      \overline{\nabla}^\rho\overline{\nabla}^\sigma
 \!+\!\frac12\overline{g}^{\rho\sigma}{\overline{\nabla}'}^\mu{\overline{\nabla}'}^\nu
                     \!+\! \overline{g}^{\mu\nu}\overline{\nabla}^\rho\overline{\nabla}^\sigma
       \!+\!\overline{g}^{\rho\sigma}{\overline{\nabla}'}^\mu{\overline{\nabla}'}^\nu
\nonumber\\
&&\hskip -0.cm  
       +\,\overline{g}^{\mu\nu}{\overline{\nabla}'}^{(\rho}\overline{\nabla}^{\sigma)}
      \!+\!\overline{g}^{\rho\sigma}{\overline{\nabla}'}^{(\mu}\overline{\nabla}^{\nu)}
         \Big)
\nonumber\\
&&\hskip -2.7cm              
             -\,\overline{g}^{\mu)(\rho} 
       \Big(2{\overline{\nabla}'}^{\sigma)}{\overline{\nabla}'}^{(\nu}
           \!+\!2\overline{\nabla}^{\sigma)}\overline{\nabla}^{(\nu}  
            \!+\!3{\overline{\nabla}'}^{\sigma)}\overline{\nabla}^{(\nu}  
           \Big)
\nonumber\\
&&\hskip -2.7cm               
          -\,\frac12\Big(\overline{g}^{\mu\nu} \overline{g}^{\rho\sigma}
           \!-\!2 \overline{g}^{\mu(\rho}\overline{g}^{\sigma)\nu}         
             \Big)
           \Big(\overline{\dAlembert}'\!+\!\overline{\dAlembert}
                      \!+\! {\overline{\nabla}'}\!\cdot\!\overline{\nabla} 
           \Big)
            \!+\!\frac12\overline{g}^{\mu(\rho}\overline{g}^{\sigma)\nu} 
                       {\overline{\nabla}'}\!\cdot\!\overline{\nabla}
                     \bigg\}  \frac{\delta^D(x\!-\!x')}{\sqrt{-\overline{g}}}            
\,.
\nonumber\\
&&\hskip -0.cm 
\label{second variation of the Hilbert-Einstein action 3}
\end{eqnarray}
This can be further simplified by moving all of the derivatives onto $x$, 
\begin{eqnarray}
\big[{\!\,}^{\mu\nu}\mathcal{\overline{L}}^{\rho\sigma}
  \big]_A(x;x')
         &\!\!=\!\!&  \bigg\{\!\!-\!\frac12\big( \overline{g}^{\mu\nu}\overline{G}^{\rho\sigma}
                           \!+\!\overline{g}^{\rho\sigma} \overline{G}^{\mu\nu}\big)
                               \!+\!2\overline{g}^{\mu)(\rho} \overline{G}^{\sigma)(\nu}
\quad
 \nonumber\\
&&\hskip 0.3cm 
                         -\, \frac14  \big(\overline{g}^{\mu\nu}\overline{g}^{\rho\sigma} 
           \!-\!2\overline{g}^{\mu(\rho}\overline{g}^{\sigma)\nu}\big)\overline{R}
\nonumber\\
&&\hskip -2.1cm  
       +\,\frac12 \Big( \overline{g}^{\mu\nu}
      \overline{\nabla}^\rho\overline{\nabla}^\sigma
             \!+\!\overline{g}^{\rho\sigma}\overline{\nabla}^\mu\overline{\nabla}^\nu
         \Big)
\!-\!\overline{g}^{\mu)(\rho} \overline{\nabla}^{\sigma)}\overline{\nabla}^{(\nu}                
          \!-\! \frac12\Big(\overline{g}^{\mu\nu} \overline{g}^{\rho\sigma} 
          \!-\! \overline{g}^{\mu(\rho}\overline{g}^{\sigma)\nu}      
             \Big)\overline{\dAlembert}
                     \bigg\}  
\nonumber\\
&&\hskip 0.5cm 
                     \times \frac{\delta^D(x\!-\!x')}{\sqrt{-\overline{g}}}            
\,.
\label{second variation of the Hilbert-Einstein action: Lichnerowicz}
\end{eqnarray}
When the Lichnerowicz operator acts on $\delta g_{\rho\sigma}$ it produces,
\begin{eqnarray}
\int\! {\rm d}^D x'\sqrt{-\overline{g}(x')}
   \big[{\!\,}^{\mu\nu}\mathcal{\overline{L}}^{\rho\sigma}\big]_A(x;x')\delta g_{\rho\sigma}(x')
         &\!\!=\!\!& -\frac12\big( \overline{g}^{\mu\nu}\overline{G}^{\rho\sigma}
       \!+\!\overline{g}^{\rho\sigma} \overline{G}^{\mu\nu}\big)\delta g_{\rho\sigma}(x)
\quad
 \nonumber\\
&&\hskip -3.3cm 
                  +\,2\overline{G}^{\mu)(\rho}\overline{g}^{\sigma)(\nu}\delta g_{\rho\sigma}
                   \!-\! \frac14  \big(\overline{g}^{\mu\nu}\overline{g}^{\rho\sigma} 
\!-\!2\overline{g}^{\mu(\rho}\overline{g}^{\sigma)\nu}\big)\overline{R}
          \delta g_{\rho\sigma}
\nonumber\\
&&\hskip -7.6cm  
       + \bigg\{\frac12 \Big( \overline{g}^{\mu\nu}
      \overline{\nabla}^\rho\overline{\nabla}^\sigma
             \!+\!\overline{g}^{\rho\sigma}\overline{\nabla}^\mu\overline{\nabla}^\nu
         \Big)\!
\!-\!\overline{g}^{\mu)(\rho} \overline{\nabla}^{\sigma)}\overline{\nabla}^{(\nu}                
          \!\!-\! \frac12\Big( \overline{g}^{\mu\nu} \overline{g}^{\rho\sigma} 
          \!-\!  \overline{g}^{\mu(\rho}\overline{g}^{\sigma)\nu}       
             \Big)\overline{\dAlembert}
                     \bigg\}  \delta g_{\rho\sigma}(x)         
\,.
\nonumber\\
&&\hskip -0.cm 
\label{second variation of the Hilbert-Einstein action: Lichnerowicz 2}
\end{eqnarray}
On the other hand we have,
$\delta G^{\mu\nu}=-2\delta g^{\mu)(\rho}\overline{G}^{\sigma)(\nu} \!+\! \overline{g}^{\mu(\alpha}\overline{g}^{\beta)\nu}\delta G_{\alpha\beta}$,
where 
\begin{eqnarray}
\delta G_{\alpha\beta}&\!\!=\!\!& \frac12
          \overline{g}_{\alpha\beta}\overline{G}^{\rho\sigma}
                    \delta g_{\rho\sigma}(x)
   \!+\!\frac14 \overline{R} \big(\overline{g}_{\alpha\beta} \overline{g}^{\rho\sigma} 
           \!-\!2\delta_\alpha^{(\rho}\delta_\beta^{\sigma)}\big)\delta g_{\rho\sigma}
\label{first variation of Einstein tensor: lower}
\\
&&\hskip -1.5cm  
       -\, \bigg[\!\!-\!\delta_{(\alpha}^{(\rho}\overline{\nabla}^{\sigma)}\overline{\nabla}_{\beta)}  
      \!+\!  \frac12 \Big( \overline{g}_{\alpha\beta}
      \overline{\nabla}^\rho\overline{\nabla}^\sigma
             \!+\overline{g}^{\rho\sigma}\overline{\nabla}_\alpha\overline{\nabla}_\beta
         \Big)              
          \!+\! \frac12\Big(\delta_\alpha^{(\rho}\delta_\beta^{\sigma)}  
          \!-\!\overline{g}_{\alpha\beta} \overline{g}^{\rho\sigma}
             \Big)\overline{\dAlembert}
                     \bigg] \delta g_{\rho\sigma}(x)         
\,,
\nonumber
\end{eqnarray}
such that
\begin{eqnarray}
\delta G^{\mu\nu} &\!\!=\!\!& \frac12
 \overline{g}^{\mu\nu}\overline{G}^{\rho\sigma}
      \delta g_{\rho\sigma}(x)
                  -\,2\overline{G}^{\mu)(\rho}\delta g^{\sigma)(\nu}
                   \!+\! \frac14  \big(\overline{g}^{\mu\nu}\overline{g}^{\rho\sigma} 
\!-\!2\overline{g}^{\mu(\rho}\overline{g}^{\sigma)\nu}\big)\overline{R}
          \delta g_{\rho\sigma}
\nonumber\\
&&\hskip -1.6cm  
       -\, \bigg[\frac12 \Big( \overline{g}^{\mu\nu}
      \overline{\nabla}^\rho\overline{\nabla}^\sigma
             \!+\!\overline{g}^{\rho\sigma}\overline{\nabla}^\mu\overline{\nabla}^\nu
         \Big)\!
\!-\!\overline{g}^{\mu)(\rho} \overline{\nabla}^{\sigma)}\overline{\nabla}^{(\nu}                
          \!\!-\! \frac12\Big(\overline{g}^{\mu\nu} \overline{g}^{\rho\sigma}
           \!-\! \overline{g}^{\mu(\rho}\overline{g}^{\sigma)\nu}        
             \Big)\overline{\dAlembert}
                     \bigg]  \delta g_{\rho\sigma}(x)    
\,.
\nonumber\\
&&\hskip -0.cm 
\label{first variation of Einstein tensor: upper}
\end{eqnarray}
Comparing this with Eq.~(\ref{second variation of the Hilbert-Einstein action: Lichnerowicz 2})
we can see that Eq.~(\ref{second variation of the Hilbert-Einstein action: Lichnerowicz 2})
can be recast as,
\begin{eqnarray}
\int\! {\rm d}^D x'\sqrt{-\overline{g}(x')}
   \big[{\!\,}^{\mu\nu}\mathcal{\overline{L}}^{\rho\sigma}\big]_A(x;x')\delta g_{\rho\sigma}(x')
         &\!\!=\!\!&-\,\delta G^{\mu\nu}
          -\frac12\overline{g}^{\rho\sigma} \overline{G}^{\mu\nu}\delta g_{\rho\sigma}(x)      
\nonumber\\
&&\hskip -0.5cm 
=\,-\frac{1}{\sqrt{-\overline{g}}}
\left(\delta\Big[\sqrt{-g} G^{\mu\nu}\Big]
\right)_{\overline{g}_{\alpha\beta}}
\,.\qquad\;\,
\label{second variation of the Hilbert-Einstein action: Lichnerowicz 3}
\end{eqnarray}
This means that the Lichnerowicz operator (if defined in terms of the second variation 
of the Hilbert-Einstein action) does not reproduce the result obtained by the
variation of the equation of motion, as it is often stated in the literature.
Instead it accords with the variation of $\sqrt{-g} G^{\mu\nu}$.


\bigskip
\noindent
{\bf Case~B.} Next we briefly consider case B defined
 in Eq.~(\ref{metric perturbations: case B}),
in which the metric perturbation is defined in terms of the inverse metric, 
\begin{eqnarray}
g^{\mu\nu} &\!\!=\!\!& \overline{g}^{\mu\nu}\!+\!\epsilon\delta g^{\mu\nu}
\,,\qquad
\label{variation gmn: case B}\\
\sqrt{-g} &\!\!=\!\!&\sqrt{-\overline{g}}\left\{1\!+\!\epsilon\left[\!-\frac12 \delta g_\alpha^{\;\alpha}\right]
                  \!+\!\epsilon^2\left[\frac18 \big(\delta g_\alpha^{\;\alpha}\big)^2
                   \!+\!  \frac14 \big(\delta g^{\mu\nu}\delta g_{\mu\nu}\big)\right]
                    \!+\!\mathcal{O}\big(\epsilon^3\big) \right\}
,\qquad\;
\label{variation root g: case B}\\
g_{\mu\nu}  &\!\!=\!\!&\overline{g}_{\mu\nu}\!-\!\epsilon\delta g_{\mu\nu}
 \!+\!\epsilon^2\delta g_{\mu\rho}\delta g^{\rho}_{\;\nu}
  \!+\!\mathcal{O}\big(\epsilon^3\big) 
 \, ,\qquad
\label{variation inverse gmn}\\
R_{\mu\nu}  &\!\!=\!\!&\overline{R}_{\mu\nu}\!+\!\epsilon\delta R_{\mu\nu}
 \!+\!\epsilon^2\delta^2 R_{\mu\nu}
  \!+\!\mathcal{O}\big(\epsilon^3\big) 
  \,,\qquad
\label{variation Rmn: case B}
\end{eqnarray}
where $\overline{R}_{\mu\nu}$ is defined in Eq.~(\ref{background Ricci tensor}).
Next step is to expand $\sqrt{-g}R$, 
\begin{eqnarray}
\sqrt{-g}R &\!\!=\!\!& \sqrt{-\overline{g}}\overline{R}
   \!+\!\epsilon \sqrt{-\overline{g}}\left[\!-\frac12 \delta g_\rho^{\;\rho} \overline{R}
        \!+\! \delta g^{\mu\nu} \overline{R}_{\mu\nu}  
             \!+\!  \overline{g}^{\mu\nu}\delta R_{\mu\nu} \right]
\nonumber\\
&&\hskip -.0cm
 +\,\epsilon^2 \sqrt{-\overline{g}}\bigg[\Big(\frac18 (\delta g_\rho^{\;\rho}\big)^2
           \!+\!\frac14\delta g^{\mu\nu} \delta g_{\mu\nu}\Big)\overline{R}
        \!-\!\frac12 \delta g_{\rho}^{\;\rho}\Big( \overline{g}^{\mu\nu}\delta R_{\mu\nu}
               \!+\!\delta g^{\mu\nu}\overline{R}_{\mu\nu}\Big)
\nonumber\\
&&\hskip 2cm  
             +\, \Big(\delta g^{\mu\nu}\delta R_{\mu\nu}
                   \!+\!\overline{g}^{\mu\nu}\delta^2 R_{\mu\nu}\Big)
                     \bigg]
\,.\quad
\label{expansion of sqrt g R: case B}
\end{eqnarray}
The first variation of the connection and Ricci tensor have now opposite signs
({\it cf.} Eqs.~(\ref{connection: first variation}) and~(\ref{Ricci tensor: first variation})), 
\begin{eqnarray}
\delta \Gamma_{\alpha\beta}^\rho &\!\!=\!\!& 
           -\overline{\nabla}_{(\alpha} \delta g_{\beta)}^\rho 
            +\frac12 \overline{\nabla}^\rho \delta g_{\alpha\beta}
\,,\qquad \delta \Gamma_{\rho\beta}^\rho = 
           -\frac12\overline{\nabla}_{\beta}\big( \delta g_{\rho}^{\;\rho}\big) 
\,,\quad 
\label{connection: first variation: case B}\\
\delta R_{\alpha\beta} &\!\!=\!\!&  
  -\overline{\nabla}_{\rho} \overline{\nabla}_{(\alpha} \delta g_{\beta)}^\rho 
            +\frac12 \big(\overline{\dAlembert} \delta g_{\alpha\beta}
              +\overline{\nabla}_{\alpha}\overline{\nabla}_{\beta}
              \delta g_{\rho}^{\;\rho}
            \big)
\,,\quad
\nonumber\\
\hskip -.0cm
\overline{g}^{\alpha\beta}\delta R_{\alpha\beta}&\!\!=\!\!&  
  -\overline{\nabla}^\beta \overline{\nabla}^{\alpha} \delta g_{\beta\alpha}
           \!+\!\overline{\dAlembert} \delta g_{\alpha}^{\;\alpha}
\,.\qquad
\label{Ricci tensor: first variation: case B}
\end{eqnarray}
To obtain the second variation we proceed as in case~A. 
The second variation of the Ricci tensor can be written as 
in Eq.~(\ref{Ricci tensor: second variation}), 
\begin{equation}
\delta^2 R_{\alpha\beta} = \overline{\nabla}_\rho \delta^2 \Gamma_{\alpha\beta}^\rho
  \!-\!\overline{\nabla}_\beta\delta^2 \Gamma_{\rho\alpha}^\rho
   \!+\!\delta\Gamma_{\rho\lambda}^\rho\delta\Gamma_{\beta\alpha}^\lambda
    \!-\!\delta\Gamma_{\beta\lambda}^\rho\delta\Gamma_{\rho\alpha}^\lambda
\,,\qquad
\label{Ricci tensor: second variation: case B}
\end{equation}
where the second variation of the connection now reads,
\begin{equation}
\delta^2 \Gamma_{\alpha\beta}^\rho = 
           \delta g_{(\alpha}^{\;\gamma}\overline{\nabla}_{\beta)} \delta g_\gamma^{\;\rho}
            \!-\!\delta g_{(\alpha}^{\;\gamma}\overline{\nabla}^\rho\delta g_{\beta)\gamma}
        \!+\!\frac12\delta g^{\rho\gamma}\overline{\nabla}_{\gamma} \delta g_{\alpha\beta}
        \,,\quad
        \delta^2 \Gamma_{\alpha\rho}^\rho 
                   = \frac12\delta g_{\delta}^{\;\gamma}
                            \overline{\nabla}_{\alpha} \delta g_{\gamma}^{\;\delta}
\,.\quad
\label{connection: second variation: case B}
\end{equation}
Inserting Eqs.~(\ref{connection: second variation: case B}) 
and~(\ref{connection: first variation: case B})
 into~(\ref{Ricci tensor: second variation: case B}) gives the desired second variation of the Ricci tensor,
\begin{eqnarray}
\delta^2 R_{\alpha\beta}  \!&\!\!=\!\!&  \!
\delta g_{(\alpha}^{\;\gamma}\overline{\nabla}^\delta
            \overline{\nabla}_{\beta)} \delta g_{\gamma\delta}
            \!-\!\delta g_{(\alpha}^{\;\gamma}\overline{\dAlembert}
                       \delta g_{\beta)\gamma}
        \!+\!\frac12\delta g^{\delta\gamma}\overline{\nabla}_\delta
                      \overline{\nabla}_{\gamma} \delta g_{\alpha\beta}
  \!-\! \frac12\delta g_{\delta}^{\;\gamma}
       \overline{\nabla}_\beta \overline{\nabla}_{\alpha} \delta g_{\gamma}^{\;\delta}
  \nonumber\\
&&\hskip 0cm    
+\,\big(\overline{\nabla}_\rho \delta g_{(\alpha}^{\;\gamma}\big)
            \big(\overline{\nabla}_{\beta)} \delta g_\gamma^{\;\rho}\big)
       \!-\! \big(\overline{\nabla}_\rho\delta g_{(\alpha}^{\;\gamma}\big)
                   \big( \overline{\nabla}^\rho\delta g_{\beta)\gamma}\big)
        \!+\!\frac12 \big(\overline{\nabla}_\rho\delta g^{\rho\gamma}\big)
             \big(\overline{\nabla}_{\gamma} \delta g_{\alpha\beta}\big)    
  \nonumber\\
&&\hskip 0cm       
        -\, \frac12\big(\overline{\nabla}_\beta\delta g_{\delta}^{\;\gamma}\big)
                            \big(\overline{\nabla}_{\alpha} \delta g_{\gamma}^{\;\delta}\big)
 \!+\!\frac14\big( \overline{\nabla}_{\lambda}\delta g_{\rho}^{\;\rho}\big)
  \Big(\overline{\nabla}_{\alpha} \delta g_{\beta}^\lambda 
  \!+\!\overline{\nabla}_{\beta} \delta g_{\alpha}^\lambda 
            \!-\!\overline{\nabla}^\lambda \delta g_{\alpha\beta}\Big)
 \nonumber\\
&&\hskip 0cm
    -\,\frac14\Big(\overline{\nabla}_{\lambda} \delta g_{\beta}^\rho 
        \!+\!\overline{\nabla}_{\beta} \delta g_{\lambda}^\rho 
            \!-\!\overline{\nabla}^\rho \delta g_{\lambda\beta}\Big)
            \Big(\overline{\nabla}_{\alpha} \delta g_{\rho}^\lambda 
            \!+\!\overline{\nabla}_{\rho} \delta g_{\alpha}^\lambda 
             \!-\!\overline{\nabla}^\lambda \delta g_{\alpha\rho}\Big) 
\,.\qquad\;\;
\label{Ricci tensor: second variation 2: case B}
\end{eqnarray}
We can now calculate the variation of the Hilbert-Einstein 
action~(\ref{Hilbert-Einstein action}),
\begin{eqnarray}
S_{\rm HE}\big[\overline{g}^{\mu\nu}\!+\!\delta g^{\mu\nu}\big] 
 &\!\!\!=\!\!\!& \frac{1}{\kappa^2}\int {\rm d}^Dy\sqrt{-\overline{g}(y)}\overline{R}(y)
 \!+\! \frac{1}{\kappa^2}\int\! {\rm d}^Dx
\left(\frac{\delta S_{\rm HE}}{\delta g^{\mu\nu}(x)}
         \right)_{\overline{g}_{\alpha\beta}}
\!\!   \delta g^{\mu\nu}(x)
\nonumber\\
\nonumber\\
&&\hskip -3.cm    
   +\,\frac12 \frac{1}{\kappa^2}\int \!{\rm d}^Dx\,{\rm d}^Dx'\delta g^{\rho\sigma}(x')\left(\frac{\delta^2 S_{\rm HE}}
     {\delta g^{\rho\sigma}(x')\delta g^{\mu\nu}(x)}
\right)_{\overline{g}_{\alpha\beta}}
             \!\!\delta g^{\mu\nu}(x)
             \!+\! \mathcal{O}\Big(\big(\delta g^{\alpha\beta}\big)^3\Big)
\,.
\nonumber\\
\label{Hilbert-Einstein action: expanded: case B}
\end{eqnarray}
Making use of Eqs.~(\ref{expansion of sqrt g R: case B}) 
and~(\ref{Ricci tensor: first variation: case B}) one obtains the first variation,
\begin{equation}
\!\left(\frac{\delta S_{\rm HE}}{\delta g^{\mu\nu}(x)}
      \right)_{\overline{g}_{\alpha\beta}}
   = \frac{1}{\kappa^2}\int\! {\rm d}^Dy\sqrt{-\overline{g}(y)}
     \Big(\overline{G}_{\mu\nu}(y)\!-\!\overline{\nabla}^y_\mu\overline{\nabla}^y_\nu
           \!+\!\overline{g}_{\mu\nu}(y)\overline{\dAlembert}_y\Big)
                  \delta^D(x\!-\!y)
\,.\!
\label{Hilbert-Einstein action: first variation: case B}
\end{equation}
Neglecting the boundary terms (see footnote~\ref{York-Gibbons-Hawking}), 
this simplifies to, 
\begin{equation}
\!\left(\frac{\delta S_{\rm HE}}{\delta g^{\mu\nu}(x)}
\right)_{\overline{g}_{\alpha\beta}}
   = \frac{\sqrt{-\overline{g}(x)}}{\kappa^2}\overline{G}_{\mu\nu}(x)
\,,\!\qquad
\label{Hilbert-Einstein action: first variation 2: case B}
\end{equation}
which is the expected result.
Next, varying the $\mathcal{O}(\epsilon^2)$ part of Eq.~(\ref{expansion of sqrt g R: case B}) 
gives the second variation of the 
Hilbert-Einstein action~(\ref{Hilbert-Einstein action}),
\begin{eqnarray}
\!\left(\frac{\delta^2 S_{\rm HE}}{\delta g^{\mu\nu}(x)\delta g^{\rho\sigma}(x')} 
\right)_{\overline{g}_{\alpha\beta}}
&\!\!=\!\!& \frac{1}{\kappa^2}\int {\rm d}^Dy 
 \sqrt{-\overline{g}}
 \nonumber\\
&&\hskip -3.8cm 
 \times\bigg\{
\!-\!\frac12\Big( \overline{g}_{\mu\nu}\overline{G}_{\rho\sigma}
                           \!+\!\overline{g}_{\rho\sigma} \overline{G}_{\mu\nu}\Big)
                            \delta^D(x\!-\!y) \delta^D(x'\!-\!y)
\nonumber\\
&&\hskip -.0cm
 -\,\frac14 \bigg(\overline{g}_{\mu\nu}\overline{g}_{\rho\sigma} 
           \!-\!2\overline{g}_{\mu(\rho}\overline{g}_{\sigma)\nu}\bigg)\overline{R}
                        \delta^D(x\!-\!y) \delta^D(x'\!-\!y)
\nonumber\\
&&\hskip -3.8cm  
        -\,\frac12\bigg(\overline{g}_{\mu\nu}\delta^D(x\!-\!y) \overline{g}^{\alpha\beta}
         \left(\frac{\delta R_{\alpha\beta}}{\delta g^{\rho\sigma}(x')}
               \right)_{\overline{g}_{\alpha\beta}}
          \!+\!\overline{g}_{\rho\sigma}\delta^D(x'\!-\!y) \overline{g}^{\alpha\beta}
        \left(\frac{ \delta R_{\alpha\beta}}{\delta g^{\mu\nu}(x)} 
             \right)_{\overline{g}_{\alpha\beta}}
             \bigg)
\nonumber\\
&&\hskip -3.8cm  
             +\, \bigg(\delta_\mu^{(\alpha}\delta_\nu^{\beta)}\delta^D(x\!-\!y)
              \left(\frac{\delta R_{\alpha\beta}}{\delta g^{\rho\sigma}(x')}
                    \right)_{\overline{g}_{\alpha\beta}}
                  \!+\!  \delta_\rho^{(\alpha}\delta_\sigma^{\beta)}\delta^D(x'\!-\!y)
              \left(\frac{\delta R_{\alpha\beta}}{\delta g^{\mu\nu}(x)}
          \right)_{\overline{g}_{\alpha\beta}}
\bigg)
\nonumber\\
&&\hskip -1.2cm  
                +\,\overline{g}^{\alpha\beta}
     \left(\frac{ \delta^2 R_{\alpha\beta}}{\delta g^{\mu\nu}(x)\delta g^{\rho\sigma}(x')}\right)_{\overline{g}_{\alpha\beta}}
\bigg\}
\,.\quad
\label{second variation of the Hilbert-Einstein action: case B}
\end{eqnarray}
Upon inserting Eqs.~(\ref{Ricci tensor: first variation: case B}) 
and~(\ref{Ricci tensor: second variation 2: case B}) 
into Eq.~(\ref{second variation of the Hilbert-Einstein action: case B})
one arrives at, 
\begin{eqnarray}
\left(\frac{\delta^2 S_{\rm HE}}{\delta g^{\mu\nu}(x)\delta g^{\rho\sigma}(x')} \right)_{\overline{g}_{\alpha\beta}}
&\!\!=\!\!& \frac{1}{\kappa^2}\int {\rm d}^Dy 
 \sqrt{-\overline{g}}
 \nonumber\\
&&\hskip -3.8cm 
 \times\bigg\{
\!-\!\frac12\Big( \overline{g}_{\mu\nu}\overline{G}_{\rho\sigma}
                           \!+\!\overline{g}_{\rho\sigma} \overline{G}_{\mu\nu}\Big)
                            \delta^D(x\!-\!y) \delta^D(x'\!-\!y)
\nonumber\\
&&\hskip -1.0cm
 -\,\frac14 \bigg(\overline{g}_{\mu\nu}\overline{g}_{\rho\sigma} 
           \!-\!2\overline{g}_{\mu(\rho}\overline{g}_{\sigma)\nu}\bigg)\overline{R}
                        \delta^D(x\!-\!y) \delta^D(x'\!-\!y)
\nonumber\\
&&\hskip -3.2cm  
        -\,\frac12\overline{g}_{\rho\sigma}\delta^D(x'\!-\!y) 
       \Big(\!\!-\!\overline{\nabla}_\mu^y \overline{\nabla}_\nu^y
            \!+\!\overline{\dAlembert}^y \overline{g}_{\mu\nu}
            \Big)\delta^D(x\!-\!y)  
 \nonumber\\
&&\hskip -3.2cm       
            -\,\frac12\overline{g}_{\mu\nu}\delta^D(x\!-\!y)
       \Big(\!\!-\!\overline{\nabla}_\rho^y \overline{\nabla}_\sigma^y
            \!+\!\overline{\dAlembert}^y \overline{g}_{\rho\sigma}
            \Big)\delta^D(x'\!-\!y)  
\nonumber\\
&&\hskip -3.2cm  
             -\,
             \delta^D(x'\!-\!y)
              \bigg(\overline{g}_{\mu)(\rho}
                \overline{\nabla}_{\sigma)}^y \overline{\nabla}_{(\nu}^y
   \!-\!\frac12\big(\overline{\dAlembert}^y  \overline{g}_{\mu(\rho}\overline{g}_{\sigma)\nu}
              +\overline{\nabla}_{\rho}^y\overline{\nabla}_{\sigma}^y
             \overline{g}_{\mu\nu}
            \big)\bigg)\delta^D(x\!-\!y)
 \nonumber\\
&&\hskip -3.2cm       
             -\,
             \delta^D(x\!-\!y)
              \bigg(\overline{g}_{\mu)(\rho}
                \overline{\nabla}_{\sigma)}^y \overline{\nabla}_{(\nu}^y
    \!-\!\frac12\big(\overline{\dAlembert}^y  \overline{g}_{\mu(\rho}\overline{g}_{\sigma)\nu}
              +\overline{\nabla}_{\mu}^y\overline{\nabla}_{\nu}^y
             \overline{g}_{\rho\sigma}
            \big)\bigg)\delta^D(x'\!-\!y)
 \nonumber\\
&&\hskip -3.2cm  
       +\,\delta^D(x\!-\!y)
       \Big(\overline{g}_{\mu)(\rho} 
       \overline{\nabla}_{\sigma)}^y\overline{\nabla}_{(\nu}^y
            \!+\!\frac12 \overline{\nabla}_{\mu}^y
             \overline{\nabla}_{\nu}^y\overline{g}_{\rho\sigma}
           \!-\! \frac32\overline{g}_{\mu(\rho}\overline{g}_{\sigma)\nu}
                        \overline{\dAlembert}^y\Big)\delta^D(x'\!-\!y)
\nonumber\\
&&\hskip -2.8cm 
                      +\,\delta^D(x'\!-\!y)
       \Big(\overline{g}_{\mu)(\rho} 
       \overline{\nabla}_{\sigma)}^y\overline{\nabla}_{(\nu}^y
    \!+\!\frac12 \overline{\nabla}_{\rho}^y\overline{\nabla}_{\sigma}^y\overline{g}_{\mu\nu}
          \!-\!\frac32 \overline{g}_{\mu(\rho}\overline{g}_{\sigma)\nu}\overline{\dAlembert}^y
              \Big)\delta^D(x\!-\!y)
\nonumber\\
&&\hskip -3.2cm  
       +\,\Big( \overline{\nabla}_{(\mu}^y\delta^D(x\!-\!y)\Big)
       \Big(\overline{g}_{\nu)(\rho} 
       \overline{\nabla}_{\sigma)}^y
            \!+\!\frac12 \overline{\nabla}_{\nu)}^y\overline{g}_{\rho\sigma}
               \Big)\delta^D(x'\!-\!y)
\nonumber\\
&&\hskip -3.2cm 
       +\,\Big( \overline{\nabla}^{(\rho}_y\delta^D(x'\!-\!y)\Big)
       \Big(\overline{g}^{\sigma)(\mu} 
       \overline{\nabla}^{\nu)}_y
            \!+\!\frac12 \overline{\nabla}^{\sigma)}_y\overline{g}_{\mu\nu}
               \Big)\delta^D(x\!-\!y)
\nonumber\\
&&\hskip -1.cm 
                -\, 3{\;\cancel 2\;}\Big(\overline{\nabla}^{\lambda}_y\delta^D(x\!-\!y)\Big)
                    \overline{g}_{\mu(\rho}\overline{g}_{\sigma)\nu}
                    \Big(\overline{\nabla}_{\lambda}^y\delta^D(x'\!-\!y)\Big)
\nonumber\\
&&\hskip -3.2cm 
                      +\,\frac12\overline{g}_{\mu\nu}
       \Big(\overline{\nabla}_{(\rho}^y\delta^D(x\!-\!y)\Big)
        \Big(\overline{\nabla}_{\sigma)}^y\delta^D(x'\!-\!y)\Big)
\nonumber\\
&&\hskip -1cm 
                      +\,\frac12\overline{g}_{\rho\sigma}
       \Big(\overline{\nabla}_{(\mu}^y\delta^D(x'\!-\!y)\Big)
        \Big(\overline{\nabla}_{\nu)}^y\delta^D(x\!-\!y)\Big)
\nonumber\\
&&\hskip -1cm 
           -\,\frac12\overline{g}_{\mu\nu}\overline{g}_{\rho\sigma}
                   \Big(\overline{\nabla}^{\lambda}_y\delta^D(x\!-\!y)\Big)
        \Big(\overline{\nabla}_{\lambda}^y\delta^D(x'\!-\!y)\Big)
\nonumber\\
&&\hskip -3.2cm 
                      -\,\overline{g}_{\mu)(\rho}
       \Big(\overline{\nabla}_{\sigma)}^y\delta^D(x\!-\!y)\Big)
        \Big(\overline{\nabla}_{(\nu}^y\delta^D(x'\!-\!y)\Big)
\nonumber\\
&&\hskip -1cm 
           +\,\frac12\overline{g}_{\mu(\rho}\overline{g}_{\sigma)\nu}
                   \Big(\overline{\nabla}^{\lambda}_y\delta^D(x\!-\!y)\Big)
        \Big(\overline{\nabla}_{\lambda}^y\delta^D(x'\!-\!y)\Big)
\bigg\}
\,.\quad
\label{second variation of the Hilbert-Einstein action 2: case B}
\end{eqnarray}
where the last ten lines come from $\overline{g}^{\alpha\beta}\delta^2R_{\alpha\beta}$.
Moving the derivatives from $y$ to $x$ or $x'$ one can perform the integral over $y$,
resulting in  the nonlocal Lichnerowicz operator,
\begin{eqnarray}
\big[{\!\,}_{\mu\nu}\mathcal{\overline{L}}_{\rho\sigma}\big]_B(x;x')
&\!\!\equiv\!\!&  \frac{\kappa^2}{\sqrt{-g}\sqrt{-g'}}
    \left(\frac{\delta^2 S_{\rm HE}}{\delta g^{\mu\nu}(x)\delta g^{\rho\sigma}(x')} \right)_{\overline{g}_{\alpha\beta}}
     \nonumber\\
&&\hskip -3.cm 
         =\,  \bigg\{\!\!-\!\frac12\big( \overline{g}_{\mu\nu}\overline{G}_{\rho\sigma}
                           \!+\!\overline{g}_{\rho\sigma} \overline{G}_{\mu\nu}\big)
                           \!-\! \frac14  \big(\overline{g}_{\mu\nu}\overline{g}_{\rho\sigma} 
           \!-\!2\overline{g}_{\mu(\rho}\overline{g}_{\sigma)\nu}\big)\overline{R}
\nonumber\\
&&\hskip -2.7cm  
       +\,\frac12 \Big( 2\overline{g}_{\mu\nu}
      {\overline{\nabla}'}_\rho{\overline{\nabla}'}_\sigma
        \!+\!\overline{g}_{\mu\nu}\overline{\nabla}_\rho\overline{\nabla}_\sigma
        \!+\!2\overline{g}_{\mu\nu}{\overline{\nabla}'}_{(\rho}\overline{\nabla}_{\sigma)}
             \!+\!\overline{g}_{\rho\sigma}\overline{\nabla}_\mu\overline{\nabla}_\nu
\nonumber\\
&&\hskip -0.cm  
       +\,2\overline{g}_{\rho\sigma}{\overline{\nabla}'}_\mu{\overline{\nabla}'}_\nu
      \!+\!2\overline{g}_{\rho\sigma}{\overline{\nabla}'}_{(\mu}\overline{\nabla}_{\nu)}
         \Big)
             +\,\overline{g}_{\mu)(\rho} {\overline{\nabla}'}_{\sigma)}\overline{\nabla}_{(\nu}  
\nonumber\\
&&\hskip -2.7cm               
          -\,\frac12\Big(\overline{g}_{\mu\nu} \overline{g}_{\rho\sigma}
           \!+\!2 \overline{g}_{\mu(\rho}\overline{g}_{\sigma)\nu}         
             \Big)
           \Big(\overline{\dAlembert}'\!+\!\overline{\dAlembert}
                      \!+\! {\overline{\nabla}'}\!\cdot\!\overline{\nabla} 
           \Big)
            \!-\!\frac32\overline{g}_{\mu(\rho}\overline{g}_{\sigma)\nu} 
                       {\overline{\nabla}'}\!\cdot\!\overline{\nabla}
                     \bigg\}  \frac{\delta^D(x\!-\!x')}{\sqrt{-\overline{g}}}            
\,.\quad
\nonumber\\
&&\hskip -0.cm 
\label{second variation of the Hilbert-Einstein action 3: case B}
\end{eqnarray}
When all of the derivatives are moved onto $x$, this simplifies further to, 
\begin{eqnarray}
\big[{\!\,}_{\mu\nu}\mathcal{\overline{L}}_{\rho\sigma}
  \big]_B(x;x')
         &\!\!=\!\!&  \bigg\{\!\!-\!\frac12\big( \overline{g}_{\mu\nu}\overline{G}_{\rho\sigma}
                           \!+\!\overline{g}_{\rho\sigma} \overline{G}_{\mu\nu}\big)
                           \!-\! \frac14  \big(\overline{g}_{\mu\nu}\overline{g}_{\rho\sigma} 
           \!-\!2\overline{g}_{\mu(\rho}\overline{g}_{\sigma)\nu}\big)\overline{R}
\nonumber\\
&&\hskip -2.cm  
       +\,\frac12 \Big( \overline{g}_{\mu\nu}
      \overline{\nabla}_\rho\overline{\nabla}_\sigma
             \!+\!\overline{g}_{\rho\sigma}\overline{\nabla}_\mu\overline{\nabla}_\nu
         \Big)
\!-\!\overline{g}_{\mu)(\rho} \overline{\nabla}_{\sigma)}\overline{\nabla}_{(\nu}                
          \!+\! \frac12\Big( \overline{g}_{\mu(\rho}\overline{g}_{\sigma)\nu}
          \!-\!\overline{g}_{\mu\nu} \overline{g}_{\rho\sigma}       
             \Big)\overline{\dAlembert}
                     \bigg\}  
\quad
\nonumber\\
&&\hskip 0.5cm 
                     \times \frac{\delta^D(x\!-\!x')}{\sqrt{-\overline{g}}}            
\,.\qquad\;\,
\label{second variation of the Hilbert-Einstein action: Lichnerowicz: case B}
\end{eqnarray}
When the Lichnerowicz 
operator~(\ref{second variation of the Hilbert-Einstein action: Lichnerowicz: case B}) 
acts on $\delta g^{\rho\sigma}$ it yields,
\begin{eqnarray}
\int \!{\rm d}^D x'\sqrt{-\overline{g}(x')}
   \big[{\!\,}_{\mu\nu}\mathcal{\overline{L}}_{\rho\sigma}\big]_B(x;x')\delta g^{\rho\sigma}(x')
         &\!\!=\!\!& -\frac12\big( \overline{g}_{\mu\nu}\overline{G}_{\rho\sigma}
       \!+\!\overline{g}_{\rho\sigma} \overline{G}_{\mu\nu}\big)\delta g^{\rho\sigma}(x)
\quad
 \nonumber\\
&&\hskip -3.3cm 
                   -\, \frac14 \overline{R} \big(\overline{g}_{\mu\nu}\overline{g}_{\rho\sigma} 
\!-\!2\overline{g}_{\mu(\rho}\overline{g}_{\sigma)\nu}\big)
          \delta g^{\rho\sigma}
\nonumber\\
&&\hskip -7.8cm  
       +\, \bigg\{\frac12 \Big( \overline{g}_{\mu\nu}
      \overline{\nabla}_\rho\overline{\nabla}_\sigma
             \!+\!\overline{g}_{\rho\sigma}\overline{\nabla}_\mu\overline{\nabla}_\nu
         \Big)\!
\!-\!\overline{g}_{\mu)(\rho} \overline{\nabla}_{\sigma)}\overline{\nabla}_{(\nu}                
          \!\!-\! \frac12\Big( \overline{g}_{\mu\nu} \overline{g}_{\rho\sigma} 
          \!-\!     \overline{g}_{\mu(\rho}\overline{g}_{\sigma)\nu}    
             \Big)\overline{\dAlembert}
                     \bigg\}  \delta g^{\rho\sigma}(x)         
\,,
\nonumber\\
&&\hskip -0.cm 
\label{second variation of the Hilbert-Einstein action: Lichnerowicz 2: case B}
\end{eqnarray}
Comparing with the variation of the Einstein tensor,  
\begin{eqnarray}
\delta G_{\mu\nu}&\!\!=\!\!& - \frac12\overline{g}_{\mu\nu}\overline{G}_{\rho\sigma}
                    \delta g^{\rho\sigma}(x)
   \!-\!\frac14 \overline{R} \big(\overline{g}_{\mu\nu} \overline{g}_{\rho\sigma} 
           \!-\!2\overline{g}_{\mu(\rho}\overline{g}_{\sigma)\nu}\big)\delta g^{\rho\sigma}
\label{first variation of Einstein tensor: lower: case B}
\\
&&\hskip -1.5cm  
       +\, \bigg[\!-\!\overline{g}_{\mu)(\rho}
                  \overline{\nabla}_{\sigma)}\overline{\nabla}_{(\nu}  
      \!+\!  \frac12 \Big( \overline{g}_{\mu\nu}
      \overline{\nabla}_\rho\overline{\nabla}_\sigma
             \!+\overline{g}_{\rho\sigma}\overline{\nabla}_\mu\overline{\nabla}_\nu
         \Big)              
          \!-\! \frac12\Big(\overline{g}_{\mu\nu} \overline{g}_{\rho\sigma}
          \!-\!
          \overline{g}_{\mu(\rho}\overline{g}_{\sigma)\nu}  
             \Big)\overline{\dAlembert}
                     \bigg] \delta g^{\rho\sigma}       
\,,
\nonumber
\end{eqnarray}
we see that 
Eq.~(\ref{second variation of the Hilbert-Einstein action: Lichnerowicz 2: case B})
we can recast as, 
\begin{eqnarray}
\int\! {\rm d}^D x'\sqrt{-\overline{g}(x')}
   \big[{\!\,}_{\mu\nu}\mathcal{\overline{L}}_{\rho\sigma}\big]_B(x;x')\delta g^{\rho\sigma}(x')
         \!&\!\!=\!\!&\delta G_{\mu\nu}
         \! -\!\frac12\overline{g}_{\rho\sigma} \overline{G}_{\mu\nu}\delta g^{\rho\sigma}(x)   
\nonumber\\
&&\hskip -0.5cm 
=\,\frac{1}{\sqrt{-\overline{g}}}
  \left(\delta\Big[\sqrt{-g} G_{\mu\nu}\Big]
      \right)_{\overline{g}_{\alpha\beta}}
\,,\qquad\;\,
\label{second variation of the Hilbert-Einstein action: Lichnerowicz 3: case B}
\end{eqnarray}
a similar result as we obtained in case~A in 
Eq.~(\ref{second variation of the Hilbert-Einstein action: Lichnerowicz 3}). 
Furthermore, comparing the non-local Lichnerowicz 
operators in Eqs.~(\ref{second variation of the Hilbert-Einstein action: Lichnerowicz})
and~(\ref{second variation of the Hilbert-Einstein action: Lichnerowicz: case B})
we see notable differences, which motivates a further study of 
the dependence of the gravitational dynamics on the representation of 
metric perturbations and of the corresponding Noether-Ward identities.

\bigskip\bigskip


\section*{Appendix B: Variation of the counterterm action}
\label{Appendix B: Variation of the counterterm action}


The gravitational counterterm action in general consists of the Hilbert-Einstein action 
(which includes the Ricci scalar and cosmological constant) and the following three terms,
\begin{eqnarray}
S_{\rm ct}[g_{\mu\nu}] &\!\!\!=\!\!\!& \alpha_{R^2}\!\int\! {\rm d}^Dx \sqrt{-g}\, R^2
 \!+\!  \alpha_{\rm Ric^2}\!\!\int\! {\rm d}^Dx \sqrt{-g}\, {\rm Ric}^2
 \!+\! \alpha_{\rm Riem^2}\!\!\int\! {\rm d}^Dx \sqrt{-g}\, {\rm Riem}^2
 \!\nonumber\\
 &\!\!\!\equiv\!\!\!& S_{R^2}[g_{\mu\nu}]
 \!+\!  S_{\rm Ric^2}[g_{\mu\nu}]
  \!+\! S_{\rm Riem^2}[g_{\mu\nu}]
 \,,\qquad
\label{counterterm action}
\end{eqnarray}
where ${\rm Ric}^2 \equiv R_{\alpha\beta}R^{\alpha\beta}$ and 
${\rm Riem}^2 \equiv R_{\alpha\beta\gamma\delta}R^{\alpha\beta\gamma\delta}$. 
Generally speaking, two of these terms are enough to renormalize
the the quantum contributions as the following linear combination, known 
as the Gauss-Bonnet term,
\begin{equation}
{\rm G\!B} \equiv R^2 \!-\! 4 {\rm Ric}^2 \!+\! {\rm Riem}^2
\,,\quad
\label{Gauss-Bonnet}
\end{equation}
is topological in $D=4$, and therefore its contribution to the variation principle is 
$\propto (D\!-\!4)$, and thus cannot be used for renormalization.
Therefore, an alternative way of writing the counterterm action~(\ref{counterterm action}) is as follows,
\begin{eqnarray}
\widetilde{S}_{\rm ct}[g_{\mu\nu}] &\!\!\!=\!\!\!& \alpha_{R^2}\!\int\! {\rm d}^Dx \sqrt{-g}\, R^2
 \!+\!  \alpha_{\rm Weyl^2}\!\!\int\! {\rm d}^Dx \sqrt{-g}\, {\rm Weyl}^2
  \!+\!  \alpha_{\rm G\!B}\!\!\int\! {\rm d}^Dx \sqrt{-g}\, {\rm G\!B}
 \!\nonumber\\
 &\!\!\!\equiv\!\!\!& S_{R^2}[g_{\mu\nu}]
  \!+\! S_{\rm Weyl^2}[g_{\mu\nu}]\!+\! S_{\rm G\!B}[g_{\mu\nu}]
 \,,\qquad
\label{counterterm action 2}
\end{eqnarray}
where ${\rm Weyl}^2 \equiv W_{\alpha\beta\gamma\delta}W^{\alpha\beta\gamma\delta}$,
where $W_{\alpha\beta\gamma\delta}$ denotes the Weyl curvature tensor (also known as 
the conformal tensor),
\begin{equation}
W_{\alpha\beta\gamma\delta} \equiv R_{\alpha\beta\gamma\delta}
   - \frac{2}{D\!-\!2}\left(g_{\alpha[\gamma}R_{\delta]\beta}
   -g_{\beta[\gamma}R_{\delta]\alpha}\right)
    +\frac{2}{(D\!-\!1)(D\!-\!2)}g_{\alpha[\gamma}g_{\delta]\beta}R
\,,\qquad
\label{Weyl curvature tensor}
\end{equation}
such that,
\begin{equation}
{\rm Weyl}^2 ={\rm Riem}^2
   - \frac{4}{D\!-\!2}{\rm Ric}^2
    +\frac{2}{(D\!-\!1)(D\!-\!2)}R^2
\,.\quad
\label{Weyl curvature tensor 2}
\end{equation}

\bigskip
\noindent
{\bf $\mathbf{R}^\mathbf{2}$ counterterm.} This counterterm action can be expanded around 
some background metric $\overline{g}_{\mu\nu}$ in perturbations $\delta g_{\mu\nu}$ as,
\begin{eqnarray}
S_{R^2}\big[\overline{g}_{\mu\nu}\!+\!\delta g_{\mu\nu}\big] 
 &\!\!=\!\!& \alpha_{R^2}\! \int\! {\rm d}^Dx\sqrt{-\overline{g}}\,\overline{R}^2
 +\alpha_{R^2}\!\! \int {\rm d}^Dx\left(\frac{\delta S_{R^2}}{\delta g_{\mu\nu}(x)}
   \right)_{\!\overline{g}_{\alpha\beta}}
             \!\!\delta g_{\mu\nu}(x)
\nonumber\\
&&\hskip -1.85cm    
   +\,\frac{\alpha_{R^2}}2\!\int \!{\rm d}^Dx\,{\rm d}^Dx'\delta g_{\rho\sigma}(x')
   \left(\frac{\delta^2 S_{R^2}}
     {\delta g_{\rho\sigma}(x')\delta g_{\mu\nu}(x)}\right)_{\!\overline{g}_{\mu\nu}}
             \!\!\delta g_{\mu\nu}(x)
             \!+\! \mathcal{O}\big(\delta g_{\alpha\beta}^3\big)
\,,
\nonumber\\
&&\hskip -2.8cm
 =\,S^{(0)}_{R^2}\big[\overline{g}_{\mu\nu}\big]
    + S^{(1)}_{R^2}\big[\overline{g}_{\mu\nu},\delta g_{\mu\nu}\big]
    + S^{(2)}_{R^2}\big[\overline{g}_{\mu\nu},\delta g_{\mu\nu}\big]
    + S^{(\geq 3)}_{R^2}\big[\overline{g}_{\mu\nu},\delta g_{\mu\nu}\big]
\,,\qquad\;\;
\label{counterterm action R2}
\end{eqnarray}
where 
\begin{eqnarray}
\delta S_{R^2} &\!\!\!=\!\!\!&\alpha_{R^2}\! \int\! {\rm d}^Dx 
\left\{\big(\delta\sqrt{-g}\,\big)\,\overline{R}^2
   \!\!+\! 2\sqrt{-\overline{g}}\,\overline{R}\Big(\!\!-\!\delta g^{\alpha\beta} \overline{R}_{\alpha\beta}
             \!+\!\overline{g}^{\alpha\beta} \delta R_{\alpha\beta}    \Big)
\right\}
,
\label{counterterm action R2 b1}\\
\delta^2 S_{R^2} &\!\!=\!\!& \alpha_{R^2}\!\! \int {\rm d}^Dx \Big\{
  2\big(\delta\sqrt{-g}\,\big)\,\overline{R}\Big(\!\!-\!\delta g^{\alpha\beta} \overline{R}_{\alpha\beta}
              \!+\!\overline{g}^{\alpha\beta} \delta R_{\alpha\beta}    \Big)
\nonumber\\
&&\hskip 1.3cm
   +\, \sqrt{-\overline{g}}
   \Big(\!\!-\!\delta g^{\alpha\beta} \overline{R}_{\alpha\beta}
                 \!+\!\overline{g}^{\alpha\beta} \delta R_{\alpha\beta}\Big)
  \Big(\!\!-\!\delta g^{\gamma\delta} \overline{R}_{\gamma\delta}
           \!+\!\overline{g}^{\gamma\delta} \delta R_{\gamma\delta}\Big)
\nonumber\\
&&\hskip -0.9cm
+\,\big(\delta^2\sqrt{-g}\,\big)\,\overline{R}^2
              \!\!+\! 2\sqrt{-\overline{g}}\,\overline{R}
   \Big(\delta g^{\alpha\gamma}\delta g_{\gamma}^{\;\beta} \overline{R}_{\alpha\beta} 
   \!-\!\delta g^{\alpha\beta} \delta{R}_{\alpha\beta} 
   \!+\!\overline{g}^{\alpha\beta}\delta^2 R_{\alpha\beta}\Big)
\Big\}
,\qquad\;
\label{counterterm action R2 b2}
\end{eqnarray}
and $\delta \sqrt{-g}$, $\delta^2  \sqrt{-g}$,  
$\delta g^{\alpha\beta}$, $\delta^2 g^{\alpha\beta}$,  $\delta R_{\alpha\beta}$ 
and $\delta^2 R_{\alpha\beta}$ are given in Eqs.~(\ref{variation root g})--(\ref{variation Rmn}),
(\ref{Ricci tensor: first variation}) and~(\ref{Ricci tensor: second variation}), respectively.

It then follows that, 
\begin{eqnarray}
\!\!\!\!\!\!\!\!
\left(\frac{\delta S_{R^2}}{\delta g_{\mu\nu}(x)}\right)
 _{\!g_{\alpha\beta}\rightarrow \overline{g}_{\alpha\beta}}\!\!
 &\!\!=\!\!&  \alpha_{R^2}\sqrt{-\overline{g}} 
    \bigg\{\!\!-\!\frac12 \overline{g}^{\mu\nu}\,\overline{R}^2
              \!-\! 2
               \big(\overline{G}^{\mu\nu}\!\!-\!\overline{\nabla}^{\mu} \overline{\nabla}^{\nu}
            \!\!+\! \overline{g}^{\mu\nu}\overline{\dAlembert}
                 \big)\overline{R}
    \bigg\}
,\quad
\label{counterterm action R2 c1}
\end{eqnarray}
and
\begin{eqnarray}
\frac{1}{\sqrt{-\overline{g}}\sqrt{-\overline{g}'}}
 \left(\frac{\delta^2 S_{R^2}}{\delta g^{\mu\nu}(x)\delta g^{\rho\sigma}(x')}
          \right)_{\!g_{\alpha\beta}\rightarrow \overline{g}_{\alpha\beta}}\!\!
 &\!\!\!=\!\!\!&
  \alpha_{R^2}\bigg\{\!
    \frac14\overline{R}^2\big(\overline{g}^{\mu\nu} \overline{g}^{\rho\sigma}
     \!\!+\!6\overline{g}^{\mu(\rho}\overline{g}^{\sigma)\nu}\big)
\nonumber\\
&&\hskip -4.7cm
     +\,4\overline{R}\,\overline{G}^{\mu)(\rho}\overline{g}^{\sigma)(\nu}
  \!\!+\!2\big(\overline{R}^{\mu\nu}
           \!\!-\!\overline{\nabla}^{\mu}\overline{\nabla}^{\nu}
          \!\!+\!\overline{g}^{\mu\nu}\overline{\dAlembert}\big)
          \big({\overline{R}'}^{\rho\sigma} 
          \!\!-\!{\overline{\nabla}'}^{\rho}{\overline{\nabla}'}^{\sigma}
           \!\!\!+\!{\overline{g}'}^{\rho\sigma}\overline{\dAlembert}'\big)
\nonumber\\
&&\hskip -4.7cm
   -\, \overline{g}^{\mu\nu}\overline{R} \big(\overline{R}^{\rho\sigma} 
          \!\!-\!\overline{\nabla}^{\rho}\overline{\nabla}^{\sigma}
           \!\!+\!\overline{g}^{\rho\sigma}\overline{\dAlembert}\big)
       -\, {\overline{g}'}^{\rho\sigma}{\overline{R}'} \big({\overline{R}'}^{\mu\nu} 
          \!\!-\!{\overline{\nabla}'}^{\mu}{\overline{\nabla}'}^{\nu}
           \!\!+\!{\overline{g}'}^{\mu\nu}\overline{\dAlembert}'\big)     
\nonumber\\
&&\hskip -4.7cm
                 +\,2\overline{R}\big(\!-2\overline{g}^{\mu)(\rho}
                         \overline{\nabla}^{\sigma)}\overline{\nabla}^{(\nu}
                         \!\!+\!\overline{g}^{\rho\sigma}
                         \overline{\nabla}^{\mu}\overline{\nabla}^{\nu}
                          \!\!+\!\overline{g}^{\mu(\rho}\overline{g}^{\sigma)\nu}
                         \overline{\dAlembert}\big)
  \nonumber\\
&&\hskip -4.7cm
   +\,2\overline{R}'\big(\!-2{\overline{g}'}^{\mu)(\rho}
                         {\overline{\nabla}'}^{\sigma)}{\overline{\nabla}'}^{(\nu}
                         \!\!+\!{\overline{g}'}^{\rho\sigma}
                         {\overline{\nabla}'}^{\mu}{\overline{\nabla}'}^{\nu}
                          \!\!+\!{\overline{g}'}^{\mu(\rho}{\overline{g}'}^{\sigma)\nu}
                         \overline{\dAlembert}'\big)
  \nonumber\\
&&\hskip -4.7cm
+\,\Big[\!-\!\big(\overline{g}^{\mu\nu}\overline{g}^{\rho\sigma}
             \!-\!3\overline{g}^{\mu(\rho}\overline{g}^{\sigma)\nu}\big)
             \overline{\nabla}\!\cdot\!{\overline{\nabla}'}
\!+\!2\big(\overline{g}^{\mu\nu}\overline{\nabla}^{(\rho}{\overline{\nabla}'}^{\sigma)}
                  \!+\!\overline{g}^{\rho\sigma}\overline{\nabla}^{(\mu}{\overline{\nabla}'}^{\nu)}\big)
  \nonumber\\
&&\hskip -3.99cm
   -\,3\overline{g}^{\mu(\rho}\big(\overline{\nabla}^{\sigma)}{\overline{\nabla}'}^{(\nu}
                  \!+\!{\overline{\nabla}'}^{\sigma)}\overline{\nabla}^{(\nu}\big)           
       \Big]\overline{R}      \bigg\}
         \frac{\delta^D(x\!-\!x')}{\sqrt{-\overline{g}}}
\,,\qquad\;
\label{counterterm action R2 c2}
\end{eqnarray}
where the Ricci scalar in front of the delta function means that it can be evaluated at $x$ or $x'$,
making the expression symmetric under $(\mu,\nu,x)\leftrightarrow (\rho,\sigma,x')$.
Notice also that the first variation in Eq.~(\ref{counterterm action R2 c1}),
\begin{eqnarray}
\overline{\nabla}_\mu
\left(\frac{1}{\sqrt{-\overline{g}} }\frac{\delta S_{R^2}}{\delta g_{\mu\nu}(x)}
           \right)_{\!\overline{g}_{\alpha\beta}}
 &\!\!=\!\!&  \alpha_{R^2}
    \Big\{
  \!-\overline{R}\overline{\nabla}^{\nu}\,\overline{R}
   \!-\! 2\overline{G}^{\mu\nu}\overline{\nabla}_\mu\overline{R}
              \!+\! 2\overline{R}^{\mu\nu}\overline{\nabla}_\mu\overline{R}
    \Big\}
\nonumber\\
&&\hskip -0.5cm
  = 0
\,,\quad
\label{counterterm action R2 d1}
\end{eqnarray}
is covariantly conserved. The conservation holds for a general background metric
 $\overline{g}_{\mu\nu}(x)$, and it is enforced by the gravitational Noether identity for the counterterm action $S_{R^2}$.

Finally, we address how the operator in Eq.~(\ref{counterterm action R2 c2}) acts on
gravitational perturbations $\delta g_{\rho\sigma}(x')$,
\begin{eqnarray}
\!\!\int\!{\rm d}^Dx'\sqrt{-\overline{g}'}\frac{1}{\sqrt{-\overline{g}}\sqrt{-\overline{g}'}}
 \left(\frac{\delta^2 S_{R^2}}{\delta g_{\mu\nu}(x)\delta g_{\rho\sigma}(x')}
          \right)_{\!\overline{g}_{\alpha\beta}} \!\!\!\delta g_{\rho\sigma}(x')
 &\!\!\!=\!\!\!&
  \alpha_{R^2}\bigg\{\!4\overline{R}\,\overline{G}^{\mu)(\rho}\overline{g}^{\sigma)(\nu}
\nonumber\\
&&\hskip -7.5cm
  -\,\frac14\overline{R}^2\big(\overline{g}^{\mu\nu} \overline{g}^{\rho\sigma}
     \!\!-\!6\overline{g}^{\mu(\rho}\overline{g}^{\sigma)\nu}\big)
\nonumber\\
&&\hskip -7.5cm
  +\,2\big(\overline{G}^{\mu\nu}
           \!\!-\!\overline{\nabla}^{\mu}\overline{\nabla}^{\nu}
          \!\!+\!\overline{g}^{\mu\nu}\overline{\dAlembert}\big)
          \big(\overline{G}^{\rho\sigma} 
          \!\!-\!\overline{\nabla}^{\rho}\overline{\nabla}^{\sigma}
           \!\!+\!\overline{g}^{\rho\sigma}\overline{\dAlembert}\big)   
\nonumber\\
&&\hskip -7.5cm
                 +\,\overline{R}\Big(\!-2\overline{g}^{\mu)(\rho}
                         \overline{\nabla}^{\sigma)}\overline{\nabla}^{(\nu}
                         \!\!+\!\big(\overline{g}^{\mu\nu}\overline{g}^{\rho\sigma}
             \!+\!\overline{g}^{\mu(\rho}\overline{g}^{\sigma)\nu}\big)\overline{\dAlembert}\Big)
  \nonumber\\
&&\hskip -7.5cm
   +\,2\Big[\big(\!-2\overline{g}^{\mu)(\rho}
                         \overline{\nabla}^{\sigma)}\overline{\nabla}^{(\nu}
                         \!\!+\!\overline{g}^{\mu\nu}
                        {\overline{\nabla}}^{\rho}\overline{\nabla}^{\sigma}
                          \!\!+\!\overline{g}^{\mu(\rho}\overline{g}^{\sigma)\nu}
                         \overline{\dAlembert}\big)\overline{R}\Big]
  \nonumber\\
&&\hskip -7.5cm
+\,(\overline{\nabla}_\lambda \overline{R})\Big[\!-\!\overline{g}^{\lambda(\mu}\overline{g}^{\nu)(\rho}
                       \overline{\nabla}^{\sigma)}
                       \!-\!\overline{g}^{\lambda(\rho}\overline{g}^{\sigma)(\mu}
                       \overline{\nabla}^{\nu)}
             \!+\!2\overline{g}^{\rho\sigma}\overline{g}^{\lambda(\mu}\overline{\nabla}^{\nu)}
  \nonumber\\
&&\hskip -5.9cm
 -\,2\overline{g}^{\mu\nu}\overline{g}^{\lambda(\rho}\overline{\nabla}^{\sigma)}
   \!\!+\!\big(\overline{g}^{\mu\nu} \overline{g}^{\rho\sigma}
     \!\!+\!\overline{g}^{\mu(\rho}\overline{g}^{\sigma)\nu}\big)\overline{\dAlembert}   
       \Big]\bigg\}
         \delta g_{\rho\sigma}(x)
,\qquad
\label{counterterm action R2 d1}\\
 &&\hskip -7.5cm
 =\, \frac{1}{\sqrt{-\overline{g}}}\left(\delta\left[\frac{\delta S_{R^2}}{\delta g_{\mu\nu}(x)}\right]\right)
 _{\!g_{\alpha\beta}\rightarrow \overline{g}_{\alpha\beta}}
 \,,\qquad
 \label{counterterm action R2 d2}
\end{eqnarray}
where to get this result we integrated by parts all the derivatives 
in Eq.~(\ref{counterterm action R2 c2}) acting on $x'$ in $\delta^D(x-x')/\sqrt{-\overline{g}}$.
On the other hand,  acting with the operator in Eq.~(\ref{counterterm action R2 c2}) 
on gravitational perturbations $\delta g_{\mu\nu}(x)$ will produce 
an expression that can be obtained from~(\ref{counterterm action R2 d1}) by exacting 
 $(\mu,\nu)\leftrightarrow (\rho,\sigma)$ and $x\rightarrow x'$
 replacements on Eq.~(\ref{counterterm action R2 d1}).
 Finally, an important check of the accuracy of the expression~(\ref{counterterm action R2 d1}) 
 is Eq.~(\ref{counterterm action R2 d2}), which can be obtained by varying 
 Eq.~(\ref{counterterm action R2 c1}).


\bigskip
\noindent
{\bf Ricci tensor squared counterterm.} Just as above we can expand the Ricci squared counterterm
action as,
\begin{eqnarray}
S_{\rm Ric^2}\big[\overline{g}_{\mu\nu}\!+\!\delta g_{\mu\nu}\big] \!\!
 &\!\!=\!\!&\!\! \alpha_{\rm Ric^2}\! \int\!{\rm d}^Dx\sqrt{-\overline{g}}\,
                \overline{R}^{\alpha\beta}\overline{R}_{\alpha\beta}
 \!+\!\alpha_{\rm Ric^2}\!\! \int\! {\rm d}^Dx\!
 \left(\frac{\delta S_{\rm Ric^2}}{\delta g_{\mu\nu}(x)}\right)_{\!\overline{g}_{\alpha\beta}}
             \!\!\!\!\delta g_{\mu\nu}(x)
\nonumber\\
&&\hskip -1.9cm    
   +\,\frac{1}2\alpha_{\rm Ric^2}\!\int \!{\rm d}^Dx\,{\rm d}^Dx'\delta g_{\rho\sigma}(x')
   \left(\frac{\delta^2 S_{\rm Ric^2}}
     {\delta g_{\rho\sigma}(x')\delta g_{\mu\nu}(x)}\right)_{\!\overline{g}_{\alpha\beta}}
             \!\!\delta g_{\mu\nu}(x)
             \!+\! \mathcal{O}\big(\delta g_{\alpha\beta}^3\big)
\nonumber\\
&&\hskip -3.3cm
 =\,S^{(0)}_{\rm Ric^2}\big[\overline{g}_{\mu\nu}\big]
    + S^{(1)}_{\rm Ric^2}\big[\overline{g}_{\mu\nu},\delta g_{\mu\nu}\big]
    + S^{(2)}_{\rm Ric^2}\big[\overline{g}_{\mu\nu},\delta g_{\mu\nu}\big]
    + S^{(\geq 3)}_{\rm Ric^2}\big[\overline{g}_{\mu\nu},\delta g_{\mu\nu}\big]
\,,\qquad\;
\label{counterterm action Ric 2}
\end{eqnarray}
where 
\begin{eqnarray}
\delta S_{\rm Ric^2} &\!\!\!=\!\!\!&\alpha_{\rm Ric^2}\! \int\! {\rm d}^Dx 
\left\{\delta\sqrt{-g}\,\overline{R}_{\alpha\beta}\overline{R}^{\alpha\beta} 
   \!\!\!+\! 2\sqrt{-\overline{g}}\,\overline{R}^{\alpha\beta}\big(\delta R_{\alpha\beta} 
   \!-\!\delta g_{\alpha\gamma}\overline{R}_{\;\beta}^{\gamma}\big)
\right\}
,\!
\label{counterterm action Ric 2 b1}\\
\delta^2 S_{\rm Ric^2} &\!\!=\!\!& \alpha_{\rm Ric^2}\!\! \int {\rm d}^Dx \Big\{
  2\delta\sqrt{-g}\,\overline{R}^{\alpha\beta}
     \Big(\delta{R}_{\alpha\beta}\!-\!\overline{R}^{\gamma}_{\;\alpha}\delta{g}_{\beta\gamma}\Big)
   \!-\! 4\sqrt{-\overline{g}}\,\overline{R}^{\alpha\beta}\delta g_{\alpha\gamma} 
                  \delta R^{\gamma}_{\;\beta}
 \nonumber\\
&&\hskip 1.2cm 
+\,\big(\delta^2\sqrt{-g}\big)\,\overline{R}^{\alpha\beta}\overline{R}_{\alpha\beta} 
   \!+\!\sqrt{-\overline{g}}\, \overline{g}^{\alpha\gamma}\overline{g}^{\beta\delta} 
      \delta R_{\alpha\beta}  \delta R_{\gamma\delta}
\nonumber\\
&&\hskip 1.2cm
+\, \sqrt{-\overline{g}}\,\overline{R}^{\alpha\beta}
   \Big(
   \delta g_{\alpha\gamma}\delta g_{\beta\delta} \overline{R}^{\gamma\delta} 
   \!+\!2\delta g_{\alpha\delta}\delta g^{\delta}_{\;\gamma} \overline{R}_{\;\beta}^{\gamma} 
   \!+\!2\delta^2 R_{\alpha\beta}\Big) 
\Big\}
,\qquad\;
\label{counterterm action Ric 2 b2}
\end{eqnarray}
and $\delta \sqrt{-g}$, $\delta^2  \sqrt{-g}$,  
$\delta g^{\alpha\beta}$, $\delta^2 g^{\alpha\beta}$,  $\delta R_{\alpha\beta}$ 
and $\delta^2 R_{\alpha\beta}$ are given in Eqs.~(\ref{variation root g})--(\ref{variation Rmn}),
(\ref{Ricci tensor: first variation}) and~(\ref{Ricci tensor: second variation}), respectively.

The first variation follows from Eq.~(\ref{counterterm action Ric 2 b1}), 
\begin{eqnarray}
\!\!\!\!\!\!\!\!
\left(\frac{\delta S_{\rm Ric^2}}{\delta g_{\mu\nu}(x)}\right)
 _{\!g_{\alpha\beta}\rightarrow \overline{g}_{\alpha\beta}}\!\!
 &\!\!=\!\!&  \alpha_{\rm Ric^2}\sqrt{-\overline{g}} 
    \bigg\{\frac12 \overline{g}^{\mu\nu}\,\overline{R}_{\alpha\beta}\overline{R}^{\alpha\beta}
              \!-\! 2\overline{R}^{\alpha(\mu}\overline{R}^{\nu)}_{\;\alpha}
              \!+\!2\overline{\nabla}_{\alpha} \overline{\nabla}^{(\mu}\overline{R}^{\nu)\alpha}
\nonumber\\
&&\hskip 2.25cm              
            -\, \overline{\dAlembert}\,\overline{R}^{\mu\nu}
          \!-\! \overline{g}^{\mu\nu}\overline{\nabla}_{\alpha} \overline{\nabla}_{\beta}
                          \overline{R}^{\alpha\beta}
    \bigg\}
,\quad
\label{counterterm action Ric 2 c1}\\
 &\!\!=\!\!&  \alpha_{\rm Ric^2}\sqrt{-\overline{g}} 
    \bigg\{\frac12 \overline{g}^{\mu\nu}\,\overline{R}_{\alpha\beta}\overline{R}^{\alpha\beta}
              \!+\! 2\overline{R}_{\alpha\beta}\overline{R}^{\alpha(\mu\nu)\beta}
              \!+\!\overline{\nabla}^{\mu} \overline{\nabla}^{\nu}\overline{R}
\nonumber\\
&&\hskip 2.25cm              
            -\, \overline{\dAlembert}\,\overline{R}^{\mu\nu}
          \!-\!\frac12 \overline{g}^{\mu\nu}\overline{\dAlembert}\overline{R}
    \bigg\}
\,,\quad
\label{counterterm action Ric 2 c1 bis}
\end{eqnarray}
where to get the last equality we made use of the identity,
\begin{equation}
\overline{\nabla}_{\alpha} \overline{\nabla}^{(\mu}\overline{R}^{\nu)\alpha}
  = \frac12\overline{\nabla}^{\mu} \overline{\nabla}^{\nu}\overline{R}
  \!+\! \overline{R}_{\alpha\beta}\overline{R}^{\alpha(\mu\nu)\beta}
           \!+\! \overline{R}^{\alpha(\mu}\overline{R}^{\nu)}_{\;\alpha}
\,.\quad
\label{commuting derivatives on Ricci tensor}
\end{equation}
The second variation follows from Eq.~(\ref{counterterm action Ric 2 b2}),
\begin{eqnarray}
\frac{1}{\sqrt{-\overline{g}}\sqrt{-\overline{g}'}}
 \left(\frac{\delta^2 S_{\rm Ric^2}}{\delta g_{\mu\nu}(x)\delta g_{\rho\sigma}(x')}
             \right)_{\!g_{\alpha\beta}\rightarrow \overline{g}_{\alpha\beta}}\!\!
 &\!\!\!=\!\!\!&
  \alpha_{\rm Ric^2}\bigg\{4\overline{R}^{\mu)\alpha} 
                    \overline{R}^{(\rho}_{\;\alpha}\overline{g}^{\sigma)(\nu}
                    \!+\!2\overline{R}^{\mu(\rho} \overline{R}^{\sigma)\nu} 
 \nonumber\\
&&\hskip -5.7cm                         
 +\,\frac14\overline{R}^{\alpha\beta}\overline{R}_{\alpha\beta}
     \big(\overline{g}^{\mu\nu} \overline{g}^{\rho\sigma}
                   \!\!-\!2\overline{g}^{\mu(\rho}\overline{g}^{\sigma)\nu}\big)
 \!-\!\big(\overline{g}^{\mu\nu} \overline{R}^{\alpha(\rho}\overline{R}_{\alpha}^{\;\sigma)}
    \!+\!\overline{g}^{\rho\sigma} \overline{R}^{\alpha(\mu}\overline{R}_{\alpha}^{\;\nu)}
    \big)
\nonumber\\
&&\hskip -5.7cm
     +\,2\Big(\overline{\nabla}_{(\alpha}\delta_{\beta)}^{\;(\mu}\overline{\nabla}^{\nu)}
          \!-\!\frac12 \delta_\alpha^{\;(\mu}\delta_\beta^{\;\nu)}\overline{\dAlembert}
            \!-\!\frac12 \overline{g}^{\mu\nu}\overline{\nabla}_{(\alpha}\overline{\nabla}_{\beta)}\Big)
\nonumber\\
&&\hskip -3.5cm
   \times\Big(\overline{\nabla}^{(\rho}\overline{g}^{\sigma)(\alpha}\overline{\nabla}^{\beta)}
          \!-\!\frac12 \overline{g}^{\alpha(\rho}\overline{g}^{\sigma)\beta}\overline{\dAlembert}
 \!-\!\frac12\overline{g}^{\rho\sigma}\overline{\nabla}^{(\alpha}\overline{\nabla}^{\beta)}\Big)
\nonumber\\
&&\hskip -5.7cm
     +\,\overline{g}^{\mu\nu}
         \Big(\overline{R}^{\alpha(\rho}\overline{\nabla}^{\sigma)}\overline{\nabla}_{\alpha}
          \!-\!\frac12 \overline{R}^{\rho\sigma}\overline{\dAlembert}
               \!-\!\frac12 \overline{g}^{\rho\sigma}\overline{R}^{\alpha\beta}\overline{\nabla}_{\alpha}
                             \overline{\nabla}_{\beta}\Big)
\nonumber\\
&&\hskip -5.7cm
     +\,{\overline{g}'}^{\rho\sigma}
         \Big({\overline{R}'}^{\alpha(\mu}{\overline{\nabla}'}^{\nu)}{\overline{\nabla}'}_{\!\alpha}
          \!-\!\frac12 {\overline{R}'}^{\mu\nu}\overline{\dAlembert}'
          \!-\!\frac12 {\overline{g}'}^{\mu\nu}{\overline{R}'}^{\alpha\beta}{\overline{\nabla}'}_{\!\alpha}
                             {\overline{\nabla}'}_{\!\!\beta}\Big)
\nonumber\\
&&\hskip -5.7cm
     -\,
         \Big(2\overline{R}^{\mu)(\rho}\overline{\nabla}^{\sigma)}\overline{\nabla}^{(\nu}           
                 \!+\!2\overline{R}^{\alpha(\mu}\overline{g}^{\nu)(\rho}\overline{\nabla}^{\sigma)}           
                         \overline{\nabla}_{\alpha}
          \!-\!2\overline{R}^{\rho)(\mu}\overline{g}^{\nu)(\sigma}\overline{\dAlembert}
\nonumber\\
&&\hskip -1.7cm
               -\,\overline{g}^{\rho\sigma}\overline{R}^{\alpha(\mu} \overline{\nabla}^{\nu)}
                          \overline{\nabla}_{\alpha}
           \!-\!\overline{g}^{\rho\sigma}\overline{R}^{\alpha(\mu}\overline{\nabla}_{\alpha}
                            \overline{\nabla}^{\nu)}
                          \Big)
\nonumber\\
&&\hskip -6.95cm
     -\,
         \Big(2{\overline{R}'}^{\rho)(\mu}{\overline{\nabla}'}^{\nu)}{\overline{\nabla}'}^{(\sigma}           
                \!+\!2{\overline{R}'}^{\alpha(\rho}{\overline{g}'}^{\sigma)(\mu}{\overline{\nabla}'}^{\nu)}           
                         {\overline{\nabla}'}_{\!\!\alpha}
          \!\!-\!2{\overline{R}'}^{\rho)(\mu}{\overline{g}'}^{\nu)(\sigma}{\overline{\dAlembert}'}
\nonumber\\
&&\hskip -1.7cm
               -\,{\overline{g}'}^{\mu\nu}{\overline{R}'}^{\alpha(\rho} {\overline{\nabla}'}^{\sigma)}
                          {\overline{\nabla}'}_{\!\!\alpha}
    \!-\!{\overline{g}'}^{\mu\nu}{\overline{R}'}^{\alpha(\rho}{\overline{\nabla}'}_{\!\!\alpha}
                     {\overline{\nabla}'}^{\sigma)}      
                          \Big) 
\nonumber\\
&&\hskip -5.7cm
     +\,
         \Big(\overline{R}^{\rho\sigma}\overline{\nabla}^{(\mu}\overline{\nabla}^{\nu)}
                         \!+\!\overline{g}^{\mu(\rho}\overline{g}^{\sigma)\nu}
                                \overline{R}^{\alpha\beta} \overline{\nabla}_{\alpha}\overline{\nabla}_{\beta}           
          \!-\!2\overline{R}^{\alpha(\rho}\overline{g}^{\sigma)(\mu}
                         \overline{\nabla}^{\nu)} \overline{\nabla}_{\alpha}\Big)
\nonumber\\
&&\hskip -5.7cm
     +\,
         \Big({\overline{R}'}^{\mu\nu}{\overline{\nabla}'}^{(\rho}           
                         {\overline{\nabla}'}^{\sigma)}
  \!+\!\overline{g}^{\mu(\rho}\overline{g}^{\sigma)\nu}
                 {\overline{R}'}^{\alpha\beta}{\overline{\nabla}'}_{\!\alpha}{\overline{\nabla}'}_{\!\beta}
             \!-\!2{\overline{R}'}^{\alpha(\mu}{\overline{g}'}^{\nu)(\rho}
                {\overline{\nabla}'}^{\sigma)}{\overline{\nabla}'}_{\!\!\alpha}\Big)
  \nonumber\\
&&\hskip -6.55cm
+\,\Big(\!\!-\!\frac12\overline{g}^{\mu(\rho}\overline{g}^{\sigma)\nu}
    \overline{\nabla}_{\alpha}\overline{R}^{\alpha\beta}\overline{\nabla}_{\beta}
\!-\!\overline{g}^{\mu)(\rho}\overline{\nabla}_\lambda\overline{R}^{\sigma)(\nu}
         \overline{\nabla}^\lambda
\!+\!\overline{\nabla}^{(\rho}\overline{R}^{\sigma)(\mu}\overline{\nabla}^{\nu)}\Big)
  \nonumber\\
&&\hskip -6.55cm
+\,\Big(\!\!-\!\frac12\overline{g}^{\mu(\rho}\overline{g}^{\sigma)\nu}
    {\overline{\nabla}'}_{\!\alpha}{\overline{R}'}^{\alpha\beta}{\overline{\nabla}'}_{\!\beta}
    \!-\!\overline{g}^{\mu)(\rho}{\overline{\nabla}'}_{\!\lambda}{\overline{R}'}^{\sigma)(\nu}
              {\overline{\nabla}'}^\lambda
\!+\!{\overline{\nabla}'}^{(\mu}{\overline{R}'}^{\nu)(\rho}{\overline{\nabla}'}^{\sigma)}
       \Big)
 \nonumber\\
&&\hskip -6.55cm
+\,\Big(2
    \overline{\nabla}^{(\mu}\overline{g}^{\nu)(\rho}\overline{R}^{\sigma)\alpha}
             \overline{\nabla}_{\alpha}
    \!-\!\overline{\nabla}^{\mu)}\overline{R}^{\rho\sigma}\overline{\nabla}^{(\nu}
\!-\!\overline{g}^{\mu\nu}\overline{\nabla}^{(\rho}\overline{R}^{\sigma)\alpha}
               \overline{\nabla}_{\alpha}
         \!+\!\frac12\overline{g}^{\mu\nu}\overline{\nabla}_\lambda \overline{R}^{\rho\sigma}
                    \overline{\nabla}^\lambda
                \Big)
  \nonumber\\
&&\hskip -6.55cm
+\,\Big(2
    {\overline{\nabla}'}^{(\rho}{\overline{g}'}^{\sigma)(\mu}{\overline{R}'}^{\nu)\alpha}{\overline{\nabla}'}_{\!\!\alpha}
    \!-\!{\overline{\nabla}'}^{\rho)}{\overline{R}'}^{\mu\nu}{\overline{\nabla}'}^{(\sigma}
\!\!-\!\overline{g}^{\rho\sigma}{\overline{\nabla}'}^{(\mu} {\overline{R}'}^{\nu)\alpha}
          {\overline{\nabla}'}_{\!\!\alpha}
    \nonumber\\
&&\hskip -3.5cm        
         +\,\frac12\overline{g}^{\rho\sigma}{\overline{\nabla}'}_{\!\!\lambda} {\overline{R}'}^{\mu\nu}
                 {\overline{\nabla}'}^\lambda
                \Big)\! \bigg\}
        \!\times\! \frac{\delta^D(x\!-\!x')}{\sqrt{-\overline{g}}}
\,,\qquad\;
\label{counterterm action Ric 2 c2}
\end{eqnarray}
where the derivative operators act on the quantities to the right,
making the expression manifestly symmetric under $(\mu,\nu,x)\leftrightarrow (\rho,\sigma,x')$.
Note that the four derivative term in the third and fourth lines can be equivalently written 
in a manifestly symmetric form as, 
\begin{eqnarray}
&&\hskip -.7cm
     \bigg\{\Big(\overline{\nabla}_{(\alpha}\overline{\nabla}^{(\mu}\delta_{\beta)}^{\;\nu)}
          \!-\!\frac12 \delta_\alpha^{\;(\mu}\delta_\beta^{\;\nu)}\overline{\dAlembert}
            \!-\!\frac12 \overline{g}^{\mu\nu}\overline{\nabla}_{(\alpha}\overline{\nabla}_{\beta)}\Big)
\nonumber\\
&&\hskip 0.5cm
   \times
  \Big(\overline{\nabla}^{(\rho}\overline{g}^{\sigma)(\alpha)}\overline{\nabla}^{\beta)}
          \!-\!\frac12 \overline{g}^{\alpha(\rho}\overline{g}^{\sigma)\beta}\overline{\dAlembert}
 \!-\!\frac12\overline{g}^{\rho\sigma}\overline{\nabla}^{(\alpha}\overline{\nabla}^{\beta)}\Big)
\nonumber\\
&&\hskip -.7cm
    +\, \Big({\overline{\nabla}'}_{(\alpha}\delta{g}_{\beta)}^{\;(\rho}{\overline{\nabla}'}^{\sigma)}
          \!-\!\frac12 \delta_{(\alpha}^{\;(\rho}\delta^{\;\sigma)}_{\beta)}\overline{\dAlembert}'
            \!-\!\frac12 {\overline{g}'}^{\rho\sigma}{\overline{\nabla}'}_{(\alpha}{\overline{\nabla}'}_{\beta)}\Big)
\nonumber\\
&&\hskip 0.5cm
   \times\Big({\overline{\nabla}'}^{(\mu}{\overline{g}'}^{\nu)(\alpha}{\overline{\nabla}'}^{\beta)}
          \!-\!\frac12 {\overline{g}'}^{\alpha(\mu}{\overline{g}'}^{\nu)\beta}\overline{\dAlembert}'
 \!-\!\frac12{\overline{g}'}^{\mu\nu}{\overline{\nabla}'}^{(\alpha}{\overline{\nabla}'}^{\beta)}\Big)
 \bigg\}\frac{\delta^D(x\!-\!x')}{\sqrt{-\overline{g}}}
 \,.\qquad\;
\label{four derivative term}
\end{eqnarray}

As in the case of the $R^2$ counterterm, 
the expression in Eq.~(\ref{counterterm action Ric 2 c1 bis}) is covariantly conserved,
\begin{eqnarray}
\overline{\nabla}_\mu
\left(\frac{1}{\sqrt{-\overline{g}} }\frac{\delta S_{\rm Ric^2}}{\delta g_{\mu\nu}(x)}\right)
_{\!\overline{g}_{\alpha\beta}}\!\!\!
 &\!\!\!=\!\!\!&  \alpha_{\rm Ric^2}
 \bigg\{\frac12 \overline{\nabla}^{\nu}\big(\overline{R}_{\alpha\beta}\overline{R}^{\alpha\beta}\,\big)
    \!+\! 2\overline{\nabla}_\mu\big(\overline{R}_{\alpha\beta}\overline{R}^{\mu(\alpha\beta)\nu}\big)
\nonumber\\
&&\hskip 1.18cm      
             +\,\overline{\dAlembert} \overline{\nabla}^{\nu}\overline{R}        
           \! -\! \overline{\nabla}_{\mu}\overline{\dAlembert}\,\overline{R}^{\mu\nu}
          \!-\!\frac12\overline{\nabla}^{\nu}\overline{\dAlembert}\overline{R}
    \bigg\}
\nonumber\\
 &\!\!=\!\!&  \alpha_{\rm Ric^2}
 \bigg\{\frac12 \overline{\nabla}^{\nu}\big(\overline{R}_{\alpha\beta}\overline{R}^{\alpha\beta}\,\big)
    \!+\! \overline{\nabla}_\mu\big(\overline{R}_{\alpha\beta}\overline{R}^{\mu(\alpha\beta)\nu}\big)
\label{counterterm action Ric 2 d1b}\\
&&\hskip 0.9cm  
            +\,\frac12 \overline{R} ^{\nu\alpha}\overline{\nabla}_{\alpha}\overline{R}          
 \!-\! \overline{\nabla}_{\alpha}\big(\overline{R}^{\alpha}_{\;\beta}\overline{R}^{\beta\nu}\big)
 \!-\!\overline{R}^{\gamma(\alpha\beta)\nu}\overline{\nabla}_{\gamma}\overline{R}_{\alpha\beta}
    \bigg\}\,
\nonumber\\
&&\hskip -0.5cm
  = 0
\,,\quad
\label{counterterm action Ric 2 d1c}
\end{eqnarray}
and the conservation holds for a general background metric $\overline{g}_{\mu\nu}(x)$.
To get the second equality~(\ref{counterterm action Ric 2 d1b}) we used,
\begin{eqnarray}
 \overline{\nabla}_{\mu}\overline{\dAlembert}\,\overline{R}^{\mu\nu}
  &\!\!=\!\!&\frac12 \overline{\dAlembert}\overline{\nabla}^{\nu}\overline{R}
 \!+\! \overline{\nabla}_{\mu}\big(\overline{R}^{\mu}_{\;\,\alpha}\overline{R}^{\alpha\nu}
 \!+\!\overline{R}^{\mu\alpha\beta\nu}\overline{R}_{\alpha\beta}\big)
 \!+\!\overline{R}^{\nu\alpha\beta\mu}\overline{\nabla}_{\mu}\overline{R}_{\alpha\beta}
\nonumber\\
 \big(\overline{\dAlembert}\overline{\nabla}^{\nu}
               \!-\!\overline{\nabla}^{\nu}\overline{\dAlembert} \big)\overline{R}    
  &\!\!=\!\!& \overline{R} ^{\nu\mu}\overline{\nabla}_{\mu}\overline{R} 
\,,\quad
\label{three derivative identities}
\end{eqnarray}
and to get the final result~(\ref{counterterm action Ric 2 d1c}) we
applied $A{\rm d}B = {\rm d}(AB) -({\rm d}A)B$ to the first and third term in 
the second line of Eq.~(\ref{counterterm action Ric 2 d1b}) and  used 
contracted Bianchi identities,
\begin{equation}
\overline{\nabla}_{\mu}\overline{R}^{\mu\alpha\beta\gamma} 
 =\overline{\nabla}^{\beta}\overline{R}^{\alpha\gamma}
  -\overline{\nabla}^{\gamma}\overline{R}^{\alpha\beta}
\,,\qquad
\overline{\nabla}_{\alpha}\overline{R}^{\alpha\beta} =\frac12\overline{\nabla}^{\beta}\overline{R}
  \,.\qquad
\label{contracted Bianchi identities}
\end{equation}

Next we address how the operator in Eq.~(\ref{counterterm action Ric 2 c2}) acts on
gravitational perturbations $\delta g_{\rho\sigma}(x')$,
\begin{eqnarray}
\!\int\!{\rm d}^Dx'\sqrt{-\overline{g}'}\left(\frac{1}{\sqrt{-\overline{g}}\sqrt{-\overline{g}'}}
 \frac{\delta^2 S_{\rm Ric^2}}{\delta g^{\mu\nu}(x)\delta g^{\rho\sigma}(x')}
    \right)_{\!\overline{g}_{\alpha\beta}} \!\!\!\delta g_{\rho\sigma}(x')
 &\!\!\!=\!\!\!&
   \alpha_{\rm Ric^2}
\nonumber\\
&&\hskip -9.7cm   
   \times\,\bigg\{\Big(4\overline{R}^{\mu)\alpha} 
                    \overline{R}^{(\rho}_{\;\alpha}\overline{g}^{\sigma)(\nu}
                    \!+\!2\overline{R}^{\mu(\rho} \overline{R}^{\sigma)\nu} \Big) \delta g_{\rho\sigma}
 \nonumber\\
&&\hskip -9.7cm                         
 +\,\frac14\overline{R}^{\alpha\beta}\overline{R}_{\alpha\beta}
     \big(\overline{g}^{\mu\nu} \overline{g}^{\rho\sigma}
                   \!\!-\!2\overline{g}^{\mu(\rho}\overline{g}^{\sigma)\nu}\big) \delta g_{\rho\sigma}
 \!-\!\big(\overline{g}^{\mu\nu} \overline{R}^{\alpha(\rho}\overline{R}_{\alpha}^{\;\sigma)}
    \!+\!\overline{g}^{\rho\sigma} \overline{R}^{\alpha(\mu}\overline{R}_{\alpha}^{\;\nu)}
    \big)  \delta g_{\rho\sigma}
\nonumber\\
&&\hskip -9.7cm
     +\,2\,\Big(\overline{\nabla}_{(\alpha}\delta_{\beta)}^{\;(\mu}\overline{\nabla}^{\nu)}
          \!-\!\frac12 \delta_\alpha^{\;(\mu}\delta_\beta^{\;\nu)}\overline{\dAlembert}
            \!-\!\frac12 \overline{g}^{\mu\nu}\overline{\nabla}_{(\alpha}\overline{\nabla}_{\beta)}\Big)
\nonumber\\
&&\hskip -6.5cm
   \times\,\Big(\overline{\nabla}_{\rho}\overline{\nabla}^{(\alpha} \delta g^{\beta)\rho}
          \!-\!\frac12 \overline{\dAlembert} \delta g^{\alpha\beta}
 \!-\!\frac12\overline{\nabla}^{\alpha}\overline{\nabla}^{\beta} \delta g_{\rho}^{\;\rho}\Big)
\nonumber\\
&&\hskip -9.7cm
     +\,\overline{g}^{\mu\nu}
         \Big(\overline{R}^{\alpha(\rho}\overline{\nabla}^{\sigma)}\overline{\nabla}_{\alpha}
          \!-\!\frac12 \overline{R}^{\rho\sigma}\overline{\dAlembert}
               \!-\!\frac12 \overline{g}^{\rho\sigma}\overline{R}^{\alpha\beta}\overline{\nabla}_{\alpha}
                             \overline{\nabla}_{\beta}\Big) \delta g_{\rho\sigma}
\nonumber\\
&&\hskip -9.7cm
     +\,\overline{g}^{\rho\sigma}
         \Big(\overline{\nabla}_{\alpha}\overline{\nabla}^{(\mu}\overline{R}^{\nu)\alpha}
          \!-\!\frac12\overline{\dAlembert}\,\overline{R}^{\mu\nu}
             \!-\!\frac12 \overline{g}^{\mu\nu}\overline{\nabla}_{\alpha}
                             \overline{\nabla}_{\beta}\overline{R}^{\alpha\beta}\Big) \delta g_{\rho\sigma}
\nonumber\\
&&\hskip -9.7cm
     -\,
         \Big(2\overline{R}^{\mu)(\rho}\overline{\nabla}^{\sigma)}\overline{\nabla}^{(\nu}           
                         \!+\!2\overline{R}^{\alpha(\mu}\overline{g}^{\nu)(\rho}\overline{\nabla}^{\sigma)}           
                         \overline{\nabla}_{\alpha}
          \!-\!2\overline{R}^{\rho)(\mu}\overline{g}^{\nu)(\sigma}\overline{\dAlembert}
\nonumber\\
&&\hskip -5.7cm
              -\,\overline{g}^{\rho\sigma}\overline{R}^{\alpha(\mu} \overline{\nabla}_{\alpha}
                        \overline{\nabla}^{\nu)}  
     \!-\!\overline{g}^{\rho\sigma}\overline{R}^{\alpha(\mu} \overline{\nabla}^{\nu)}
                          \overline{\nabla}_{\alpha}\Big) \delta g_{\rho\sigma}
\nonumber\\
&&\hskip -9.7cm
     -\,
         \Big(
         2\overline{\nabla}^{\rho)}\overline{\nabla}^{(\mu}\overline{R}^{\nu)(\sigma}  
                  \!+\!2\overline{\nabla}_{\alpha}\overline{\nabla}^{(\mu} \overline{g}^{\nu)(\rho}          
                         \overline{R}^{\sigma)\alpha} 
          \!\!-\!2\overline{g}^{\mu)(\rho}\overline{\dAlembert}\,\overline{R}^{\sigma)(\nu}
\nonumber\\
&&\hskip -5.7cm
              -\,\overline{g}^{\mu\nu}\overline{\nabla}_{\alpha}\overline{\nabla}^{(\rho}
                          \overline{R}^{\sigma)\alpha} 
               \!-\!\overline{g}^{\mu\nu}\overline{\nabla}^{(\rho}\overline{\nabla}_{\alpha}
                          \overline{R}^{\sigma)\alpha} 
                          \Big) \delta g_{\rho\sigma}
\nonumber\\
&&\hskip -9.7cm
     +\,
         \Big(\overline{R}^{\rho\sigma}\overline{\nabla}^{(\mu}\overline{\nabla}^{\nu)}
  \!-\!2\overline{R}^{\alpha(\rho}\overline{g}^{\sigma)(\mu}
                         \overline{\nabla}^{\nu)} \overline{\nabla}_{\alpha}
                              \!+\!\overline{g}^{\mu(\rho}\overline{g}^{\sigma)\nu}                                       
                              \overline{R}^{\alpha\beta} \overline{\nabla}_{\alpha}\overline{\nabla}_{\beta}           
        \Big) \delta g_{\rho\sigma}
\nonumber\\
&&\hskip -9.7cm
     +\,
         \Big(\overline{\nabla}^{(\rho}           
                         \overline{\nabla}^{\sigma)}\overline{R}^{\mu\nu}
  \!-\!2\overline{\nabla}_{\alpha}\overline{\nabla}^{(\rho}
                     \overline{g}^{\sigma)(\mu}\overline{R}^{\nu)\alpha}
  \!+\!\overline{g}^{\mu(\rho}\overline{g}^{\sigma)\nu}
                \overline{\nabla}_{\alpha}\overline{\nabla}_{\beta}\overline{R}^{\alpha\beta}
           \Big) \delta g_{\rho\sigma}
  \nonumber\\
&&\hskip -9.7cm
+\,\Big(\!\!-\!\overline{g}^{\mu(\rho}\overline{g}^{\sigma)\nu}
    \overline{\nabla}_{\alpha}\overline{R}^{\alpha\beta}\overline{\nabla}_{\beta}
    \!-\!2\overline{g}^{\mu)(\rho}\overline{\nabla}_\lambda \overline{R}^{\sigma)(\nu}
                             \overline{\nabla}^\lambda
\!\!+\!2\overline{\nabla}^{(\rho}\overline{R}^{\sigma)(\mu}\overline{\nabla}^{\nu)}
         \Big)
           \delta g_{\rho\sigma}
 \nonumber\\
&&\hskip -9.7cm
+\,\Big(2\overline{\nabla}^{(\mu}\overline{g}^{\nu)(\rho}
   \overline{R}^{\sigma)\alpha} \overline{\nabla}_{\alpha}
    \!-\!\overline{\nabla}^{\mu)}\overline{R}^{\rho\sigma}\overline{\nabla}^{(\nu}
\!\!-\!\overline{g}^{\mu\nu}\overline{\nabla}^{(\rho}
              \overline{R}^{\sigma)\alpha}\overline{\nabla}_{\alpha}
         \!+\!\frac12\overline{g}^{\mu\nu}\overline{\nabla}_\lambda
                 \overline{R}^{\rho\sigma}\overline{\nabla}^\lambda\Big) \delta g_{\rho\sigma}
  \nonumber\\
&&\hskip -9.7cm
+\,\Big(2\overline{\nabla}_{\alpha}
    \overline{R}^{\alpha(\mu}\overline{g}^{\nu)(\rho}\overline{\nabla}^{\sigma)}
    \!\!-\!\overline{\nabla}^{\rho)}\overline{R}^{\mu\nu}\overline{\nabla}^{(\sigma}
      \! \!\!-\!\overline{g}^{\rho\sigma}\overline{\nabla}_{\alpha}
              \overline{R}^{\alpha(\mu}\overline{\nabla}^{\nu)}
         \!\!+\!\frac12\overline{g}^{\rho\sigma}\overline{\nabla}_\lambda
                 \overline{R}^{\mu\nu}\overline{\nabla}^\lambda\Big) \delta g_{\rho\sigma}\! \bigg\},\!
\nonumber\\
 \label{counterterm action Ric 2 d2}
\end{eqnarray}
where the derivatives act on the quantities to the right. 
This can be simplified to, 
\begin{eqnarray}
\!\int\!{\rm d}^Dx'\sqrt{-\overline{g}'}\left(\frac{1}{\sqrt{-\overline{g}}\sqrt{-\overline{g}'}}
 \frac{\delta^2 S_{\rm Ric^2}}{\delta g^{\mu\nu}(x)\delta g^{\rho\sigma}(x')}
    \right)_{\!\overline{g}_{\alpha\beta}} \!\!\!\delta g_{\rho\sigma}(x')
&&
\nonumber\\
 &&\hskip -9.7cm
   =\,\alpha_{\rm Ric^2}
   \times\,\bigg\{\Big(4\overline{R}^{\mu)\alpha} 
                    \overline{R}^{(\rho}_{\;\alpha}\overline{g}^{\sigma)(\nu}
                    \!+\!2\overline{R}^{\mu(\rho} \overline{R}^{\sigma)\nu} \Big) \delta g_{\rho\sigma}
 \nonumber\\
&&\hskip -9.7cm                         
 +\,\frac14\overline{R}^{\alpha\beta}\overline{R}_{\alpha\beta}
     \big(\overline{g}^{\mu\nu} \overline{g}^{\rho\sigma}
                   \!\!-\!2\overline{g}^{\mu(\rho}\overline{g}^{\sigma)\nu}\big) \delta g_{\rho\sigma}
 \!-\!\big(\overline{g}^{\mu\nu} \overline{R}^{\alpha(\rho}\overline{R}_{\alpha}^{\;\sigma)}
    \!+\!\overline{g}^{\rho\sigma} \overline{R}^{\alpha(\mu}\overline{R}_{\alpha}^{\;\nu)}
    \big)  \delta g_{\rho\sigma}
\nonumber\\
&&\hskip -9.7cm
     +\,2\,\Big(\overline{\nabla}_{(\alpha}\delta_{\beta)}^{\;(\mu}\overline{\nabla}^{\nu)}
          \!-\!\frac12 \delta_\alpha^{\;(\mu}\delta_\beta^{\;\nu)}\overline{\dAlembert}
            \!-\!\frac12 \overline{g}^{\mu\nu}\overline{\nabla}_{(\alpha}\overline{\nabla}_{\beta)}\Big)
\nonumber\\
&&\hskip -6.5cm
   \times\,
    \Big(\overline{\nabla}_{\rho}\overline{\nabla}^{(\alpha} \delta g^{\beta)\rho}
          \!-\!\frac12 \overline{\dAlembert} \delta g^{\alpha\beta}
 \!-\!\frac12\overline{\nabla}^{\alpha}\overline{\nabla}^{\beta} \delta g_{\rho}^{\;\rho}\Big)
\nonumber\\
&&\hskip -9.7cm
     +\,
         \Big(\!\!-\!2\overline{R}^{\mu)(\rho}\overline{\nabla}^{\sigma)}\overline{\nabla}^{(\nu}           
                 \!\!-\!2\overline{R}^{\alpha(\mu}\overline{g}^{\nu)(\rho}\overline{\nabla}^{\sigma)}           
                         \overline{\nabla}_{\alpha}
                  \!\!-\!2\overline{R}^{\alpha(\rho}\overline{g}^{\sigma)(\mu}\overline{\nabla}_{\alpha}
                  \overline{\nabla}^{\nu)}  
          \!\!+\!2\overline{R}^{\rho)(\mu}\overline{g}^{\nu)(\sigma}\overline{\dAlembert}
              \Big) \delta g_{\rho\sigma}
\nonumber\\
&&\hskip -9.7cm
     +\,
         \Big(\overline{g}^{\mu\nu}\overline{R}^{\alpha(\rho}\overline{\nabla}_{\alpha}
                     \overline{\nabla}^{\sigma)}
           \!+\!\overline{g}^{\mu\nu}\overline{R}^{\alpha(\rho}
                     \overline{\nabla}^{\sigma)}\overline{\nabla}_{\alpha}
                      \!+\!2\overline{g}^{\rho\sigma}\overline{R}^{\alpha(\mu} \overline{\nabla}^{\nu)}
                          \overline{\nabla}_{\alpha}
              \overline{\nabla}^{\nu)}
\nonumber\\
&&\hskip -5.7cm
               -\,\big(\overline{g}^{\mu\nu} \overline{g}^{\rho\sigma}
               \!-\!\overline{g}^{\mu(\rho}\overline{g}^{\sigma)\nu}\big)\overline{R}^{\alpha\beta}           
                        \overline{\nabla}_{\alpha}\overline{\nabla}_{\beta}
                        \Big) \delta g_{\rho\sigma}
\nonumber\\
&&\hskip -9.7cm
     +\,\big(\overline{\nabla}_{\lambda}\overline{R}_{\alpha\beta}\big)
         \Big(
            \!\!-\!2\overline{g}^{\lambda(\alpha}\overline{g}^{\beta)(\rho}\overline{g}^{\sigma)(\mu} 
                    \overline{\nabla}^{\nu)}  
            \!\!-\!2\overline{g}^{\lambda(\mu}\overline{g}^{\nu)(\alpha}\overline{g}^{\beta)(\rho} 
                      \overline{\nabla}^{\sigma)}        
 \! -\!2\overline{g}^{\lambda(\rho}\overline{g}^{\sigma)(\mu}
      \overline{g}^{\nu)(\alpha}\overline{\nabla}^{\beta)}
\nonumber\\
&&\hskip -8.7cm
          +\,2\overline{g}^{\mu)(\rho}\overline{g}^{\sigma)(\alpha}\overline{g}^{\beta)(\nu}
              \overline{\nabla}^\lambda
         \!+\,\frac12\big(\overline{g}^{\mu\nu}\overline{g}^{\rho(\alpha}\overline{g}^{\beta)\sigma}
         \!-\!\overline{g}^{\rho\sigma} \overline{g}^{\mu(\alpha} \overline{g}^{\beta)\nu}\big)
                 \overline{\nabla}^\lambda
  \nonumber\\
&&\hskip -8.7cm       
       +\,  \overline{g}^{\rho\sigma}\overline{g}^{\lambda(\mu}\overline{g}^{\nu)(\alpha}\overline{\nabla}^{\beta)}  
               \!\!+\!2\overline{g}^{\mu\nu}\overline{g}^{\lambda(\alpha}\overline{g}^{\beta)(\rho}
                         \overline{\nabla}^{\sigma)}
               \!\!+\!\overline{g}^{\mu\nu}\overline{g}^{\lambda(\rho}
                          \overline{g}^{\sigma)(\alpha}\overline{\nabla}^{\beta)}
 \nonumber\\
&&\hskip -8.7cm
      +\,\overline{g}^{\alpha(\mu}\overline{g}^{\nu)\beta}    
                         \overline{g}^{\lambda(\rho}\overline{\nabla}^{\sigma)}  
     \!\!-\!\overline{g}^{\alpha(\rho}\overline{g}^{\sigma)\beta}
        \overline{g}^{\lambda (\mu}\overline{\nabla}^{\nu)}                          
       \!\!-\!\big(\overline{g}^{\mu\nu}\overline{g}^{\rho\sigma}
         \!-\!\overline{g}^{\mu(\rho}\overline{g}^{\sigma)\nu}
         \big)
               \overline{g}^{\lambda(\alpha} \overline{\nabla}^{\beta)}   
            \Big) \delta g_{\rho\sigma}
\nonumber\\
&&\hskip -9.7cm
     +\,\delta g_{\rho\sigma}\overline{g}^{\rho\sigma}
         \Big(\overline{\nabla}_{\alpha}\overline{\nabla}^{(\mu}\overline{R}^{\nu)\alpha}
          \!-\!\frac12\overline{\dAlembert}\,\overline{R}^{\mu\nu}
             \!-\!\frac12 \overline{g}^{\mu\nu}\overline{\nabla}_{\alpha}
                             \overline{\nabla}_{\beta}\overline{R}^{\alpha\beta}\Big) 
\nonumber\\
&&\hskip -9.7cm
     +\,\delta g_{\rho\sigma}
         \Big( 
         \!\!-\!2\overline{\nabla}^{\rho)}\overline{\nabla}^{(\mu}\overline{R}^{\nu)(\sigma}  
   \!-\!2\overline{\nabla}_{\alpha}\overline{\nabla}^{(\rho}
                     \overline{g}^{\sigma)(\mu}\overline{R}^{\nu)\alpha}
                  \!-\!2\overline{\nabla}_{\alpha}\overline{\nabla}^{(\mu} \overline{g}^{\nu)(\rho}          
                         \overline{R}^{\sigma)\alpha}    
                        \!\!+\! \overline{\nabla}^{(\rho}           
                         \overline{\nabla}^{\sigma)}\overline{R}^{\mu\nu}
                           \Big)
\nonumber\\
&&\hskip -9.7cm
     +\, \delta g_{\rho\sigma}
         \Big(
       2\overline{g}^{\mu)(\rho}\overline{\dAlembert}\,\overline{R}^{\sigma)(\nu}                    
  \!\!+\overline{g}^{\mu(\rho}\overline{g}^{\sigma)\nu}
                \overline{\nabla}_{\alpha}\overline{\nabla}_{\beta}\overline{R}^{\alpha\beta}
               \!\!\!+\!\overline{g}^{\mu\nu}\overline{\nabla}_{\alpha}\overline{\nabla}^{(\rho}
                          \overline{R}^{\sigma)\alpha}
               \!\!\!+\!\overline{g}^{\mu\nu}\overline{\nabla}^{(\rho}\overline{\nabla}_{\alpha}
                          \overline{R}^{\sigma)\alpha}
           \Big)
 \bigg\}.\!\!\!\!
\nonumber\\
 \label{counterterm action Ric 2 simple 1}
\end{eqnarray}
The same expression is obtained by varying Eq.~(\ref{counterterm action Ric 2 c1});
here we give an intermediate step,
\begin{eqnarray}
\frac{1}{\sqrt{-\overline{g}}}
\delta\left(\frac{\delta S_{\rm Ric^2}}{\delta g_{\mu\nu}(x)}\right)
 _{\!\overline{g}_{\alpha\beta}}\!\!
 &\!\!=\!\!&  \alpha_{\rm Ric^2}
    \bigg\{\delta\Big(\frac12 g^{\mu\nu}g^{\alpha(\gamma}g^{\delta)\beta}
                    \!-\! 2g^{\gamma(\alpha}g^{\beta)(\mu}g^{\nu)\delta}
                      \Big)    
               \overline{R}_{\alpha\beta}\overline{R}_{\gamma\delta}
\nonumber\\
&&\hskip -4.0cm  
+\,\delta\Big(2g^{\alpha\gamma}g^{\delta(\mu}g^{\nu)\beta}
                    \!-\! g^{\gamma\delta}g^{\alpha(\mu}g^{\nu)\beta}
                     \!-\! g^{\mu\nu}g^{\alpha(\gamma}g^{\delta)\beta}
                      \Big)    
              \overline{\nabla}_{\gamma}\overline{\nabla}_{\delta}\overline{R}_{\alpha\beta}
   \nonumber\\
&&\hskip -4.0cm                        
   +\,\delta g_\rho^{\;\rho}\Big(\frac14 \overline{g}^{\mu\nu}\,\overline{R}_{\alpha\beta}\overline{R}^{\alpha\beta}
              \!\!-\! \overline{R}^{\alpha(\mu}\overline{R}^{\nu)}_{\;\alpha}
              \!+\!\overline{\nabla}_{\alpha} \overline{\nabla}^{(\mu}\overline{R}^{\nu)\alpha}              
            \!\!-\! \frac12\overline{\dAlembert}\,\overline{R}^{\mu\nu}
          \!\!-\!\frac12 \overline{g}^{\mu\nu}\overline{\nabla}_{\alpha} \overline{\nabla}_{\beta}
                          \overline{R}^{\alpha\beta}
                          \Big)
   \nonumber\\
&&\hskip -4.0cm                        
    +\,\big(\overline{g}^{\mu\nu}\overline{R}^{\alpha\beta}
              \!\!\!-\! 4\overline{R}^{\alpha(\mu}\overline{g}^{\nu)\beta}\big)\delta R_{\alpha\beta} 
\nonumber\\
&&\hskip -1.0cm      
              +\,2\Big(\overline{\nabla}^{\alpha} \overline{\nabla}^{(\mu}\overline{g}^{\nu)\beta}
    \!\!-\! \frac12\overline{\dAlembert}\overline{g}^{\mu(\alpha}\overline{g}^{\beta)\nu}
         \! \!-\! \frac12\overline{g}^{\mu\nu}\overline{\nabla}^{\alpha} \overline{\nabla}^{\beta}\Big)
                          \delta R_{\alpha\beta}
\nonumber\\
&&\hskip -4.0cm 
              -\,\big(2\overline{g}^{\alpha\gamma} \overline{g}^{\delta(\mu}\overline{g}^{\nu)\beta}
              \!-\!\overline{g}^{\gamma\delta} \overline{g}^{\mu(\alpha}\overline{g}^{\beta)\nu}
              \!-\! \overline{g}^{\mu\nu}\overline{g}^{\alpha\gamma}\overline{g}^{\delta\beta}
               \big)
\Big[\overline{\nabla}_{\gamma}\big( 
                          \delta\Gamma_{\delta\alpha}^\omega\overline{R}_{\omega\beta}
                          \!+\!\delta\Gamma_{\delta\beta}^\omega\overline{R}_{\alpha\omega}\big)
\nonumber\\
&&\hskip -1.0cm 
        +\,\big(
            \delta\Gamma_{\gamma\delta}^\omega\overline{\nabla}_\omega\overline{R}_{\alpha\beta}
      \!+\!  \delta\Gamma_{\gamma\alpha}^\omega\overline{\nabla}_\delta\overline{R}_{\omega\beta}
      \!+\!  \delta\Gamma_{\gamma\beta}^\omega\overline{\nabla}_\delta\overline{R}_{\omega\beta}
                \big)\Big]
    \bigg\}
,\quad
\label{counterterm action Ric 2 c1 delta}
\end{eqnarray}
which, when worked out, represents a nontrivial check of the result in Eq.~(\ref{counterterm action Ric 2 simple 1}).

By varying expression~(\ref{counterterm action Ric 2 c1 bis}) one gets a slightly 
simpler expression,
\begin{eqnarray}
\frac{1}{\sqrt{-\overline{g}}}
\left[\delta\left(\frac{\delta S_{\rm Ric^2}}{\delta g_{\mu\nu}(x)}\right)
\right]
 _{\!\overline{g}_{\alpha\beta}}\!\!
 &\!\!=\!\!& \alpha_{\rm Ric^2}
    \bigg\{\delta g_\rho^{\;\rho}
    \Big(\frac14 \overline{g}^{\mu\nu}\,\overline{R}_{\alpha\beta}\overline{R}^{\alpha\beta}
              \!+\!\overline{R}_{\alpha\beta}\overline{R}^{\alpha(\mu\nu)\beta}\Big)
\nonumber\\
&&\hskip -0cm              
         +\,\delta g_\rho^{\;\rho}
    \Big(\frac12\overline{\nabla}^{\mu} \overline{\nabla}^{\nu}\overline{R}
                        \!-\!\frac12\overline{\dAlembert}\,\overline{R}^{\mu\nu}  
          \!-\!\frac14 \overline{g}^{\mu\nu}\overline{\dAlembert}\overline{R}\Big)
\nonumber\\
&&\hskip -4.0cm           
   +\,\big(\overline{g}^{\mu\nu}\overline{R}^{\alpha\beta}
              \!+\! 2\overline{R}^{\alpha(\mu\nu)\beta}\big)\delta R_{\alpha\beta}
              \!+\! 2\overline{R}_{\alpha}^{\;\beta}
              \overline{g}^{\mu(\gamma} \overline{g}^{\delta)\nu}
                      \delta R^{\alpha}_{\;\gamma\delta\beta}
\nonumber\\
&&\hskip -4.0cm       
         +\,  \Big(\overline{g}^{\alpha\beta}\overline{\nabla}^{(\mu}\overline{\nabla}^{\nu)}
                    \!-\! \overline{g}^{\mu(\alpha}\overline{g}^{\beta)\nu}\overline{\dAlembert}
                    \!-\! \frac12 \overline{g}^{\mu\nu} \overline{g}^{\alpha\beta}\overline{\dAlembert} \Big)    
               \delta R_{\alpha\beta} 
\nonumber\\
&&\hskip -4.0cm           
   +\, \frac12\delta\big(g^{\mu\nu} g^{\alpha(\gamma}g^{\delta)\beta}\big)
                 \overline{R}_{\alpha\beta}\overline{R}_{\gamma\delta}
                    \!+\!2 \delta\Big(g^{\gamma(\mu}g^{\nu)\delta}g^{\beta\eta}\big)
                  \overline{R}_{\alpha\beta}\overline{R}^{\alpha}_{\;\gamma\delta\eta}
\nonumber\\
&&\hskip -4.0cm       
         +\,  \delta\Big(g^{\mu(\gamma}g^{\delta)\nu} g^{\alpha\beta}
                    \!-\! g^{\gamma\delta} g^{\mu(\alpha}g^{\beta)\nu}
                    \!-\! \frac12 g^{\mu\nu} g^{\gamma\delta} g^{\alpha\beta} \Big)    
               \overline{\nabla}_{\gamma}\overline{\nabla}_{\delta}\overline{R}_{\alpha\beta}        
\nonumber\\
&&\hskip -4.0cm                   
      -\,
\Big(g^{\mu(\gamma}g^{\delta)\nu} g^{\alpha\beta}
                    \!-\! g^{\gamma\delta} g^{\mu(\alpha}g^{\beta)\nu}
                    \!-\! \frac12 g^{\mu\nu} g^{\gamma\delta} g^{\alpha\beta} \Big) 
\Big[\overline{\nabla}_{\gamma}\big( 
                          \delta\Gamma_{\delta\alpha}^\omega\overline{R}_{\omega\beta}
                          \!+\!\delta\Gamma_{\delta\beta}^\omega\overline{R}_{\alpha\omega}\big)
\nonumber\\
&&\hskip -1.2cm 
        +\,\big(
            \delta\Gamma_{\gamma\delta}^\omega\overline{\nabla}_\omega\overline{R}_{\alpha\beta}
      \!+\!  \delta\Gamma_{\gamma\alpha}^\omega\overline{\nabla}_\delta\overline{R}_{\omega\beta}
      \!+\!  \delta\Gamma_{\gamma\beta}^\omega\overline{\nabla}_\delta\overline{R}_{\omega\beta}
                \big)\Big]
    \bigg\}
\,,\quad
\label{counterterm action Ric 2 c1 bis delta}
\end{eqnarray}
which when evaluated yields, 
\begin{eqnarray}
\!\int\!{\rm d}^Dx'\sqrt{-\overline{g}'}\left(\frac{1}{\sqrt{-\overline{g}}\sqrt{-\overline{g}'}}
 \frac{\delta^2 S_{\rm Ric^2}}{\delta g^{\mu\nu}(x)\delta g^{\rho\sigma}(x')}
    \right)_{\!\overline{g}_{\alpha\beta}} \!\!\!\delta g_{\rho\sigma}(x')
&&
\nonumber\\
 &&\hskip -9.8cm
   =\,\alpha_{\rm Ric^2}
   \!\times\!\bigg\{\!\!-\!2\Big(\overline{R}^{\alpha\beta} 
                    \big({{\overline{R}_{\alpha}}^{\!\rho)(\mu}}_{\!\beta}\overline{g}^{\nu)(\sigma}
                    \!+\!{{\overline{R}_{\alpha}}^{\!\mu)(\rho}}_{\!\beta}\overline{g}^{\sigma)(\nu}\big)
                    \!+\!\overline{R}_{\alpha}^{\;\,(\rho}\overline{R}^{\sigma)(\mu\nu)\alpha}\Big) \delta g_{\rho\sigma}
 \nonumber\\
&&\hskip -9.8cm                         
 +\,\frac14\overline{R}^{\alpha\beta}\overline{R}_{\alpha\beta}
     \big(\overline{g}^{\mu\nu} \overline{g}^{\rho\sigma}
                   \!\!-\!2\overline{g}^{\mu(\rho}\overline{g}^{\sigma)\nu}\big) \delta g_{\rho\sigma}
 \!-\!\big(\overline{g}^{\mu\nu} \overline{R}^{\alpha(\rho}\overline{R}_{\alpha}^{\;\sigma)}
    \!-\!\overline{g}^{\rho\sigma} \overline{R}_{\alpha\beta}\overline{R}^{\rho(\mu\nu)\sigma}
    \big)  \delta g_{\rho\sigma}
\nonumber\\
&&\hskip -9.8cm
     +\,\Big(\overline{g}^{\alpha\beta}\overline{\nabla}^{\mu}\overline{\nabla}^{\nu)}
          \!-\!\overline{g}^{\alpha(\mu}\overline{g}^{\nu)\beta}\overline{\dAlembert}
            \!-\!\frac12 \overline{g}^{\mu\nu}\overline{g}^{\alpha\beta}\overline{\dAlembert}\Big)
\nonumber\\
&&\hskip -6.5cm
   \times\,
    \Big(\overline{\nabla}^{(\rho}\delta^{\sigma)}_{\;(\alpha}\overline{\nabla}_{\beta)} 
          \!-\!\frac12 \delta_{\alpha}^{\;(\rho}\delta_{\beta}^{\;\sigma)}\overline{\dAlembert} 
 \!-\!\frac12 \overline{g}^{\rho\sigma}\overline{\nabla}_{\alpha}\overline{\nabla}_{\beta}\Big)\delta g_{\rho\sigma}
\nonumber\\
&&\hskip -9.8cm
     +\,
         \Big(2\overline{R}^{\alpha(\mu\nu)(\rho}\overline{\nabla}^{\sigma)}\overline{\nabla}_{\alpha}           
                 \!-\!\overline{g}^{\rho\sigma}\overline{R}^{\alpha(\mu\nu)\beta}
                     \overline{\nabla}_{\alpha}\overline{\nabla}_{\beta}
                  \!-\!\overline{R}^{\rho(\mu\nu)\sigma}\overline{\dAlembert}
              \Big) \delta g_{\rho\sigma}
\nonumber\\
&&\hskip -9.8cm
     +\,\Big(\overline{R}^{\mu)(\rho}\overline{\nabla}^{\sigma)}\overline{\nabla}^{(\nu}           
                 \!\!-\!\overline{R}^{\alpha(\mu}\overline{g}^{\nu)(\rho}\overline{\nabla}^{\sigma)}           
                         \overline{\nabla}_{\alpha}
                  \!\!-\!\overline{R}^{\rho\sigma}\overline{\nabla}^{(\mu}\overline{\nabla}^{\nu)} 
          \!\!+\!\overline{R}^{\rho)(\mu}\overline{g}^{\nu)(\sigma}\overline{\dAlembert}
\nonumber\\
&&\hskip -5.8cm
     +\,\overline{g}^{\mu\nu}\overline{R}^{\alpha(\rho}\overline{\nabla}^{\sigma)}\overline{\nabla}_{\alpha}
              \!-\!\frac12\overline{g}^{\mu\nu} \overline{g}^{\rho\sigma}\overline{R}^{\alpha\beta}           
                        \overline{\nabla}_{\alpha}\overline{\nabla}_{\beta}
                        \Big) \delta g_{\rho\sigma}
\nonumber\\
&&\hskip -9.8cm
\nonumber\\
&&\hskip -9.8cm
     +\,\big(\overline{\nabla}_{\lambda}\overline{R}_{\alpha\beta}\big)
         \Big[2\overline{g}^{\lambda(\rho}\overline{g}^{\sigma)(\alpha}
      \overline{g}^{\beta)(\mu}\overline{\nabla}^{\nu)}  
         \! -\!2\overline{g}^{\lambda(\rho}\overline{g}^{\sigma)(\mu}
      \overline{g}^{\nu)(\alpha}\overline{\nabla}^{\beta)}
      \!-\!\overline{g}^{\alpha\beta}\overline{g}^{\lambda(\rho}\overline{g}^{\sigma)(\mu}\overline{\nabla}^{\nu)}
\nonumber\\
&&\hskip -9.8cm
          +\,\Big(\!2\overline{g}^{\alpha(\mu}\overline{g}^{\nu)(\rho}\overline{g}^{\sigma)\beta}
         \!+\,\frac12\big(\overline{g}^{\alpha\beta}\overline{g}^{\mu(\rho}\overline{g}^{\sigma)\nu}
         \!-\!\overline{g}^{\rho\sigma} \overline{g}^{\alpha(\mu} \overline{g}^{\nu)\beta}\big)
         \!+\!\overline{g}^{\mu\nu}\overline{g}^{\alpha(\rho}\overline{g}^{\sigma)\beta}
           \Big)\overline{\nabla}^\lambda
  \nonumber\\
&&\hskip -9.8cm      
           -\, \frac14 \overline{g}^{\alpha\beta}\overline{g}^{\mu\nu}\overline{g}^{\rho\sigma}
          \overline{\nabla}^\lambda 
     \!\!+\! \overline{g}^{\alpha(\mu}\overline{g}^{\nu)\beta}
                        \overline{g}^{\lambda(\rho} \overline{\nabla}^{\sigma)}
     \!\!\!-\! 2\overline{g}^{\alpha(\rho}\overline{g}^{\sigma)\beta}
                        \overline{g}^{\lambda(\mu} \overline{\nabla}^{\nu)}
               \!\!+\!\frac12\overline{g}^{\alpha\beta}\overline{g}^{\mu\nu}\overline{g}^{\lambda(\rho}
                         \overline{\nabla}^{\sigma)} 
            \Big] \delta g_{\rho\sigma}\!
\nonumber\\
&&\hskip -9.8cm
\nonumber\\
&&\hskip -9.8cm
     +\,\delta g_{\rho\sigma}
         \Big(\!\!-\!2\overline{\nabla}^{(\rho}\overline{g}^{\sigma)(\mu}\overline{\nabla}^{\nu)}\overline{R}
         \!\!+\!\overline{\nabla}^{(\rho}\overline{\nabla}^{\sigma)} \overline{R}^{\mu\nu}
          \!\!+\!2 \overline{\dAlembert}\overline{g}^{\mu)(\rho}\overline{R}^{\sigma)(\nu}
        \!\!+\!\frac12\overline{g}^{\mu(\rho}\overline{g}^{\sigma)\nu}
        \overline{\dAlembert}\overline{R}
\nonumber\\
&&\hskip -9.1cm
          +\,\frac12\big(\overline{g}^{\mu\nu}\overline{\nabla}^{\rho}\overline{\nabla}^{\sigma}
               \!\!+\!\overline{g}^{\rho\sigma}\overline{\nabla}^{\mu}\overline{\nabla}^{\nu}\big)\overline{R}
            \!+\!\frac12\overline{\dAlembert}\big(\overline{g}^{\mu\nu}\overline{R}^{\rho\sigma}
               \!\!-\!\overline{g}^{\rho\sigma}\overline{R}^{\mu\nu}\big)
               \!-\!\frac14\overline{g}^{\mu\nu}\overline{g}^{\rho\sigma}\overline{\dAlembert} \overline{R}
            \Big) 
 \bigg\}\,.
 \nonumber\\
 \label{counterterm action Ric 2 alternative}
\end{eqnarray}
%


\bigskip
\noindent
{\bf Riemann squared counterterm.} Next we expand the Riemann squared counterterm
action as,
\begin{eqnarray}
S_{\rm Riem^2}\big[\overline{g}_{\mu\nu}\!+\!\delta g_{\mu\nu}\big] \!\!
 &\!\!=\!\!&\!\! \alpha_{\rm Riem^2}\! \int\!{\rm d}^Dx\sqrt{-\overline{g}}\,
                \overline{R}^{\alpha\beta\gamma\delta}\overline{R}_{\alpha\beta\gamma\delta}
\label{counterterm action Riem 2}\\
&&\hskip -.5cm 
 +\,\alpha_{\rm Riem^2}\!\! \int\! {\rm d}^Dx\!
 \left(\frac{\delta S_{\rm Riem^2}}{\delta g_{\mu\nu}(x)}\right)_{\!\overline{g}_{\alpha\beta}}
             \!\!\!\!\delta g_{\mu\nu}(x)
\nonumber\\
&&\hskip -3.3cm    
   +\,\frac{1}2\alpha_{\rm Riem^2}\!\int \!{\rm d}^Dx\,{\rm d}^Dx'\delta g_{\rho\sigma}(x')
   \left(\frac{\delta^2 S_{\rm Riem^2}}
     {\delta g_{\rho\sigma}(x')\delta g_{\mu\nu}(x)}\right)_{\!\overline{g}_{\alpha\beta}}
             \!\!\delta g_{\mu\nu}(x)
             \!+\! \mathcal{O}\big(\delta g_{\alpha\beta}^3\big)
\nonumber\\
&&\hskip -3.3cm
 =\,S^{(0)}_{\rm Riem^2}\big[\overline{g}_{\mu\nu}\big]
    + S^{(1)}_{\rm Riem^2}\big[\overline{g}_{\mu\nu},\delta g_{\mu\nu}\big]
    + S^{(2)}_{\rm Riem^2}\big[\overline{g}_{\mu\nu},\delta g_{\mu\nu}\big]
    + S^{(\geq 3)}_{\rm Riem^2}\big[\overline{g}_{\mu\nu},\delta g_{\mu\nu}\big]
\,,
\nonumber
\end{eqnarray}
where~\footnote{Notice that one may be tempted to rewrite the second line in Eq.~(\ref{counterterm action Riem 2 b1}) in a simpler form as, 
\begin{equation}
  -\,2\sqrt{-\overline{g}}\,\overline{R}^{\alpha}_{\;\,\beta\gamma\delta}\overline{R}^{\rho\beta\gamma\delta}
                          \delta g_{\alpha\rho}
\,.\qquad
\label{counterterm action Riem 2 b1: simpler}
\end{equation}
This is a correct expression at the first order in $\delta g_{\mu\nu}(x)$. However, as one can explicitly check,
Eq.~(\ref{counterterm action Riem 2 b1: simpler}) would  lead to an incorrect expression 
at the second order in the metric perturbations, which is for what we use it below.
The reason is that not all identities (symmetries) that hold for $\overline{R}^\alpha_{\;\,\beta\gamma\delta}$
also hold for $\delta{R}^\alpha_{\;\,\beta\gamma\delta}$. This can be traced back to the fact that 
$\overline{R}^\alpha_{\;\,\beta\gamma\delta}$ is a function of the background metric only,
but  $\delta{R}^\alpha_{\;\,\beta\gamma\delta}$ is a function of both the background metric and 
of the perturbed metric.
} 
\begin{eqnarray}
\delta S_{\rm Riem^2} &\!\!\!=\!\!\!&\alpha_{\rm Riem^2}\! \int\! {\rm d}^Dx 
\Big\{\big(\delta\sqrt{-g}\,\big)\overline{R}_{\alpha\beta\gamma\delta}\overline{R}^{\alpha\beta\gamma\delta} 
   \!\!\!+\! 2\sqrt{-\overline{g}}\,\overline{R}_{\alpha}^{\;\,\beta\gamma\delta}
                 \delta R^{\alpha}_{\;\,\beta\gamma\delta} 
\nonumber\\
&&\hskip -1.6cm
   +\,\sqrt{-\overline{g}}\,\Big[
            \overline{R}^{\mu}_{\;\,\beta\gamma\delta}\overline{R}^{\nu\beta\gamma\delta}
            \!-\!\overline{R}^{\;\;\,\mu}_{\alpha\;\;\gamma\delta}\overline{R}^{\alpha\nu\gamma\delta}
              \!-\!\overline{R}^{\;\;\;\;\,\mu}_{\alpha\beta\;\;\delta}\overline{R}^{\alpha\beta\nu\delta}
              \!-\!\overline{R}^{\;\;\;\;\;\;\mu}_{\alpha\beta\gamma}\overline{R}^{\alpha\beta\gamma\nu}
                          \Big]\delta g_{\mu\nu}(x)
\Big\}
\,,
\nonumber\\
\label{counterterm action Riem 2 b1}\\
\delta^2 S_{\rm Riem^2} &\!\!=\!\!&\alpha_{\rm Riem^2}\!\! \int {\rm d}^Dx \sqrt{-\overline{g}}\Big\{
  \Big[\frac18\big(\delta g_\rho^{\;\rho}\big)^2 \!-\! \frac14\delta g^{\rho\sigma} \delta g_{\rho\sigma} \Big]
    \overline{R}^{\alpha\beta\gamma\delta} \overline{R}_{\alpha\beta\gamma\delta}
\nonumber\\
 &&\hskip 3.8cm
     +\,\overline{R}^{\alpha\beta\gamma\delta} \overline{R}^\lambda_{\;\,\beta\gamma\delta}
                          \big(3 \delta g_{\alpha\omega} \delta g_{\;\,\lambda}^{\omega} 
                                   \!-\!\delta g_\rho^{\;\rho}\delta g_{\alpha\lambda}\big)
\nonumber\\
&&\hskip -.cm
+\,  \big[2\overline{R}^{\lambda\beta\gamma\delta}\delta g_{\lambda\alpha}
               \!-\! 2\overline{R}_\alpha^{\;\,\lambda\gamma\delta}\delta g_{\lambda}^{\;\beta}
               \!-\!4 \overline{R}_\alpha^{\;\,\beta\lambda\delta}\delta g_{\lambda}^{\;\gamma}
               \!+\!\overline{R}_{\alpha}^{\;\beta\gamma\delta}\delta g_\rho^{\;\rho}\big]
               \delta{R}^{\alpha}_{\;\,\beta\gamma\delta}
\nonumber\\
&&\hskip -.cm
    +\,\overline{g}_{\alpha\lambda}\overline{g}^{\beta\eta}\overline{g}^{\gamma\omega}
                  \overline{g}^{\delta\zeta}\delta R^\lambda_{\;\,\eta\omega\zeta}
               \delta{R}^{\alpha}_{\;\,\beta\gamma\delta}
        \!+\! 2\overline{R}_{\alpha}^{\;\,\beta\gamma\delta}
                 \delta^2 R^{\alpha}_{\;\,\beta\gamma\delta} 
\Big\}
\,,\quad
\label{counterterm action Riem 2 b2}
\end{eqnarray}
with 
\begin{eqnarray}
\delta R^\rho_{\;\,\sigma\mu\nu} &\!\!=\!\!& \frac12\overline{\nabla}_\mu\overline{\nabla}_\nu
                          \delta g^{\rho}_{\;\,\sigma}
                          \!+\! \frac12\overline{\nabla}_\mu\overline{\nabla}_\sigma
                          \delta g^{\rho}_{\;\,\nu}
                          \!-\! \frac12\overline{\nabla}_\mu\overline{\nabla}^\rho
                          \delta g_{\nu\sigma} - (\mu\leftrightarrow \nu)
\,,\qquad
\label{Riemann tensor: first variation}\\
\delta^2 R^\rho_{\;\,\sigma\mu\nu} &\!\!=\!\!& \overline{\nabla}_\mu \delta^2 \Gamma^\rho_{\sigma\nu}
       \!+\!\delta \Gamma^\rho_{\mu\lambda}\delta \Gamma^\lambda_{\sigma\nu}
        \!-\!\big(\mu\leftrightarrow \nu\big)
\nonumber\\
 &\!\!=\!\!& - \frac12 \delta g^{\rho\lambda}\overline{\nabla}_\mu     
                         \big[\,\overline{\nabla}_\nu \delta g_{\sigma\lambda}
                         \!+\!\overline{\nabla}_\sigma \delta g_{\nu\lambda}
                         \!-\!\overline{\nabla}_\lambda\delta g_{\sigma\nu} \big]\!-\!\big(\mu\leftrightarrow \nu\big)
\nonumber\\
          &&-\, \frac14\big[\,\overline{\nabla}_\mu \delta g^{\rho\lambda} 
               \!+\! \overline{\nabla}^\rho \delta g_{\mu}^{\;\,\lambda}  
               \!-\! \overline{\nabla}^\lambda \delta g^{\rho}_{\;\,\mu} \big]
                         \big[\,\overline{\nabla}_\nu \delta g_{\sigma\lambda}
                         \!+\!\overline{\nabla}_\sigma \delta g_{\nu\lambda}
                         \!-\!\overline{\nabla}_\lambda\delta g_{\sigma\nu} \big]
  \nonumber\\
&&\hskip 1.99cm                       
     -\,\big(\mu\leftrightarrow \nu\big)
\,.\qquad\;\,
\label{Riemann tensor: second variation}
\end{eqnarray}
When Eqs.~(\ref{Riemann tensor: first variation})
and~(\ref{Riemann tensor: second variation}) are contracted with $\delta_\rho^{\;\,\mu}$ they reproduce
Eqs.~(\ref{Ricci tensor: first variation}) and~(\ref{Ricci tensor: second variation 2}), respectively.
The first variation follows immediately from Eq.~(\ref{counterterm action Riem 2 b1}), 
\begin{eqnarray}
\!\!
\left(\frac{\delta S_{\rm Riem^2}}{\delta g_{\mu\nu}(x)}\right)
 _{\! \overline{g}_{\alpha\beta}}\!\!
 &\!\!=\!\!&  \alpha_{\rm Riem^2}\sqrt{-\overline{g}} 
    \bigg\{\frac12 \overline{g}^{\mu\nu}\,\overline{R}_{\alpha\beta\gamma\delta}
                \overline{R}^{\alpha\beta\gamma\delta}
\nonumber\\
&&\hskip -2.25cm              
            +\, \Big(\overline{g}^{\delta\zeta}\overline{g}^{\gamma\xi}\overline{g}^{\eta(\mu}
           \!-\!\overline{g}^{\gamma\zeta}\overline{g}^{\delta\xi}\overline{g}^{\eta(\mu}
                  \!+\!\overline{g}^{\gamma\eta}\overline{g}^{\delta\zeta}\overline{g}^{\xi(\mu}
                  \!-\!\overline{g}^{\gamma\eta}\overline{g}^{\delta\xi}\overline{g}^{\zeta(\mu}
                  \Big)\overline{\nabla}_{\gamma} \overline{\nabla}_{\delta}\overline{R}^{\nu)}_{\;\;\eta\zeta\xi}
\nonumber\\
&&\hskip -2.cm
          \!+\, \Big(\!\!-\!\overline{g}^{\gamma\zeta}\overline{g}^{\eta(\mu}\overline{g}^{\nu)\xi}
                  \!+\!\overline{g}^{\gamma\xi}\overline{g}^{\eta(\mu}\overline{g}^{\nu)\zeta}
                  \Big)
                  \overline{\nabla}_{\alpha} \overline{\nabla}_{\gamma}\overline{R}^{\alpha}_{\;\;\eta\zeta\xi}
\nonumber\\
&&\hskip -2.25cm
 +\,\Big(\delta_\alpha^{\;(\mu}\delta^{\nu)}_{\;\omega}\overline{g}^{\beta\eta}
                        \overline{g}^{\gamma\zeta}\overline{g}^{\delta\xi}
        \!-\!\overline{g}_{\alpha\omega}\overline{g}^{\beta(\mu}\overline{g}^{\nu)\eta}
                           \overline{g}^{\gamma\zeta}\overline{g}^{\delta\xi}
                            \!-\!\overline{g}_{\alpha\omega}\overline{g}^{\beta\eta}
                           \overline{g}^{\gamma(\mu}\overline{g}^{\nu)\zeta}\overline{g}^{\delta\xi}
\nonumber\\
&&\hskip -1.5cm
                            \!-\,\overline{g}_{\alpha\omega}\overline{g}^{\beta\eta}
                           \overline{g}^{\gamma\zeta}\overline{g}^{\delta(\mu}\overline{g}^{\nu)\xi}
 \Big)
   \!\times\!\Big(\overline{R}^{\alpha}_{\;\,\beta\gamma\delta}\overline{R}^{\omega}_{\;\;\eta\zeta\xi}
             \Big)
    \bigg\}
,\quad
\label{counterterm action Riem 2 c1}\\
 &\!\!=\!\!&  \alpha_{\rm Riem^2}\sqrt{-\overline{g}} 
    \bigg\{\frac12 \overline{g}^{\mu\nu}\,\overline{R}_{\alpha\beta\gamma\delta}
                \overline{R}^{\alpha\beta\gamma\delta}
         \!\!-\! 2\overline{R}^{(\mu}_{\;\,\alpha\beta\gamma}\overline{R}^{\nu)\alpha\beta\gamma}
         \!+\! 4\overline{R}_{\alpha\beta}\overline{R}^{\alpha(\mu\nu)\beta}
\nonumber\\
&&\hskip 2.4cm                 
           +\, 4\overline{R}^{\alpha(\mu}\overline{R}^{\nu)}_{\;\alpha}
            + 2\overline{\nabla}^{\mu} \overline{\nabla}^{\nu}\overline{R}
            - 4\overline{\dAlembert}\overline{R}^{\mu\nu}
    \bigg\}
\,,\qquad
\label{counterterm action Riem 2 c1 bis}
\end{eqnarray}
where to get the last equality we made use of 
Eq.~(\ref{contracted Bianchi identities}), from which the following identity can be derived,
\begin{equation}
\overline{\nabla}_{\alpha} \overline{\nabla}_{\beta}\overline{R}^{\alpha(\mu\nu)\beta}
  = \frac12\overline{\nabla}^{\mu} \overline{\nabla}^{\nu}\overline{R}
     + \overline{R}_{\alpha\beta}\overline{R}^{\alpha(\mu\nu)\beta}
           \!+\! \overline{R}^{\alpha(\mu}\overline{R}^{\nu)}_{\;\alpha}
            - \overline{\dAlembert}\overline{R}^{\mu\nu}
\,.\quad
\end{equation}
When one takes account of $\delta g_{\mu\nu} 
= - \overline{g}_{\mu\alpha}\delta g^{\alpha\beta}  \overline{g}_{\beta\nu}$
(and one corrects some sign errors)
one sees that Eqs.~(\ref{counterterm action R2 c1}), (\ref{counterterm action Ric 2 c1 bis})
and~(\ref{counterterm action Riem 2 c1 bis})
agree with Eqs.~(6.53), (6.54) and~(6.55) of Ref.~\cite{Birrell:1982ix}, respectively.

Just as above, and as dictated by the gravitational Noether theorem, one can prove that 
Eq.~(\ref{counterterm action Riem 2 c1 bis}) implies a conserved energy-momentum tensor.
Indeed, acting $\overline{\nabla}_\mu$ on Eq.~(\ref{counterterm action Riem 2 c1 bis}) gives,
\begin{eqnarray}
\overline{\nabla}_\mu\! \left[\frac{1}{\sqrt{-\overline{g}}}
   \left(\frac{\delta S_{\rm Riem^2}}{\delta g_{\mu\nu}(x)}\right)
 _{\! \overline{g}_{\alpha\beta}}\right]\!\!
 &\!\!=\!\!&   \alpha_{\rm Riem^2}
    \bigg\{\frac12\overline{\nabla}^\nu\big(\overline{R}_{\alpha\beta\gamma\delta}
                \overline{R}^{\alpha\beta\gamma\delta}\big)
         \!-\! 2\overline{\nabla}_\mu\big(\overline{R}^{\mu}_{\;\,\alpha\beta\gamma}
                            \overline{R}^{\nu\alpha\beta\gamma}\big)
\nonumber\\
&&\hskip 1.5cm 
    -\, 4\overline{R}^{\nu\alpha\beta\mu}\overline{\nabla}_{\mu}\overline{R}_{\alpha\beta}
    \bigg\}
\nonumber\\
 &&\hskip -3.6cm    =\,\alpha_{\rm Riem^2}
    \bigg\{\frac12\overline{\nabla}^\nu\big(\overline{R}_{\alpha\beta\gamma\delta}
                \overline{R}^{\alpha\beta\gamma\delta}\big)
         \!-\! 2\overline{\nabla}_\mu\big(\overline{R}^{\mu}_{\;\,\alpha\beta\gamma}
                            \overline{R}^{\nu\alpha\beta\gamma}\big)
   \!-\!  4\overline{\nabla}_{\mu}\big(\overline{R}^{\mu\alpha\beta\nu}\overline{R}_{\alpha\beta}\big)
\nonumber\\
&&\hskip -3.6cm 
   +\,4\overline{\nabla}_{\mu}\big(\overline{R}^{\mu\alpha}\overline{R}_{\alpha}^{\;\,\nu}\big)
   \!-\! 2\overline{\nabla}^{\nu}\big(\overline{R}^{\alpha\beta}\overline{R}_{\alpha\beta}\big)
   \!-\! 2\overline{\nabla}_{\mu}\big(\overline{R}^{\mu\nu}\overline{R}\big)
   \!+\! \frac12\overline{\nabla}^{\nu}\big(\overline{R}^2\big)  
    \bigg\}
\,,\qquad
\label{counterterm action Riem 2: conservation}
\end{eqnarray}
where to get the first equality we used Eqs.~(\ref{three derivative identities}), and to get the second 
equality we applied $A{\rm d}B = {\rm d}(AB) - ({\rm d}A)B$ several times on the last term. 
Finally, one can use the antisymmetry property of the Riemann tensor 
on the first term in Eq.~(\ref{counterterm action Riem 2: conservation}), 
$\overline{\nabla}^{\nu} \overline{R}^{\alpha\beta\gamma\delta}
=-\overline{\nabla}^{\alpha} \overline{R}^{\beta\nu\gamma\delta}
-\overline{\nabla}^{\beta} \overline{R}^{\nu\alpha\gamma\delta}$ to derive the following identity, 
\begin{eqnarray}
\!\frac12\overline{\nabla}^\nu\big(\overline{R}_{\alpha\beta\gamma\delta}
                \overline{R}^{\alpha\beta\gamma\delta}\big)
     &\!\!=\!\!&    2\overline{\nabla}_\mu\big(\overline{R}^{\mu}_{\;\,\alpha\beta\gamma}
                            \overline{R}^{\nu\alpha\beta\gamma}\big)
   \!+\!  4\overline{\nabla}_{\mu}\big(\overline{R}^{\mu\alpha\beta\nu}\overline{R}_{\alpha\beta}\big)
\nonumber\\
&&\hskip -2.cm 
   -\,4\overline{\nabla}_{\mu}\big(\overline{R}^{\mu\alpha}\overline{R}_{\alpha}^{\;\,\nu}\big)
   \!+\! 2\overline{\nabla}^{\nu}\big(\overline{R}^{\alpha\beta}\overline{R}_{\alpha\beta}\big)
   \!+\! 2\overline{\nabla}_{\mu}\big(\overline{R}^{\mu\nu}\overline{R}\big)
   \!-\! \frac12\overline{\nabla}^{\nu}\big(\overline{R}^2\big) 
\,,\qquad 
\label{identity for contracted Riem squared}
\end{eqnarray}
which when inserted into Eq.~(\ref{counterterm action Riem 2: conservation}) produces the desired conservation law,
\begin{eqnarray}
\!\!
\overline{\nabla}_\mu\! \left[\frac{1}{\sqrt{-\overline{g}}}
   \left(\frac{\delta S_{\rm Riem^2}}{\delta g_{\mu\nu}(x)}\right)
 _{\! \overline{g}_{\alpha\beta}}\right]\!\!
 &\!\!=\!\!&   0
\,.\qquad
\label{counterterm action Riem 2: conservation 2}
\end{eqnarray}
All of these conservation laws associated with individual counterterms are, of course, not coincidental. 
In fact, any
gravitational action which is a scalar functional of $g_{\mu\nu}$
$R$, $R_{\alpha\beta}$, $R_{\alpha\beta\gamma\delta}$
will produce upon variation with respect to  $g_{\mu\nu}$
 a conserved gravitational energy-momentum tensor, as required by the Noether theorem.
 This conservation holds true for any (off-shell) metric tensor.

\bigskip

Making use of Eqs.~(\ref{counterterm action Riem 2 b2}), (\ref{Riemann tensor: first variation}) and~(\ref{Riemann tensor: second variation})
one can obtain the second variation of Eq.~(\ref{counterterm action Riem 2}),
\begin{eqnarray}
\frac{1}{\sqrt{-\overline{g}(x)}\sqrt{-\overline{g}(x')}}
     \left(\frac{\delta^2 S_{\rm Riem^2}}{\delta g_{\mu\nu}(x)\delta g_{\rho\sigma}(x')}
      \right)_{\overline{g}_{\alpha\beta}} 
      &\!\!=\!\!&\alpha_{\rm Riem^2}
\nonumber\\
 &&\hskip -7.5cm      
     \times \bigg\{
 \frac14\big(\overline{g}^{\mu\nu}\overline{g}^{\rho\sigma}
  \!-\!2\overline{g}^{\mu(\rho}\overline{g}^{\sigma)\nu} \big)
   \overline{R}^{\alpha\beta\gamma\delta} \overline{R}_{\alpha\beta\gamma\delta}
\nonumber\\
 &&\hskip -6.9cm
    +\,  \Big(6\overline{R}^{(\mu}_{\;\,\,\beta\gamma\delta} \overline{g}^{\nu)(\rho} 
                                     \overline{R}^{\sigma)\beta\gamma\delta}
                            \!-\!\overline{g}^{\mu\nu} \overline{R}^{(\rho}_{\;\,\,\beta\gamma\delta}
                               \overline{R}^{\sigma)\beta\gamma\delta}
                        \!-\!\overline{g}^{\rho\sigma}\overline{R}^{(\mu}_{\;\,\,\beta\gamma\delta}
                         \overline{R}^{\nu)\beta\gamma\delta} \Big)
\nonumber\\
&&\hskip -7.5cm
    +\,\Big(\frac32\overline{g}^{\mu(\rho}\overline{g}^{\sigma)\nu}
                               \overline{\nabla}^\alpha\overline{\dAlembert}\overline{\nabla}_\alpha
                             \!-\!\overline{g}^{\mu)(\rho}
                              \overline{\nabla}^{\sigma)}\overline{\dAlembert}\overline{\nabla}^{(\nu}
\nonumber\\
&&\hskip -6.5cm                     
    -\,\frac12\overline{g}^{\mu(\rho}\overline{g}^{\sigma)\nu}
    \overline{\nabla}^\alpha\overline{\nabla}^\beta\overline{\nabla}_\alpha\overline{\nabla}_\beta
    \!-\!  \overline{g}^{\mu)(\rho}
    \overline{\nabla}^\alpha\overline{\nabla}^{\sigma)}\overline{\nabla}^{(\nu}\overline{\nabla}_\alpha            
  \!+\!\overline{\nabla}^{(\rho}\overline{\nabla}^{\sigma)}\overline{\nabla}^{(\mu}\overline{\nabla}^{\nu)}
             \Big)
\nonumber\\
&&\hskip -7.5cm
    +\,\Big(\frac32{\overline{g}'}^{\mu(\rho}{\overline{g}'}^{\sigma)\nu}
                               {\overline{\nabla}'}^\alpha\overline{\dAlembert}'{\overline{\nabla}'}_{\!\alpha}
                             \!-\!{\overline{g}'}^{\rho)(\mu}
                              {\overline{\nabla}'}^{\nu)}\overline{\dAlembert}'{\overline{\nabla}'}^{(\sigma}
\nonumber\\
&&\hskip -6.7cm                     
    -\,\frac12 {\overline{g}'}^{\mu(\rho}{\overline{g}'}^{\sigma)\nu}
    {\overline{\nabla}'}^\alpha{\overline{\nabla}'}^\beta{\overline{\nabla}'}_{\!\alpha}
       {\overline{\nabla}'}_{\!\beta}
    \!-\!{\overline{g}'}^{\rho)(\mu}
    {\overline{\nabla}'}^\alpha{\overline{\nabla}'}^{\nu)}{\overline{\nabla}'}^{(\sigma}
       {\overline{\nabla}'}_{\!\alpha}
  \!+\!{\overline{\nabla}'}^{(\mu}{\overline{\nabla}'}^{\nu)}{\overline{\nabla}'}^{(\rho}
       {\overline{\nabla}'}^{\sigma)}
             \Big)
\nonumber\\
 &&\hskip -7.5cm 
+\, 4
\Big(\overline{g}^{\rho)(\mu}\overline{R}^{\nu)(\sigma|\alpha\beta} \overline{\nabla}_\alpha
                               \overline{\nabla}_\beta
                        \!-\!\overline{R}^{\alpha(\rho\sigma)(\mu} \overline{\nabla}^{\nu)}\overline{\nabla}_\alpha
               \!-\!\overline{g}^{\mu)(\rho} \overline{R}^{\sigma)\alpha\beta(\nu}\overline{\nabla}_\beta
                          \overline{\nabla}_\alpha
            \Big)
 \nonumber\\
&&\hskip -5.5cm
               +\,2\overline{g}^{\mu\nu}
              \overline{R}^{\alpha(\rho\sigma)\beta} \overline{\nabla}_\beta\overline{\nabla}_\alpha
\nonumber\\             
&&\hskip -7.5cm
+\, 4
\Big({\overline{g}'}^{\mu)(\rho}{\overline{R}'}^{\sigma)(\nu|\alpha\beta}{\overline{\nabla}'}_{\!\alpha}
                               {\overline{\nabla}'}_{\!\beta}
                      \!-\!{\overline{R}'}^{\alpha(\mu\nu)(\rho}{\overline{\nabla}'}^{\sigma)}
                       {\overline{\nabla}'}_{\!\alpha}
            \!-\!{\overline{g}'}^{\rho)(\mu}{\overline{R}'}^{\nu)\alpha\beta(\sigma}{\overline{\nabla}'}_{\!\beta}
                          {\overline{\nabla}'}_{\!\alpha}
            \Big)
 \nonumber\\
&&\hskip -5.5cm
               +\,2{\overline{g}'}^{\rho\sigma}
              {\overline{R}'}^{\alpha(\mu\nu)\beta}{\overline{\nabla}'}_{\!\beta}{\overline{\nabla}'}_{\!\alpha}
\nonumber\\
 &&\hskip -7.5cm 
    -\,2\Big[\overline{g}^{\rho)(\mu}\overline{R}^{\nu)(\sigma|\alpha\beta}
                          \overline{\nabla}_\alpha\overline{\nabla}_\beta
                         \!+\!\overline{g}^{\rho)(\mu} \overline{R}^{\nu)\alpha\beta(\sigma}
                         \overline{\nabla}_\beta\overline{\nabla}_\alpha
                         \!+\! \overline{R}^{\alpha(\rho\sigma)(\mu}\overline{\nabla}_\alpha
                         \overline{\nabla}^{\nu)}
                              \Big]
                              \nonumber\\
 &&\hskip -7.5cm 
    -\,2\Big[{\overline{g}'}^{\mu)(\rho}{\overline{R}'}^{\sigma)(\nu|\alpha\beta}
                          {\overline{\nabla}'}_{\!\alpha}{\overline{\nabla}'}_{\!\beta}
                         \!+\!{\overline{g}'}^{\mu)(\rho}{\overline{R}'}^{\sigma)\alpha\beta(\nu}
                         {\overline{\nabla}'}_{\!\beta}{\overline{\nabla}'}_{\!\alpha}
                         \!+\! {\overline{R}'}^{\alpha(\mu\nu)(\rho}{\overline{\nabla}'}_{\!\alpha}
                         {\overline{\nabla}'}^{\sigma)}
                              \Big]
\nonumber\\
 &&\hskip -7.5cm 
    +\, \Big[ 
         2\overline{\nabla}_\alpha
         \Big(\overline{g}^{\rho)(\mu}\overline{R}^{\nu)(\sigma|\alpha\beta}\overline{\nabla}_\beta 
                         \!+\!\overline{g}^{\rho)(\mu}\overline{R}^{\nu)\beta\alpha(\sigma}\overline{\nabla}_\beta
                         \!+\!\overline{R}^{\alpha(\rho\sigma)(\mu}\overline{\nabla}^{\nu)}
                         \Big)
\nonumber\\
 &&\hskip -5.5cm 
              +\,2\overline{\nabla}^{(\rho} \overline{R}^{\sigma)(\mu\nu)\alpha}\overline{\nabla}_\alpha
              \!-\!\overline{\nabla}^\lambda \overline{R}^{\rho)(\mu\nu)(\sigma}\overline{\nabla}_\lambda             
     \Big]
\nonumber\\
 &&\hskip -7.5cm 
    +\, \Big[ 
         2{\overline{\nabla}'}_{\!\alpha}
         \Big({\overline{g}'}^{\mu)(\rho}{\overline{R}'}^{\sigma)(\nu|\alpha\beta}{\overline{\nabla}'}_{\!\beta}      
            \!+\!{\overline{g}'}^{\mu)(\rho}{\overline{R}'}^{\sigma)\beta\alpha(\nu}{\overline{\nabla}'}_{\!\beta}
                         \!+\!{\overline{R}'}^{\alpha(\mu\nu)(\rho}{\overline{\nabla}'}^{\sigma)}
                         \Big)
\nonumber\\
 &&\hskip -5.5cm 
     +\,2{\overline{\nabla}'}^{(\mu} {\overline{R}'}^{\nu)(\rho\sigma)\alpha}{\overline{\nabla}'}_{\!\alpha}
         \!-\!{\overline{\nabla}'}^\lambda{\overline{R}'}^{\mu)(\rho\sigma)(\nu}{\overline{\nabla}'}_{\!\lambda}
     \Big]
\bigg\}
\!\times\!
\frac{\delta^D(x\!-\!x')}{\sqrt{-\overline{g}}}
\,,\qquad
\label{second variation: counterterm action Riem2}
\end{eqnarray}

where the derivatives act to the right. When this acts on a metric perturbation $\delta g_{\rho\sigma}(x')$ one obtains,
\begin{eqnarray}
\int {\rm d}^Dx'\sqrt{-\overline{g}(x')}
          \frac{1}{\sqrt{-\overline{g}(x)}\sqrt{-\overline{g}(x')}}
     \left(\frac{\delta^2 S_{\rm Riem^2}}{\delta g_{\mu\nu}(x)\delta g_{\rho\sigma}(x')}
      \right)_{\overline{g}_{\alpha\beta}} \delta g_{\rho\sigma}(x')
      &\!\!=\!\!&\alpha_{\rm Riem^2}
\nonumber\\
 &&\hskip -12.cm      
     \times \bigg\{
 \frac14\big(\overline{g}^{\mu\nu}\overline{g}^{\rho\sigma}
  \!-\!2\overline{g}^{\mu(\rho}\overline{g}^{\sigma)\nu} \big)
   \overline{R}^{\alpha\beta\gamma\delta} \overline{R}_{\alpha\beta\gamma\delta}
\nonumber\\
 &&\hskip -11cm
    +\,  \Big(6\overline{R}^{(\mu}_{\;\,\,\beta\gamma\delta} \overline{g}^{\nu)(\rho} 
                                     \overline{R}^{\sigma)\beta\gamma\delta}
                            \!-\!\overline{g}^{\mu\nu} \overline{R}^{(\rho}_{\;\,\,\beta\gamma\delta}
                               \overline{R}^{\sigma)\beta\gamma\delta}
                        \!-\!\overline{g}^{\rho\sigma}\overline{R}^{(\mu}_{\;\,\,\beta\gamma\delta}
                         \overline{R}^{\nu)\beta\gamma\delta} \Big)
\nonumber\\
&&\hskip -12cm
    +\,\Big(3\overline{g}^{\mu(\rho}\overline{g}^{\sigma)\nu}
                               \overline{\nabla}^\alpha\overline{\dAlembert}\overline{\nabla}_\alpha
                             \!-\!2\overline{g}^{\mu)(\rho}
                              \overline{\nabla}^{\sigma)}\overline{\dAlembert}\overline{\nabla}^{(\nu}
\nonumber\\
&&\hskip -11cm                     
    -\,\overline{g}^{\mu(\rho}\overline{g}^{\sigma)\nu}
    \overline{\nabla}^\alpha\overline{\nabla}^\beta\overline{\nabla}_\alpha\overline{\nabla}_\beta
    \!-\! 2 \overline{g}^{\mu)(\rho}
    \overline{\nabla}^\alpha\overline{\nabla}^{\sigma)}\overline{\nabla}^{(\nu}\overline{\nabla}_\alpha            
  \!+\!2\overline{\nabla}^{(\rho}\overline{\nabla}^{\sigma)}\overline{\nabla}^{(\mu}\overline{\nabla}^{\nu)}
             \Big)
\nonumber\\
 &&\hskip -12cm 
+\, 4
\Big(\overline{g}^{\rho)(\mu}\overline{R}^{\nu)(\sigma|\alpha\beta} \overline{\nabla}_\alpha
                               \overline{\nabla}_\beta
                        \!-\!\overline{R}^{\alpha(\rho\sigma)(\mu} \overline{\nabla}^{\nu)}\overline{\nabla}_\alpha
               \!-\!\overline{g}^{\mu)(\rho} \overline{R}^{\sigma)\alpha\beta(\nu}\overline{\nabla}_\beta
                          \overline{\nabla}_\alpha
            \Big)
 \nonumber\\
&&\hskip -10cm
               +\,2\overline{g}^{\mu\nu}
              \overline{R}^{\alpha(\rho\sigma)\beta} \overline{\nabla}_\beta\overline{\nabla}_\alpha
\nonumber\\             
&&\hskip -12cm
+\, 4
\Big(\overline{\nabla}_\beta\overline{\nabla}_\alpha\overline{g}^{\mu)(\rho}\overline{R}^{\sigma)(\nu|\alpha\beta}
                      \!-\!\overline{\nabla}_\alpha\overline{\nabla}^{\sigma)}
                       \overline{R}^{\alpha(\mu\nu)(\rho}
            \!-\!\overline{\nabla}_\alpha\overline{\nabla}_\beta
                          \overline{g}^{\rho)(\mu}\overline{R}^{\nu)\alpha\beta(\sigma}
            \Big)
 \nonumber\\
&&\hskip -10cm
               +\,2\overline{\nabla}_\alpha\overline{\nabla}_\beta
               \overline{g}^{\rho\sigma}\overline{R}^{\alpha(\mu\nu)\beta}
\nonumber\\
 &&\hskip -12cm 
    -\,2\Big[\overline{g}^{\rho)(\mu}\overline{R}^{\nu)(\sigma|\alpha\beta}
                          \overline{\nabla}_\alpha\overline{\nabla}_\beta
                         \!+\!\overline{g}^{\rho)(\mu} \overline{R}^{\nu)\alpha\beta(\sigma}
                         \overline{\nabla}_\beta\overline{\nabla}_\alpha
                         \!+\! \overline{R}^{\alpha(\rho\sigma)(\mu}\overline{\nabla}_\alpha
                         \overline{\nabla}^{\nu)}
                              \Big]
                              \nonumber\\
 &&\hskip -12cm 
    -\,2\Big[\overline{\nabla}_\beta\overline{\nabla}_\alpha
                         \overline{g}^{\mu)(\rho}\overline{R}^{\sigma)(\nu|\alpha\beta}
                         \!+\!\overline{\nabla}_\alpha\overline{\nabla}_\beta
                         \overline{g}^{\mu)(\rho} \overline{R}^{\sigma)\alpha\beta(\nu}
                         \!+\!\overline{\nabla}^{\sigma)}\overline{\nabla}_\alpha
                          \overline{R}^{\alpha(\mu\nu)(\rho}
                              \Big]
\nonumber\\
 &&\hskip -12cm 
    +\, \Big[ 
         2\overline{\nabla}_\alpha
         \Big(\overline{g}^{\rho)(\mu}\overline{R}^{\nu)(\sigma|\alpha\beta}\overline{\nabla}_\beta 
                         \!+\!\overline{g}^{\rho)(\mu}\overline{R}^{\nu)\beta\alpha(\sigma}\overline{\nabla}_\beta
                         \!+\!\overline{R}^{\alpha(\rho\sigma)(\mu}\overline{\nabla}^{\nu)}
                         \Big)
\nonumber\\
 &&\hskip -10cm 
              +\,2\overline{\nabla}^{(\rho} \overline{R}^{\sigma)(\mu\nu)\alpha}\overline{\nabla}_\alpha
              \!-\!\overline{\nabla}^\lambda \overline{R}^{\rho)(\mu\nu)(\sigma}\overline{\nabla}_\lambda             
     \Big]
\nonumber\\
 &&\hskip -12cm 
    +\,\Big[ 
  2\Big(\overline{\nabla}_\beta\overline{g}^{\mu)(\rho}\overline{R}^{\sigma)(\nu|\alpha\beta}
            \!+\!\overline{\nabla}_\beta\overline{g}^{\mu)(\rho}\overline{R}^{\sigma)\beta\alpha(\nu}
                         \!+\!\overline{\nabla}^{\sigma)}\overline{R}^{\alpha(\mu\nu)(\rho}
                         \Big)\overline{\nabla}_\alpha
\nonumber\\
 &&\hskip -10cm 
     +\,2\overline{\nabla}_\alpha\overline{R}^{\nu)(\rho\sigma)\alpha}\overline{\nabla}^{(\mu} 
         \!-\!\overline{\nabla}^\lambda\overline{R}^{\mu)(\rho\sigma)(\nu}\overline{\nabla}_\lambda
     \Big]
\bigg\}\delta g_{\rho\sigma}(x)
\,.\qquad
\label{second variation: counterterm action Riem2: perturbation}
\end{eqnarray}
This can be further simplified to,
\begin{eqnarray}
\int {\rm d}^Dx'\sqrt{-\overline{g}(x')}
          \frac{1}{\sqrt{-\overline{g}(x)}\sqrt{-\overline{g}(x')}}
     \left(\frac{\delta^2 S_{\rm Riem^2}}{\delta g_{\mu\nu}(x)\delta g_{\rho\sigma}(x')}
      \right)_{\overline{g}_{\alpha\beta}} \delta g_{\rho\sigma}(x')
      &\!\!=\!\!&\alpha_{\rm Riem^2}
\nonumber\\
 &&\hskip -12.cm      
     \times \bigg\{
\bigg[ \frac14\big(\overline{g}^{\mu\nu}\overline{g}^{\rho\sigma}
  \!-\!2\overline{g}^{\mu(\rho}\overline{g}^{\sigma)\nu} \big)
   \overline{R}^{\alpha\beta\gamma\delta} \overline{R}_{\alpha\beta\gamma\delta}
\nonumber\\
 &&\hskip -11cm
    +\,  \Big(6\overline{R}^{(\mu}_{\;\,\,\beta\gamma\delta} \overline{g}^{\nu)(\rho} 
                                     \overline{R}^{\sigma)\beta\gamma\delta}
                            \!-\!\overline{g}^{\mu\nu} \overline{R}^{(\rho}_{\;\,\,\beta\gamma\delta}
                               \overline{R}^{\sigma)\beta\gamma\delta}
                        \!-\!\overline{g}^{\rho\sigma}\overline{R}^{(\mu}_{\;\,\,\beta\gamma\delta}
                         \overline{R}^{\nu)\beta\gamma\delta} \Big)
                         \bigg]\delta g_{\rho\sigma}
\nonumber\\
&&\hskip -12cm
    +\,\Big(3\overline{g}^{\mu(\rho}\overline{g}^{\sigma)\nu}
                               \overline{\nabla}^\alpha\overline{\dAlembert}\overline{\nabla}_\alpha
                             \!-\!2\overline{g}^{\mu)(\rho}
                              \overline{\nabla}^{\sigma)}\overline{\dAlembert}\overline{\nabla}^{(\nu}
\nonumber\\
&&\hskip -11.2cm                     
    -\,\overline{g}^{\mu(\rho}\overline{g}^{\sigma)\nu}
    \overline{\nabla}^\alpha\overline{\nabla}^\beta\overline{\nabla}_\alpha\overline{\nabla}_\beta
    \!-\! 2 \overline{g}^{\mu)(\rho}
    \overline{\nabla}^\alpha\overline{\nabla}^{\sigma)}\overline{\nabla}^{(\nu}\overline{\nabla}_\alpha            
  \!+\!2\overline{\nabla}^{(\rho}\overline{\nabla}^{\sigma)}\overline{\nabla}^{(\mu}\overline{\nabla}^{\nu)}
             \Big)\delta g_{\rho\sigma}
\nonumber\\             
&&\hskip -12.2cm 
+\, 2\delta g_{\rho\sigma}
\Big(\overline{\nabla}_\beta\overline{\nabla}_\alpha
             \overline{g}^{\mu)(\rho}\overline{R}^{\sigma)(\nu|\alpha\beta}
            \!-\!2\overline{\nabla}_\alpha\overline{\nabla}_\beta
                          \overline{g}^{\rho)(\mu}\overline{R}^{\nu)\alpha\beta(\sigma}
          \!-\!\overline{\nabla}_\alpha\overline{\nabla}_\beta
                          \overline{g}^{\rho)(\mu}\overline{R}^{\nu)\beta\alpha(\sigma}
 \nonumber\\
&&\hskip -10.5cm
                +\,\overline{\nabla}_\alpha\overline{\nabla}_\beta
                          \overline{g}^{\rho\sigma}\overline{R}^{\alpha(\mu\nu)\beta}
              \!-\!2\overline{\nabla}_\alpha\overline{\nabla}^{(\rho}
                         \overline{R}^{\sigma)(\mu\nu)\alpha}
              \!-\!\overline{\nabla}^{(\rho}\overline{\nabla}_\alpha
                          \overline{R}^{\sigma)(\mu\nu)\alpha}
                           \Big) 
 \nonumber\\
 &&\hskip -12.1cm 
+\, 2
\Big( 4\overline{g}^{\rho)(\mu}\overline{R}^{\nu)(\sigma|\alpha\beta} \overline{\nabla}_\alpha
                               \overline{\nabla}_\beta
       \!-\!4\overline{g}^{\rho)(\mu} \overline{R}^{\nu)\alpha\beta(\sigma}\overline{\nabla}_\alpha
                          \overline{\nabla}_\beta
    \!-\!2\overline{R}^{\alpha(\mu\nu)(\rho}\overline{\nabla}_\alpha\overline{\nabla}^{\sigma)}
 \nonumber\\
&&\hskip -11.6cm
     +\,\overline{R}^{\alpha(\mu\nu)(\rho}\overline{\nabla}^{\sigma)}\overline{\nabla}_\alpha
        \!-\!2\overline{R}^{\alpha(\rho\sigma)(\mu} \overline{\nabla}^{\nu)}\overline{\nabla}_\alpha
\!+\!\overline{R}^{\alpha(\rho\sigma)(\mu} \overline{\nabla}_\alpha\overline{\nabla}^{\nu)}
 \nonumber\\
&&\hskip -11.6cm
           +\,\overline{g}^{\mu\nu} \overline{R}^{\alpha(\rho\sigma)\beta}
                           \overline{\nabla}_\beta\overline{\nabla}_\alpha
               \!+\!\overline{g}^{\rho\sigma}\overline{R}^{\alpha(\mu\nu)\beta}
                \overline{\nabla}_\alpha\overline{\nabla}_\beta
                \!-\!\overline{R}^{\rho)(\mu\nu)(\sigma}\overline{\dAlembert}
                            \Big)\delta g_{\rho\sigma}      
 \nonumber\\             
&&\hskip -12.2cm 
+\, 2
\Big[2\overline{g}^{\rho)(\mu}\big(\overline{\nabla}_\alpha\overline{R}^{\nu)(\sigma|\alpha\beta}\big)
                          \overline{\nabla}_\beta
         \!-\!3 \overline{g}^{\rho)(\mu}\big(\overline{\nabla}_\alpha\overline{R}^{\nu)\alpha\beta(\sigma}\big)
                            \overline{\nabla}_\beta  
         \!-\!\overline{g}^{\rho)(\mu}\big(\overline{\nabla}_\alpha \overline{R}^{\nu)\beta\alpha(\sigma}\big)
                            \overline{\nabla}_\beta   
\nonumber\\
&&\hskip -11.3cm
       -\,3\big(\overline{\nabla}_\alpha\overline{R}^{\alpha(\mu\nu)(\rho}\big)\overline{\nabla}^{\sigma)}
          \!+\!2\big(\overline{\nabla}_\alpha\overline{R}^{\alpha(\rho\sigma)(\mu}\big)\overline{\nabla}^{\nu)}
         \!-\!\big(\overline{\nabla}^{(\rho}\overline{R}^{\sigma)(\mu\nu)\alpha}\big)\overline{\nabla}_\alpha            
\nonumber\\
&&\hskip -11.3cm
      -\,\big(\overline{\nabla}^\alpha\overline{R}^{\mu)(\rho\sigma)(\nu}\big)\overline{\nabla}_\alpha
 \!+\!2\overline{g}^{\rho\sigma}\big(\overline{\nabla}_\alpha\overline{R}^{\alpha(\mu\nu)\beta}\big)
                             \overline{\nabla}_\beta
                             \Big]\delta g_{\rho\sigma}(x)
\bigg\}
\,.\qquad
\label{second variation: counterterm action Riem2: perturbation2}
\end{eqnarray}
%


\bigskip
On the other hand, varying Eq.~(\ref{counterterm action Riem 2 c1}) results in,
\begin{eqnarray}
\!\!
\frac{1}{\sqrt{-\overline{g}}}\left[\delta
\left(\frac{\delta S_{\rm Riem^2}}{\delta g_{\mu\nu}(x)}\right)\right]_{\! \overline{g}_{\alpha\beta}}\!\!
\!\! &\!\!=\!\!&  \alpha_{\rm Riem^2}
    \bigg\{\overline{g}^{\rho\sigma}\delta g_{\rho\sigma}
       \Big(\frac14 \overline{g}^{\mu\nu}\,\overline{R}_{\alpha\beta\gamma\delta}
                \overline{R}^{\alpha\beta\gamma\delta}
         \!\!-\!\overline{R}^{(\mu}_{\;\,\alpha\beta\gamma}\overline{R}^{\nu)\alpha\beta\gamma}
\nonumber\\
&&\hskip 2.25cm 
          +\, 2\overline{\nabla}_{\alpha} \overline{\nabla}_{\beta}\overline{R}^{\alpha(\mu\nu)\beta}\Big)
 \nonumber\\
&&\hskip -3.4cm           
+\,\frac12 \delta \big(g^{\mu\nu}g_{\omega\alpha}g^{\eta\beta}g^{\zeta\gamma}g^{\xi\delta}\big) \,
                \overline{R}^{\omega}_{\;\,\eta\zeta\xi}\overline{R}^{\alpha}_{\;\,\beta\gamma\delta}
             \!+\! \overline{g}^{\mu\nu}\,\overline{R}_{\alpha}^{\;\,\beta\gamma\delta}
                \delta{R}^{\alpha}_{\;\;\beta\gamma\delta}
\nonumber\\
&&\hskip  -3.5cm
   +\,\delta\big(\delta^{\;(\mu}_{\alpha}\delta^{\nu)}_{\;\omega}g^{\beta\eta}g^{\gamma\zeta}g^{\delta\xi}
                     \!-\!g_{\alpha\omega}g^{\beta(\mu}g^{\nu)\eta}g^{\gamma\zeta}g^{\delta\xi}
                      \!-\!g_{\alpha\omega}g^{\beta\eta}g^{\gamma(\mu}g^{\nu)\zeta}g^{\delta\xi}
\nonumber\\
&&\hskip  .0cm                      
                       -\,g_{\alpha\omega}g^{\beta\eta}g^{\gamma\zeta}g^{\delta(\mu}g^{\nu)\xi}\big)
             \overline{R}^{\alpha}_{\;\,\beta\gamma\delta}\overline{R}^{\omega}_{\;\,\eta\zeta\xi}
\nonumber\\
&&\hskip -3.5cm
 +\,\Big(\delta_\alpha^{\;(\mu}\delta^{\nu)}_{\;\omega}\overline{g}^{\beta\eta}
                        \overline{g}^{\gamma\zeta}\overline{g}^{\delta\xi}
        \!-\!\overline{g}_{\alpha\omega}\overline{g}^{\beta(\mu}\overline{g}^{\nu)\eta}
                           \overline{g}^{\gamma\zeta}\overline{g}^{\delta\xi}
                            \!-\!\overline{g}_{\alpha\omega}\overline{g}^{\beta\eta}
                           \overline{g}^{\gamma(\mu}\overline{g}^{\nu)\zeta}\overline{g}^{\delta\xi}
\nonumber\\
&&\hskip -1.6cm
                            \!-\,\overline{g}_{\alpha\omega}\overline{g}^{\beta\eta}
                           \overline{g}^{\gamma\zeta}\overline{g}^{\delta(\mu}\overline{g}^{\nu)\xi}
 \Big)
   \!\times\!\Big(\overline{R}^{\alpha}_{\;\,\beta\gamma\delta}\delta R^{\omega}_{\;\;\eta\zeta\xi}
                          \!+\!  \delta R^{\alpha}_{\;\,\beta\gamma\delta}\overline{R}^{\omega}_{\;\;\eta\zeta\xi}
             \Big)             
\nonumber\\
&&\hskip  -3.5cm       
    +\,\delta\Big(g^{\beta\zeta}g^{\alpha\xi}g^{\eta(\mu}\delta^{\nu)}_{\;\omega}
                 \!-\!g^{\alpha\zeta}g^{\beta\xi}g^{\eta(\mu}\delta^{\nu)}_{\;\omega}
                  \!+\!g^{\alpha\eta}g^{\beta\zeta}g^{\xi(\mu}\delta^{\nu)}_{\;\omega}
                  \!-\!g^{\alpha\eta}g^{\beta\xi}g^{\zeta(\mu}\delta^{\nu)}_{\;\omega}
\nonumber\\
&&\hskip  .0cm
    -\,g^{\beta\zeta}g^{\eta(\mu}g^{\nu)\xi}\delta^{\alpha}_{\;\omega}
                  \!+\!g^{\beta\xi}g^{\eta(\mu}g^{\nu)\zeta}\delta^{\alpha}_{\;\omega}
                  \Big) \overline{\nabla}_{\alpha} \overline{\nabla}_{\beta}\overline{R}^{\omega}_{\;\;\eta\zeta\xi}
\nonumber\\
&&\hskip  -3.5cm       
            +\, \Big(\overline{g}^{\beta\zeta}\overline{g}^{\alpha\xi}\overline{g}^{\eta(\mu}\delta^{\nu)}_{\;\omega}
                 \!-\!\overline{g}^{\alpha\zeta}\overline{g}^{\beta\xi}\overline{g}^{\eta(\mu}\delta^{\nu)}_{\;\omega}
                  \!+\!\overline{g}^{\alpha\eta}\overline{g}^{\beta\zeta}\overline{g}^{\xi(\mu}\delta^{\nu)}_{\;\omega}
                  \!-\!\overline{g}^{\alpha\eta}\overline{g}^{\beta\xi}\overline{g}^{\zeta(\mu}\delta^{\nu)}_{\;\omega}
\nonumber\\
&&\hskip  .0cm
    -\,\overline{g}^{\beta\zeta}\overline{g}^{\eta(\mu}\overline{g}^{\nu)\xi}\delta^{\alpha}_{\;\omega}
                  \!+\!\overline{g}^{\beta\xi}\overline{g}^{\eta(\mu}\overline{g}^{\nu)\zeta}\delta^{\alpha}_{\;\omega}
                  \Big) \overline{\nabla}_{\alpha} \overline{\nabla}_{\beta}\delta\overline{R}^{\omega}_{\;\;\eta\zeta\xi}
\nonumber\\
&&\hskip  -3.5cm       
            +\, \Big(\overline{g}^{\beta\zeta}\overline{g}^{\alpha\xi}\overline{g}^{\eta(\mu}\delta^{\nu)}_{\;\omega}
                 \!-\!\overline{g}^{\alpha\zeta}\overline{g}^{\beta\xi}\overline{g}^{\eta(\mu}\delta^{\nu)}_{\;\omega}
                  \!+\!\overline{g}^{\alpha\eta}\overline{g}^{\beta\zeta}\overline{g}^{\xi(\mu}\delta^{\nu)}_{\;\omega}
                  \!-\!\overline{g}^{\alpha\eta}\overline{g}^{\beta\xi}\overline{g}^{\zeta(\mu}\delta^{\nu)}_{\;\omega}
\nonumber\\
&&\hskip  .0cm
    -\,\overline{g}^{\beta\zeta}\overline{g}^{\eta(\mu}\overline{g}^{\nu)\xi}\delta^{\alpha}_{\;\omega}
                  \!+\!\overline{g}^{\beta\xi}\overline{g}^{\eta(\mu}\overline{g}^{\nu)\zeta}\delta^{\alpha}_{\;\omega}
   \Big)
\nonumber\\
&&\hskip  -3.5cm 
\times\Big[\overline{\nabla}_{\alpha}\Big(
\delta\Gamma_{\beta\gamma}^\omega \overline{R}^{\gamma}_{\;\;\eta\zeta\xi} 
\!-\!\delta\Gamma_{\beta\eta}^\gamma\overline{R}^{\omega }_{\;\;\gamma\zeta\xi}  
\!-\!\delta\Gamma_{\beta\zeta}^\gamma\overline{R}^{\omega }_{\;\;\eta\gamma\xi}  
\!-\!\delta\Gamma_{\beta\xi}^\gamma\overline{R}^{\omega }_{\;\;\eta\zeta\gamma}  \Big)
\nonumber\\
&&\hskip  -3.cm       
+\,\Big(
          \!-\!\delta\Gamma_{\alpha\beta}^\gamma\overline{\nabla}_\gamma\overline{R}^{\omega}_{\;\;\eta\zeta\xi}
          \!+\!\delta\Gamma_{\alpha\gamma}^\omega\overline{\nabla}_\beta\overline{R}^{\gamma}_{\;\;\eta\zeta\xi}
          \!-\!\delta\Gamma_{\alpha\eta}^\gamma\overline{\nabla}_\beta\overline{R}^{\omega}_{\;\;\gamma\zeta\xi}
          \!-\!\delta\Gamma_{\alpha\zeta}^\gamma\overline{\nabla}_\beta\overline{R}^{\omega}_{\;\;\eta\gamma\xi}
\nonumber\\
&&\hskip 0.0cm 
          \!-\!\delta\Gamma_{\alpha\xi}^\gamma\overline{\nabla}_\beta\overline{R}^{\omega}_{\;\;\eta\zeta\gamma}
\Big) 
\Big]
    \bigg\}
\,,\quad
\label{counterterm action Riem 2: variation of variation 0}
\end{eqnarray}
which when evaluated gives Eq.~(\ref{second variation: counterterm action Riem2: perturbation2}).
This rather complicated form of the variation of the first variation is required as the symmetries of 
$\overline{R}^{\alpha}_{\;\,\beta\gamma\delta}$ and $\delta R^{\alpha}_{\;\,\beta\gamma\delta}$
are not the same. This also means that in the case of the Riemann tensor squared counterterm,
varying the first variation is not simpler than calculating directly its second variation. Nevertheless, 
doing the calculation both ways is useful for checking the result.


\bigskip
\noindent
{\bf Weyl squared counterterm.} The results for the Weyl squared counterterm 
action~(\ref{Weyl curvature tensor 2}) can be obtained from the defining relation,
\begin{equation}
{\rm Weyl}^2 ={\rm Riem}^2
   - \frac{4}{D\!-\!2}{\rm Ric}^2
    +\frac{2}{(D\!-\!1)(D\!-\!2)}R^2
\,.\quad
\label{Weyl curvature tensor 2 OLD}
\end{equation}
For example, making use of Eqs.~(\ref{counterterm action R2 c1}), 
(\ref{counterterm action Ric 2 c1}) and~(\ref{counterterm action Riem 2 c1}) 
one obtains for the first order variation,
\begin{eqnarray}
\!\!\!\!\!
\frac{1}{\sqrt{-\overline{g}}}\left(\frac{\delta S_{\rm Weyl^2}}{\delta g_{\mu\nu}(x)}\right)
 _{\!\overline{g}_{\alpha\beta}}\!\!\!\!
 &\!\!=\!\!& \! \alpha_{\rm Weyl^2} \bigg\{
\frac12 \overline{g}^{\mu\nu}\,\overline{R}_{\alpha\beta\gamma\delta}
                \overline{R}^{\alpha\beta\gamma\delta}
        \!\!\!-\! 2\overline{R}^{(\mu}_{\;\,\alpha\beta\gamma}\overline{R}^{\nu)\alpha\beta\gamma}
          \!\!+\! 4\overline{\nabla}_{\alpha} \overline{\nabla}_{\beta}\overline{R}^{\alpha(\mu\nu)\beta}\!
\nonumber\\
&&\hskip -3.cm
  -\,\frac{4}{D\!-\!2}  \bigg[\frac12 \overline{g}^{\mu\nu}\,\overline{R}_{\alpha\beta}\overline{R}^{\alpha\beta}
              \!-\! 2\overline{R}^{\alpha(\mu}\overline{R}^{\nu)}_{\;\alpha}
              \!+\!2\overline{\nabla}_{\alpha} \overline{\nabla}^{(\mu}\overline{R}^{\nu)\alpha}
            \!-\! \overline{\dAlembert}\,\overline{R}^{\mu\nu}
          \!-\! \overline{g}^{\mu\nu}\overline{\nabla}_{\alpha} \overline{\nabla}_{\beta}
                          \overline{R}^{\alpha\beta}
    \bigg]
\nonumber\\
&&\hskip -3cm   
+\,\frac{2}{(D\!-\!1)(D\!-\!2)}    \bigg[\frac12 \overline{g}^{\mu\nu}\,\overline{R}^2
              \!-\! 2
              \big(\overline{R}^{\mu\nu}\!-\!\overline{\nabla}^{\mu} \overline{\nabla}^{\nu}
            \!\!+\! \overline{g}^{\mu\nu}\overline{\dAlembert}
                 \big)\overline{R}
    \bigg]
      \bigg\}
\,.\quad
\label{counterterm action Weyl2: 1st order}
\end{eqnarray}
%

%

\bigskip
\noindent
{\bf Gauss-Bonnet counterterm.} Even though the Gauss-Bonnet action contributes
to the energy-momentum tensor and self-energy as 
$\propto (D\!-\!4)$, it is sometimes used
for obtaining finite contributions to the energy-momentum tensor from a divergent action.
Just as in the case of the Weyl squared counterterm, the defining relation can be 
used to compute the resulting counterterm from the results above.
We leave as an exercise to the reader to show that both the energy-momentum tensor 
and the graviton self-energy are proportional to $(D\!-\!4)\alpha_{\rm G\!B}$.
When worked out explicitly, one obtains,
\begin{eqnarray}
\!\!\!\!\!
\frac{1}{\sqrt{-\overline{g}} }\left(\frac{\delta S_{\rm G\!B}}{\delta g_{\mu\nu}(x)}\right)
 _{\!\overline{g}_{\alpha\beta}}\!\!\!\!
 &\!\!=\!\!&  \alpha_{\rm G\!B}
    \bigg\{\frac12 \overline{g}^{\mu\nu}\,\overline{\rm G\!B}
              \!-\! 2
               \big(\overline{R}^{\mu\nu}\!\!-\!\overline{\nabla}^{\mu} \overline{\nabla}^{\nu}
            \!\!+\! \overline{g}^{\mu\nu}\overline{\dAlembert}
                 \big)\overline{R}
\nonumber\\
&&\hskip -3.7cm  
 -\, 4\bigg[
              \!-\! 2\overline{R}^{\alpha(\mu}\overline{R}^{\nu)}_{\;\alpha}
              \!+\!2\overline{\nabla}_{\alpha} \overline{\nabla}^{(\mu}\overline{R}^{\nu)\alpha}           
           \!-\! \overline{\dAlembert}\,\overline{R}^{\mu\nu}
          \!-\! \overline{g}^{\mu\nu}\overline{\nabla}_{\alpha} \overline{\nabla}_{\beta}
                          \overline{R}^{\alpha\beta}
    \bigg]
\nonumber\\
&&\hskip -3.7cm      
    +\,\bigg[
        \!\!-\! 2\overline{R}^{(\mu}_{\;\,\alpha\beta\gamma}\overline{R}^{\nu)\alpha\beta\gamma}
          \!+\! 4\overline{\nabla}_{\alpha} \overline{\nabla}_{\beta}\overline{R}^{\alpha(\mu\nu)\beta}
    \bigg]
    \bigg\}
\nonumber\\ 
&&\hskip -3.7cm 
 =\, \alpha_{\rm G\!B}
    \bigg\{\frac12 \overline{g}^{\mu\nu}\Big(\overline{R}_{\alpha\beta\gamma\delta}
                 \overline{R}^{\alpha\beta\gamma\delta}
              \!-\!4\overline{R}_{\alpha\beta}\overline{R}^{\alpha\beta}
               \!+\!\overline{R}^2\Big)
\nonumber\\ 
&&\hskip -2.7cm 
              -\, 2\Big(2\overline{R}^{(\mu}_{\;\,\alpha\beta\gamma}\overline{R}^{\nu)\alpha\beta\gamma}
              \!+\!2\overline{R}_{\alpha\beta}\overline{R}^{\alpha(\mu\nu)\beta}
              \!-\!2\overline{R}_{\alpha}^{\;\;(\mu}\overline{R}^{\nu)\alpha}
               \!+\!\overline{R}^{\mu\nu}\overline{R}\Big)
    \bigg\}
,\qquad
\label{counterterm action GB: 1st order}
\end{eqnarray}
which is variation of a topological counterterm, and therefore ought to be proportional to $(D\!-\!4)$~\cite{Chern:1945,Yale:2011usf}.


\end{document}